\documentclass[12pt]{article}
\pdfoutput=1
\usepackage{amsmath,amssymb,amsfonts,amsthm,bm,bbm,cancel,wasysym,mathtools,bm}
\usepackage{epsfig,graphics,graphicx,epstopdf,tikz-feynman}
\usepackage{feynmf}
\graphicspath{{Charts/}}
\usepackage{array,booktabs,colortbl,colordvi,multirow}
\usepackage{colordvi,color,xcolor}
\usepackage{hyperref}
\usepackage{rotating}

\usepackage{verbatim}
\usepackage{cite}
\usepackage{caption}
\usepackage{subcaption}
\usepackage{setspace}
\usepackage{url}
\usepackage[percent]{overpic}
\usepackage{slashed}
\usepackage{authblk}
\usepackage{xspace}
\usepackage{fullpage}

\def\ie{{\it i.e.}}
\def\eg{{\it e.g.}}

\newskip\zatskip \zatskip=0pt plus0pt minus0pt
\def\matth{\mathsurround=0pt}

\def\gsim{\mathrel{\mathpalette\atversim>}}
\def\atversim#1#2{\lower0.7ex\vbox{\baselineskip\zatskip\lineskip\zatskip
  \lineskiplimit 0pt\ialign{$\matth#1\hfil##\hfil$\crcr#2\crcr\sim\crcr}}}

\begin{document}


\begin{flushright}
SLAC-PUB-17476 \\
\today
\end{flushright}
\vspace*{5mm}

\renewcommand{\thefootnote}{\fnsymbol{footnote}}
\setcounter{footnote}{1}

\begin{center}

{\Large {\bf Local $SU(2)\times U(1)$ Quark Flavor Symmetry in the RS Bulk }}\\

\vspace*{0.75cm}
{\bf G. N. Wojcik}~\footnote{gwojci03@stanford.edu} and {\bf T. G. Rizzo}~\footnote{rizzo@slac.stanford.edu},

\vspace{0.5cm}

{SLAC National Accelerator Laboratory, 2575 Sand Hill Rd, Menlo Park, CA, 94025, USA}

\end{center}
\vspace{.5cm}

\begin{abstract}
 
\noindent
We propose a model of quark flavor based on an additional $SU(2)\times U(1)$ local symmetry in a warped extra dimensional bulk. In contrast to other works, we break the additional gauge symmetry in the bulk via two complex scalars which acquire bulk vevs, rather than relying on brane-localized symmetry breaking. A gauge-covariant Kaluza-Klein decomposition of a theory with a bulk spontaneously broken gauge symmetry is performed, and exact expressions for the bulk profiles of all physical particles in such systems are given. The SM quark masses and mixings are then recreated using gauge-covariant bulk quark mass terms and Yukawa-like couplings to the new bulk scalars. A numerical sampling of points in the model parameter space that recreate the quark masses and mixings is performed at a KK scale of $M_{KK} =5$ TeV. We then compute the $\Delta F = 2$ 4-quark operators arising from our new flavor gauge bosons and scalars, and those arising from Kaluza-Klein modes of SM gauge bosons. By decoupling one of our bulk scalar fields to all quark fields except the right-handed up-like sector, we find that it is possible to greatly suppress tree-level contributions to the highly constrained Kaon mixing parameters. Instead, the dominant constraints on the model emerge from neutral $B_d$ and $D$ meson mixing. These constraints are explored with our numerical sampling of the model parameter space, and the specific contribution of the new flavor gauge bosons and scalars is discussed. We find that for a significant range of realistic flavor gauge couplings, the new gauge bosons compete with the normally dominant gluon flavor-changing currents, but flavor-changing operators emerging from the bulk scalar fields are highly suppressed. Finally, we briefly comment on flavor constraints that are independent of the flavor gauge sector arising from the $Z \bar{b}_L b_L$ coupling and rare top decays.
\end{abstract}

\renewcommand{\thefootnote}{\arabic{footnote}}
\setcounter{footnote}{0}
\thispagestyle{empty}
\vfill
\newpage
\setcounter{page}{1}



\section{Introduction}
\label{Sections/Section_1_Intro}

As a theoretical tool, the Randall-Sundrum (RS) model \cite{Randall:1999ee,eDims,Davoudiasl:1999tf} has proven itself remarkably effective at generating large hierarchies in Standard Model (SM) parameters from $O(1)$ differences in fundamental parameters. In essence, the scheme proposes that rather than existing in a 4-dimensional spacetime, we can extend our theory by introducing an additional warped extra dimension, parameterized by $-\pi<\phi<\pi$, that is compactified on an $S^1/Z_2$ orbifold with boundaries at the branes $\phi=0$ (the Planck brane) and $\phi=\pi$ (the TeV-brane). The full metric of spacetime is given by

\begin{equation}\label{RSMetric}
    ds^2 = e^{-2 \sigma (\phi)} \eta_{\mu \nu} dx^{\mu} dx^{\nu}-r_c^2 d \phi^2,
\end{equation}
where $\sigma(\phi)\equiv k r_c |\phi|$, $\pi r_c$ is the size of the extra dimension, and $k \sim M_{\textrm{Pl}}$ is a parameter that describes the curvature of the space. If the Higgs field is localized at (or very near) the TeV-brane, then even if the Higgs vev's value in the fundamental five-dimensional theory is roughly equivalent to the 5-dimensional Planck scale (as naturalness would suggest), the vev in the effective 4-dimensional theory should be suppressed by a factor of $\epsilon \equiv e^{-k r_c \pi}$ relative to the 4-dimensional Planck scale. As a result, for $kr_c \approx 11-12$, the $O(10^{15})$ discrepancy between the weak scale and the Planck scale (the so-called gauge hierarchy problem) can be resolved; furthermore, it has been shown that the size of the extra dimension can be naturally stabilized such that this value is realized \cite{Goldberger:1999uk}. For the numerical components of our analyses, we assume that $kr_c = 11.3$

In addition to addressing the gauge hierarchy, it has been noted that if the SM fermion and gauge boson fields are allowed to propagate in the bulk, $O(1)$ differences in bulk mass parameters and TeV-brane localized Yukawa couplings to the Higgs field among different fermion species can be used to naturally explain the large hierarchies in observed fermion masses, as well as the small degree of flavor mixing observed in the quark sector \cite{casagrande,huberHierarchy,ghergetta,ahmed,blanke,Grossman:2000,Burdman:2002gr,Moreau:2006np}. However, the promotion of the entire SM field content (except for the Higgs) to bulk fields leads to the introduction of copious new physics at the scale of $M_{KK} \equiv k \epsilon$, and in particular to new tree-level flavor-changing neutral currents \cite{casagrande,gori,ahmed,fitzpatrick,agasheRSGIM,csakiShining,csakiCompositeHiggs,perez,agasheBFactory,blanke,burdman,Burdman:2002gr,Moreau:2006np} at this energy scale. In spite of the fact that naive operator analysis indicates that any new tree-level flavor changing processes introduced to the SM must be due to physics at a much greater scale than $M_{KK}$, these effects are naturally suppressed in the RS framework due to the so-called RS-GIM mechanism \cite{agasheRSGIM}: New flavor violation effects in the RS model arise primarily from non-universality of different generations of fermion fields' couplings to Kaluza-Klein (KK) modes of gauge bosons, however, the light SM fermions are localized close to the Planck brane while KK modes of gauge bosons are localized close to the TeV brane, so these new flavor-changing couplings are suppressed for the light fermions. The RS-GIM mechanism, however, is not sufficient to suppress flavor-changing processes to within experimental tolerances alone without either imposing draconian constraints such as requiring the lightest KK modes of the gauge bosons to have a mass $\gsim O(20 \; \textrm{TeV})$ \cite{csakiShining,csakiCompositeHiggs,gori,blanke} or requiring fine tuning of the quark Yukawa couplings \cite{casagrande2,gori}. As such, experimental evidence suggests additional flavor symmetries must be present in order to render the RS model phenomenologically viable. There have been a number of proposals in this direction previously \cite{csakiShining,csakiU1,frank,kadosh,hernandez,fitzpatrick,cacciapaglia,santiago,bauer}. In particular, it has been previously noted that the AdS/CFT correspondence which relates 5-dimensional Randall-Sundrum theories to 4-dimensional strongly coupled CFTs \cite{Maldacena:1997re,PerezVictoria:2001pa} suggests the introduction of $\textit{global}$ flavor symmetries in the 4-dimensional CFT in order to suppress dangerous new flavor-changing effects, which correspond in turn to $\textit{local}$ bulk flavor symmetries in the 5-dimensional theory \cite{csakiShining,csakiU1}. As a result, we might naturally expect that any continuous bulk flavor symmetry in the RS model would be local, and therefore give rise to new gauge bosons and require controlled breaking to produce a phenomenologically realistic model.

Previous discussions of gauged RS bulk flavor symmetries, however, have generally limited their exploration of the effect of these new gauge forces. For example, in \cite{csakiU1} and \cite{cacciapaglia}, at least one of the the coupling constants associated with the new flavor gauge group is assumed to be very weak, in order to avoid unacceptably large contributions to either FCNC's (in \cite{csakiU1}) or electroweak precision bounds (in \cite{cacciapaglia}) arising from new gauge bosons in the 4-D theory with masses substantially below the KK scale. In the setup of \cite{csakiShining}, no restriction needs to be imposed on the flavor gauge couplings, however this is because the model is specifically constructed so that the new flavor gauge bosons don't mediate any flavor-changing interactions in the mass eigenstate basis for the SM-like quarks (the authors note here that there exist generalizations of their model in which the flavor gauge bosons $\textit{do}$ mediate additional flavor-changing interactions, but the effects of these interactions are not quantitatively explored). To build on this body of work, then, it is interesting to consider if it is possible to construct a model in which a new bulk flavor gauge sector in RS can contribute significantly to flavor observables, but the model as a whole maintains phenomenological viability. To that end, we construct a model with a gauged bulk flavor symmetry based on the group $SU(2)_F\times U(1)_F$ and perform a detailed calculation of the tree-level flavor-changing effects from all sources, including from the new flavor gauge bosons. In contrast to earlier work, in which symmetry breaking effects were localized in the fifth dimension (generally on the UV or IR branes), we find it useful to enact entirely non-localized bulk symmetry breaking, which can be realized dynamically by two bulk complex scalar doublets which attain vacuum expectation values that completely break $SU(2)_F \times U(1)_F$, imparting bulk masses to all gauge bosons of the new flavor symmetry. This method of symmetry breaking in particular possesses certain advantages over other proposed methods. First, as long as the gauge symmetry is completely broken in the bulk, the model does not predict any additional light gauge bosons, as emerge in \cite{csakiU1,cacciapaglia}. Furthermore, unlike \cite{csakiShining}, in which the authors present a highly general set of UV-brane localized flavor violating terms that transmit this symmetry breaking to the bulk via a bulk scalar (this mechanism is known as shining), the minimization of a bulk scalar potential presents a specific dynamical realization of flavor symmetry breaking.

In order to effect the flavor mixing we observe in the SM, we couple the scalar doublets to the bulk fields in bulk Yukawa-like terms, so that after these scalars achieve vacuum expectation values we can recreate the observed SM quark masses and mixings in the effective 4-dimensional theory. Notably, we find that through careful arrangement of our vacuum expectation values and couplings to the quark fields, we can entirely eliminate any $\Delta F = 2$ tree-level FCNCs featuring the $s$ quark, except those that arise from the exchange of bulk scalars. Furthermore, because we find that couplings of the bulk scalars to the SM-like quarks in our effective 4-dimensional theory are highly suppressed, we can protect certain sensitive flavor observables, most notably the Kaon indirect CP violation parameter $\epsilon_K$, from unacceptably large corrections.

Our paper is laid out as follows. In Section \ref{BulkSSBSection}, we consider the effective 4-dimensional theory arising from a toy $U(1)$ gauge symmetry spontaneously broken by a bulk complex scalar field that acquires a constant vacuum expectation value. In particular, we derive the bulk profiles of the Kaluza-Klein modes of $\textit{all}$ physical particles emerging in the theory. In this toy model, we find that there are three distinct KK towers of physical states: One of 4-dimensional vector gauge bosons, one of scalars that arises from the massive component of the bulk scalar field, and one of pseudoscalars that arises from a mixture of the fifth component of the gauge boson and the bulk Goldstone boson. This is consistent with the analogous calculation in flat space \cite{Flacke:2008ne}, and the treatment of this scenario in a more general warped background in \cite{Falkowski:2008fz}. In Section \ref{BulkEWSection}, we outline our treatment of the SM gauge fields in the bulk, giving expressions for the bulk profiles of each gauge field's KK tower modes and including an exact treatment of TeV-brane localized electroweak symmetry breaking. In Section \ref{BulkFermionSection}, we outline our treatment of the bulk quark fields, including exact and approximate expressions for the resultant quark KK towers in the mass eigenstate basis.

In Section \ref{BulkModelSection}, we develop our $SU(2)_F \times U(1)_F$ model of flavor explicitly. First, we discuss how the bulk flavor gauge symmetry is broken by two complex $SU(2)_F$ doublet scalars in Section \ref{BulkScalarModelSection}, deriving the bulk masses for the physical bulk scalar fields and the bulk $SU(2)_F \times U(1)_F$ gauge bosons. Then, in Section \ref{BulkMatterModelSection}, we place the bulk quark fields into representations of $SU(2)_F\times U(1)_F$, and in Section \ref{CKMSection}, we discuss how the observed quark masses and the CKM matrix are realized in our model, with our methodology for numerically sampling our model's parameter space given in Section \ref{NumericalParamPointsSection}.

In Section \ref{4QuarkOperatorSection}, we derive low-energy effective 4-quark operators deriving from the tree-level exchange of both the KK modes of the SM gauge bosons and the new gauge bosons and scalars arising from our flavor symmetry sector. In Section \ref{FlavorObservablesSection}, we then compute the effect of these 4-quark operators on flavor observables; in Sections \ref{KKBarMixingSection}-\ref{DDBarMixingSection}, we discuss neutral meson mixing processes, finding that the dominant constraints on our model arise from $\bar{B}^{0}-B^{0}$ and $\bar{D}^{0}-D^{0}$ mixing, while in Section \ref{OtherFlavorObsSection} we address several other significant flavor observables which are independent of the new flavor gauge sector, finding them to be insignificant within our model. Finally, in Section \ref{ConclusionSection}, we summarize our findings and discuss potential future directions of research.


\section{Spontaneously Broken Bulk Gauge Symmetry}\label{BulkSSBSection}
Because our flavor symmetry has been promoted to a local symmetry broken in the bulk by scalar fields, it is instructive to explore the new gauge and scalar fields created by this symmetry in the effective 4-dimensional theory. To begin, we consider the case of a single $U(1)$ bulk gauge symmetry, broken by a bulk scalar vev. The case for other symmetries, such as the $SU(2)\times U(1)$ bulk symmetry discussed in this paper, is straightforwardly generalizable from this treatment. To begin, we write the action for the gauge boson $A$, with coupling constant $g_A$, and the complex scalar field $\Phi$ (with $U(1)$ charge $Y_\Phi$) prior to symmetry breaking in the $U(1)$ toy theory as

\begin{eqnarray}
 S_A & = & \int d^4 x \int r_c d\phi \, \sqrt {-G} \,
\left\{-\frac{1}{4}G^{MK}G^{NL} (\partial_M A_N-\partial_N A_M)(\partial_K A_L-\partial_L A_K) \right. \nonumber\\ 
& + &
\left. G^{MN}(D_M \Phi)^* (D_N \Phi)-V(\Phi) \right\}\,,\\
D_N & \equiv & \partial_N-\sqrt{2 \pi r_c}i g_A Y_\Phi A_N, \nonumber
\end{eqnarray}
where the metric $G^{AB}$ is given by Eq.(\ref{RSMetric}), and $\sqrt{-G} =\sqrt{|det(G_{AB})|}= e^{-4 \sigma(\phi)}$. Note that we have defined the gauge coupling constant $g_A$ such that it is dimensionless, by including an extra factor of $\sqrt{2 \pi r_c}$ in the coupling term. The specific choice of this factor is to mimic the convention normally used to interrelate bulk gauge coupling constants to their SM equivalents in RS; in the absence of bulk symmetry breaking, the effective 4-dimensional coupling constant of the $A$ field's massless zero-mode would be $g_A$. Additionally, $V(\Phi)$ is simply some potential such that $\Phi$ achieves a vacuum expectation value in the bulk.  For the sake of simplicity, for the remainder of the section we shall assume that this vev is constant, and not a function of the fifth-dimensional spacetime coordinate $\phi$. This is easily achieved, for example, by requiring that the potential has no explicit dependence on $\phi$. An extension to the case of a non-flat bulk vev would significantly complicate the analysis, and is therefore beyond the scope of this paper.

Assuming that $v/\sqrt{2 \pi r_c}$ is the bulk vacuum expectation value of $\Phi$ (note that in our definition of $v$, we factor out $\sqrt{2 \pi r_c}$ so that the scale $v$ has mass dimension 1, similar to our redefinition of $g_A$), we rewrite $\Phi=\frac{1}{\sqrt{2}}(h+i \varphi +v/\sqrt{2 \pi r_c})$, where $h$ and $\varphi$ are real scalar fields. The action then becomes

\begin{eqnarray}
 S_A & = & \int d^4 x \int r_c d\phi \,
\left\{-\frac{1}{4}(\partial_\mu A_\nu-\partial_\nu A_\mu)^2+\frac{e^{-2 \sigma}}{2 r_c^2}(\partial_\phi A_\mu)^2+\frac{g_A^2 Y_\Phi^2 v^2 e^{-2 \sigma} }{2}(A_\mu)^2 \right. \nonumber\\
& + & \left. \frac{e^{-2 \sigma}}{2 r_c^2}(\partial_\mu A_\phi)^2-\frac{e^{-2 \sigma}}{r_c^2} (\partial_\mu A_\phi)(\partial_\phi A_\mu) -\frac{g_A^2 Y_\Phi^2 v^2 e^{-4 \sigma}}{2 r_c^2} (A_\phi)^2\right. \nonumber\\ 
& + &
\left. \frac{e^{-2 \sigma}}{2}(\partial_\mu h)^2+\frac{e^{-2 \sigma}}{2}(\partial_\mu \varphi)^2-\frac{e^{-4 \sigma}}{2 r_c^2}(\partial_\phi h)^2-\frac{e^{-4 \sigma}}{2 r_c^2}(\partial_\phi \varphi)^2-\frac{b^2 k^2 e^{-4 \sigma}}{2} h^2 \right.\\
& - & g_A Y_\Phi v e^{-2 \sigma} A^\mu \partial_\mu \varphi+\frac{g_A Y_\Phi v e^{-4 \sigma} }{r_c^2} A_\phi \partial_\phi \varphi + \textrm{ interaction terms} \nonumber \bigg\}.
\end{eqnarray}

Here, the dimensionless parameter $b$ represents the bulk mass for $h$ which derives from the potential $V(\Phi)$, while $\varphi$ remains without a bulk mass, as a Goldstone boson. To eliminate mixing between $A_\mu$ and the other two fields, we then add the gauge fixing Lagrangian
\begin{equation}\label{GaugeFlavorFixing}
     \mathcal{L}_{\textrm{gf}}=-\frac{1}{2 \xi} \bigg( \partial^\mu A_\mu-\xi \frac{\partial_\phi e^{-2 \sigma} A_\phi}{r_c^2}+\xi e^{-2 \sigma}  g_A Y_\Phi v \varphi \bigg)^2,
\end{equation}
which corresponds to a modified $R_\xi$ gauge, similar to the choice made in \cite{casagrande,csakiTASI,Randall:2001gb}. Adding this gauge fixing term and integrating by parts (assuming that, on the orbifold, $A_\mu$, $h$, and $\varphi$ are even, while $A_\phi$ is odd) finally yields the action

\begin{eqnarray}\label{GaugeFlavorAction}
 S_A & = & \int d^4 x \int r_c d\phi \,
\left\{-\frac{1}{4}(\partial_\mu A_\nu-\partial_\nu A_\mu)^2-\frac{1}{2 \xi}(\partial_\mu A^{\mu})^2+\frac{e^{-2 \sigma}}{2 r_c^2}(\partial_\phi A_\mu)^2+\frac{\gamma^2 k^2 e^{-2 \sigma}}{2}(A_\mu)^2 \right. \nonumber\\
& + & \left. \frac{e^{-2 \sigma}}{2 r_c^2}(\partial_\mu A_\phi)^2+\frac{\xi}{2} \bigg( \frac{e^{-2 \sigma} \partial_\phi^2 e^{-2 \sigma} A_\phi}{r_c^4}\bigg) A_\phi-\frac{\gamma^2 k^2 e^{-4 \sigma}}{2 r_c^2} (A_\phi)^2\right. \nonumber\\ 
& + &
\left.\frac{e^{-2 \sigma}}{2}(\partial_\mu \varphi)^2-\frac{e^{-4 \sigma}}{2 r_c^2}(\partial_\phi \varphi)^2-\frac{\xi}{2}\gamma^2 k^2 e^{-4 \sigma} \varphi^2\right.\\
& + & \left.\frac{\gamma k e^{-4 \sigma} }{r_c^2}A_\phi \partial_\phi \varphi + \xi \gamma k e^{-2 \sigma} \bigg(\frac{\partial_\phi e^{-2 \sigma} A_\phi}{r_c^2} \bigg) \varphi\right. \nonumber \\
& + & \left. \frac{e^{-2 \sigma} }{2}(\partial_\mu h)^2-\frac{e^{-4 \sigma}}{2 r_c^2}(\partial_\phi h)^2-\frac{b^2 k^2 e^{-4 \sigma}}{2} h^2 + \textrm{ interaction terms} \nonumber \right\} \nonumber,\\
\gamma & \equiv & \left. \frac{g_A Y_\Phi v}{k} \nonumber.\right.
\end{eqnarray}
Notably, while the $A_\mu$ field does not mix with any other field in this gauge, the $A_\phi$ and $\varphi$ fields do still mix with one another. We shall discuss this mixing, and a method for exactly determining the Kaluza-Klein spectrum for the $A_\phi$ and $\varphi$ fields, in Section \ref{BulkSSBPseudoscalarsSection}.

\subsection{The Vector Gauge Field $A_\mu$}
First, we determine the Kaluza-Klein spectrum of the field $A_\mu$, as has previously been done in \cite{huberBulkGauges}. To begin, we perform the KK decomposition,

\begin{eqnarray}\label{GaugeFlavorVectorKK}
   & & \left. A_\mu (x,\phi) = \frac{1}{\sqrt{r_c}}\sum_n A^{(n)}_\mu (x)\chi^A_n (\phi),\right.\\
    & & \left. \int_{-\pi}^{\pi} d \phi \chi^{A}_m (\phi) \chi^{A}_n (\phi) = \delta_{m n}. \nonumber\right.
\end{eqnarray}
Applying this to Eq.(\ref{GaugeFlavorAction}) yields the action
\begin{eqnarray}\label{VectorGaugeFlavorAction}
   S_{\textrm{vector}} & = & \left. \sum_{n,m} \int d^4 x \int_{-\pi}^{\pi} d \phi \bigg\{ \bigg(-\frac{1}{4}(F^{(n)}_{\mu \nu})(F^{(m)}_{\mu \nu})-\frac{1}{2 \xi}(\partial_\mu A^{(n)}_\mu)(\partial_\nu A^{(m)}_\nu)\bigg)\chi^A_n(\phi)\chi^A_m(\phi)\right. \nonumber\\
   & + & \left. \chi^A_{n}(\phi) \bigg(-\frac{1}{2 r_c^2}(\partial_\phi e^{-2 \sigma} \partial_\phi \chi^{A}_{m}(\phi))+\frac{1}{4}\gamma^2 k^2 e^{-2 \sigma} \chi^{A}_m(\phi) \bigg)A^{(n)}_\mu A^{(m)}_\mu\bigg\}.\right.
\end{eqnarray}
The orthonormality condition in Eq.(\ref{GaugeFlavorVectorKK}) automatically produces diagonalized, canonically normalized kinetic terms for the flavor gauge field's KK modes in the effective 4-dimensional theory. To diagonalize the mass terms in Eq.(\ref{VectorGaugeFlavorAction}), we find $\chi^A_n(\phi)$ that satisfies the equation,
\begin{equation}\label{VectorGaugeFlavorWavefunction}
    -\frac{1}{r_c^2} \partial_\phi (e^{-2 \sigma} \partial_\phi \chi^{A}_n(\phi))+\gamma^2 k^2 e^{-2 \sigma} \chi^A_n(\phi)=(m^A_n)^2 \chi^A_n(\phi).
\end{equation}
This equation can be solved using Bessel functions $J$ and $Y$ (of the first and second kind, respectively), yielding the solution,
\begin{align}\label{VectorGaugeFlavorSolution}
    &\chi^A_n(\phi) = \frac{e^\sigma}{N^A_n} \zeta^A_{\delta}(z^A_n)\equiv \frac{e^{\sigma}}{N^A_n}(\beta^A_n J_{\delta}(z^A_n)+\alpha^A_n Y_{\delta}(z^A_n)) \nonumber \\
    &N^A_n \equiv \bigg(\frac{1}{kr_c \epsilon^2 (x^A_n)^2}[((x^A_n)^2-(\delta^2-1))(\zeta^A_\delta (x^A_n))^2-((x^A_n)^2 \epsilon^2-(\delta^2-1))(\zeta^A_\delta (x^A_n \epsilon))^2] \bigg)^{\frac{1}{2}}\\
    &\alpha^A_n \equiv -(x^A_n \epsilon J_{\delta-1}(x^A_n \epsilon)+(1-\delta)J_\delta (x^A_n \epsilon)), \; \; \beta^A_n \equiv x^A_n \epsilon Y_{\delta-1}(x^A_n \epsilon)+(1-\delta)Y_\delta (x^A_n \epsilon) \nonumber\\
    &z^A_n \equiv \frac{m^A_n}{k} e^{\sigma(\phi)}, \; \; x^A_n \equiv \frac{m^A_n}{k \epsilon}, \; \; \delta \equiv \sqrt{1+\gamma^2}. \nonumber
\end{align}
Here, the constants $\alpha^A_n$ and $\beta^A_n$ are derived from the orbifold boundary condition,
$ \partial_\phi \chi^{A}_{n}(0)=0 $
, while the normalization constant $N^A_n$ is derived from the orthonormality condition in Eq.(\ref{GaugeFlavorVectorKK}). The mass eigenvalues $m^A_n$ can then be derived from the orbifold boundary condition at $\phi=\pm \pi$, which requires that $x^A_n$ satisfies the equation,
\begin{equation}
    (1-\delta)\zeta^A_\delta(x^A_n)+x^A_n \zeta^A_{\delta-1}(x^A_n)=0.
\end{equation}
Notably, unlike the case without a bulk mass term, Eq.(\ref{VectorGaugeFlavorWavefunction}) does not admit a solution when $m^A_n=0$ that satisfies the orbifold boundary conditions, $\partial_\phi \chi^A_n(0)=0$ and $\partial_\phi \chi^A_n(\pm \pi)=0$. As a result, excluding cases of extreme fine-tuning \cite{huberBulkGauges,chang1999}, any gauge field with a non-zero bulk mass in an RS model will have its lightest states be of mass $O(M_{KK})$.

\subsection{Bulk Scalar $h$}

Next, we address the bulk scalar field $h$ in Eq.(\ref{GaugeFlavorAction}). Like $A_\mu$, this field does not mix with any others in Eq.(\ref{GaugeFlavorAction}), and can therefore be addressed separately. Meanwhile, we assume that since the vev of $\Phi$ is even on the orbifold, $h$ must be even as well, so we impose even orbifold boundary conditions on this field. We then begin by performing the KK expansion,
\begin{eqnarray}
    & & \left. h(x,\phi) = \frac{1}{\sqrt{r_c}}\sum_n h^{(n)}(x) \chi^{h}_n (\phi),\right.\\
    & & \left. \int_{-\pi}^{\pi} d\phi e^{-2 \sigma} \chi^{h}_n(\phi) \chi^{h}_m(\phi) = \delta_{m n}, \nonumber\right.
\end{eqnarray}
which yields the action,
\begin{eqnarray}
    S_h & = & \left. \int d^4 x \int_{-\pi}^\pi d \phi \bigg\{ \frac{1}{2}(\partial_\mu h^{(n)})(\partial_\mu h^{(m)})e^{-2 \sigma} \chi^h_n(\phi) \chi^h_m(\phi)\right.\\
    & + & \left. \frac{1}{2}\chi^h_n(\phi) \bigg(\frac{1}{r_c^2}\partial_\phi e^{-4 \sigma}\partial_\phi \chi^h_m (\phi)-b^2 k^2 e^{-4 \sigma} \chi^h_m (\phi) \bigg) h^{(n)}h^{(m)} \bigg\}. \nonumber\right.
\end{eqnarray}
To diagonalize the mass matrix in the effective 4-dimensional theory, then, we need $\chi^h(\phi)$ to satisfy the equation
\begin{equation}\label{ScalarGaugeFlavorWavefunction}
    -\frac{1}{r_c^2}\partial_\phi e^{-4 \sigma} \partial_\phi \chi^h_n(\phi)+b^2 k^2 e^{-4 \sigma}\chi^h_n(\phi)=(m^h_n)^2 e^{-2 \sigma} \chi^h_n(\phi).
\end{equation}
The solution to this equation, with even orbifold boundary conditions applied, is
\begin{align}
    &\chi^h_n(\phi) = \frac{e^{2 \sigma}}{N^h_n}\zeta^h_\rho(z^h_n) \equiv \beta^h_n J_\rho (z^h_n)+\alpha^h_n Y_\rho(z^h_n), \nonumber\\
    &N^h_n \equiv \bigg( \frac{1}{\epsilon^2 kr_c (x^h_n)^2} [((x^h_n)^2+4-\rho^2)(\zeta^h_\rho (x^h_n))^2-(\epsilon^2 (x^h_n)^2 +4-\rho^2)(\zeta^h_\rho (x^h_n \epsilon))^2]\bigg)^{\frac{1}{2}}\\
    &\alpha^h_n \equiv (2-\rho)J_\rho (x^h_n \epsilon)+x^h_n \epsilon J_{\rho-1}(x^h_n \epsilon), \;\; \beta^h_n \equiv -((2-\rho)Y_\rho (x^h_n \epsilon)+x^h_n \epsilon Y_{\rho-1}(x^h_n \epsilon)), \nonumber \\
    &z^h_n \equiv \frac{m^h_n}{k}e^\sigma, \; x^h_n \equiv \frac{m^h_n}{k \epsilon}, \; \rho \equiv \sqrt{4+b^2}. \nonumber
\end{align}
The mass eigenvalues $m^h_n$ are found, just as in the case of $A_\mu$, using the even orbifold boundary condition at $\phi=\pm \pi$, yielding the condition
\begin{equation}
    (2-\rho)\zeta^h_\rho(x^h_n)+x^h_n \zeta^h_{\rho-1}(x^h_n)=0.
\end{equation}
As in the case of the vector gauge field $A_\mu$, we note that Eq.(\ref{ScalarGaugeFlavorWavefunction}) lacks a solution when $m^h_n=0$ that satisfies the orbifold boundary conditions. As a result, any physical scalars arising from this mode in the effective 4-dimensional theory will have a mass of at least $O(M_{KK})$.

\subsection{The Scalars $A_\phi$ and $\varphi$}\label{BulkSSBPseudoscalarsSection}

We next consider the fifth component of the gauge field, $A_\phi$, and the bulk Goldstone boson $\varphi$, which after the addition of the gauge fixing terms given in Eq.(\ref{GaugeFlavorFixing}), have kinetic and mass terms given by

\begin{eqnarray}\label{GaugeFlavorGoldstoneAction}
    S_{A_\phi, \varphi} & = & \left. \int d^4 x \int_{-\pi}^{\pi} r_c \, d \phi \bigg\{ \frac{1}{2}\frac{e^{-2 \sigma}}{r_c^2}(\partial_\mu A_\phi)^2+\frac{1}{2}e^{-2 \sigma} (\partial_\mu \varphi)^2 \nonumber\right. \\
    & - & \left. \frac{\xi}{2}\bigg( \frac{\partial_\phi e^{-2 \sigma} A_\phi}{r_c^2}\bigg)^2-\frac{\xi}{2} \gamma^2 k^2 e^{-4 \sigma} \varphi^2+\xi \gamma k e^{-2 \sigma} \varphi \bigg( \frac{\partial_\phi e^{-2 \sigma} A_\phi}{r_c^2}\bigg)\right.\\
    & - & \left. \frac{1}{2} \gamma^2 k^2 e^{-4 \sigma}\bigg( \frac{A_\phi}{r_c} \bigg)^2-\frac{e^{-4 \sigma}}{2} \bigg( \frac{\partial_\phi \varphi}{r_c} \bigg)^2+\gamma k e^{-4 \sigma} \bigg(\frac{A_\phi}{r_c}\bigg) \bigg( \frac{\partial_\phi \varphi}{r_c} \bigg) \bigg\} \nonumber.\right.
\end{eqnarray}
Here, we have placed the mass terms in a form that suggests we define new bulk fields, $G$ and $a$, as
\begin{eqnarray}\label{GoldstoneNewFields}
    G & \equiv & \left. \gamma k e^{-2 \sigma} \varphi-\frac{\partial_\phi e^{-2 \sigma} A_\phi}{r_c^2},\right. \\
    a & \equiv & \left. \frac{e^{-2 \sigma}}{r_c}(\partial_\phi \varphi -\gamma k \, A_\phi) \nonumber.\right.
\end{eqnarray}
In terms of these fields, the bulk mass terms of Eq.(\ref{GaugeFlavorGoldstoneAction}) simply become $-(\xi/2) G^2$ and $-(1/2) a^2$. The task of determining the wavefunctions for the KK towers of $a$ and $G$ then becomes diagonalizing the kinetic terms of Eq.(\ref{GaugeFlavorGoldstoneAction}) in terms of the new fields. To accomplish this, we first solve Eq.(\ref{GoldstoneNewFields}) for $\varphi$ and $A_\phi$. For $A_\phi$, we arrive at

\begin{align}\label{GoldstoneAPhiEq}
    \bigg( \gamma^2 k^2-\frac{1}{r_c^2} \partial_\phi e^{2 \sigma} \partial_\phi e^{-2 \sigma} \bigg) A_\phi &= -\gamma k r_c e^{2 \sigma} a+\partial_\phi e^{2 \sigma} G.
\end{align}
Invoking the orbifold odd boundary conditions, $A_\phi|_{\phi=0}=0$ and $A_\phi|_{\phi=\pm \pi}=0$, we can then find an integral expression for $A_\phi$ from Eq.(\ref{GoldstoneAPhiEq}), yielding
\begin{align}
    A_\phi = \frac{-\sigma' e^\sigma}{k^2 \delta \sinh[\delta kr_c \pi]}\int_{0}^{\pi} d \phi_0 &(e^{-\sigma_0}\partial_{\phi_0} e^{2 \sigma_0} G(\phi_0)-\gamma k r_c e^{\sigma_0} a(\phi_0)) \label{GoldstoneAPhi}\\
    \times \{ \sinh[\delta \sigma]\sinh[(\sigma_0-k r_c \pi) \delta] &\theta(|\phi_0|-|\phi|)+ \sinh[\delta \sigma_0]\sinh[(\sigma-k r_c \pi)\delta] \theta(|\phi|-|\phi_0|) \}, \nonumber
\end{align}
where $\sigma_0 \equiv \sigma(\phi_0)$ and $\delta$ is defined as it was in Eq.(\ref{VectorGaugeFlavorSolution}). In Eq.(\ref{GoldstoneAPhi}), we have explicitly written the dependence of $G$ and $a$ on the fifth-dimensional coordinate $\phi_0$ (while continuing to suppress its dependence on the four Minkowski coordinates), to reduce confusion due to the presence of two five-dimensional coordinates in the expression. For $\varphi$, we arrive at
\begin{align}
    &\bigg( -\frac{1}{r_c^2}e^{2 \sigma}\partial_\phi e^{-2 \sigma} \partial_\phi+\gamma^2 k^2 \bigg) \varphi = -\frac{e^{2 \sigma}}{r_c} \partial_\phi a + e^{2 \sigma} \gamma k G. \label{GoldstonevarPhiEq}
\end{align}
Once we impose the even orbifold boundary conditions $\partial_\phi \varphi|_{\phi=0}=0$ and $\partial_\phi \varphi|_{\phi=\pm \pi}=0$, Eq.(\ref{GoldstonevarPhiEq}) then yields the result
\begin{align}
    \varphi &= \frac{-k r_c e^\sigma}{(1-\delta^2)k^2 \delta \sinh[\delta k r_c \pi]} \int_0^{\pi} d \phi_0 \bigg( -\frac{e^{\sigma_0}}{r_c}\partial_{\phi_0}a(\phi_0) + \gamma k e^{\sigma_0} G(\phi_0) \bigg) \label{GoldstonevarPhi} \\
    &\times \{ \lambda_\delta (\sigma_0-k r_c \pi)\lambda_\delta(\sigma)\theta(|\phi_0|-|\phi|) + \lambda_\delta (\sigma_0)\lambda_\delta(\sigma-k r_c \pi) \theta(|\phi|-|\phi_0|), \nonumber \\
    \nonumber
\end{align}
where
\begin{align}
    &\lambda_\delta(\sigma) \equiv \sinh[\delta \sigma]- \delta \cosh[\delta \sigma].
\end{align}
We also use the definition, $\delta=\sqrt{1+\gamma^2}$, in order to eliminate all appearances of $\gamma$ in favor of $\delta$. Eqs.(\ref{GoldstoneAPhi}) and (\ref{GoldstonevarPhi}) can then be inserted into the kinetic terms of Eq.(\ref{GaugeFlavorGoldstoneAction}). Exploiting the overall evenness of the action under the transformation $\phi\rightarrow-\phi$, we determine that the $G$ and $a$ fields' kinetic terms in the action are given by
\begin{align}
&\frac{k^2 r_c^2}{k^4 \delta^2 \sinh^2[\delta k r_c \pi]}\int d^4 x \int_0^\pi \int_0^\pi \int_0^\pi d\phi \, d\phi_1 \, d\phi_2 \, r_c \, \partial_\mu G(\phi_1) \partial_\mu G(\phi_2) \times \nonumber\\
&\left\{ -\frac{k^2 e^{(\sigma_1+\sigma_2)}}{(1-\delta^2)}\Delta_1(\sigma;\sigma_1)\Delta_1(\sigma; \sigma_2)+\frac{1}{r_c^2}e^{2(\sigma_1+\sigma_2)}(\partial_{\phi_1}e^{-\sigma_1}\Delta_2(\sigma;\sigma_1))(\partial_{\phi_2}e^{-\sigma_2}\Delta_2(\sigma; \sigma_2))\right\} \label{GoldstoneKinetic}\\
&+\frac{k^2 r_c^2}{k^4 \delta^2 \sinh^2[\delta kr_c \pi]} \int d^4 x \int_0^{\pi} \int_0^{\pi} \int_0^{\pi} d \phi \, d \phi_1 \, d \phi_2 \, r_c \, \partial_\mu a(\phi_1) \partial_\mu a(\phi_2) \times \nonumber \\
&\left\{ \frac{1}{r_c^2 (1-\delta^2)^2} (\partial_{\phi_1}e^{\sigma_1}\Delta_1(\sigma; \sigma_1))(\partial_{\phi_2} e^{\sigma_2} \Delta_1(\sigma; \sigma_2))+(\delta^2-1) k^2 e^{(\sigma_1+\sigma_2)} \Delta_2 (\sigma;\sigma_1) \Delta_2 (\sigma;\sigma_2) \right\}, \nonumber
\end{align}
where
\begin{align}
    &\Delta_1(\sigma_j; \sigma_i) \equiv \begin{cases}
    \lambda_\delta (\sigma_i-kr_c \pi)\lambda_\delta (\sigma_j), & |\phi_i| \geq |\phi_j| \\
    \lambda_\delta (\sigma_i) \lambda_\delta (\sigma_j-kr_c \pi), & |\phi_i| < |\phi_j|
    \end{cases} \\
    &\Delta_2 (\sigma_j; \sigma_i) \equiv \begin{cases}
    \sinh[\sigma_j \delta] \sinh[(\sigma_i-kr_c \pi) \delta], & |\phi_i| \geq |\phi_j| \\
    \sinh[\sigma_i \delta] \sinh[(\sigma_j-kr_c \pi) \delta], & |\phi_i| < |\phi_j|
    \end{cases} \nonumber
\end{align}
Notably, cross-terms of the form $\partial_\mu a \, \partial_\mu G$ vanish when integrated over $\phi$, and are hence not included in Eq.(\ref{GoldstoneKinetic}). The proof of this vanishing is straightforward, albeit lengthy and unenlightening, so we omit it here. 

Since both the mass terms and the kinetic terms of the action now lack any mixing terms between $a$ and $G$, we may move to addressing the two fields separately. We begin with $G$, by performing a KK expansion, 
\begin{align}
    G(x,\phi) &= \frac{1}{\sqrt{r_c}}\sum_n m^G_n G^{(n)}(x) \chi^G_n(\phi), & \int_{-\pi}^{\pi} d\phi \, \chi^G_n(\phi)\chi^G_m(\phi)=\delta_{m n}.
\end{align}
To achieve a diagonal KK tower of canonically-normalized fields, then, $\chi^G_n$ must satisfy the integral equation,
\begin{align}
    &\chi^G_n(\phi_1)=\frac{k^2 r_c^2 (m^G_n)^2}{k^4 \delta^2 \sinh^2[\delta k r_c \pi]} \int_0^{\pi} \int_0^{\pi} d\phi \, d\phi_2 \, \chi^G_n(\phi_2) \times \nonumber \\
    &\left\{ -\frac{k^2}{(1-\delta^2)} e^{(\sigma_1+\sigma_2)}\Delta_1(\sigma;\sigma_1)\Delta_1(\sigma;\sigma_2)+\frac{1}{r_c^2}e^{2(\sigma_1+\sigma_2)}(\partial_{\phi_1}e^{-\sigma_1}\Delta_2(\sigma;\sigma_1))(\partial_{\phi_2}e^{-\sigma_2}\Delta_2(\sigma;\sigma_2)) \right\}.
\end{align}
Integrating over $\phi$, this equation becomes
\begin{align}\label{GoldstoneGaugeFlavorWavefunction}
    &\chi^G_n(\phi_1) = \frac{(m^G_n)^2 r_c^2}{k r_c \delta \sinh[\delta k r_c \pi](\delta^2-1)} \int_0^{\pi} d\phi_2 \, \chi^G_n(\phi_2) e^{(\sigma_1+\sigma_2)} \Delta_1(\sigma_1;\sigma_2).
\end{align}
This integral equation can be expressed as
\begin{align}\label{GoldstoneGaugeFlavorWavefunctionDiff}
    &\bigg( -\frac{1}{r_c^2} e^{2 \sigma} \partial_\phi e^{-2 \sigma}\partial_\phi+\gamma^2 k^2 \bigg) \chi^G_n(\phi)=(m^G_n)^2 e^{2 \sigma} \chi^G_n(\phi), \textrm{ with }  \partial_\phi \chi^G_n(\phi)|_{\phi=0,\pm \pi}=0,
\end{align}
where we have converted the integral equation into a differential one (taking care to properly differentiate $\Delta_1$, which includes step functions), and invoked the definition $\delta=\sqrt{1+\gamma^2}$. Notably, the differential equation form of Eq.(\ref{GoldstoneGaugeFlavorWavefunction}) is identical to Eq.(\ref{VectorGaugeFlavorWavefunction}), as are their normalization conditions. So, the KK expansion for $G$ may be rewritten as
\begin{equation}
    G(x,\phi)=\frac{1}{\sqrt{r_c}} \sum_n m^A_n G^{(n)}(x) \chi^A_n(\phi),
\end{equation}
with $\chi^A_n(\phi)$ given by Eq.(\ref{VectorGaugeFlavorSolution}). This produces an effective 4-dimensional action for the $G^{(n)}$ fields of
\begin{equation}
    S_G = \sum_n \int d^4 x \left\{ \frac{1}{2}(\partial_\mu G^{(n)}(x))^2-\frac{\xi}{2} (m^A_n)^2 (G^{(n)}(x))^2 \right\},
\end{equation}
indicating that the $G$ fields may be eliminated from the spectrum of physical particles by making the gauge choice $\xi \rightarrow \infty$, corresponding to the unitary gauge. Hence, the fields $G^{(n)}(x)$, a mixture of $A_\phi$ and $\varphi$ fields, correspond to the Goldstone bosons of the massive $A_\mu$ gauge bosons.

We now move on to a similar treatment for the $a$ fields, beginning with a KK expansion,
\begin{align}\label{PseudoGaugeFlavorKK}
    &a(x,\phi)=\frac{1}{\sqrt{r_c}} \sum_n m^a_n a^{(n)}(x) \chi^a_n(\phi), & \int_{-\pi}^{\pi} d\phi \, \chi^a_n(\phi)\chi^a_m(\phi)=\delta_{m n}.
\end{align}
To ensure that the $a^{(n)}$ fields are canonically normalized, then, we require
\begin{align}
    &\chi^a_n(\phi_1)=\frac{k^2 r_c^2 (m^a_n)^2}{k^4 \delta^2 \sinh^2[\delta k r_c \pi]} \int_0^{\pi} \int_0^{\pi} d\phi \, d\phi_2 \, \chi^a_n(\phi_2) \times \nonumber \\
    & \left\{ \frac{1}{(1-\delta^2)^2 r_c^2}(\partial_{\phi_1}e^{\sigma_1}\Delta_1(\sigma;\sigma_1))(\partial_{\phi_2}e^{\sigma_2} \Delta_1(\sigma;\sigma_2))+\gamma^2 k^2 e^{(\sigma_1+\sigma_2)}\Delta_2(\sigma;\sigma_1)\Delta_2(\sigma;\sigma_2) \right\}.
\end{align}
Integrating over $\phi$ then yields the equation, 
\begin{align}\label{PseudoGaugeFlavorWavefunction}
    &\chi^a_n(\phi_1) = -\frac{1}{k r_c}\frac{(m^a_n)^2 r_c^2}{\delta \sinh[\delta k r_c \pi]} \int_0^{\pi} d \phi_2 \, \chi^a_n(\phi_2) e^{(\sigma_1+\sigma_2)} \Delta_2(\sigma_2;\sigma_1) \\
    &\rightarrow \bigg( -\frac{1}{r_c^2} \partial_\phi e^{2 \sigma} \partial_\phi e^{-2 \sigma}+\gamma^2 k^2\bigg)\chi^a_n(\phi) = (m^a_n)^2 e^{2 \sigma} \chi^a_n(\phi), & \chi^a_n(0)=\chi^a_n(\pm \pi)=0. \nonumber
\end{align}
The differential equation form of Eq.(\ref{PseudoGaugeFlavorWavefunction}) can be solved, as in the case for the other KK towers arising in this model, using Bessel functions. The solution is 
\begin{align}\label{PseudoGaugeFlavorSolution}
    \chi^a_n(\phi) &= \frac{e^\sigma}{N^a_n} \sigma' \zeta^a_\delta (z^a_n) \equiv \frac{e^\sigma}{N^a_n}\sigma'(\alpha^a_n J_\delta (z^a_n)+\beta^a_n Y_\delta (z^a_n)), \nonumber \\
    N^a_n &\equiv \frac{\sqrt{k r_c}}{\epsilon}[(\zeta^a_\delta(x^a_n))^2-\epsilon^2(\zeta^a_\delta(x^a_n \epsilon))^2]^{\frac{1}{2}}, \\
    &\alpha^a_n \equiv -Y_\delta (x^a_n \epsilon), \; \beta^a_n \equiv J_\delta (x^a_n \epsilon), \nonumber \\
    &z^a_n \equiv \frac{m^a_n}{k}e^{\sigma}, \; x^a_n \equiv \frac{m^a_n}{k \epsilon}. \nonumber
\end{align}
By applying the boundary condition at $\phi=\pi$ to Eq.(\ref{PseudoGaugeFlavorSolution}), we arrive at the equation that each mass eigenvalue $m^a_n = M_{KK} x^a_n$ must satisfy, namely
\begin{equation}
    \zeta^a_\delta (x^a_n)=0.
\end{equation}
The final 4-dimensional effective action for the $a^{(n)}$ fields now takes the form,
\begin{equation}
    S_a = \sum_n \int d^4 x \left\{ \frac{1}{2}(\partial_\mu a^{(n)}(x))^2-\frac{1}{2} (m^a_n)^2 (a^{(n)}(x))^2 \right\}.
\end{equation}
Notably, just as in the case for $A_\mu$ and $h$, the bulk equation of motion for $a$, Eq.(\ref{PseudoGaugeFlavorWavefunction}), lacks a solution that satisfies the boundary conditions when $x^a_n=0$, again indicating the absence of a massless zero-mode state in the KK tower. Additionally, in contrast to the fields $G^{(n)}$, the masses of the fields $a^{(n)}$ are independent of $\xi$, and represent a tower of physical pseudoscalar particles arising from bulk SSB.

Having derived expressions for the $G$ and $a$ fields' bulk profiles, it is useful to now possess expressions for  the original bulk scalar fields we considered, $A_\phi$ and $\varphi$, since these fields appear elsewhere in our action (for example, in couplings to quark fields). For simplicity, we move to the unitary gauge ($\xi\rightarrow \infty$), in which the $G$ fields are infinitely massive and hence decoupled from the theory. It should be noted that the unitary gauge here corresponds to the selection, $\gamma k e^{-2 \sigma} \varphi=(1/r_c^2) \partial_\phi e^{-2 \sigma} A_\phi$, as opposed to the equivalent in the absence of any bulk symmetry breaking, in which case the gauge choice is simply $A_\phi=0$. Given this gauge choice, we have, from inserting Eqs.(\ref{PseudoGaugeFlavorKK}) and (\ref{PseudoGaugeFlavorSolution}) into Eq.(\ref{GoldstoneAPhiEq}), the expressions
\begin{align}
    & A_\phi(x,\phi)=\frac{1}{\sqrt{r_c}} \sum_n \frac{\gamma k r_c}{k \epsilon x^a_n} \chi^a_n(\phi) a^{(n)}(x), \\
    & \varphi(x,\phi)=\frac{1}{\sqrt{r_c}} \sum_n \frac{1}{k r_c \epsilon x^a_n} (e^{2 \sigma} \partial_\phi e^{-2 \sigma} \chi^a_n(\phi)) a^{(n)}(x),
\end{align}
for the KK expansions of the fields $A_\phi$ and $\varphi$ in the unitary gauge, in terms of the physical pseudoscalar fields $a^{(n)}(x)$.

It bears mentioning that certain results in this analysis, in particular the equality (up to proportionality) of the Kaluza-Klein modes of the Goldstone field $G$ and those of the vector gauge boson $A_\mu$, as well as the vanishing of mixing terms between $a$ and $G$, do eventually emerge in our analysis, but are difficult to see from the outset. However, we note that our route to these results is not unique; during the preparation of this manuscript, we became aware of an alternative treatment of this problem (applied to a more general warped metric) that appears in \cite{Falkowski:2008fz}. In that work's formalism, these characteristics are more immediately apparent, in part due to the starting assumption (which must be true, since each massive KK mode of the vector gauge boson has a longitudinal degree of freedom that must emerge from somewhere) that one of the two KK towers must be a tower of Goldstone bosons for the vector gauge fields. We therefore refer the reader to that work for a different treatment of this problem which may clarify seemingly accidental results of our formalism.

\subsection{Summing Over KK Modes}\label{BulkSSBKKModeSum}
In probing the phenomenology of our model, we shall find it useful to evaluate sums of the form $\sum_n \mathcal{F}(\chi^{A,a,h}_n (\phi_1), \chi^{A,a,h}_n (\phi_2))$ over all KK modes $n$, where $\mathcal{F}$ is some function (following the treatment of SM fields given in \cite{casagrande}). This is in order, for example, to estimate effective four-fermion operators arising from the exchange of all KK modes of a given field in the low-energy limit. To accomplish this, we exploit the orthonormality of the various functions $\chi^{A,a,h}_n (\phi)$ in order to derive several convenient summation identities. To do so, we take a modified version of the approach of \cite{casagrande, hirn}, exploiting orthonormality relations for the various wavefunctions $\chi^{A,a,h}_n$. In particular, we note that \cite{casagrande, hirn}
\begin{equation}\label{VectorSumOrthonormality}
    \int_{-\pi}^{\pi} d \phi \chi_m(\phi) \chi_n(\phi) = \delta_{m n} \rightarrow \sum_{n} \chi_n(\phi_1) \chi_n(\phi_2) = \frac{1}{2}[\delta(\phi_1-\phi_2)+\delta(\phi_1+\phi_2)].
\end{equation}
First, we evaluate the sum $\sum_{n} \chi^A_n(\phi_1)\chi^A_n(\phi_2)/(m^A_n)^2$, which appears in the evaluation of four-fermion operators from the exchange of the entire $A_\mu$ KK tower. From Eq.(\ref{VectorGaugeFlavorAction}) and the orbifold boundary conditions $\partial_\phi \chi^A_n(\phi)|_{\phi=0, \pm \pi} = 0$, we can write
\begin{align}\label{FlavorGaugeIntegralEq}
    \chi^A_n(\phi) &= \frac{kr_c e^\sigma}{k^2 \delta \sinh[\delta kr_c \pi]} \int_0^\pi d\phi_0 \frac{e^{\sigma_0} (m^A_n)^2 \chi^A_n (\phi_0)}{\delta^2-1}\\
    &\times[\lambda_\delta (\sigma_0-kr_c \pi) \lambda_\delta(\sigma)\theta(|\phi_0|-|\phi|)+\lambda_\delta(\sigma_0)\lambda_\delta(\sigma-kr_c \pi) \theta(|\phi|-|\phi_0|)], \nonumber\\
\end{align}
where
\begin{align}
        \delta \equiv \sqrt{1+\frac{g^2 Y^2_A v^2}{k^2}}, \;\; \lambda_\delta(\sigma) \equiv \sinh[\delta \sigma]-\delta \cosh[\delta \sigma], \;\; \sigma_i \equiv \sigma(\phi_i).
\end{align}
Eq.(\ref{FlavorGaugeIntegralEq}) then allows us to write
\begin{align}\label{FlavorGaugeSum}
    \sum_n \frac{\chi^A_n (\phi_1)\chi^A_n(\phi_2)}{(m^A_n)^2} &= \frac{kr_c e^{\sigma_2}}{k^2 \delta \sinh[\delta k r_c \pi](\delta^2-1)}\int_0^\pi d\phi_0 \, \bigg( \sum_n \chi^A_n(\phi_1)\chi^A_n(\phi_0) \bigg) e^{\sigma_0}\\
    &\times[\lambda_\delta(\sigma_0-kr \pi) \lambda_\delta (\sigma_2) \theta(|\phi_0|-|\phi_2|)+\lambda_\delta(\sigma_0)\lambda_\delta(\sigma_2-kr \pi)\theta(|\phi_2|-|\phi_0|)] \nonumber.
\end{align}
Using Eq.(\ref{VectorSumOrthonormality}), we can then write
\begin{align}\label{FlavorGaugeTowerExchange}
    \sum_n \frac{\chi^A_n (\phi_1)\chi^A_n(\phi_2)}{(m^A_n)^2} = \frac{kr_c e^{(\sigma_1+\sigma_2)}\epsilon^2}{2 M_{KK}^2 (\delta^2-1) \delta \sinh[\delta k r_c \pi]} \begin{cases}
    \lambda_\delta (\sigma_1-kr_c \pi)\lambda_\delta (\sigma_2) & |\phi_1| \geq |\phi_2| \\
    \lambda_\delta (\sigma_1)\lambda_\delta (\sigma_2-kr_c \pi) & |\phi_1|<|\phi_2|
    \end{cases}.
\end{align}

Next, we consider the analogous sum for the tower exchange of the scalar field $h$, namely, $\sum_{n} \chi^h_n(\phi_1)\chi^h_n(\phi_2)/(m^h_n)^2$. From Eq.(\ref{ScalarGaugeFlavorWavefunction}) (and the even orbifold boundary conditions of the $h$ field), we obtain
\begin{align}\label{FlavorScalarIntegralEq}
    \chi^h_n(\phi) = \frac{kr_c e^{2 \sigma}}{k^2\sinh[\rho kr_c \pi]} \int_0^\pi \frac{\chi^h_n(\phi_0)}{\rho^2-4} &[\omega_\rho (\sigma_0-kr_c \pi) \omega_\rho(\sigma)\theta(|\phi_0|-|\phi|)\\
    &+\omega_\rho (\sigma_0) \omega_\rho (\sigma-kr_c \pi)\theta(|\phi|-|\phi_0|) \nonumber,
\end{align}
where
\begin{align}
    \rho \equiv \sqrt{4+b^2}, \; \; \omega_\rho(\sigma) \equiv 2 \sinh[\rho \sigma] &-\rho \cosh[\rho \sigma].
\end{align}
We can then insert this identity into the sum we wish to evaluate, and proceed identically to our treatment of the vector gauge boson sum, the only exception being that the orthonormality relation among the bulk profiles is now given by
\begin{align}
    \sum_{n} e^{-2 \sigma_2} \chi^h_n (\phi_1) \chi^h_n (\phi_2) = \frac{1}{2}[\delta(\phi_1-\phi_2)+\delta(\phi_1+\phi_2)].
\end{align}
Applying this relation, we arrive at
\begin{align}\label{FlavorScalarTowerExchange}
    \sum_n \frac{\chi^h_n(\phi_1)\chi^h_n(\phi_2)}{(m^h_n)^2} = \frac{kr_c e^{2 (\sigma_1+\sigma_2)} \epsilon^2}{2 M_{KK}^2 (\rho^2-4) \rho \sinh[\rho kr_c \pi]} \begin{cases}
    \omega_{\rho}(\sigma_1-kr_c \pi) \omega_{\rho}(\sigma_2) & |\phi_1| \geq |\phi_2| \\
    \omega_{\rho}(\sigma_1) \omega_{\rho}(\sigma_2-kr_c \pi) & |\phi_1| < |\phi_2|
    \end{cases}.
\end{align}

The sum identities which are required for the exchange of the pseudoscalar boson, $a$, are somewhat more complex than the prior cases, due to the highly non-trivial mixing between the $A_\phi$ and $\varphi$ fields which produce it. We shall see that, in order to evaluate the effective four-fermion operators for exchange of the tower of $a$ fields, we shall need to evaluate several sums, namely,
\begin{align}\label{NeededSums}
    &\sum_n \frac{(\delta^2-1)k^2 \chi^a_n (\phi_1) \chi^a_n (\phi_2)}{(m^a_n)^4}, \nonumber\\
    &\sum_n \frac{k^2 e^{2(\sigma_1 +\sigma_2)}}{(k r_c)^2(m^a_n)^4}(\partial_{\phi_1} e^{-2 \sigma_1} \chi^a_n (\phi_1))(\partial_{\phi_2} e^{-2 \sigma_2}\chi^a_n(\phi_2)), \\
    &\sum_n \frac{k^2 \sqrt{\delta^2-1}}{kr_c (m^a_n)^4}(\partial_{\phi_1}e^{-2 \sigma_1}\chi^a_n(\phi_1))\chi^a_n (\phi_2), \nonumber
\end{align}
where $\delta$ is defined in the same way as it is in Eq.(\ref{FlavorGaugeIntegralEq}). It should be noted that even with the $(m^a_n)^{-4}$ dependence of these sums, they still represent dimension-6 operators; the $k^2$ terms that multiply each sum in Eq.(\ref{NeededSums}) cancel out the extra factors $m^a_n$ in the denominator. To actually evaluate the sums of Eq.(\ref{NeededSums}), we first use the integral form of Eq.(\ref{PseudoGaugeFlavorWavefunction}), the equation of motion for $\chi^a_n(\phi)$, to evaluate the sum $\sum_n \chi^a_n (\phi_1) \chi^a_n (\phi_2)/(m^a_n)^2$ in a manner directly analogous to our discussion leading up to Eq.(\ref{FlavorGaugeTowerExchange}), arriving at
\begin{align}\label{InitialPseudoscalarSum}
    \sum_n \frac{\chi^a_n (\phi_1) \chi^a_n (\phi_2)}{(m^a_n)^2} = -\frac{kr_c e^{(\sigma_1+\sigma_2)} \epsilon^2}{2 M_{KK}^2 \delta \sinh[\delta k r_c \pi]} \begin{cases}
    \sinh[\delta \sigma_1] \sinh[(\sigma_2-kr_c \pi)\delta] & |\phi_2| \geq |\phi_1| \\
    \sinh[(\sigma_1-kr_c \pi)\delta] \sinh[\delta \sigma_2] & |\phi_2| < |\phi_1|
    \end{cases}.
\end{align}
We can then use Eq.(\ref{InitialPseudoscalarSum}) to evaluate the summations listed in Eq.(\ref{NeededSums}). First, we note that using Eq.(\ref{FlavorGaugeIntegralEq}), we can write
\begin{align}
    \sum_n \frac{\chi^a_n (\phi_1) \chi^a_n (\phi_2)}{(m^a_n)^4} &= -\frac{kr_c e^{\sigma_2}}{\delta \sinh[\delta kr_c \pi]} \int_0^\pi d\phi_0 \bigg( \sum_n \frac{\chi^a_n(\phi_0)\chi^a_n(\phi_1)}{(m^a_n)^2} \bigg) \\
    &\times e^{\sigma_0} (\sinh[\delta \sigma_2]\sinh[(\sigma_0-kr_c \pi)\delta]\theta(\sigma_0-\sigma_2)+\sinh[\delta \sigma_0] \sinh[(\sigma_2-kr_c \pi)\delta]).\nonumber
\end{align}
Substituting Eq.(\ref{InitialPseudoscalarSum}) into this expression and performing the integration yields the expression,
\begin{align}\label{NeededSum1}
    \sum_n &\frac{\chi^a_n (\phi_1) \chi^a_n(\phi_2)}{(m^a_n)^4}= \frac{-k r_c e^{(\sigma_1+\sigma_2)}}{8 k^4 (\delta^2-1)\delta \sinh^2[\delta kr_c \pi]}\\
    &\times \bigg( e^{2 \sigma_>}\sinh[\delta kr_c \pi] \sinh[\delta \sigma_<]\lambda_\delta(\sigma_>-kr_c \pi)+e^{2 \sigma_<}\sinh[\delta kr_c \pi] \sinh[(\sigma_>-kr_c \pi)\delta]\lambda_\delta (\sigma_<) \nonumber\\
    &-\delta \sinh[(\sigma_1-kr_c \pi) \delta] \sinh[(\sigma_2-kr_c \pi)\delta]+\epsilon^{-2} \delta \sinh[\delta \sigma_1]\sinh[\delta \sigma_2]\bigg) \nonumber,
\end{align}
where $\sigma_{>(<)}$ denotes the larger (smaller) value of the pair, $\sigma_1$ and $\sigma_2$, and $\lambda_\delta(\sigma)$ has the same definition as it does in Eq.(\ref{FlavorGaugeIntegralEq}). By differentiating functions of the integral expression in Eq.(\ref{PseudoGaugeFlavorWavefunction}), we can similarly derive the other sums in Eq.(\ref{NeededSums}). We arrive at
\begin{align}\label{NeededSum2}
    \sum_n \frac{e^{2 \sigma_1}}{(m^a_n)^4}(\partial_{\phi_1} e^{-2 \sigma_1} \chi^a_n(\phi_1))\chi^a_n(\phi_2) = \frac{-(kr_c)^2 e^{(\sigma_1+\sigma_2)}}{8 k^4 (\delta^2-1)\delta \sinh^2[\delta kr_c \pi]}& \nonumber \\
    \times \bigg( -e^{2 \sigma_2} \sinh[\delta kr_c \pi] \lambda_\delta(\sigma_>-kr_c \pi) \lambda_\delta(\sigma_<)&\\
    -e^{2 \sigma_1}(\delta^2-1)\sinh[\delta kr_c \pi] \sinh[\delta \sigma_<] \sinh[(\sigma_>-kr_c \pi) \delta]& \nonumber\\
    +\delta \sinh[(\sigma_2-kr_c \pi)\delta]\lambda_\delta(\sigma_1-kr_c \pi) - \delta \epsilon^{-2} \sinh[\delta \sigma_2] \lambda_\delta(\sigma_1) \bigg)& \nonumber
\end{align}
and
\begin{align}\label{NeededSum3}
    \sum_n \frac{e^{2 (\sigma_1+\sigma_2)}}{(m^a_n)^4} (\partial_{\phi_1} e^{-2 \sigma_1} \chi^a_n (\phi_1))(\partial_{\phi_2} e^{-2 \sigma_2}\chi^a_n(\phi_2))  = \frac{(k r_c)^3 e^{(\sigma_1+\sigma_2)}}{8 k^4 (\delta^2-1)\delta \sinh^2[\delta kr_c \pi]}& \nonumber\\
    \times\bigg( -e^{2 \sigma_>}(\delta^2-1)\sinh[\delta k r_c \pi] \sinh[(\sigma_>-kr_c \pi)] \lambda_\delta(\sigma_<)&\\
    -e^{2 \sigma_<} (\delta^2-1)\sinh[\delta kr_c \pi] \lambda_\delta(\sigma_>-kr_c \pi)\sinh[\delta \sigma_<]& \nonumber\\
    +\delta \lambda_\delta(\sigma_1-kr_c \pi) \lambda_\delta(\sigma_2-kr_c \pi)-\epsilon^{-2} \delta \lambda_\delta(\sigma_1)\lambda_\delta(\sigma_2) \bigg).& \nonumber
\end{align}

\section{SM In the Bulk: Gauge Bosons}\label{BulkEWSection}
In addition to the introduction of new gauge symmetries broken in the bulk, our model must of course include the SM gauge group $SU(3)_C \times SU(2)_L \times U(1)_Y$. While a discussion on precisely how to realize the bulk SM in warped spacetime is readily available \cite{casagrande,Davoudiasl:2000wi,carenaEW}, for definiteness and clarity we quote relevant results here, particularly regarding the electroweak sector. In the absence of any brane-localized symmetry breaking, the spectrum of physical particles (and their bulk wave functions) arising in the effective 4-D theory of a gauge field is trivially derivable from our treatment of bulk symmetry breaking in Section \ref{BulkSSBSection}; in the unitary gauge, it's simply given by the KK tower of vector gauge bosons $A_\mu$ with the bulk mass set to zero. While this suffices for the gluons, the electroweak sector is more complex. We employ the treatment of \cite{casagrande}, and for greater detail, we encourage the reader to consult that work. The quadratic terms of the action of this sector of the theory (in the unitary gauge, which eliminates the fifth component of the gauge fields) is given by \cite{casagrande}
\begin{align}
    S_{A,Z,W,h} = \int d^4 x \int d\phi \, r_c \bigg\{ &-\frac{1}{4} F_{\mu \nu}F^{\mu \nu} -\frac{1}{4} Z_{\mu \nu}Z^{\mu \nu} -\frac{1}{2} W^{+}_{\mu \nu} W^{-\mu \nu}\\
    & + \frac{e^{-2 \sigma}}{2 r_c^2}(\partial_\phi A_\mu)^2  + \frac{e^{-2 \sigma}}{2 r_c^2}(\partial_\phi Z_\mu)^2 +\frac{e^{-2 \sigma}}{r_c^2}\partial_\phi W^{+}_\mu \partial_\phi W^{-\mu} \nonumber\\
    &+\frac{\delta(|\phi|-\pi)}{r_c} \bigg[\frac{1}{2}\partial_\mu H \partial^\mu H-\lambda v^2 H^2+\frac{M_Z^2}{2} Z_\mu Z^\mu + M_W^2 W^{+}_{\mu} W^{-\mu}\bigg] \bigg\} \nonumber,
\end{align}
where $A_\mu$ is the bulk vector photon field (and $F_{\mu \nu}$ its corresponding field strength tensor), $Z_\mu$ the bulk vector $Z$ field (with its field strength tensor $Z_{\mu \nu}$, $W^{\pm}_\mu$ the vector $W$ bosons (with their field strength tensors $W^{\pm}_\mu$), and $H$ the conventional SM Higgs boson, localized on the TeV-brane. Here, all the listed bulk fields are even on the orbifold. Kaluza-Klein decomposition of the bulk fields is performed in the usual way, where for each vector field we write \cite{casagrande}
\begin{align}\label{SMGaugeKKExpansion}
    B_\mu (x,\phi) = \frac{1}{\sqrt{r_c}}\sum_{n}B^{(n)}_\mu (x) \chi^{B}_{n}(\phi)
\end{align}
with $B=A,Z,W^{+},W^{-}$. In order to produce a diagonalized, canonically-normalized effective 4-dimensional theory, each vector field must then satisfy \cite{casagrande}
\begin{align}\label{SMGaugeEq}
    &-\frac{1}{r_c^2}(\partial_\phi e^{-2 \sigma} \partial_\phi \chi^B_n(\phi))=(m^B_n)^2 \chi^B_n(\phi) - \frac{\delta(|\phi|-\pi)}{r_c} M_B^2 \chi^B_n(\phi), \\
    &\partial_\phi \chi^B_n(0)=0, \; \partial_\phi \chi^B_n(\phi)|_{\phi\rightarrow \pi^-} = -\frac{r_c M_B^2}{2 \epsilon^2} \chi^B_n (\pi), \nonumber
\end{align}
where again $B=A,Z,W^{+},W^{-}$, and $m^B_n$ is the mass eigenvalue of the $n^{th}$ Kaluza Klein mode of the field. The bulk wave functions that satisfy this equation are of the form,
\begin{align}\label{SMGaugeWaveFunction}
    &\chi^{B}_n(\phi) = \frac{e^\sigma}{N^B_n} \bar{\zeta}^B_1 (z^B_n), \nonumber\\
    &\bar{\zeta}^B_q (z^B_n) \equiv Y_0(x^B_n \epsilon) J_q(z^B_n) - J_0 (x^B_n \epsilon) Y_q (z^B_n), \nonumber\\
    &N^B_n \equiv \frac{1}{\sqrt{\epsilon^2 kr_c}}\bigg( \bar{\zeta}^B_1(x^B_n)^2-\bar{\zeta}^B_0(x^B_n)^2-\frac{2}{x^B_n}\bar{\zeta}^B_1(x^B_n)\bar{\zeta}^B_0(x^B_n)-\epsilon^2 \bar{\zeta}^B_1(x^B_n \epsilon)\bigg)^{\frac{1}{2}},
\end{align}
where
\begin{align}
    z^B_n \equiv x^B_n e^{(\sigma-kr_c \pi)}, \; x^B_n \equiv \frac{m^B_n}{k \epsilon}.
\end{align}
We can find the allowed eigenvalues of $x^B_n$, as usual, by finding the roots of the TeV-brane boundary condition in Eq.(\ref{SMGaugeEq}). In the case of the photon field (and the gluon field, as it also lacks a brane mass), there's an additional massless zero-mode that satisfies Eq.(\ref{SMGaugeEq}) with $x^A_n=0$, in particular, this is given by the field $\chi^A_0(\phi) = (2 \pi)^{-\frac{1}{2}}$. The $W$ and $Z$ fields lack this zero-mode, however they do each possess a light mode that corresponds to a SM $W$ or $Z$ boson, with a wavefunction approximately given by \cite{casagrande}
\begin{align}
    \chi^{W,Z}_0 (\phi) \approx \frac{1}{\sqrt{2 \pi}} \bigg[ 1+ \frac{(m_{W,Z})^2}{4 M_{KK}^2}\bigg( 1-\frac{1}{kr_c \pi}+\epsilon^2 e^{2 \sigma}(1-2 \sigma)\bigg) + O\bigg( \frac{(m_{W,Z})^4}{M_{KK}^4}\bigg)\bigg],
\end{align}
where $m_{W(Z)}$ is the mass of the light $W(Z)$ mode in the effective 4-D theory. It should be noted that the mass relations between $W$, $Z$, and the electroweak couplings and Higgs VEV are altered slightly in the RS framework from their SM forms \cite{casagrande, hewett, casagrande:2010,Burdman:2002gr}, however, as these modifications are independent at tree-level from the flavor structure we are exploring in this work, we shall not reproduce them in detail here, instead simply quoting the tree-level corrections to the precision electroweak observables $S$ and $T$, which provide tight constraints on any non-custodial RS model with SM fields propagating in the bulk \cite{casagrande,hewett,carenaEW,agashe,Burdman:2002gr}. The tree-level corrections to these parameters are simply \cite{casagrande,goertz}
\begin{align}
    S=\frac{2 \pi v^2}{M_{KK}^2}\bigg(1-\frac{1}{kr_c \pi} \bigg), \; \; T=\frac{\pi v^2}{2 \cos^2\theta_W M_{KK}^2} \bigg( kr_c \pi - \frac{1}{2 kr_c \pi} \bigg).
\end{align}
Because these corrections (in particular the correction to $T$) are independent of the flavor structure, they provide a strong constraint on the KK scale of our model; we shall discuss the numerical implications of these constraints when we perform a numerical probe of our model space in Section \ref{NumericalParamPointsSection}.

Finally, we complete our discussion of the SM gauge sector here with a brief review of some summation identities we shall use when discussing effective 4-fermion operators that arise from an exchange over the massive particles in the gauge boson Kaluza-Klein towers. In particular, we have \cite{casagrande}
\begin{align}
    &\sum_{n=0}^{\infty} \frac{\chi^{W,Z}_n(\phi_1)\chi^{W,Z}_n(\phi_2)}{(m^{W,Z}_n)^2} = \frac{1}{2 \pi (m_{W,Z})^2} \label{WZGaugeSum}\\
    &+\frac{1}{4 \pi M_{KK}^2} \bigg[ kr_c \pi \epsilon^2 e^{2\sigma_<}-kr_c \pi \epsilon^2 (e^{2 \sigma_1}+e^{2 \sigma_2})+1-\frac{1}{2 kr_c \pi} + O\bigg( \frac{(m_{W,Z})^2}{M_{KK}^2}\bigg) \bigg], \nonumber
\end{align}
for the exchange of $W$ or $Z$ towers, where $\sigma_<$ is the smaller of the pair, $\sigma_1$ and $\sigma_2$, and
\begin{align}
    &\sum_{n=1}^{\infty} \frac{\chi^\gamma_n(\phi_1)\chi^\gamma_n(\phi_2)}{(m^\gamma_n)^2} = \label{PhotonGaugeSum}\\
    &\frac{1}{4 \pi M_{KK}^2} \bigg[ kr_c \pi \epsilon^2 e^{2\sigma_<}-\epsilon^2 e^{2 \sigma_1} \bigg(\frac{1}{2}+kr_c \pi-\sigma_1 \bigg)-\epsilon^2 e^{2 \sigma_2}\bigg( \frac{1}{2}+kr_c \pi -\sigma_2 \bigg) + \frac{1}{2 kr_c \pi}\bigg], \nonumber
\end{align}
for the exchange of photon or gluon towers.

\section{Fermions in Warped Spacetime}\label{BulkFermionSection}
Before discussing the particulars of our model of flavor, it is useful to outline the general treatment of chiral fermions in an RS framework. The exact solutions for quark bulk profiles in the case of multiple generations has been well explored in \cite{casagrande,huberHierarchy,huber}, and rather than repeating that work, we shall simply restrict ourselves to quoting important results. In particular, we follow the notation of \cite{casagrande,casagrande2,goertz}, and for a more detailed discussion of this treatment, we refer the reader to those works. Generically, the quark fields (arranged into multiplets as in the SM) will have the action

\begin{align}\label{FermionAction}
    S_{\textrm{quarks}} &= \int d^4 x \int_{-\pi}^{\pi} d \phi \, r_c \bigg\{ e^{-3 \sigma} (\frac{i}{2} \bar{\mathbf{Q}}\slashed{\partial}\mathbf{Q}+h.c.)+e^{-3 \sigma} (\frac{i}{2} \bar{\mathbf{u}}\slashed{\partial}\mathbf{u}+h.c.)+e^{-3 \sigma} (\frac{i}{2} \bar{\mathbf{d}}\slashed{\partial}\mathbf{d}+h.c.) \nonumber \\
    &-\frac{e^{-2 \sigma}}{r_c}[\bar{\mathbf{Q}}_{L} \partial_\phi \mathbf{Q}_{R}-\bar{\mathbf{Q}}_{R} \partial_\phi e^{-2 \sigma} \mathbf{Q}_{L}+\sum_{\mathbf{q}=\mathbf{u},\mathbf{d}}(\bar{\mathbf{q}}_{L} \partial_\phi \mathbf{q}_{R}-\bar{\mathbf{q}}_{R} \partial_\phi e^{-2 \sigma} \mathbf{q}_{L})] \nonumber\\
    &-e^{-4 \sigma} k \,\textrm{sgn}(\phi)(\bar{\mathbf{Q}} \bm{\omega}_Q \mathbf{Q}-\bar{\mathbf{u}}\bm{\omega}_u \mathbf{u}-\bar{\mathbf{d}} \bm{\omega}_d \mathbf{d}) \\
    &-\frac{2}{kr_c} \delta(|\phi|-\pi)e^{-3 \sigma}\frac{v}{\sqrt{2}}[\bar{\mathbf{Q}}_{L} \mathbf{Y}^u \mathbf{u}_{R}+\bar{\mathbf{Q}}_{L} \mathbf{Y}^d \mathbf{d}_{R}+h.c.] \nonumber \bigg\} .
\end{align}
Here, $\mathbf{Q}$ represents the $SU(2)_L$ doublet fields, while $\mathbf{u}$ and $\mathbf{d}$ represent the up-like and down-like $SU(2)_L$ singlets occuring in the SM, while $v \approx 246$ GeV is the standard electroweak Higgs vev. To produce the appropriate spectrum of SM fermions, we impose the condition that $\mathbf{Q}_L$, $\mathbf{u}_R$, and $\mathbf{d}_R$ are even on the orbifold, while $\mathbf{Q}_R$, $\mathbf{u}_L$, and $\mathbf{d}_L$ are odd. Here, $\mathbf{Q}_{L,R}$, $\mathbf{u}_{L,R}$, and $\mathbf{d}_{L,R}$ are all 3-component vectors in generation space, while $\mathbf{Y}^{u,d}$ and $\bm{\omega}_{Q,u,d}$ denote $3\times 3$ matrices in this space. The matrices $\bm{\omega}^{Q,u,d}$ are simply bulk mass matrices (rendered dimensionless by factoring out $k$); note the inclusion of a factor of $\textrm{sgn}(\phi)$ in these terms due to the opposite $Z_2$ parity of the quarks' left- and right-handed fields. This sign term can arise, for example, from the vacuum expectation value of an odd gauge singlet bulk scalar field \cite{Grossman:2000,ahmed}. In this section, we shall assume that the $\bm{\omega}$ matrices are all real and diagonal; generically this can always be made true by performing rotations on the various quark fields, and we shall do so explicitly when discussing our model of flavor. For simplicity's sake, we have foregone any extensions of the SM gauge group here (beyond our eventual inclusion of a flavor gauge symmetry), in particular the introduction of a custodial symmetry, which is often used to mitigate the draconian constraint on RS models from the $T$ parameter, as in \cite{casagrande:2010,goertz,carenaEW,agashe,bouchart}. In any event, we find that the constraint on $M_{KK}=k \epsilon$ from the $T$ parameter is ultimately secondary to flavor constraints in most regions of parameter space for our model, so the omission of such a symmetry does not overly constrain our model.

Armed with our action in Eq.(\ref{FermionAction}), we perform the KK expansions (following the notation of \cite{casagrande})
\begin{align}\label{FermionKKExpansions}
    \mathbf{Q}_L &=\frac{e^{2 \sigma}}{\sqrt{r_c}} \sum_n \mathbf{C}^{(Q)}_n (\phi) \begin{pmatrix}
    \Vec{a}^{(U)}_n u^{(n)}_L(x) \\
    \Vec{a}^{(D)}_n d^{(n)}_L(x)
    \end{pmatrix}, & \mathbf{Q}_R &=\frac{e^{2 \sigma}}{\sqrt{r_c}} \sum_n \mathbf{S}^{(Q)}_n (\phi) \begin{pmatrix}
    \Vec{b}^{(U)}_n u^{(n)}_R(x) \\
    \Vec{b}^{(D)}_n d^{(n)}_R(x)
    \end{pmatrix}, \nonumber\\
    \mathbf{u}_L &=\frac{e^{2 \sigma}}{\sqrt{r_c}} \sum_n \mathbf{S}^{(u)}_n(\phi) \Vec{b}^{(u)}_n u^{(n)}_L(x), & \mathbf{u}_R &=\frac{e^{2 \sigma}}{\sqrt{r_c}} \sum_n \mathbf{C}^{(u)}_n(\phi) \Vec{a}^{(u)}_n u^{(n)}_R(x), \\
    \mathbf{d}_L &=\frac{e^{2 \sigma}}{\sqrt{r_c}} \sum_n \mathbf{S}^{(d)}_n(\phi) \Vec{b}^{(d)}_n d^{(n)}_L(x), & \mathbf{d}_R &=\frac{e^{2 \sigma}}{\sqrt{r_c}} \sum_n \mathbf{C}^{(d)}_n(\phi) \Vec{a}^{(d)}_n d^{(n)}_R(x). \nonumber
\end{align}
Here, the various $\Vec{a}$ and $\Vec{b}$ vectors are three-dimensional complex vectors in generation space, while $\mathbf{C}^{(Q,u,d)}_n(\phi)$ and $\mathbf{S}^{(Q,u,d)}_n(\phi)$ are diagonal matrices, with each non-zero entry given by a real function of $\phi$. Given our boundary condition choices for the quark fields, we see that $\mathbf{C}^{(Q,u,d)}_n(\phi)$ consists of functions even under $\phi\rightarrow -\phi$, while $\mathbf{S}^{(Q,u,d)}_n(\phi)$ consists of odd functions. It is straightforward to show that, in order for these KK modes to be mass eigenstates in the effective 4-dimensional theory, $\mathbf{S}^{(Q,u,d)}_n(\phi)$ and $\mathbf{C}^{(Q,u,d)}_n(\phi)$ must satisfy the equations of motion,
\begin{align}\label{QuarkDiffEqs}
    - &\frac{1}{r_c}\partial_\phi \mathbf{C}^{(u,d)}_n \Vec{a}^{(u,d)}_n+\frac{\sigma'}{r_c}\bm{\omega}_{u,d} \mathbf{C}^{(u,d)}_n \Vec{a}^{(u,d)}_n &=& -m^{u,d}_n e^{\sigma} \mathbf{S}^{(u,d)}_n \Vec{b}^{(u,d)}_n, \nonumber\\
    &\frac{1}{r_c} \partial_\phi \mathbf{C}^{(U,D)}_n \Vec{a}^{(U,D)}_n-\frac{\sigma'}{r_c} \bm{\omega}_Q \mathbf{C}^{(U,D)}_n \Vec{a}^{(U,D)}_n &=& -m^{u,d}_n e^{\sigma}\mathbf{S}^{(U,D)}_n \Vec{b}^{(U,D)}_n,\\
    - &\frac{1}{r_c}\partial_\phi \mathbf{S}^{(U,D)}_n \Vec{b}^{(U,D)}_n-\frac{\sigma'}{r_c}\bm{\omega}_Q \mathbf{S}^{(U,D)}_n \Vec{b}^{(U,D)}_n &=& -m^{u,d}_n e^\sigma \mathbf{C}^{(U,D)}_n \Vec{a}^{U,D}_n+\delta(|\phi|-\pi)e^{\sigma} \mathbf{M}^{(U,D)}, \nonumber\\
    &\frac{1}{r_c} \partial_\phi \mathbf{S}^{(u,d)}_n \Vec{b}^{(u,d)}_n+\frac{\sigma'}{r_c} \bm{\omega}_{u,d}\mathbf{S}^{(u,d)}_n \Vec{b}^{(u,d)}_n &=& -m^{u,d}_n e^{\sigma}\mathbf{C}^{(u,d)}_n \Vec{a}^{(u,d)}_n+\delta(|\phi|-\pi)\mathbf{M}^{(u,d)}, \nonumber
\end{align}
where
\begin{align}
    &\mathbf{M}^{(U,D)} \equiv \frac{2 v}{\sqrt{2}kr_c}e^{\sigma} \mathbf{Y}^{u,d} \mathbf{C}^{(u,d)}_n \Vec{a}^{(u,d)}_n, \\
    &\mathbf{M}^{(u,d)} \equiv \frac{2 v}{\sqrt{2}kr_c}e^{\sigma} (\mathbf{Y}^{u,d})^\dagger \mathbf{C}^{(U,D)}_n \Vec{a}^{(U,D)}_n, \nonumber
\end{align}
and the normalization conditions,
\begin{align}\label{FermionNormalizationConditions}
    &\int_{-\pi}^{\pi} d\phi \, e^{\sigma} \left\{ \Vec{a}^{(U,D)\dagger}_n \mathbf{C}^{(U,D)}_n \mathbf{C}^{(U,D)}_m \Vec{a}^{(U,D)}_m+\Vec{b}^{(u,d) \dagger}_n \mathbf{S}^{(u,d)}_n \mathbf{S}^{(u,d)}_m \Vec{b}^{(u,d)}_m \right\} = \delta_{nm},\\
    &\int_{-\pi}^{\pi} d\phi \, e^{\sigma} \left\{ \Vec{b}^{(U,D)\dagger}_n \mathbf{S}^{(U,D)}_n \mathbf{S}^{(U,D)}_m \Vec{b}^{(U,D)}_m+\Vec{a}^{(u,d) \dagger}_n \mathbf{C}^{(u,d)}_n \mathbf{C}^{(u,d)}_m \Vec{a}^{(u,d)}_m \right\} = \delta_{nm}. \nonumber
\end{align}
Here, $m^{u(d)}_n$ is the mass of the $n^{\textrm{th}}$ Kaluza-Klein mode of the up(down)-like quark KK tower. These conditions yield the solutions \cite{casagrande}
\begin{align}\label{QuarkSolutions}
    &\mathbf{C}^{(U,D)}_n(\phi) = e^{\sigma/2} diag \bigg(\Vec{N}^{U,D}_n \circ \Vec{\zeta}^Q_{+}(z^{u,d}_n)\bigg), & \mathbf{S}^{(U,D)}_n = &\frac{-\sigma' e^{\sigma/2}}{kr_c} diag \bigg(\Vec{N}^{U,D}_n \circ \Vec{\zeta}^Q_{-}(z^{u,d}_n) \bigg),\\
    &\mathbf{C}^{(u,d)}_n(\phi) = e^{\sigma/2} diag \bigg(\Vec{N}^{u,d}_n \circ \Vec{\zeta}^Q_{+}(z^{u,d}_n)\bigg), & \mathbf{S}^{(u,d)}_n = &\frac{\sigma' e^{\sigma/2}}{kr_c} diag \bigg(\Vec{N}^{u,d}_n \circ \Vec{\zeta}^{u,d}_{-}(z^{u,d}_n) \bigg), \nonumber \\
    \nonumber \\
    & \Vec{a}^{(U,u,D,d)}_n=\Vec{b}^{(U,u,D,d)}_n, \; \; \Vec{a}^{(U,D)\dagger}_n \Vec{a}^{(U,D)}_n+\Vec{a}^{(u,d)\dagger}_n \Vec{a}^{(u,d)}_n = 1, \nonumber
\end{align}
where $\circ$ is the Hadamard product, that is $\Vec{x} \circ \Vec{y} = (x_1 y_1, x_2 y_2, ..., x_N y_N)$ for two $N$-dimensional vectors $\Vec{x}$ and $\Vec{y}$. The terms $x^{u}_n$, $z^{u}_n$, and the three-dimensional (in generation space) vectors $\Vec{\zeta^{Q,u,d}_{+}}$, $\Vec{\zeta^{Q,u,d}_{-}}$, and $\Vec{N}^{U}_n$ are given by the expressions
\begin{align}\label{QuarkZetaDefinition}
    & \Vec{\zeta}^{Q,u,d}_{+}(z^{u,d}_n) \equiv J_{-\frac{1}{2}-\Vec{\bm{\omega}}_{Q,u,d}}(x^{u,d}_n \epsilon) J_{-\frac{1}{2}+\Vec{\bm{\omega}}_{Q,u,d}}(z^{u,d}_n)+J_{\frac{1}{2}+\Vec{\bm{\omega}}_{Q,u,d}}(x^{u,d}_n \epsilon) J_{\frac{1}{2}-\Vec{\bm{\omega}}_{Q,u,d}}(z^{u,d}_n), \nonumber\\
    \nonumber\\
    & \Vec{\zeta}^{Q,u,d}_{-}(z^{u,d}_n) \equiv J_{\frac{1}{2}+\Vec{\bm{\omega}}_{Q,u,d}}(x^{u,d}_n \epsilon) J_{-\frac{1}{2}-\Vec{\bm{\omega}}_{Q,u,d}}(z^{u,d}_n)-J_{-\frac{1}{2}-\Vec{\bm{\omega}}_{Q,u,d}}(x^{u,d}_n \epsilon) J_{\frac{1}{2}+\Vec{\bm{\omega}}_{Q,u,d}}(z^{u,d}_n),\\
    &(\Vec{N}^{U}_n)_{i} \equiv \sqrt{kr_c}\epsilon x^{u}_n \bigg( (x^{u}_n)^2 ((\Vec{\zeta}^{U}_{+}(x^{u}_n))_i)^2+(x^{u}_n)^2 ((\Vec{\zeta}^{U}_{-}(x^{u}_n))_i)^2 +2 x^{u}_n (\Vec{\bm{\omega}}_Q)_i (\Vec{\zeta}^{U}_{+}(x^{u}_n))_i (\Vec{\zeta}^{U}_{-}(x^{u}_n))_i \bigg)^{-\frac{1}{2}}, \nonumber\\
    & z^{u}_n \equiv \frac{m^u_n}{k}e^{\sigma}, \; \; \; x^{u}_n \equiv \frac{m^u_n}{k \epsilon}, \nonumber
\end{align}
with the expressions for $\Vec{N}^{D,u,d}_n$, $z^d_n$, and $x^d_n$ given by corresponding definitions. Note that the definition of $\zeta^{Q,u,d}_{\pm}(z^{u,d}_n)$ depends on the index $n$ of the Kaluza-Klein mode; there is a unique $\zeta_+$ and $\zeta_-$ for each mass eigenstate of the quark KK tower. Here, we shall always have the index of the argument of these functions $z^{(u,d)}_n$, be the index $n$ also employed in all other cases in Eq.(\ref{QuarkZetaDefinition}). So, $\zeta^{Q}_{+}(z^{u,d}_n)$ and $\zeta^{Q}_{+}(z^{u,d}_m)$ both uniquely specify different functions given by Eq.(\ref{QuarkZetaDefinition}) as long as $m \neq n$. For convenience, it is now useful to define the matrices
\begin{align}
    \bm{\zeta}^{(U,u,D,d)}_{\pm}(z^{u,d}_n) \equiv \textrm{diag} \bigg( \frac{1}{\Vec{N}^{U,u,D,d}_n} \zeta^{(U,u,D,d)}_{\pm}(z^{u,d}_n) \bigg).
\end{align}
The mass eigenvalues $m^{u,d}_n=x^{u,d}_n M_{KK}$ for the up-like and down-like quark fields, as well as their eigenvalues, are then given by solutions to the boundary value equations \cite{casagrande}
\begin{align}
    &-\bm{\zeta}^{(U,D)}_{-}(x^{u,d}_n) \Vec{a}^{(U,D)}_n = \frac{v}{\sqrt{2} k} \mathbf{Y}^{u,d} \bm{\zeta}^{(u,d)}_{+}(x^{(u,d)}_n) \Vec{a}^{(u,d)}_n, \\
    &-\bm{\zeta}^{(u,d)}_{-}(x^{u,d}_n) \Vec{a}^{(u,d)}_n = \frac{v}{\sqrt{2} k} (\mathbf{Y}^{u,d})^\dagger \bm{\zeta}^{(U,D)}_{+}(x^{(u,d)}_n) \Vec{a}^{(U,D)}_n.
\end{align}
This can be reimagined as a block eigenvector equation,
\begin{align}\label{FermionEigenvectorsExact}
    &\begin{pmatrix}
    \Vec{a}^{(U,D)}_n \\
    \Vec{a}^{(u,d)}_n
    \end{pmatrix}
    = -\frac{v}{\sqrt{2} k} \bm{\mathcal{M}}_{u,d} 
    \begin{pmatrix}
    \Vec{a}^{(U,D)}_n \\
    \Vec{a}^{(u,d)}_n
    \end{pmatrix},\\
    &\bm{\mathcal{M}}_{u,d} \equiv
    \begin{pmatrix}
     0 & (\bm{\zeta}^{(U,D)}_{-}(x^{u,d}_n))^{-1}\mathbf{Y}^{u,d} \bm{\zeta}^{(u,d)}_{+}(x^{u,d}_n) \\
    (\bm{\zeta}^{(u,d)}_{-}(x^{u,d}_n))^{-1}(\mathbf{Y}^{u,d})^\dagger \bm{\zeta}^{(U,D)}_{+}(x^{u,d}_n) & 0
    \end{pmatrix}. \nonumber
\end{align}
The allowed values of $x^{u,d}_n$ are then simply the solutions to the equation
\begin{align}
    \textbf{Det} \bigg( \mathbb{I}_{6\times 6}+\frac{v}{\sqrt{2} k} \bm{\mathcal{M}}_{u,d} \bigg)=0,
\end{align}
where $\mathbb{I}_{6\times6}$ is the $6\times6$ identity matrix, while the eigenvectors $\Vec{a}^{(U,u,D,d)}_n$ can be found as components of the corresponding eigenvector, $(\Vec{a}^{(U,D)}_n,\Vec{a}^{(u,d)}_n)$, which we recall from Eq.(\ref{QuarkSolutions}) has a magnitude equal to unity.

It is helpful at this point to discuss some subtleties regarding the notation above. First, rather than treating each fermion generation individually, the treatment we have employed instead performs KK decompositions on the three-dimensional (in flavor space) objects $\mathbf{Q}_{L,R}$, $\mathbf{u}_{L,R}$, and $\mathbf{d}_{L,R}$. As a result, each of the Kaluza-Klein towers given in Eq.(\ref{FermionKKExpansions}) can be thought of as \textit{three} towers rolled into one. In the absence of brane-localized mass terms that mix the generations, each of the KK decompositions in Eq.(\ref{FermionKKExpansions}) could be cleanly separated into different towers for the first, second, and third generations. In the presence of these terms, as in our current case, each KK mode is a mixture of all three fermion generations, and such a separation is impossible. Furthermore, we note that without these brane-localized mass terms, each of the KK towers in Eq.(\ref{FermionKKExpansions}) would contain three massless chiral zero modes, corresponding to the zero modes of the three different generations' KK towers. With the introduction of the brane mass terms arising from the TeV-brane localized Higgs field, however, these modes not only mix with one another, they also acquire three different masses, each well below the scale $M_{KK}$. We identify these light modes with the SM quarks: The three lightest KK modes of the up(down)-like quarks, with masses we denote as $m^{u(d)}_{1}$, $m^{u(d)}_{2}$, and $m^{u(d)}_{3}$, will be identified with the SM $u(d)$, $c(s)$, and $t(b)$ quarks respectively.

Using our analytical exact expressions for the Kaluza-Klein decompositions of the quark fields, it is now useful to also give approximate expressions for the bulk profiles of these light SM-like modes. Unlike the other KK modes of the quark towers, these fields have masses much lower than $M_{KK}$, and as such, it is reasonable to approximate them in the limit as their masses approach zero. These approximations, valid to leading order in $x^{u,d}_{1,2,3} \equiv m^{u,d}_{1,2,3}/M_{KK}$, are given by \cite{casagrande}
\begin{align}\label{QuarkApproxProfiles}
    &\mathbf{C}^{(Q,u,d)}_n(\phi) \approx \sqrt{k r_c \epsilon}  \; \textrm{diag} \bigg( F(\Vec{\eta}_{Q,u,d}) e^{\frac{1}{2}(\Vec{\eta}_{Q,u,d}-1)(\sigma-k r_c \pi)} \bigg), \nonumber \\
    &\mathbf{S}^{(U,D)}_n(\phi) \approx \textrm{sgn}(\phi) \sqrt{k r_c \epsilon} x^{u,d}_n \; \textrm{diag} \bigg( \frac{\epsilon^{\frac{1}{2}(\Vec{\eta}_{Q}+1)}F(\Vec{\eta}_{Q})}{\Vec{\eta}_{Q}} (e^{\frac{1}{2}(\Vec{\eta}_{Q}+1)\sigma}-e^{-\frac{1}{2}(\Vec{\eta}_Q-1)\sigma}) \bigg),\\
    &\mathbf{S}^{(u,d)}_n(\phi) \approx -\textrm{sgn}(\phi) \sqrt{k r_c \epsilon} x^{u,d}_n \; \textrm{diag} \bigg( \frac{\epsilon^{\frac{1}{2}(\Vec{\eta}_{u,d}+1)}F(\Vec{\eta}_{u,d})}{\Vec{\eta}_{u,d}} (e^{\frac{1}{2}(\Vec{\eta}_{u,d}+1)\sigma}-e^{-\frac{1}{2}(\Vec{\eta}_{u,d}-1)\sigma}) \bigg), \nonumber \\
    &\Vec{\eta}_{Q,u,d} \equiv 1+2 \Vec{\bm{\omega}}_{Q,u,d}, \; \; \; F(\eta) \equiv \textrm{sgn}[ \cos \big( \frac{\pi}{2} (\eta-1) \big)]\sqrt{\frac{\eta}{1-\epsilon^{\eta}}} \nonumber
\end{align}
It should be noted that strictly speaking, the above approximate expressions are not entirely well-controlled; in particular for the (1,1) and (2,2) components of $\mathbf{C}^{(u)}_3(\phi)$, the bulk profile corresponding to the left-handed $t$ quark, there are terms proportional to $(x^u_3)^2$ which are rendered quite large. In practice, however, the effect of these corrections is minimal on final numerical results for quark masses and couplings, so we ignore them here. For a more detailed discussion of these correction terms, and why their effects on the flavor physics of the theory are suppressed, we refer the reader once again to \cite{casagrande}.

\section{Imposing a Local Flavor Gauge Symmetry: $SU(2)_F \times U(1)_F$}\label{BulkModelSection}

Having now discussed how the various elements of our model are realized separately, it is now necessary to synthesize them and discuss the explicit form of our model of flavor. Our model extends the SM gauge group in the bulk by adding the additional flavor symmetry, $SU(2)_F \times U(1)_F$. Our selection of this specific flavor symmetry is motivated by several factors. First, due to work exploring two-Higgs doublet models, the vacua of systems with two complex scalar doublets are well-understood \cite{diazcruz} (at least given our assumption, already mentioned in Section \ref{BulkSSBSection}, that the vacuum expectation values of our bulk fields have flat bulk profiles), and we find that it is possible for our two-scalar potential to reach an absolute minimum in an arrangement that completely breaks the gauge symmetry. Second, with such a gauge symmetry we can find vacuum expectation values for the two bulk scalars such that, even when the gauge symmetry is completely broken, the gauge boson couplings to fermions continue to respect a new conserved flavor charge as if there were a remaining unbroken $U(1)$ group; this charge conservation is only violated by the scalar interactions \cite{diazcruz}. We shall use this latter quirk in order to protect the highly sensitive flavor observable $\epsilon_K$ in Section \ref{KKBarMixingSection}. We also should mention that the dual description of this particular model as a strongly coupled CFT may be of some theoretical interest, although a detailed investigation of this correspondence, in particular regarding the  is non-trivial and beyond the scope of this paper. Instead, we content ourselves with mentioning in passing that, in analogy with the SU(5) RS GUT of \cite{Pomarol:2000hp} (the holographic dual description of which is given in \cite{ArkaniHamed:2000ds}), which has its GUT gauge group broken in the bulk by a scalar vev, the dual 4D description of our model likely involves a strongly-coupled CFT with a global $SU(2)_F \times U(1)_F$ symmetry in addition to the weakly gauged SM gauged symmetry. Additional exploration of the CFT interpretation of this model, including the gauge fields' flavor charge conservation, the role of the bulk scalars in imparting bulk fermion masses, and the flavor mixing that the model predicts are left for future work.

In summary, we posit a full theory with a gauge group of $SU(3)_C \times SU(2)_L \times SU(2)_F \times U(1)_Y \times U(1)_F$, which is then broken in the bulk via two scalars in the fundamental representation of $SU(2)_F$, down to the SM gauge group $SU(3)_C \times SU(2)_L \times U(1)_Y$. In the following sections, we now embark on a detailed discussion of the group structure of the model's matter content and the specific structure of the bulk symmetry breaking.

\subsection{Bulk Scalars in $SU(2)_F \times U(1)_F$}\label{BulkScalarModelSection}
The bulk scalar sector of our model plays two important roles: First, the scalar vacuum expectation values we posit must fully break the flavor gauge group $SU(2)_F \times U(1)_F$ in the bulk to avoid the emergence of new light flavor-changing gauge bosons, and second, these vacuum expectation values must provide adequate flavor symmetry violating quark bulk mass terms in order to recreate the observed SM quark flavor structure (that is, the quark masses and the CKM matrix). As the ability of any collection of fields to satisfy the second of these criteria is strongly dependent on the structure of the quark sector itself, we begin our model building by focusing on the first, namely, that our flavor gauge symmetry must be completely broken. We propose a scalar sector consisting of two identical fields, $\Phi_1$ and $\Phi_2$, which are given in the fundamental representation of $SU(2)_F$ and possess (the same) charge of $\frac{1}{2}Y_H$ under $U(1)_F$ (we shall discuss the value of $Y_H$ in Sec. \ref{BulkMatterModelSection}, since it is ultimately determined by its interactions with the bulk matter fields). This is in some senses a ``minimal'' scalar sector, since in order to fully break the gauge symmetry, one requires no fewer than two complex doublets with vacuum expectation values. The action of these bulk scalars, including their interaction terms with the $SU(2)_F$ gauge bosons $A^1$, $A^2$, and $A^3$ and the $U(1)_F$ gauge boson $B$, is then explicitly given by
\begin{align}\label{ModelScalarAction}
    S_{\Phi} = & \int d^4 x \int d \phi \, r_c \left\{\sum_{i=1,2} e^{-2 \sigma} (D_\mu \Phi_i)^\dagger (D_\mu \Phi_i) -\frac{e^{-4\sigma}}{r_c^2} (D_\phi \Phi_i)^\dagger (D_\phi \Phi_i)-V(\Phi_1,\Phi_2) \right\}, \nonumber\\
    &D_M=(\partial_M-i \sqrt{2 \pi r_c} g_A A^a_M \tau^a-\frac{i}{2}\sqrt{2 \pi r_c} g_B Y_H B_M).
\end{align}
Here, $\tau^a=\frac{1}{2}\sigma^a$ are the standard generators of $SU(2)$, $g_A$ and $g_B$ are the $SU(2)_F$ and $U(1)_F$ coupling constants, respectively, and $Y_H/2$ is the $U(1)_F$ charge of the scalars $\Phi_1$ and $\Phi_2$ (notably, they possess identical charges). The $\sqrt{2 \pi r_c}$ term inserted into terms with $g_A$ and $g_B$ are added in order to keep the coupling constants themselves dimensionless. Furthermore, they are selected to make $g_A$ and $g_B$ directly comparable to $SM$ gauge coupling constants; if we set $g_A$ equal to the electroweak $SU(2)_L$ gauge coupling constant, for example, the two forces will have identical bulk gauge coupling strengths. The potential term $V(\Phi_1,\Phi_2)$ is given by
\begin{align}\label{BulkV}
    V(\Phi_1,\Phi_2) &= -\mu^2(\Phi^\dagger_1 \Phi_1+\Phi^\dagger_2 \Phi_2)+2 \pi r_c \lambda_1((\Phi^\dagger_1 \Phi_1)^2+(\Phi^\dagger_2 \Phi_2)^2)+2 \pi r_c \lambda_3 (\Phi^\dagger_1 \Phi_1)(\Phi^\dagger_2 \Phi_2)\\
    &+2 \pi r_c \lambda_4 (\Phi^\dagger_1 \Phi_2)(\Phi^\dagger_2 \Phi_1) + \frac{1}{2}(2 \pi r_c) \lambda_5 ((\Phi^\dagger_1 \Phi_2)^2+(\Phi^\dagger_2 \Phi_1)^2). \nonumber
\end{align}
Here, all parameters are assumed to be real, and for simplicity, we have required that $V$ is symmetric under $\Phi_1 \rightarrow -\Phi_1$, $\Phi_2 \rightarrow -\Phi_2$, and $\Phi_1 \leftrightarrow \Phi_2$. In addition, the first two of these conditions are important to the structure of our model; were they relaxed, an additional term of the form, $\mu_{12}^2 (\Phi_1^\dagger \Phi_2) + h.c.$ would be permitted. This term vitiates \cite{diazcruz} the flavor charge near-conservation (discussed briefly in the introduction to Section \ref{BulkModelSection}) which we shall use to protect the observable $\epsilon_K$. The factors of $2 \pi r_c$ included in some terms are inserted for convenience to keep the coefficients $\lambda_i$ dimensionless in a five-dimensional spacetime, as was done for the gauge couplings. We also note that, because this potential lacks any explicit dependence on the 5-dimensional coordinate $\phi$, its minimum-energy configuration will be flat in the bulk. In order to completely break $SU(2)_F \times U(1)_F$, we require a vacuum configuration of 
\begin{align} \label{Bulkvevs}
    & \langle \Phi_1 \rangle = \sqrt{\frac{1}{4 \pi r_c}}\begin{pmatrix}
    0 \\
    v_F
    \end{pmatrix}, & \langle \Phi_2 \rangle = \sqrt{\frac{1}{4 \pi r_c}} \begin{pmatrix}
    v_F \\
    0
    \end{pmatrix},
\end{align}
where we have factored out $\sqrt{2 \pi r_c}$ from the vacuum expectation value to give the quantity $v_F$ a mass dimension of 1, and factored out a further $\sqrt{2}$ for notational convenience later on. Note that both scalar vevs have the same magnitude, $v_F$; this is an unsurprising consequence of our imposition of a $\Phi_1 \leftrightarrow \Phi_2$ symmetry on the potential. We parameterize $\Phi_{1,2}$ after spontaneous symmetry breaking as
\begin{align}\label{PhiParameterization}
    \Phi_1 = \langle \Phi_1 \rangle + \frac{1}{\sqrt{2}}\begin{pmatrix}
    h_1+i h_2 \\
    h_3+i h_4
    \end{pmatrix}, & & \Phi_2 = \langle \Phi_2 \rangle + \frac{1}{\sqrt{2}}\begin{pmatrix}
    h_5+i h_6 \\
    h_7+i h_8
    \end{pmatrix},
\end{align}
where all $h$ fields are real scalars. In agreement with \cite{diazcruz}, we find that $V(\Phi_1,\Phi_2)$ achieves an absolute minimum in such a configuration, with $v_F^2=2 \mu^2 (2 \lambda_1+\lambda_3)^{-1}$ provided that $\lambda_4 \pm \lambda_5 \geq 0$, $\lambda_1 \geq 0$, and $2 \lambda_1 \pm \lambda_3 \geq 0$. After diagonalizing the mass matrices, we rewrite the bulk scalar field components in terms of mass eigenstates. Four of the scalar fields are given bulk masses, namely
\begin{align}\label{HBulkMasses}
    &H_1 \equiv \frac{1}{\sqrt{2}}(h_1+h_7), &m_{H_1}^2 = (\lambda_4+\lambda_5) v_F^2, \\
    &H_2 \equiv \frac{1}{\sqrt{2}}(h_2+h_8), &m_{H_2}^2 = (\lambda_4-\lambda_5) v_F^2, \nonumber\\
    &H_3 \equiv \frac{1}{\sqrt{2}}(h_3-h_5), &m_{H_3}^2 = (2\lambda_1-\lambda_3) v_F^2, \nonumber\\
    &H_4 \equiv \frac{1}{\sqrt{2}}(h_3+h_5), &m_{H_4}^2 = (2\lambda_1+\lambda_3) v_F^2. \nonumber
\end{align}
We note that $H_1$, $H_2$, and $H_4$ are $CP$-even, while $H_3$ is $CP$-odd. Additionally, four Goldstone-like (that is, lacking a bulk mass) fields emerge, which we parameterize as
\begin{align}\label{HGoldstones}
    &H_5 \equiv \frac{1}{\sqrt{2}}(h_4+h_6), &H_6 \equiv \frac{1}{\sqrt{2}}(h_4-h_6),\\
    &H_7 \equiv \frac{1}{\sqrt{2}}(h_1-h_7), &H_8 \equiv \frac{1}{\sqrt{2}}(h_2+h_8). \nonumber
\end{align}
Among these fields, $H_5$ is $CP$-even, while $H_6$, $H_7$, and $H_8$ are $CP$-odd.

From Eq.(\ref{ModelScalarAction}), it is also straightforward to derive the bulk mass terms of the gauge bosons. In terms of the $\gamma$ variable we used to express bulk mass in \ref{BulkSSBSection}, originally defined in Eq.(\ref{GaugeFlavorAction}), the bulk mass parameters for the flavor gauge bosons are
\begin{align}
    & \gamma_{B}=\frac{g_B Y_H v_F}{\sqrt{2} k}, & \gamma_{A^a}=\frac{g_A v_F}{\sqrt{2} k}, \; \; a=1,2,3.
\end{align}
Notably, unlike the SM group $SU(2)_L\times U(1)_Y$ broken by the Higgs field, there is no mixing between any of the $SU(2)_F$ gauge bosons and the $U(1)_F$ boson; such mixing terms in the gauge bosons' mass matrix cancel due to the specific configuration of our bulk vevs. With these bulk masses, it is then straightforward to derive the full spectrum of physical particles emerging in the unitary gauge for this system, in direct analogy with Section \ref{BulkSSBSection}. In particular, we find that in addition to four KK towers of scalar states emerging from $H_{1-4}$, and four towers of vector bosons emerging from $A^{1,2,3}_\mu$, and $B_\mu$, four KK towers of scalar particles arise from mixtures of $H_8$ and $A^1_\phi$, $H_7$ and $A^2_\phi$, $H_6$ and $A^3_\phi$, and $H_5$ and $B_\phi$,  which we shall refer to as $a_1$, $a_2$, $a_3$, and $a_4$ respectively. 

\subsection{Bulk Matter in $SU(2)_F \times U(1)_F$}\label{BulkMatterModelSection}
We begin our discussion with the matter fields of the theory. We begin by arranging the quark fields into $SU(2)_F \times U(1)_F$ multiplets, such that three generations of a given quark field are given by one $SU(2)_F$ doublet and one $SU(2)_F$ singlet. Explicitly, we have the bulk fermion fields
\begin{align}\label{QuarkMultiplets}
    &U=\begin{pmatrix}
    (U_1)_{+\frac{1}{2}} \\
    (U_2)_{-\frac{1}{2}}
    \end{pmatrix}_{-\frac{1}{2}}, &T= (T_0)_{+1}, \\
    &u=\begin{pmatrix}
    (u_1)_{+\frac{1}{2}} \\
    (u_2)_{-\frac{1}{2}}
    \end{pmatrix}_{-\frac{1}{2}}, &t= (t_0)_{+1}, \\
    &d=\begin{pmatrix}
    (d_1)_{+\frac{1}{2}} \\
    (d_2)_{-\frac{1}{2}}
    \end{pmatrix}_{-\frac{1}{2}}, &b= (b_0)_{+1}.
\end{align}
Here, $U$ and $T$ are $SU(2)_L$ doublets corresponding to the left-handed quark doublets in the SM, while $u(d)$ and $t(b)$ are $SU(2)_L$ singlets, corresponding to the SM up(down)-like right-handed quark singlets. The subscripts $\textit{within}$ the multiplet matrices above refer to the $T^3_F$ quantum numbers of the various elements of the multiplet, while the subscripts $\textit{outside}$ of the matrices refer to the $U(1)_F$ charges of the fields. Motivated by the fact the third generation quarks, particularly the top quark, are substantially more massive than the quarks in first two generations, our notation anticipates that up to mixing due to couplings to the bulk scalar fields and the brane-localized Higgs field, the $SU(2)_F$ singlet quarks shall roughly correspond to the third generation of quarks (the $t$ and $b$ quarks), allowing these quarks to have bulk masses that differ substantially from those of the $SU(2)_F$ doublet quarks even without the contributions of the bulk scalars. The various fields' representations and charge assignments under the remaining SM gauge group, $SU(3)_C \times SU(2)_L \times U(1)_Y$ are assumed to be identical to their realization in the SM. Notably, there are only two unique $U(1)_F$ charge assignments within this model, $Q^F_{U}=Q^F_{u}=Q^{F}_{d}$, the $U(1)_F$ charge shared by all $SU(2)_F$ doublet quarks, and $Q^F_{T}=Q^F_{t}=Q^F_{b}$, the charge shared by all $SU(2)_F$ singlet quarks. As we shall see shortly, this is by design, in order to permit the bulk and brane mass terms we require in our theory. Additionally, our specific $U(1)_F$ charge assignments, $Q^F_{U}=-1/2$, $Q^F_{T}=+1$, are selected specifically to avoid chiral anomalies in the effective 4-dimensional theory stemming from the light SM-like quarks; in a 5D warped theory on an $S_1/Z_2$ orbifold, as we have here, this condition is sufficient to avoid anomalies in the full 5-dimensional theory \cite{Hirayama:2003kk}. In general, in order to avoid anomalies while simultaneously satisfying our requirement that $Q^F_{U}=Q^F_{u}=Q^F_{d}$ and $Q^F_{T}=Q^F_{t}=Q^F_{b}$, we find that $Q^F_{T,t,b}= -2 Q^F_{U,u,d}$. In Eq.(\ref{QuarkMultiplets}), we have fixed $Q^F_{T,t,b}=+1$, and shall continue to do so for the remainder of this paper.

The bulk mass terms of the quark fields then arise from fundamental terms which respect the gauge symmetry $SU(2)_F \times U(1)_F$ and symmetry-breaking terms which emerge from Yukawa-like interactions with the scalars $\Phi_{1,2}$ introduced in Section \ref{BulkScalarModelSection}. Supplemented by brane mass terms arising from a TeV-brane localized Higgs field, we arrive at the action
\begin{align}\label{QuarkAction}
    S_{q} = &\int d^4 x \int d \phi \, r_c e^{-4 \sigma} \bigg\{ \sum_{Q=U,T,u,t,d,b} \bigg( e^{\sigma} \frac{i}{2} \bar{Q} \slashed{D} Q +\frac{1}{r_c} \frac{i}{2}\bar{Q} (i \gamma_5) D_\phi Q \bigg) + h.c. \nonumber\\
    - &\sum_{Q=U,T} \textrm{sgn}(\phi) \, \nu_Q k \bar{Q} Q \; \; + \sum_{q=u,t,d,b} \textrm{sgn}(\phi) \, \nu_q k \bar{q} q\\
    - &\textrm{sgn}(\phi) \sqrt{2 \pi r_c}\bigg( y^Q \Phi_2 \bar{U} T + y^d \Phi_2 \bar{d} b + y^u_1 \Phi_1 \bar{u} t+y^u_2 \Phi_2 \bar{u} t \bigg) + h.c. \nonumber\\
    - &\frac{2}{kr_c} \delta(|\phi|-\pi) e^{\sigma}\frac{v}{\sqrt{2}}[ Y_u \bar{U}_L u_R + Y_t \bar{T_L} t_R + Y_d \bar{U}_L d_R + Y_b \bar{T}_L b_R + h.c.] \bigg\}  \nonumber.
\end{align}
Here, $\nu_{U,T,u,t,d,b}$ are all assumed to be real numbers, and for the sake of simplicity we likewise require $y^{Q,d}$ and $y^{u}_{1,2}$ to be real. This latter requirement does not actually result in a loss of generality, since any nontrivial phases in the $y$ parameters can be absorbed into redefinitions of the complex brane-localized Yukawa couplings, $Y_{u,t,d,b}$. We also have kept $y^{Q,d}$ and $y^u_{1,2}$ dimensionless, by factoring out $\sqrt{2 \pi r_c}$, as we have done with bulk gauge couplings. Additionally, note that there are $\textit{no}$ terms of the form, $\epsilon^{ab} (\Phi_i^{\dagger})_b \bar{Q}_a q$, where $a$ and $b$ are $SU(2)_F$ indices, $Q$ is an $SU(2)_F$ doublet quark, and $q$ is an $SU(2)_F$ singlet quark, since although these terms are singlets under $SU(2)_F$, this term must $\textit{not}$ be invariant under $U(1)_F$ if the Yukawa terms given in Eq.(\ref{QuarkAction}) respect $U(1)_F$ symmetry, since the scalars $\Phi_{1,2}$ must have non-zero $U(1)_F$ charges to cancel out the non-zero $U(1)_F$ charge of the product $\bar{Q} q$.

In the action of Eq.(\ref{QuarkAction}), we have made several non-trivial assumptions about the form of the bulk Yukawa couplings. Most notably, we have assumed that $\Phi_1$ is only coupled to the quark sector through the fields $u$ and $t$, and NOT to the fields $U$, $T$, $d$, and $b$. This is an $\textit{ad hoc}$ assumption motivated by phenomenological considerations; we shall find that in the absence of any couplings of $\Phi_1$ to the $SU(2)_L$ doublet and down-like singlet fields, we can numerically reproduce the observed spectrum of quark masses and mixings, and that dangerous contributions to flavor observables, in particular $\epsilon_K$, can be highly suppressed. In a more general realization of this model, we anticipate that harsh limits on couplings of $\Phi_1$ to $U$, $T$, $d$, and $b$ could be derived from the emergence of additional flavor-changing interactions at tree level, however it may be possible to formulate a slightly modified model that explicitly forbids the $\Phi_1$ couplings to $U$, $T$, $d$, and $b$: For example, one might impose an additional discrete symmetry on the action such that both $\Phi_1$ and $\Phi_2$ can only couple to $u$ and $t$, and then introduce a third scalar $\Phi_3$ such that its vev is the same (up to a real proportionality constant) as that of $\Phi_2$, but its representation under the discrete symmetry only permits couplings $U$, $T$, $d$, and $b$ (notably, such a vev configuration, if possible, would also still preserve the flavor charge near-conservation discussed in Section \ref{BulkScalarModelSection} that protects certain flavor observables). The $\Phi_3$ bulk Yukawa couplings would replace the now-forbidden couplings between $\Phi_2$ and $U$, $T$, $d$, and $b$, and the additional discrete symmetry would forbid the troublesome $\Phi_1$ couplings that we have set to zero in our original framework. Regardless, either elaborating the model in order to preclude the omitted bulk Yukawa couplings or performing a full exploration of the experimental constraints on these couplings is beyond the scope of the present work. So, we shall henceforth simply assume that these terms are identically zero, noting that there very likely exist constructions which can effect such a result, and therefore that these assumptions do not seriously impact the potential naturalness of our model.

It should also be noted that the $\textrm{sgn}(\phi)$ term included in front of the bulk Yukawa coupling terms is also somewhat $\textit{ad hoc}$, since the bulk scalars themselves are orbifold even fields in order to allow a flat bulk vev. The simplest explanation for this additional sign term would be that it arises due to the vev of an additional orbifold odd bulk scalar, which produces the bulk Yukawa coupling terms as effective vertices in some limit. However, the sign function could also be eliminated if the bulk scalar fields $\Phi_1$ and $\Phi_2$ were themselves orbifold odd. A full exploration of this latter option is beyond the scope of this work, since it introduces substantial complications to the analysis (for example, the bulk vevs of $\Phi_1$ and $\Phi_2$ would no longer be constant in the bulk), but we can briefly speculate on the effects of this choice on flavor observables. In particular, we expect qualitatively little change from the results we derive with even  bulk scalars. For example, we find that an orbifold odd scalar can still produce bulk masses (albeit no longer constant in the extra dimension coordinate $\phi$) for the bulk gauge fields, and that the field definitions given in Eq.(\ref{GoldstoneNewFields}) can still be used to diagonalize the bulk mass terms Eq.(\ref{GaugeFlavorGoldstoneAction}). As such, we anticipate that requiring the bulk scalars $\Phi_1$ and $\Phi_2$ to be orbifold odd, and possess non-flat bulk vevs, would substantially complicate the analysis of our model, but is not likely to effect any greater change than altering the spectrum  and bulk profiles of KK flavor gauge and scalar bosons; spontaneous symmetry breaking and the emergence of the new towers of scalar modes arising as mixtures of components of $\Phi_{1,2}$ and the fifth component of the bulk flavor gauge bosons will still take place. It should also be noted that in \cite{ahmed}, much of the parameter space of a single orbifold odd bulk scalar potential resulted in bulk vevs closely approximated by expressions of the form, $v_F \; \textrm{sgn}(\phi)$, and if similar results hold for a multi-scalar potential such as we have presented here (namely, that our resulting bulk vevs are identical to those given in Eq.(\ref{Bulkvevs}), only multiplied by $\textrm{sgn}(\phi)$), we would expect that fermion and flavor gauge boson bulk profiles and spectra would be given to excellent approximation by the results we derive from our flat vev, and among flavor-changing couplings only those involving the bulk scalars, which we shall see are suppressed relative to the other new physics couplings in the model, would undergo any significant alteration. Therefore, to avoid introducing additional complications into our model, we satisfy ourselves with orbifold even bulk scalars and introduce the additional $\textrm{sgn}(\phi)$ terms in our bulk Yukawa couplings by hand.

From this action, we also see that the requirement that each $SU(2)_F$ doublet share the same $U(1)_F$ charge, and each $SU(2)_F$ singlet do likewise, is essential to avoid unnecessarily complicating the SM-like Higgs sector. Were this condition not satisfied, then the SM Higgs would have to be charged under $U(1)_F$ in order to produce the brane-localized mass terms in Eq.(\ref{QuarkAction}). This would in turn dramatically complicate a treatment of the Higgs, including possibly resulting in additional brane mass terms emerging for the $U(1)_F$ flavor gauge boson. Furthermore, depending on these charge assignments, some of the brane-localized mass terms of Eq.(\ref{QuarkAction}) might be forbidden unless additional Higgses, each with different $U(1)_F$ charges, were added to the theory. By instead making the charge assignments of $U$, $u$, and $d$, identical and doing the same for the charge assignments of $T$, $t$, and $b$, we sidestep all of these issues and allow the SM Higgs to remain in the trivial representation of $SU(2)_F \times U(1)_F$. We also note that, given our charge assignments in Eq.(\ref{QuarkMultiplets}), we can now deduce the $U(1)_F$ charge of the $\Phi_{1,2}$ fields from the form of the interaction terms with $\Phi_{1,2}$ in Eq.(\ref{QuarkAction})-- it must be $-3/2$, and hence the quantity $Y_H$ in Section \ref{BulkScalarModelSection} is $Y_H=-3$ here.

At this point, in order to bring our model in line with our discussion of RS fermions in Section \ref{BulkFermionSection}, we need to diagonalize the bulk masses of our various bulk quark fields. We begin the process of doing so by rewriting the bulk mass terms in Eq.(\ref{QuarkAction}) as mass matrices, yielding
\begin{align}\label{SU2U1MassMatrices}
    S_{q,mass} = &\int d^4 x \int d\phi r_c e^{-4 \sigma} \bigg\{ k \, \textrm{sgn}(\phi)(\bar{u}_1,\bar{u}_2,\bar{t}) \begin{pmatrix}
    \nu_u & 0 & -\frac{y^u_2}{\sqrt{2}}\frac{v_F}{k} \\
    0 & \nu_u & -\frac{y^u_1}{\sqrt{2}}\frac{v_F}{k} \\
    -\frac{y^u_2}{\sqrt{2}} \frac{v_F}{k} & -\frac{y^u_1}{\sqrt{2}}\frac{v_F}{k} & \nu_t
    \end{pmatrix} \begin{pmatrix}
    u_1 \\
    u_2 \\
    t
    \end{pmatrix} \nonumber \\
    &+ k \, \textrm{sgn}(\phi)(\bar{d}_1,\bar{d}_2,\bar{b}) \begin{pmatrix}
    \nu_d & 0 & -\frac{y^d}{\sqrt{2}} \frac{v_F}{k} \\
    0 & \nu_d & 0 \\
    -\frac{y^d}{\sqrt{2}} \frac{v_F}{k} & 0 & \nu_b
    \end{pmatrix} \begin{pmatrix}
    d_1 \\
    d_2 \\
    b
    \end{pmatrix} \\
    &- k \, \textrm{sgn}(\phi)(\bar{U}_1,\bar{U}_2,\bar{T}) \begin{pmatrix}
    \nu_U & 0 & \frac{y^Q}{\sqrt{2}} \frac{v_F}{k} \\
    0 & \nu_U & 0 \\
    \frac{y^Q}{\sqrt{2}} \frac{v_F}{k} & 0 & \nu_T
    \end{pmatrix} \begin{pmatrix}
    U_1 \\
    U_2 \\
    T
    \end{pmatrix} \bigg\}. \nonumber
\end{align}

Because of the later significance that the various zero entries in the mass matrices of Eq.(\ref{SU2U1MassMatrices}) play for flavor physics, it is useful here to carefully explain their origins in the action of Eq.(\ref{QuarkAction}). To start, we note that the scalars $\Phi_1$ and $\Phi_2$ are only capable of coupling $SU(2)_F$ doublet quarks ($u_i$, $d_i$, and $U_i$) to $SU(2)_F$ singlet quarks, since the product of three fundamental representations of $SU(2)$ does not contain a singlet, and any such couplings can only be between quarks with the same representations in the SM gauge group, since $\Phi_1$ and $\Phi_2$ are SM singlets. So, the (1,2) and (2,1) entries of all quark mass matrices in Eq.(\ref{SU2U1MassMatrices}) are necessarily zero, and there are no entries in any of the mass matrices that mix quarks of different SM representations. Furthermore, we note that due to the form of the bulk scalar vevs given in Eq.(\ref{Bulkvevs}), bulk Yukawa terms in Eq.(\ref{QuarkAction}) of the form, $\Phi_2 \bar{Q} q$, where $Q$ is an $SU(2)_F$ doublet quark and $q$ is an $SU(2)_F$ singlet quark, will only result in non-zero entries in the (1,3) and (3,1) entries of the mass matrices of Eq.(\ref{SU2U1MassMatrices}), while terms of the form $\Phi_1 \bar{Q} q$ will only result in non-zero (2,3) and (3,2) entries. Recalling that, as noted previously, $SU(2)_F$-invariant bulk Yukawa terms featuring the antisymmetric symbol $\epsilon^{ab}$ are forbidden by the $U(1)_F$ gauge symmetry, the fact that the (2,3) and (3,2) entries of the electroweak doublet and down-like electroweak singlet quark mass matrices are zero then stems directly from the lack of any couplings of these quarks to the bulk scalar $\Phi_1$.

Having written down our quark mass matrices directly from our action, we now express these mass matrices as rotated diagonal matrices, writing
\begin{align}\label{DiagonalBulkMasses}
    &\begin{pmatrix}
    \nu_u & 0 & \frac{-y^u_2}{\sqrt{2}}\frac{v_F}{k} \\
    0 & \nu_u & \frac{-y^u_1}{\sqrt{2}}\frac{v_F}{k} \\
    \frac{-y^u_2}{\sqrt{2}} \frac{v_F}{k} & \frac{-y^u_1}{\sqrt{2}}\frac{v_F}{k} & \nu_t
    \end{pmatrix} &=& &\mathbf{W}^T_u \begin{pmatrix}
    \frac{1}{2}(\eta^u_1-1) & 0 & 0 \\
    0 & \frac{1}{2}(c_u^2 \eta^u_1 + s_u^2 \eta^u_3-1) & 0 \\
    0 & 0 & \frac{1}{2}(\eta^u_3-1)
    \end{pmatrix} \mathbf{W}_u, \nonumber \\
    &\;\;\;\,\begin{pmatrix}
    \nu_d & 0 & \frac{-y^d}{\sqrt{2}}\frac{v_F}{k} \\
    0 & \nu_d & 0 \\
    \frac{-y^d}{\sqrt{2}} \frac{v_F}{k} & 0 & \nu_b
    \end{pmatrix} &=& &\mathbf{W}^T_d \begin{pmatrix}
    \frac{1}{2}(\eta^d_1-1) & 0 & 0 \\
    0 & \frac{1}{2}(c_d^2 \eta^d_1 + s_d^2 \eta^d_3-1) & 0 \\
    0 & 0 & \frac{1}{2}(\eta^d_3-1)
    \end{pmatrix} \mathbf{W}_d, \\
    &\;\;\;\;\,\begin{pmatrix}
    \nu_U & 0 & \frac{y^Q}{\sqrt{2}}\frac{v_F}{k} \\
    0 & \nu_U & 0 \\
    \frac{y^Q}{\sqrt{2}} \frac{v_F}{k} & 0 & \nu_T
    \end{pmatrix} &=& &\mathbf{W}^T_Q \begin{pmatrix}
    \frac{1}{2}(\eta^Q_1-1) & 0 & 0 \\
    0 & \frac{1}{2}(c_Q^2 \eta^Q_1 + s_Q^2 \eta^Q_3-1) & 0 \\
    0 & 0 & \frac{1}{2}(\eta^Q_3-1)
    \end{pmatrix} \mathbf{W}_Q. \nonumber
\end{align}
We define the rotation matrices as
\begin{align}\label{RotMatrices}
    \mathbf{W}_u \equiv \begin{pmatrix}
    c_H c_u & c_u s_H & -s_u \\
    -s_H & c_H & 0 \\
    c_H s_u & s_H s_u & c_u
    \end{pmatrix}, \;\; \mathbf{W}_d \equiv \begin{pmatrix}
    c_d & 0 & -s_d \\
    0 & 1 & 0 \\
    s_d & 0 & c_d
    \end{pmatrix}, \;\; \mathbf{W}_Q \equiv \begin{pmatrix}
    c_Q & 0 & -s_Q \\
    0 & 1 & 0 \\
    s_Q & 0 & c_Q
    \end{pmatrix}.
\end{align}
In Eqs.(\ref{DiagonalBulkMasses}) and (\ref{RotMatrices}), we have introduced 6 real localization parameters, $\eta^{Q,u,d}_{1,3}$ (our variable selection here is informed by the usefulness of the quantity $\eta^{Q,u,d}_i$ in the approximate expressions for the bulk profiles of the SM-like fermions, given in Eq.(\ref{QuarkApproxProfiles})), and 4 angles, $\theta_{u,d,Q,H}$, where we have defined $s_i \equiv \sin(\theta_i)$ and $c_i \equiv \cos(\theta_i)$. It shall also be useful to express our original $y$ and $\nu$ parameters in terms of these new parameters; we derive the relations
\begin{align}\label{BulkLocalizationExpressions}
    &\nu_u = \frac{1}{2}(c_u^2 \eta^u_1 + s_u^2 \eta^u_3-1), \;\;\;\;\;\, \nu_t = \frac{1}{2}(s_u^2 \eta^u_1 + c_u^2 \eta^u_3-1),& \nonumber\\
    &\nu_d = \frac{1}{2}(c_d^2 \eta^d_1 + s_d^2 \eta^d_3-1), \;\;\;\;\;\,\,\nu_b = \frac{1}{2}(s_d^2 \eta^d_1 + c_d^2 \eta^d_3-1),& \\
    &\nu_U = \frac{1}{2}(c_Q^2 \eta^Q_1 + s_Q^2 \eta^Q_3-1), \;\;\nu_T = \frac{1}{2}(s_Q^2 \eta^Q_1 + c_Q^2 \eta^Q_3-1),& \nonumber
\end{align}
for the $\nu$ parameters, and
\begin{align}\label{BulkYukawaExpressions}
    &\frac{y^d}{\sqrt{2}}\frac{v_F}{k}= -\frac{1}{2} (\eta^d_3-\eta^d_1)s_d c_d, &\frac{y^Q}{\sqrt{2}}\frac{v_F}{k} = \frac{1}{2}(\eta^Q_3-\eta^Q_1)s_Q c_Q,\\
    &\frac{y^u_1}{\sqrt{2}} \frac{v_F}{k} = -\frac{1}{2}(\eta^u_3-\eta^u_1) c_H c_u s_u, &\frac{y^u_2}{\sqrt{2}}\frac{v_F}{k}= -\frac{1}{2}(\eta^u_3-\eta^u_1)s_H c_u s_u. \nonumber
\end{align}
for the bulk Yukawa-like couplings $y^{Q,d}$ and $y^{u}_{1,2}$. The bulk quark fields can then be put in the form of Section \ref{BulkFermionSection}, with diagonal bulk mass matrices and non-diagonal brane mass terms, by defining new fields rotated by the $\mathbf{W}_{Q,u,d}$ matrices,
\begin{align}
    \begin{pmatrix}
    u'_1 \\
    u'_2 \\
    t'
    \end{pmatrix} = \mathbf{W}_u \begin{pmatrix}
    u_1 \\
    u_2 \\
    t
    \end{pmatrix}, \;\; \begin{pmatrix}
    d'_1 \\
    d'_2 \\
    b'
    \end{pmatrix} = \mathbf{W}_d \begin{pmatrix}
    d_1 \\
    d_2 \\
    b
    \end{pmatrix}, \;\; \begin{pmatrix}
    U'_1 \\
    U'_2 \\
    T'
    \end{pmatrix} = \mathbf{W}_Q \begin{pmatrix}
    U_1 \\
    U_2 \\
    T
    \end{pmatrix}.
\end{align}
In terms of these new fields, we arrive at the brane-localized Yukawa matrices in terms of the matrices $\mathbf{Y}^u$ and $\mathbf{Y}^d$ given in Section \ref{BulkFermionSection}, which have the form
\begin{align}\label{YukawaMatrices}
    &\mathbf{Y}^{u}=\begin{pmatrix}
    Y_u c_H c_Q c_u + Y_t s_Q s_u & -Y_u c_Q s_H & Y_u c_H c_Q s_u - Y_t c_u s_Q \\
    Y_u c_u s_H & Y_u c_H & Y_u s_H s_u \\
    Y_u c_H c_u s_Q - Y_t c_Q s_u & -Y_u s_H s_Q & Y_t c_Q c_u + Y_u c_H s_Q s_u
    \end{pmatrix}, \\
    &\mathbf{Y}^{d}=\begin{pmatrix}
    Y_d c_d c_Q + Y_b s_d s_Q & 0 & Y_d c_Q s_d-Y_b c_d s_Q \\
    0 & Y_d & 0 \\
    Y_d c_d s_Q-Y_b c_Q s_d & 0 & Y_d s_d s_Q +Y_b c_d c_Q
    \end{pmatrix}. \nonumber
\end{align}
It should be noted that the rotation matrices $\mathbf{W}_{Q,u,d}$ do not only affect the brane-localized Yukawa terms; thanks to our field redefinition, these same rotation matrices now also appear in any term in the action that features a non-universal coupling to different matter generations. In particular, these same rotation matrices shall appear again when computing couplings of various quark fields to the flavor gauge bosons, scalars, and pseudoscalars discussed in Section \ref{BulkScalarModelSection}. It is also important to note that, because of the form of $\mathbf{W}_Q$ and $\mathbf{W}_d$, the down-like Yukawa matrix in Eq.(\ref{YukawaMatrices}) only contains mixing between the first and third generations. In particular, then, we see that the second generation down-like quarks therefore have \textit{no} mixing with other generations in the effective 4-D theory; for all second-generation quarks, the eigenvectors $\Vec{a}^{D,d}_n = (0,1/\sqrt{2},0)$, up to an overall complex phase. This is exceedingly consequential for flavor physics within our model, in particular, we shall see later that this design \textit{prevents any tree-level flavor-changing neutral currents featuring} $s$ \textit{quarks}, unless mediated by the new vector bosons and scalars introduced with the additional gauge group $SU(2)_F \times U(1)_F$.

With our fermion action now in the form we have addressed in Section \ref{BulkFermionSection}, we now can use the results given there in order to derive the flavor mixing properties of the SM-like quarks emerging in this model, in particular determining the conditions under which our model recreates the CKM matrix.

\subsection{The CKM Matrix in $SU(2)_F \times U(1)_F$}\label{CKMSection}
Following common practice in such analyses (see \cite{casagrande,huber,blanke}), we determine the conditions under which the CKM quark mixing matrix is recreated under the so-called zero-mass approximation (ZMA), in which the bulk profiles of the SM-like quark fields first defined in Eq.(\ref{FermionKKExpansions}) are given by Eq.(\ref{QuarkApproxProfiles}). Again here, as in Section \ref{BulkFermionSection}, our treatment is based heavily on that of \cite{casagrande}, and for greater detail regarding the nature of flavor mixing matrices in the RS we refer the reader to that work. To begin, we implement the ZMA by making the substitutions,
\begin{align}\label{ZMAProfileSubstitutions}
    &\mathbf{C}^{(Q)}_{1,2,3} (\phi) \Vec{a}^{(U,D)}_{1,2,3} \rightarrow \sqrt{\frac{kr_c \epsilon}{2}} \textrm{diag} \bigg( F(\Vec{\eta}_Q) e^{\frac{1}{2}(\Vec{\eta}_Q-1)(\sigma - kr_c \pi)}\bigg) \hat{a}^{(U,D)}_{1,2,3}. \nonumber\\
    &\mathbf{C}^{(u,d)}_{1,2,3} (\phi) \Vec{a}^{(u,d)}_{1,2,3} \rightarrow \sqrt{\frac{kr_c \epsilon}{2}} \textrm{diag} \bigg( F(\Vec{\eta}_{u,d}) e^{\frac{1}{2}(\Vec{\eta}_{u,d}-1)(\sigma - kr_c \pi)}\bigg) \hat{a}^{(u,d)}_{1,2,3},\\
    &\mathbf{S}^{U,D}_{1,2,3}(\phi) \Vec{a}^{(U,D)}_{1,2,3} \rightarrow \textrm{sgn}(\phi) \sqrt{\frac{k r_c \epsilon}{2}} x^{u,d}_n \; \textrm{diag} \bigg( \frac{\epsilon^{\frac{1}{2}(\Vec{\eta}_{Q}+1)}F(\Vec{\eta}_{Q})}{\Vec{\eta}_{Q}} (e^{\frac{1}{2}(\Vec{\eta}_{Q}+1)\sigma}-e^{-\frac{1}{2}(\Vec{\eta}_Q-1)\sigma}) \bigg) \hat{a}^{(U,D)}_{1,2,3}, \nonumber\\
    &\mathbf{S}^{u,d}_{1,2,3}(\phi) \Vec{a}^{(u,d)}_{1,2,3} \rightarrow -\textrm{sgn}(\phi) \sqrt{\frac{k r_c \epsilon}{2}} x^{u,d}_n \; \textrm{diag} \bigg( \frac{\epsilon^{\frac{1}{2}(\Vec{\eta}_{u,d}+1)}F(\Vec{\eta}_{u,d})}{\Vec{\eta}_{u,d}} (e^{\frac{1}{2}(\Vec{\eta}_{u,d}+1)\sigma}-e^{-\frac{1}{2}(\Vec{\eta}_{u,d}-1)\sigma}) \bigg) \hat{a}^{(u,u)}_{1,2,3}. \nonumber
\end{align}
The function $F(\eta)$ is defined in Eq.(\ref{QuarkApproxProfiles}), while we have rescaled the complex vectors $\Vec{a}^{U,D,u,d}_n$ appearing in Section \ref{BulkFermionSection}, which normally have magnitudes equal to $\frac{1}{\sqrt{2}}$ (this is straightforward to derive), into complex unit vectors $\hat{a}^{U,D,u,d}_n$. The mass terms for the SM-like quarks appearing in the effective four-dimensional theory are then
\begin{align}\label{ZMAMassMatrix1}
    &S_{q,mass}=-\frac{v}{\sqrt{2}} \Vec{\bar{\mathbf{u}}}_L \bm{\mathcal{M}}_{u} \Vec{\mathbf{u}}_R-\frac{v}{\sqrt{2}} \Vec{\bar{\mathbf{d}}}_L \bm{\mathcal{M}}_{d} \Vec{\mathbf{d}}_R + h.c.,\\
    &(\bm{\mathcal{M}}_{u,d})_{ij}=(\mathbf{Y}^{u,d})_{ij} F(\eta^Q_i)F(\eta^{u,d}_j),
\end{align}
where in our model, $\mathbf{Y}^{u,d}$ are given by Eq.(\ref{YukawaMatrices}). Here, $\Vec{\mathbf{u}}_{L,R}$ and $\Vec{\mathbf{d}}_{L,R}$ are three-dimensional vectors in flavor space, representing the three generations of up- and down-like quarks. Additionally, $\bm{\mathcal{M}}_{u,d}$ should not be confused with the variable of the same name defined in Eq.(\ref{FermionEigenvectorsExact}); they are unrelated. The crux of the ability for the RS model to generate fermion mass hierarchies lies in the function $F(\eta)$; even if the elements of $\mathbf{Y}^{u,d}$ are roughly of the same order of magnitude, the exponential dependence of $F(\eta)$ on bulk localization parameters means that the \emph{effective} Yukawa couplings experienced by each field can vary significantly; enough even to recreate the substantial mass hierarchy and suppressed mixing present among SM quarks, as has been mentioned and explored in \cite{casagrande,Grossman:2000,huberHierarchy,hewett,ahmed,blanke}. To find the quark mixing matrices here, we simply need to diagonalize the matrices $\bm{\mathcal{M}}_{u,d}$, by finding unitary matrices $\mathbf{U}^{u,d}_{L,R}$ such that
\begin{align}\label{UnitaryMatrices}
    \bm{\mathcal{M}}_{u,d} = \frac{\sqrt{2} k \epsilon}{v} \mathbf{U}^{u,d}_L \begin{pmatrix}
    x^{u,d}_{1,0} & 0 & 0 \\
    0 & x^{u,d}_{2,0} & 0 \\
    0 & 0 & x^{u,d}_{3,0}
    \end{pmatrix} (\mathbf{U}^{u,d}_R)^\dagger,
\end{align}
where $x^{u,d}_{i,0}$ are simply matrix eigenvalues here, related by a proportionality constant to the eigenvalues of $\bm{\mathcal{M}}_{u,d}$. In the language of the ZMA substitutions of Eq.(\ref{ZMAProfileSubstitutions}), we have
\begin{equation}\label{ZMAMixingSubstitutions}
    (\hat{a}^{u,d}_n)_i = (\mathbf{U}^{u,d}_R)_{in}, \; (\hat{a}^{U,D}_n)_i = (\mathbf{U}^{u,d}_L)_{in},
\end{equation}
where $i$ and $n$ are indices in three-dimensional generation space. Note that we use the index $n$ here, reminiscent of our notation for the indices of a Kaluza-Klein tower. This is by design; in the exact theory, $\hat{a}^{U,u,D,d}_n$ are just rescaled versions of the mass eigenvectors $\Vec{a}^{U,u,D,d}_n$ of the fermion fields discussed in Section \ref{BulkFermionSection}, and the indices $n=1,2,3$ correspond to the SM-like quarks, with masses far below $M_{KK}$. Meanwhile, the CKM matrix in this approximation is simply
\begin{align}
    \mathbf{V}_{CKM}= (\mathbf{U}^{u}_L)^{\dagger} \mathbf{U}^{d}_L.
\end{align}
In order for our model to be consistent with experiment, therefore, we must find sets of theoretical parameters such that the values $x^{u,d}_{1,2,3}$ are consistent with measured quark masses, and $\mathbf{U}^{u,d}_{L}$ recreate the CKM matrix. To evaluate whether or not the latter requirement is satisfied, we use the Wolfenstein parametrization of the CKM matrix \cite{wolfenstein},
\begin{equation}
    \lambda = \frac{|V_{us}|}{\sqrt{|V_{ud}|^2+|V_{us}|^2}}, \; \; A= \frac{1}{\lambda} \bigg\lvert \frac{V_{cb}}{V_{us}} \bigg\rvert, \; \; \bar{\rho} - i \bar{\eta} = -\frac{V_{ud}^* V_{ub}}{V_{cd}^* V_{cb}},
\end{equation}
and require that the mixing matrices $\mathbf{U}^{u,d}_L$ produce results consistent with the experimentally measured values \cite{pdgTopQuark},
\begin{align}
    &\lambda = 0.22453 \pm 0.00044, &A=0.836\pm0.015, \\
    &\bar{\rho}=0.122^{+0.018}_{-0.017}, &\bar{\eta}=0.355^{+0.012}_{-0.011}.
\end{align}
In order to actually probe the parameter space of our model, we now take advantage of hierarchical differences in the magnitude of $F(\eta)$ terms in $\bm{\mathcal{M}}_{u,d}$ in order to produce approximate analytical expressions for both the mass eigenvalues $x^{u,d}_{1,2,3}$ and the Wolfenstein parameters $A$, $\lambda$, $\bar{\rho}$, and $\bar{\eta}$. In particular, we assume (following the work of \cite{casagrande}) that $|F(\eta^{Q,u,d}_1| << |F(\eta^{Q,u,d}_2)| << |F(\eta^{Q,u,d}_{3})|$, and evaluate the approximate diagonalization matrices $\mathbf{U}^{u,d}_{L,R}$ and eigenvalues $x^{u,d}_{1,2,3}$ in this limit. For the quark masses evaluated at the scale $M_{KK}=k \epsilon$, we obtain the expressions,
\begin{align}\label{CKMMasses}
    &\frac{2 m_d^2}{v^2} \approx \bigg\lvert \frac{Y_b Y_d}{Y_b c_Q c_d + Y_d s_Q s_d} \bigg\rvert^2 |F(\eta^Q_1) F(\eta^d_1)|^2, &\frac{2 m_u^2}{v^2} \approx \frac{|\bar{Y}_t-Y_u|^2 |Y_u|^2}{|\bar{Y}_t|^2 c_u^2 c_Q^2 c_H^2}|F(\eta^Q_1)F(\eta^u_1)|^2, \nonumber\\
    &\frac{2 m_s^2}{v^2} = |Y_d|^2|F(\eta^Q_2)F(\eta^d_2)|^2, &\frac{2 m_c^2}{v^2} \approx \frac{|Y_u \bar{Y}_t|^2 c_H^2}{|\bar{Y}_t-Y_u s_H^2|^2}|F(\eta^Q_2)F(\eta^u_2)|^2,\\
    & \frac{2 m_b^2}{v^2} \approx | Y_b c_Q c_d + Y_d s_Q s_d |^2 |F(\eta^Q_3)F(\eta^d_3)|^2 &\frac{2 m_t^2}{v^2} \approx \frac{s_Q^2 s_u^2}{c_H^2}|\bar{Y}_t -Y_u s_H^2|^2 |F(\eta^Q_3)F(\eta^u_3)|^2, \nonumber \\
    &\bar{Y}_t \equiv \frac{1}{s_Q s_u}(Y_t c_u c_Q c_H+Y_u s_Q s_u) \nonumber,
\end{align}
in which small correction terms terms proportional to $|F(\eta^{Q,u,d}_i)|^2/|F(\eta^{Q,u,d}_j)|^2$ with $i<j$ have been dropped. Naively, these ratios should have roughly the same order of magnitude as the square of the ratio of the $i^{th}$ generation quark mass to the $j^{th}$ generation quark mass, which is invariably at most of $O(10^{-2})$. In practice, we find that these corrections are at most at the $1\%$ level, and usually are closer to the $O(10^{-3})$ level. As such, we find it well-motivated to ignore these subleading contributions. In Eq.(\ref{CKMMasses}), we have introduced a new variable $\bar{Y}_t$, which we shall find to be convenient later when solving for the appropriate bulk localizations to reproduce quark masses and mixings. We next apply a similar treatment to the Wolfenstein parameters, arriving at

\begin{align}\label{WolfensteinExpansion}
    &\lambda^2 \approx \bigg\lvert \frac{F(\eta^Q_1)}{F(\eta^Q_2)} \bigg\rvert^2 \frac{|\bar{Y}_t-Y_u|^2 s_H^2}{|\bar{Y}_t|^2 c_Q^2 c_H^2}(1+\Delta_{\lambda^2}), \nonumber\\
    &A^2 \lambda^4 \approx \bigg\lvert \frac{F(\eta^Q_2)}{F(\eta^Q_3)} \bigg\rvert^2 \frac{|Y_u|^2 s_H^2 c_H^2}{|\bar{Y}_t-Y_u s_H^2|^2 s_Q^2}(1+\Delta_{A^2 \lambda^4}),\\
    &\bar{\rho}-i \bar{\eta} \approx \frac{(Y_b Y_u c_d c_Q+Y_d Y_u s_d s_Q-\bar{Y}_t Y_d s_d s_Q)(\bar{Y}_t - Y_u s_H^2)}{s_H^2(Y_b c_d c_Q +Y_d s_d s_Q)(\bar{Y}_t-Y_u)Y_u}(1+\Delta_{\bar{\rho}-i \bar{\eta}}),  \nonumber
\end{align}
Here, we have included subleading terms (labelled $\Delta$) in our expressions for the Wolfenstein parameters. These quantities are sums of terms proportional to $|F(\eta^{Q,u,d}_1)|^2/|F(\eta^{Q,u,d}_2)|^2$ or $|F(\eta^{Q,u,d}_2)|^2/|F(\eta^{Q,u,d}_3)|^2$, \textit{i.e.} terms suppressed by the hierarchies between successive generations of quark bulk profiles. Although it is possible to omit such terms in the approximate expressions for quark masses within our model, we find that in order for the analytical expressions of Eq.(\ref{WolfensteinExpansion}) to be accurate enough to reproduce the Wolfenstein parameters to within experimental uncertainties, these correction terms must be included. The full expressions for the $\Delta$ terms are lengthy and unenlightening (they are included in full in Appendix \ref{DeltaTermsAppendix}), however we may use them, combined with the leading-order solutions to Eqs.(\ref{CKMMasses}) and (\ref{WolfensteinExpansion}), to approximately re-express the $\Delta$ terms using the quark masses and Wolfenstein parameters themselves. Doing so yields the relations,

\begin{align}\label{WolfensteinExpansion2}
    &\bigg\lvert \frac{F(\eta^Q_1)}{F(\eta^Q_2)} \bigg\rvert^2 \approx \lambda^2 \frac{|\bar{Y}_t|^2 c_Q^2 c_H^2}{|\bar{Y}_t-Y_u|^2 s_H^2} ( 1 -\Delta_{\lambda^2} ) \nonumber\\
    &\bigg\lvert \frac{F(\eta^Q_2)}{F(\eta^Q_3)} \bigg\rvert^2 \approx A^2 \lambda^4 \frac{s_Q^2 |\bar{Y}_t-Y_u s_H^2|^2}{|Y_u|^2 c_H^2 s_H^2} ( 1 - \Delta_{A^2 \lambda^4} ) \nonumber\\
    &Y_b \approx Y_d \frac{(\bar{Y}_t-Y_u) s_d s_Q}{Y_u c_d c_Q}\times\\
    &\bigg[ 1 + \nonumber\frac{\bar{Y}_t (\bar{\rho}-i \bar{\eta}) s_H^2}{\bar{Y}_t-Y_u s_H^2-(\bar{Y}_t-Y_u)(\bar{\rho}-i \bar{\eta})s_H^2 } \bigg( 1-\frac{(\bar{Y}_t-Y_u s_H^2)\Delta_{\bar{\rho}-i\bar{\eta}}}{\bar{Y}_t-Y_u s_H^2-(\bar{Y}_t-Y_u)(\bar{\rho}-i \bar{\eta})s_H^2}\bigg) \bigg] \nonumber \\
    &\Delta_{\lambda^2} \approx -\lambda^2 -\frac{2 Re[(\bar{Y}_t^*-Y_u^* s_H^2)(\bar{Y}_t-Y_u)]s_H^2}{|\bar{Y}_t-Y_u s_H^2|^2}\frac{m_u^2}{m_c^2} \frac{1}{\lambda^2}, \nonumber\\
    &\Delta_{A^2 \lambda^4} \approx -(2 \bar{\rho}-1)\lambda^2 - A^2 \lambda^4 - 2 Re \bigg[ \frac{Y_u^*}{\bar{Y}_t^*} \bigg] \frac{m_c^2}{m_t^2} \frac{s_H^2}{A^2 \lambda^4}, \nonumber\\
    &\Delta_{\bar{\rho}-i \bar{\eta}} \approx \frac{Y_u^*}{\bar{Y}_t^*} \frac{m_c^2}{m_t^2} \frac{s_H^2}{A^2 \lambda^4}+\frac{\bar{Y}_t^*-Y_u^*}{\bar{Y}_t^*-Y_u^* s_H^2} \frac{\bar{\rho}-i\bar{\eta}-1}{\bar{\rho}-i\bar{\eta}}\frac{m_u^2}{m_c^2}\frac{s_H^2}{\lambda^2}+(\bar{\rho}-i \bar{\eta}-1)(\lambda^2-A^2 \lambda^4). \nonumber
\end{align}

It should be noted that for the sake of consistency, similar correction terms should be included for the quark masses computed in Eq.(\ref{CKMMasses}). However, we find numerically that these corrections are substantially smaller than those for the Wolfenstein parameters (as we have noted earlier, the mass eigenvalue corrections are usually at the $O(10^{-3})$ level, while the Wolfenstein parameter corrections are all at the level of approximately $5\%$), and so using the leading order expressions for the quark masses does not appreciably affect our results. Therefore, we omit the next-to-leading order corrections to the quark masses here for the sake of simplicity. At this point, we now possess all the tools necessary to identify points in our model's parameter space which recreate the observed quark masses and mixing parameters. Explicitly, we must find brane Yukawa couplings $Y_u$, $\bar{Y}_t$, $Y_d$, and $Y_b$, bulk localization parameters $\eta^{Q,u,d}_{1,2,3}$, and rotation angles $\theta_{H,Q,u,d}$ that satisfy Eqs.(\ref{CKMMasses}) and (\ref{WolfensteinExpansion2}), as well as the requirement (from our parameterization in Section \ref{BulkMatterModelSection}) that $\eta^{Q,u,d}_2 = c_{Q,u,d}^2\eta^{Q,u,d}_1+s_{Q,u,d}^2 \eta^{Q,u,d}_3$. To find these solutions, we first identify that we have in total 18 real parameters (4  complex brane Yukawa couplings, 6 independent real fermion localizations $\eta^{Q,u,d}_{1,3}$, and 4 real rotation angles $\theta_{H,Q,u,d}$), with which we must satisfy 10 constraints (6 quark masses and four Wolfenstein parameters). Naively then, we should anticipate that specifying 8 of our model parameters as inputs should uniquely specify all other model parameters, if we impose the requirement that the observed SM quark masses and CKM parameters are recreated. Following this intuition, we identify points in our parameter space by specifying the complex parameters $Y_u$, $\bar{Y}_t$, and $Y_d$, and the real parameters $\theta_H$ and $\theta_Q$, and assuming that the remainder of the model parameters must be found as solutions to Eqs.(\ref{CKMMasses}) and (\ref{WolfensteinExpansion2}). We begin with the localizations $\eta^Q_{1,2,3}$. The expression for $|F(\eta^Q_1)|^2/|F(\eta^Q_2)|^2$ in Eq.(\ref{WolfensteinExpansion2}) readily gives the function $|F(\eta^Q_1)|^2$ as a function of $|F(\eta^Q_2)|^2$ and fixed model inputs (namely $\bar{Y}_t$, $Y_u$, $\theta_Q$, and $\theta_H$). From here, it is possible to get $\eta^Q_1$, the localization parameter itself, using the Lambert product log function, defined as the function $W(x)$ such that $W(x) \textrm{exp} [W(x)] = x$. In general, we find that
\begin{align}\label{ProductLogID}
    &|F(\eta)|^2=Q \rightarrow \eta=Q+\frac{1}{k r_c \pi} \bar{W}(- k r_c \pi Q),\\
    &\bar{W}(x) \equiv \begin{cases}
    W_{0}(x e^x ) & x \leq -1 \\
    W_{-1}(x e^x) & x > -1
    \end{cases} \nonumber
\end{align}
where the subscripts on $W$ refer to different branches of the product log function, following conventional notation, and we have defined the function $\bar{W}$ to select the proper branch for our purposes at all points. Explicitly, we have
\begin{align}\label{Q1Func}
    &\eta^Q_1 = |F(\eta^Q_1)|^2 + \frac{1}{k r_c \pi} \bar{W}(- k r_c \pi |F(\eta^Q_1)|^2), \\
    &|F(\eta^Q_1)|^2 \approx |F(\eta^Q_2)|^2 \lambda^2 \frac{|\bar{Y}_t|^2 c_Q^2 c_H^2}{|\bar{Y}_t-Y_u|^2 s_H^2} (1- \Delta_{\lambda^2}). \nonumber
\end{align}
Using the relation, $\eta^{Q}_2 = c_{Q}^2\eta^{Q}_1+s_{Q}^2 \eta^{Q}_3$, we can then express $\eta^Q_3$ as a function of $\eta^Q_1$, $\eta^Q_2$, and $\theta_Q$. In doing so, the expression for $|F(\eta^Q_2)|^2/|F(\eta^Q_3)|^2$ in Eq.(\ref{WolfensteinExpansion2}) now becomes
\begin{align}\label{Eta2Func}
    &\bigg\lvert \frac{F(\eta^Q_2)}{F(\csc^2(\theta_Q)\eta^Q_2-\cot^2(\theta_Q)\eta^Q_1)} \bigg\rvert^2 \approx A^2 \lambda^4 \frac{s_Q^2 |\bar{Y}_t-Y_u s_H^2|^2}{|Y_u|^2 c_H^2 s_H^2} (1-\Delta_{A^2 \lambda^4}).
\end{align}
Folding Eq.(\ref{Q1Func}) into Eq.(\ref{Eta2Func}) then gives us an expression for $\eta^Q_2$ that can be solved numerically, using only the fixed input parameters $\bar{Y}_t$, $Y_u$, $\theta_Q$, and $\theta_u$.

With $\eta^Q_2$ (and by extension, $\eta^Q_1$ and $\eta^Q_3$) fixed by Eq.(\ref{Eta2Func}), we can move on to addressing the localizations $\eta^u_{1,2,3}$. First, we use the expressions for $|F(\eta^Q_1)|^2/|F(\eta^Q_2)|^2$ and $|F(\eta^Q_2)|^2/|F(\eta^Q_3)|^2$ from Eq.(\ref{WolfensteinExpansion2}) to rewrite the expressions for the up-like quark masses in Eq.(\ref{CKMMasses}) in a more convenient form, namely,
\begin{align}\label{UpCKMMasses}
    &|F(\eta^u_1)|^2 \approx \frac{2 m_u^2}{v^2} \frac{c_u^2 c_H^2}{\lambda^2|F(\eta^Q_2)|^2}(1+\Delta_{\lambda^2}), \nonumber \\
    &|F(\eta^u_2)|^2 \approx \frac{2 m_c^2}{v^2} \frac{|\bar{Y}_t-Y_u s_H^2|^2}{|Y_u \bar{Y}_t|^2 c_H^2}\frac{1}{|F(\eta^Q_2)|^2}, \\
    &|F(\eta^u_3)|^2 \approx \frac{2 m_t^2}{v^2} \frac{A^2 \lambda^4}{|F(\eta^Q_2)|^2 s_u^2 s_H^2} (1-\Delta_{A^2\lambda^4}). \nonumber
\end{align}
The expressions for $|F(\eta^u_1)|^2$ and $|F(\eta^u_3)|^2$ in Eq.(\ref{UpCKMMasses}) can then be solved to yield expressions for $\eta^u_1$ and $\eta^u_3$ using the identity in Eq.(\ref{ProductLogID}), yielding expressions for these two bulk localization parameters in terms of $\eta^Q_2$, $\theta_u$, $\bar{Y}_t$, $Y_u$, $\theta_H$, and $\theta_Q$. Since $\eta^Q_2$ can be found by numerically solving Eq.(\ref{Eta2Func}), we therefore have both $\eta^u_1$ and $\eta^u_3$ solely in terms of $\theta_u$ and fixed input parameters. We can now use $\eta^u_2 = c_u^2 \eta^u_1 + s_u^2 \eta^u_3$ with the expression for $|F(\eta^u_2)|^2$ given in Eq.(\ref{UpCKMMasses}), which then yields the equation,
\begin{align}\label{cuFunc}
    |F(c_u^2 \eta^u_1 + s_u^2 \eta^u_3)|^2 \approx \frac{2 m_c^2}{v^2} \frac{|\bar{Y}_t-Y_u s_H^2|^2}{|Y_u \bar{Y}_t|^2 c_H^2}\frac{1}{|F(\eta^Q_2)|^2}.
\end{align}
When the expressions for $\eta^u_1$ and $\eta^u_3$ from Eq.(\ref{UpCKMMasses}) are then used here, we note that Eq.(\ref{cuFunc}) contains only $\theta_u$ and fixed input parameters. So, we can numerically solve Eq.(\ref{cuFunc}) for $\theta_u$, giving us another parameter of our model. Then, we can derive $\eta^u_1$ and $\eta^u_3$ (and by extension, $\eta^u_2$) by inserting our numerical result for $\theta_u$ into the relations of Eq.(\ref{UpCKMMasses}).

Having solved our system for $\eta^{Q}_{1-3}$, $\eta^u_{1-3}$, and $\theta_u$, the only parameters which remain unfixed are the down-like quark localizations $\eta^d_{1}$ and $\eta^d_{3}$ (recall that $\eta^d_{2}$ is fixed by the relation $\eta^d_2 = c_d^2 \eta^d_1+ s_d^2 \eta^d_3$), the mixing angle $\theta_d$, and the brane Yukawa coupling $Y_b$. To solve for these variables, we look to the expressions for the masses of the down-like quarks given in Eq.(\ref{CKMMasses}). Inserting the expressions from Eq.(\ref{WolfensteinExpansion2}) into these equations, we arrive at
\begin{align}\label{DownCKMMasses}
    &|F(\eta^d_1)|^2 \approx \frac{2 m_d^2}{v^2} \frac{|\bar{Y}_t-Y_u s_H^2|^2 c_d^2 s_H^2}{\lambda^2 |Y_d|^2 |F(\eta^Q_2)|^2 c_H^2} \frac{(1+ \Delta_{\lambda^2})}{|\bar{Y}_t-Y_u s_H^2 +(\bar{\rho}-i \bar{\eta})(1+\Delta_{\bar{\rho}-i \bar{\eta}})Y_u s_H^2 |^2}, \nonumber\\
    &|F(\eta^d_2)|^2 = \frac{2 m_s^2}{v^2} \frac{1}{|Y_d|^2|F(\eta^Q_2)|^2},\\
    &|F(\eta^d_3)|^2 \approx \frac{2 m_b^2}{v^2} \frac{A^2 \lambda^4 (1-\Delta_{A^2 \lambda^4})}{|Y_d \bar{Y}_t|^2 s_d^2 s_H^2 c_H^2 |(\bar{Y}_t-Y_u) s_H^2 (\bar{\rho}-i \bar{\eta})(1+\Delta_{\bar{\rho}-i \bar{\eta}})-(\bar{Y}_t-Y_u s_H^2)|^2}. \nonumber
\end{align}
The expressions for $|F(\eta^d_1)|^2$ and $|F(\eta^d_3)|^2$ in Eq.(\ref{DownCKMMasses}) can then, just as for the up-like quarks, be solved to yield algebraic expressions for $\eta^d_1$ and $\eta^d_3$ in terms of $c_d$ and already fixed parameters. Then, we merely need to fold these expressions into
\begin{equation}
    |F(c_d^2 \eta^d_1+s_d^2 \eta^d_3)|^2 = \frac{2 m_s^2}{v^2} \frac{1}{|Y_d|^2|F(\eta^Q_2)|^2},
\end{equation}
with the $|F(\eta^d_2)|^2$ expression in Eq.(\ref{DownCKMMasses}), in order to get a function in which the only remaining variable is $\theta_d$. We can solve this equation numerically, fixing $\theta_d$, which then allows us to use the expressions for $|F(\eta^d_1)|^2$ and $|F(\eta^d_3)|^2$ to fix $\eta^d_1$ and $\eta^d_3$ (and therefore also $\eta^d_2$). Finally, in order to complete our set of parameters, we only need to insert our value for $\theta_d$ into the expression for $Y_b$ in Eq.(\ref{WolfensteinExpansion2}), at which point we have achieved our goal: Given a specified $\bar{Y}_t$, $Y_u$, $Y_d$, $\theta_H$, and $\theta_Q$, we now have a process to solve for the remaining model parameters relating to the fermion bulk localizations and Yukawa couplings such that the observed quark masses and CKM mixing parameters are satisfied.

\subsection{Numerical Sampling of Parameter Space}\label{NumericalParamPointsSection}

Using the methods of Section \ref{CKMSection}, we can now generate a large sample of points in parameter space which recreate the observed quark masses and the CKM mixing matrix. To begin, we select a KK scale $M_{KK}$ for our analysis, and run the quark masses up to that scale. We select $M_{KK}=k \epsilon = 5$ TeV, as this is quite close to the minimum scale that is permissible from the significant constraint on the precision electroweak parameter $T$ \cite{peskin}, which represents one of the most restrictive precision constraints for RS models in the absence of a custodial protection for it \cite{casagrande,goertz,gori,csakiCompositeHiggs,ahmed}. At tree level \cite{casagrande,goertz},
\begin{equation}
    T=\frac{\pi v^2}{2 c_W^2 M_{KK}^2} \bigg( kr_c \pi - \frac{1}{2 kr_c \pi} \bigg),
\end{equation}
where $v = 246$ GeV is the Higgs vacuum expectation value, and $c_W$ is the cosine of the weak mixing angle. Using this formula, we find that in order to keep the RS contribution to $T$ within the 68\% CL range \cite{pdgTopQuark} of $T=0.07 \pm 0.12$, we need $M_{KK} > 4.78$ TeV (in rough agreement with the results of \cite{Davoudiasl:2009cd}, however their definition of $M_KK$ differs from our own by a factor of 2.45), hence our selection of $M_{KK}=5$ TeV. We note that this method of constraining $M_{KK}$ is hardly thorough; in this computation of $T$, for example, the effects of non-oblique operators that emerge in RS are not included \cite{Davoudiasl:2009cd}. However, we are focused specifically on quark flavor physics, constraints on which will only weaken as $M_{KK}$ is increased, and so a precise determination of precision electroweak constraints on non-custodial RS models is well beyond the scope of this paper. Indeed, we anticipate that reconstructing a version of this model in a framework with a bulk custodial symmetry should be straightforward and give similar results for flavor physics, with dramatically weakened electroweak precision constraints, so we do not consider the possibility that electroweak precision constraints are more severe here to significantly affect our later discussion. In the end, we use the RunDec Mathematica package \cite{rundec} to determine the quark masses at the scale $\mu=M_{KK}= 5$ TeV to be
\begin{align}
    &m_u(5 \textrm{ TeV}) = 0.972 \, \textrm{MeV}, & m_c(5 \textrm{ TeV}) = 0.486 \, \textrm{GeV}, \;\;\;\;\;\;\;\;\;\;& m_t(5 \textrm{ TeV}) = 131.0 \, \textrm{GeV}, \\
    &m_d(5 \textrm{ TeV}) = 2.08 \, \textrm{MeV}, & m_s(5 \textrm{ TeV}) = 42.4 \, \textrm{MeV}, \;\;\;\;\;\;\;\;\;\;& m_b(5 \textrm{ TeV}) = 2.19 \, \textrm{GeV}. \nonumber
\end{align}

With our KK scale selected and our quark masses determined, we perform our numerical probe in two stages. First, we generate a random sample of 8000 sets of brane Yukawa couplings $Y_u$, $\bar{Y}_t$, and $Y_d$, each with magnitudes between 1/3 and 3 and with random complex phases. Once we have these sets of couplings, we then need to determine which input angles $\theta_H$ and $\theta_Q$ permit solutions to Eqs.(\ref{WolfensteinExpansion2}),(\ref{UpCKMMasses}), and (\ref{DownCKMMasses}). To do this, we move on to the second stage of our probe, in which we perform a scan over $\theta_Q$ and $\theta_H$ parameters for each set of input Yukawa couplings. Specifically, for each set of Yukawa couplings, we test each value of $c_H^2$ and $c_Q^2$ (the squares of the cosines of the angles $\theta_H$ and $\theta_Q$) in the range from 0.05 to 0.95 for both parameters, in increments of 0.05, generating a list of $c_H^2$ and $c_Q^2$ values that permit solutions that yield the correct quark masses and Wolfenstein parameters for that set of couplings. By probing a broad range of these input parameters, our scan therefore produces, for each of our 8000 sets of Yukawa coupling parameters, the approximate full range of the model's parameter space accessible to each set. On average, each randomly generated set of Yukawa couplings has $\sim$40 $(c_H^2, c_Q^2)$ pairs which yield such solutions (found following the method of Section \ref{CKMSection}), yielding a total sample size of 323610 points in parameter space that we have sampled. However, we note that in keeping with the guiding principle that each brane Yukawa coupling constant should not have magnitude much greater than or less than $O(1)$ \cite{casagrande}, we then dismiss sampled points that have $|Y_t|$ or $|Y_b|$ greater than 10 or less than 0.1, leaving us with 229691 points in parameter space to explore.

Regarding our methodology for numerically generating points in our model's parameter space, two important observations must be made. First, although our setup in Section \ref{CKMSection} depends only on the $\textit{squares}$ of trigonometric functions of the angles $\theta_{H,Q,u,d}$, and is hence independent of the sign of these trigonometric functions, the values of $Y_t$, $Y_b$, and various coupling matrices in our model are sensitive to these signs. For simplicity, we assume that each angle is within the range of 0 and $\pi/2$, that is to say, all trigonometric functions of the angle parameters are positive. While this assumption is non-trivial (and amounts, in effect, to a set of assumptions about the signs of the bulk Yukawa couplings $y^{Q,d}$ and $y^u_{1,2}$), we do not expect this choice having a significant effect on our final results (outside of the signs of the bulk Yukawa couplings themselves). Our primary analysis will be in the form of exploring effective four-quark vertices emerging from the exchange of heavy KK tower states of various gauge bosons and bulk scalars, and we anticipate that altering the signs of the trigonometric functions of $\theta_{H,Q,u,d}$ should generically just result in $O(1)$ changes to the coefficients of these four-quark vertices. Second, we note that our probe of the parameter space of our model is by no means guaranteed to be complete or unbiased; this study is not designed to probe the fine-tuning of this model as much as it is to establish the existence of a number of points in parameter space that are at present phenomenologically viable.

\begin{figure}
    \centerline{\includegraphics[width=3.5in]{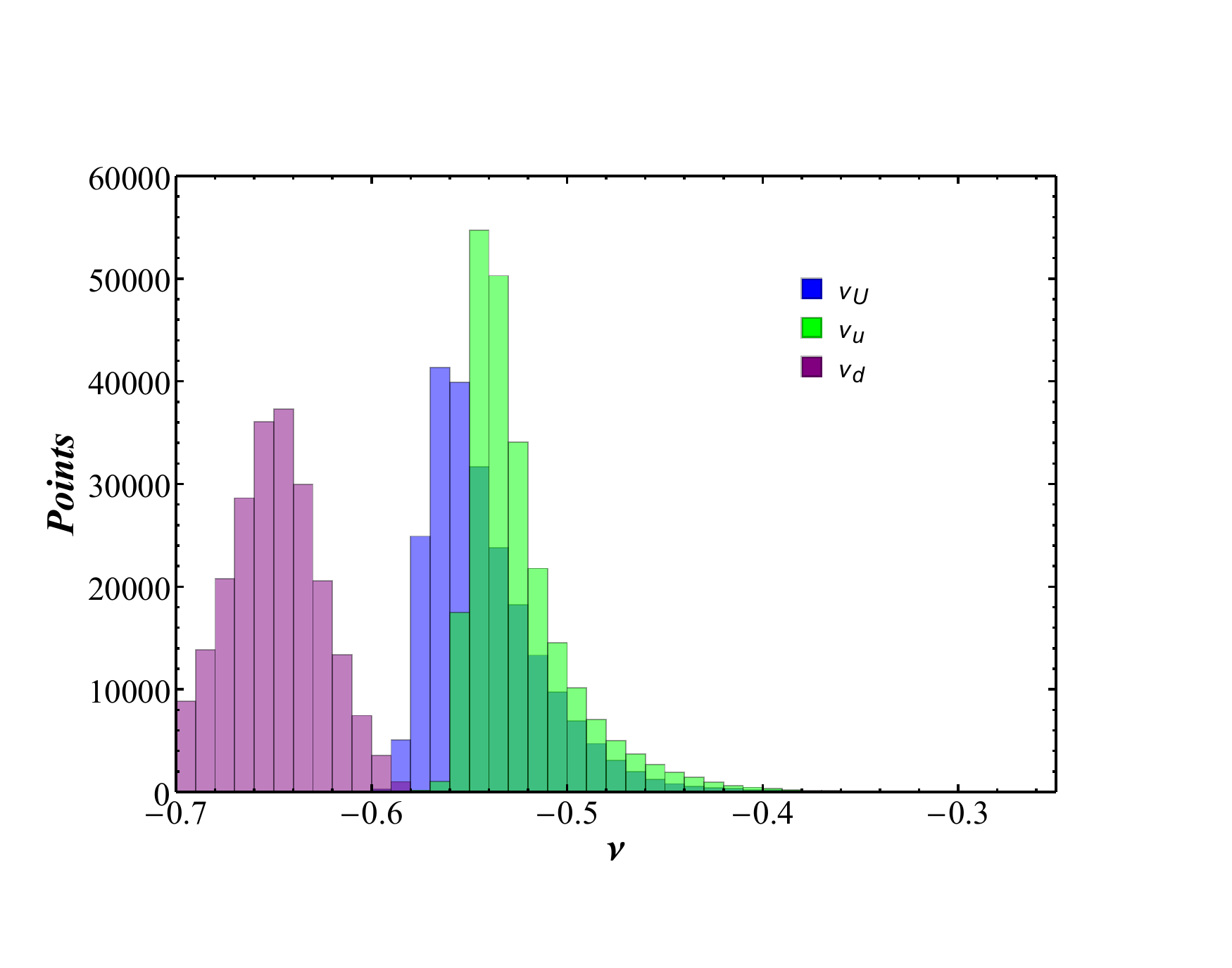}
    \hspace{-0.75cm}
    \includegraphics[width=3.5in]{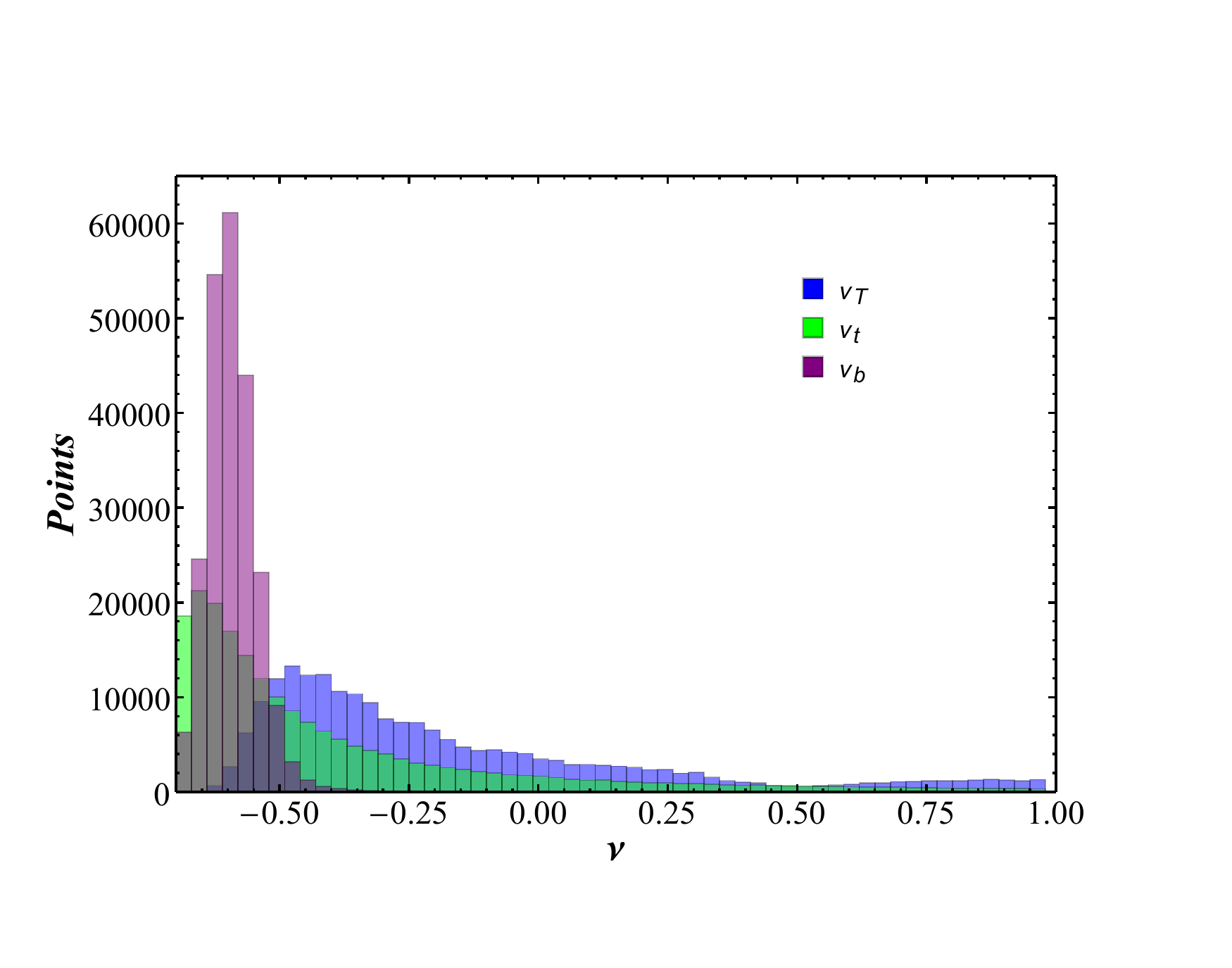}}
    \vspace*{-0.25cm}
    \centering
    \includegraphics[width=3.5in]{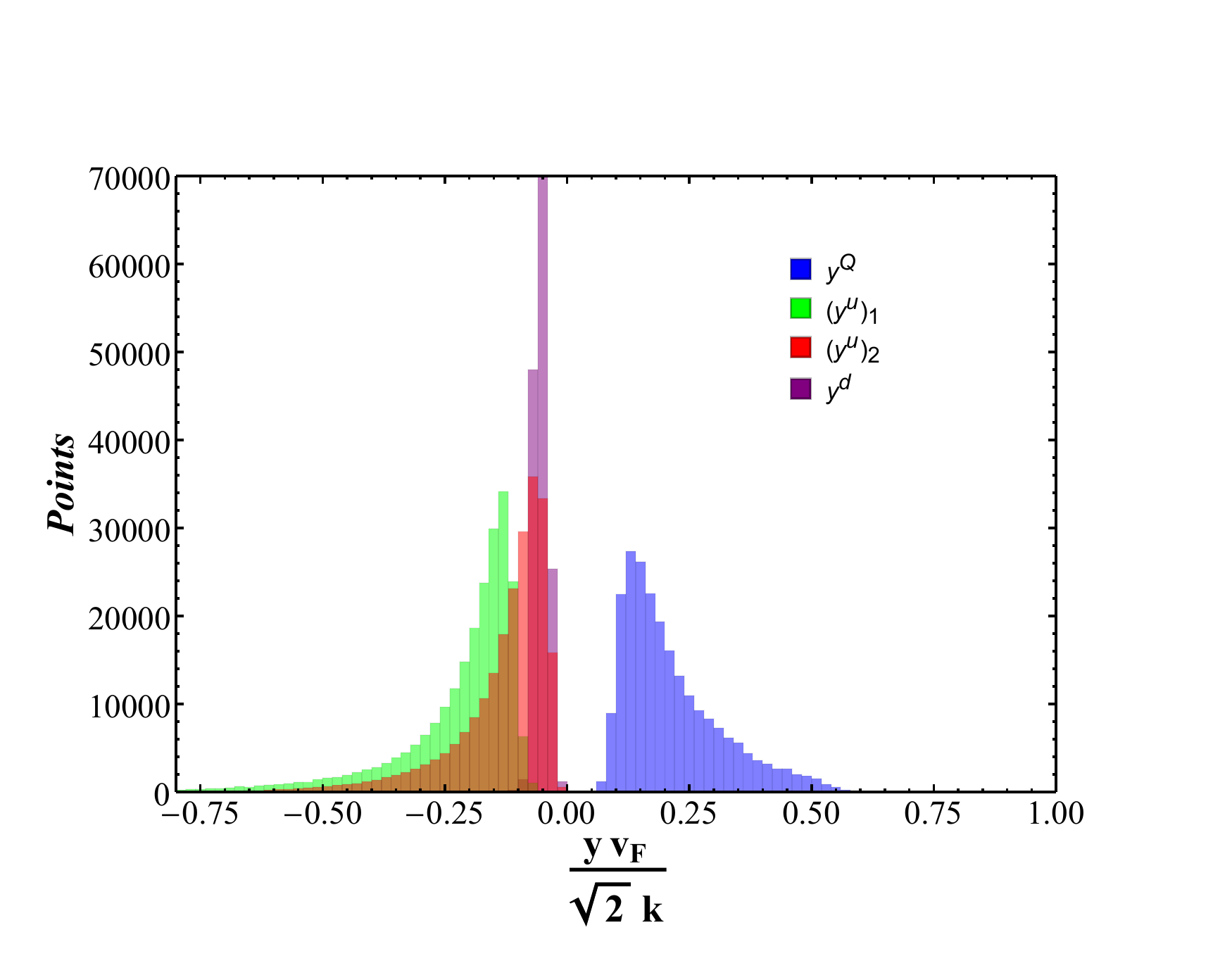}
    \caption{Bulk Localization (top) and Yukawa Coupling (bottom) parameters for sampled points in parameter space}
    \label{fig:ObservablePoints}
\end{figure}

It is instructive at this point to present certain aspects of our sample pool graphically. In Figure \ref{fig:ObservablePoints}, we depict histograms of the various model parameters $\nu_{U,u,d}$, $\nu_{T,t,b}$, $y^{Q,d}$, and $y^{u}_{1,2}$. The most salient interesting features of these distributions lie in the behavior of the $SU(2)_F$ singlet quark field localization parameters $\nu_{T,t,b}$ and the bulk Yukawa couplings. First, we observe that the localization parameters $\nu_T$ and $\nu_t$ exhibit a much broader distribution than that of $\nu_b$ and the $SU(2)_F$ doublet fields; in particular, a number of solutions exist that push either $\nu_T$ or $\nu_t$ into the large and positive regime. This is to be expected, given that in order to realize the large top quark mass, we would anticipate either or both of the third generation's $SU(2)_L$ doublet and up-like singlet bulk profiles to be highly TeV-brane localized, which in turn favors pushing $\nu_T$ and $\nu_t$ more positive. The same cannot be said of $\nu_b$, which (largely due to the $b$ quark's substantially smaller mass compared to that of the top) is favored to have a value quite similar to the localization of the $SU(2)_F$ doublet quarks. Meanwhile, among the Yukawa couplings, we note that qualitatively the magnitudes of $y^{Q}$ and $y^{u}_{1,2}$ parameters take on quite similar values between 0 and 1 (although $y^Q$ is positive while the $y^u$ parameters are both negative, which is unsurprising given the difference in sign conventions for the $SU(2)_L$ doublet and singlet quarks' bulk mass parameters, discussed in Section \ref{BulkMatterModelSection}). However, the parameter $y^d$, corresponding to the bulk Yukawa coupling for the down-like $SU(2)_F$ singlet quarks, favors dramatically smaller magnitudes, such that $|y^d v_F/(\sqrt{2}k)| \sim O(10^{-1})$. This is also intuitively reasonable. The parameter $y^d$ can be thought of as an off-diagonal mixing parameter in the bulk mass matrix that generates a discrepancy between the bulk masses of the first- and second-generation down-like $SU(2)_L$ singlet quark fields. As $y^d$ approaches zero, the bulk masses of both of these fields become equal, namely, both fields' bulk masses become $\nu_d$ as the $SU(2)_F$ symmetry demands. With this in mind, we also recall that it has been noted in other studies of RS flavor \cite{casagrande,huberHierarchy} that the localization parameters for the down-like $SU(2)_L$ singlet quarks tend to be quite similar, and hence we would expect that only a rather small mixing parameter $y^d$ would be needed to effect the minimal discrepancy between the first- and second-generation down-like quarks' bulk localizations. Notably, due to the requirement that $|y^d v_F/(\sqrt{2}k)| \sim O(10^{-1})$, we anticipate that, in order to avoid an unreasonably small $y^d$ parameter (naively assuming that our bulk Yukawa couplings should themselves be approximately $O(1)$, which while not necessarily strongly motivated from naturalness arguments has a considerable aesthetic appeal), $v_F$ should not be substantially greater than the curvature constant $k$; in our further analysis we restrict our study to values of $v_F<k$.

\section{Four-Quark Operators at Tree Level}\label{4QuarkOperatorSection}
Having constructed our scenario, we now possess all the necessary tools to explore quark couplings within the model, and specifically, additional tree-level flavor changing processes introduced by the introduction of a warped extra dimension and our flavor gauge symmetry $SU(2)_F \times U(1)_F$. In particular, within this section we shall discuss the 4-quark operators which emerge in the low-energy EFT from both the new bulk gauge bosons and scalars we propose and the SM gauge fields allowed in the bulk.

\subsection{Four-Quark Operators: Flavor Gauge Bosons}\label{FlavorGauge4FermionSection}
We begin our discussion of quark couplings in our model with currents emerging from the vector gauge bosons of the new bulk local symmetry $SU(2)_F \times U(1)_F$. In the bulk theory, this interaction arises from terms of the form (for some vector gauge boson $E_\mu$),
\begin{equation}\label{BulkGaugeCouplingTerm}
    \int d^4 x \sqrt{2 \pi r_c} g_E \int_{-\pi}^{\pi} d \phi \, r_c e^{-3 \sigma} \bar{\mathbf{q}} \gamma^\mu \mathbf{X}^q_E \mathbf{q} E_\mu,
\end{equation}
where, as in Section \ref{BulkFermionSection}, $\mathbf{q}$ represents a 3-dimensional vector of quark fields in generation space (in this case either the $SU(2)_L$ doublet $\mathbf{Q}$, the up-like $SU(2)_L$ singlet $\mathbf{u}$, or the down-like $SU(2)_L$ singlet $\mathbf{d}$. The term $\mathbf{X}^q_E$ is a $3\times3$ matrix in generation space, the precise form of which varies depending on the identities of the gauge boson $E_\mu$ and the quark field $q$. Applying the KK expansions of Eq.(\ref{FermionKKExpansions}) to the fermion fields, and Eq.(\ref{SMGaugeKKExpansion}) to the vector gauge field (it should be noted that this is identical to Eq.(\ref{GaugeFlavorVectorKK}), and so is applicable regardless of which of our model's vector bosons we are considering in our model), we arrive at interaction terms of the form,
\begin{align}\label{GaugeKKCouplingTerms}
    &\int d^4 x \sqrt{2 \pi} g_E \int_{-\pi}^{\pi} d \phi \, e^\sigma \chi^E_p(\phi) \bigg\{ \nonumber\\
    &\bigg( (\Vec{a}^{(U)}_n)^\dagger \mathbf{C}^{(U)}_n(\phi) \mathbf{X}^U_E \mathbf{C}^{(U)}_m(\phi) \Vec{a}^{(U)}_m + (\Vec{a}^{(u)}_n)^\dagger \mathbf{S}^{(u)}_n(\phi) \mathbf{X}^u_E \mathbf{S}^{(u)}_m(\phi) \Vec{a}^{(u)}_m\bigg)(\bar{u}^{(n)}_L \gamma^\mu E^{(p)}_\mu u^{(m)}_L)\\
    &+\bigg( (\Vec{a}^{(U)}_n)^\dagger \mathbf{S}^{(U)}_n(\phi) \mathbf{X}^U_E \mathbf{S}^{(U)}_m(\phi) \Vec{a}^{(U)}_m + (\Vec{a}^{(u)}_n)^\dagger \mathbf{C}^{(u)}_n(\phi) \mathbf{X}^u_E \mathbf{C}^{(u)}_m(\phi) \Vec{a}^{(u)}_m\bigg)(\bar{u}^{(n)}_R \gamma^\mu E^{(p)}_\mu u^{(m)}_R) \bigg\}, \nonumber
\end{align}
where we've given the expression for up-like quarks, but the expression for down-like quarks is readily derivable by making the substitutions $U \rightarrow D$ and $u \rightarrow d$. For the charged currents emerging from $W$ boson exchange some modification to this expression is necessary, which we shall detail in Section \ref{SMGauge4FermionSection}. Eq.(\ref{GaugeKKCouplingTerms}) contains two separate terms, one for couplings between the $p^{th}$ mode of the $B_\mu$ field and the $m^{th}$ and $n^{th}$ mode of the left-handed up-like quark fields, and one for couplings between the $p^{th}$ mode of the $E_\mu$ field and the $m^{th}$ and $n^{th}$ mode of the right-handed up-like quark fields. It is convenient at this point to define the quantities,
\begin{align}\label{OmegaDefs}
    &(\bm{\Omega}_{L}^E)_{n m}^{(U,D)}(\phi) \equiv (\Vec{a}^{(U,D)}_n)^\dagger \mathbf{C}^{(U,D)}_n(\phi) \mathbf{X}^{U,D}_E \mathbf{C}^{(U,D)}_m(\phi) \Vec{a}^{(U,D)}_m, \nonumber\\
    &(\bm{\Omega}_{R}^E)_{n m}^{(U,D)}(\phi) \equiv (\Vec{a}^{(U,D)}_n)^\dagger \mathbf{S}^{(U,D)}_n(\phi) \mathbf{X}^{U,D}_E \mathbf{S}^{(U,D)}_m(\phi) \Vec{a}^{(U,D)}_m,\\
    &(\bm{\Omega}_{L}^E)_{n m}^{(u,d)}(\phi) \equiv (\Vec{a}^{(u,d)}_n)^\dagger \mathbf{S}^{(u,d)}_n(\phi) \mathbf{X}^{u,d}_E \mathbf{S}^{(u,d)}_m(\phi) \Vec{a}^{(u,d)}_m, \nonumber \\
    &(\bm{\Omega}_{R}^E)_{n m}^{(u,d)}(\phi) \equiv (\Vec{a}^{(u,d)}_n)^\dagger \mathbf{C}^{(u,d)}_n(\phi) \mathbf{X}^{u,d}_E \mathbf{C}^{(u,d)}_m(\phi) \Vec{a}^{(u,d)}_m, \nonumber
\end{align}
so that we may, for example, write that the effective coupling constant between the $p^{th}$ mode of $B_\mu$ and the $n^{th}$ and $m^{th}$ modes of $u_L$ can be given as
\begin{equation}
    \sqrt{2 \pi} g_E \int_{-\pi}^{\pi} e^{\sigma} \chi^E_p(\phi) [(\bm{\Omega}^E_L)^{(U)}_{nm}(\phi)+(\bm{\Omega}^E_L)^{(u)}_{nm}(\phi)].
\end{equation}

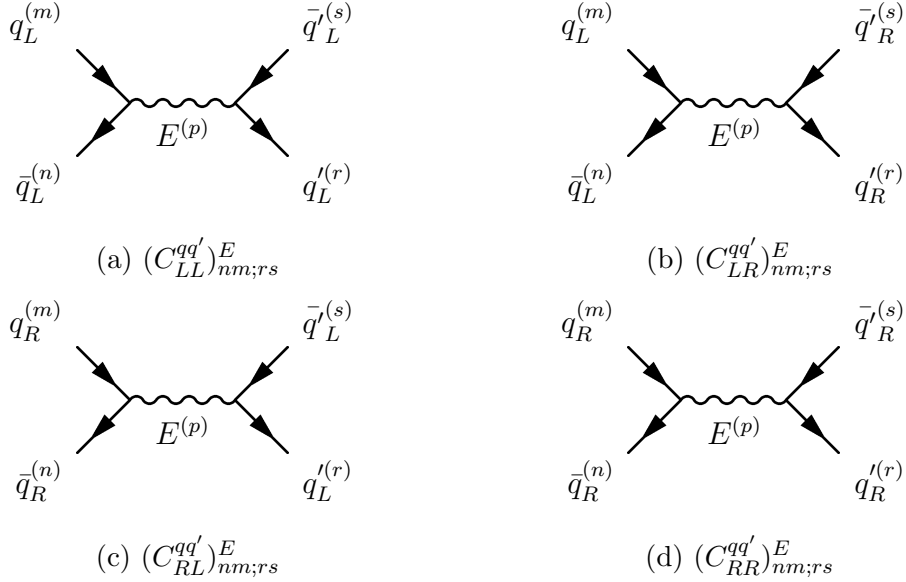
\begin{figure}
  \centering
  \begin{subfigure}[b]{0.45\textwidth}
    \centering
    \fmfframe(10,25)(10,25){
    \begin{fmfgraph*}(100,40)
        \fmfleft{i1,i2}
        \fmfright{o1,o2}
        \fmf{fermion}{i2,v1,i1}
        \fmf{boson,label=$E^{(p)}$}{v1,v2}
        \fmf{fermion}{o2,v2,o1}
        \fmflabel{$q^{(m)}_L$}{i2}
        \fmflabel{$\bar{q}^{(n)}_L$}{i1}
        \fmflabel{$\bar{q'}^{(s)}_L$}{o2}
        \fmflabel{$q'^{(r)}_L$}{o1}
    \end{fmfgraph*}
    }
    \caption{$(C^{q q'}_{LL})^E_{nm;rs}$}
  \end{subfigure}
  \begin{subfigure}[b]{0.45\textwidth}
    \centering
    \fmfframe(10,25)(10,25){
    \begin{fmfgraph*}(100,40)
        \fmfleft{i1,i2}
        \fmfright{o1,o2}
        \fmf{fermion}{i2,v1,i1}
        \fmf{boson,label=$E^{(p)}$}{v1,v2}
        \fmf{fermion}{o2,v2,o1}
        \fmflabel{$q^{(m)}_L$}{i2}
        \fmflabel{$\bar{q}^{(n)}_L$}{i1}
        \fmflabel{$\bar{q'}^{(s)}_R$}{o2}
        \fmflabel{$q'^{(r)}_R$}{o1}
    \end{fmfgraph*}
    }
    \caption{$(C^{q q'}_{LR})^E_{nm;rs}$}
  \end{subfigure}
  \begin{subfigure}[b]{0.45\textwidth}
    \centering
    \fmfframe(10,25)(10,25){
    \begin{fmfgraph*}(100,40)
        \fmfleft{i1,i2}
        \fmfright{o1,o2}
        \fmf{fermion}{i2,v1,i1}
        \fmf{boson,label=$E^{(p)}$}{v1,v2}
        \fmf{fermion}{o2,v2,o1}
        \fmflabel{$q^{(m)}_R$}{i2}
        \fmflabel{$\bar{q}^{(n)}_R$}{i1}
        \fmflabel{$\bar{q'}^{(s)}_L$}{o2}
        \fmflabel{$q'^{(r)}_L$}{o1}
    \end{fmfgraph*}
    }
    \caption{$(C^{q q'}_{RL})^E_{nm;rs}$}
  \end{subfigure}
  \begin{subfigure}[b]{0.45\textwidth}
    \centering
    \fmfframe(10,25)(10,25){
    \begin{fmfgraph*}(100,40)
        \fmfleft{i1,i2}
        \fmfright{o1,o2}
        \fmf{fermion}{i2,v1,i1}
        \fmf{boson,label=$E^{(p)}$}{v1,v2}
        \fmf{fermion}{o2,v2,o1}
        \fmflabel{$q^{(m)}_R$}{i2}
        \fmflabel{$\bar{q}^{(n)}_R$}{i1}
        \fmflabel{$\bar{q'}^{(s)}_R$}{o2}
        \fmflabel{$q'^{(r)}_R$}{o1}
    \end{fmfgraph*}
    }
    \caption{$(C^{q q'}_{RR})^E_{nm;rs}$}
  \end{subfigure}
    \caption{The Feynman diagrams that give rise to the effective 4-fermion interactions discussed in Eq.(\ref{Vector4Fermion}). Here, $E^{(p)}$ denotes the $p^{th}$ KK mode of any electrically neutral vector gauge boson, namely the flavor gauge bosons, and the SM $Z$, $\gamma$, and gluon fields.}
    \label{fig:GaugeFeynmanDiagrams}
\end{figure}

We are particularly interested in the effect of flavor-changing processes in the low-energy limit, the effects of which should appear, for example, in flavor observables such as the $B^0-\bar{B^0}$ mass splittings. As such, we find it useful to compute low-energy effective four-fermion operators which emerge in the 4-dimensional theory from exchanges over entire towers of KK bosons. In this case, an exchange over the entire $E$ boson KK tower will contribute Hamiltonian interaction terms of the form, 
\begin{align}\label{Vector4Fermion}
    &\sim (C^{q q'}_{XY})^E_{nm;rs} (\bar{q}^{(n)}_X \gamma^\mu q^{(m)}_X)(\bar{q'}^{(r)}_Y\gamma_\mu q'^{(s)}_Y),
\end{align}
where
\begin{align}
    &(C^{q q'}_{XY})^E_{nm;rs}\equiv 2 \pi g_E^2 \int_{-\pi}^{\pi} d\phi_1 \, \int_{-\pi}^{\pi} d \phi_2 \, \bigg\{ \bigg( \sum_p \frac{\chi^E_p(\phi_1)\chi^E_p(\phi_2)}{(m^E_p)^2} \bigg) \nonumber \\
    &\times e^{(\sigma_1+\sigma_2)}[(\bm{\Omega}^E_X)^{(Q)}_{nm}(\phi_1)+(\bm{\Omega}^E_X)^{(q)}_{nm}(\phi_1)][(\bm{\Omega}^E_Y)^{(Q')}_{rs}(\phi_2)+(\bm{\Omega}^E_Y)^{(q')}_{rs}(\phi_2)]\bigg\}.
\end{align}
Here, $X$ and $Y$ denote chiralities (and so may take on values of $L$ or $R$), $q$ and $q'$ denote either up- or down-like quarks (that is, $q=u,d$ and $q'=u,d$), and $Q=U$ if $q=u$, and $Q=D$ if $q=d$ (with analogous relations holding between $Q'$ and $q'$). We remind the reader that $n$, $m$, $r$, and $s$ all denote different KK tower modes for the quark fields, with the SM-like quarks being given by tower indices 1, 2, and 3, while the index $p$ denotes the $p^{th}$ KK mode of the tower of fields from some gauge boson $E$. For the sake of clarity, we have also included the Feynman diagrams that depict the tree-level processes which contribute to the four-fermion interactions of Eq.(\ref{Vector4Fermion}) in Figure \ref{fig:GaugeFeynmanDiagrams}.
\begin{fmffile}{diagram1}
\end{fmffile}

Substituting in Eq.(\ref{FlavorGaugeSum}) for the summation over the gauge boson bulk profiles, we are then able to use Eq.(\ref{Vector4Fermion}) in order to compute any four-quark operator arising from the exchange of the entire tower of any species of gauge boson. In practice, we find that for the SM-like quark fields, the four-quark couplings relevant to our analysis are well-approximated by using the expressions of Eqs.(\ref{ZMAProfileSubstitutions}) and (\ref{ZMAMixingSubstitutions}) in lieu of exact expressions for the quark bulk profiles and mixing matrices, that is, working in the previously discussed zero-mass approximation, or ZMA. Using these approximations, it is possible to produce analytical expressions for the four-quark couplings, only resorting to numerical evaluation for the mixing matrices $\mathbf{U}^{u,d}_{L,R}$ in the ZMA. The precise form of these coupling expressions varies significantly depending on the gauge field $E$ under discussion, and as such, we now specialize our discussion to  specific gauge fields present in our model.

In this case, there are four distinct gauge fields, three for the $SU(2)_F$ symmetry (which we shall denote as $A^{1,2,3}_\mu$) and one for the $U(1)_F$ symmetry, which we shall denote as $B_\mu$. We find, using the angle parameters introduced in Section \ref{BulkMatterModelSection}, expressions for the various coupling matrices $\mathbf{X}^{U,u,D,d}_{A^1, A^2, A^3,B}$ to be
\begin{align}\label{FlavorGaugeXQ}
    &\mathbf{X}^{U,D}_{A^1} = \frac{1}{2}\begin{pmatrix}
    0 & c_Q & 0 \\
    c_Q & 0 & s_Q \\
    0 & s_Q & 0
    \end{pmatrix}, & \mathbf{X}^{U,D}_{A^2} = \frac{1}{2}\begin{pmatrix}
    0 & -i c_Q & 0 \\
    i c_Q & 0 & i s_Q \\
    0 & -i s_Q & 0
    \end{pmatrix}, \\
    &\mathbf{X}^{U,D}_{A^3} = \frac{1}{2}\begin{pmatrix}
    c_Q^2 & 0 & c_Q s_Q \\
    0 & -1 & 0 \\
    c_Q s_Q & 0 & s_Q^2
    \end{pmatrix}, & \mathbf{X}^{U,D}_{B} = \frac{1}{2}\begin{pmatrix}
    2-3 c_Q^2 & 0 & -3 s_Q c_Q \\
    0 & -1 & 0 \\
    -3 s_Q c_Q & 0 & -1+3c_Q^2
    \end{pmatrix} \nonumber
\end{align}
for the $SU(2)_L$ doublet quarks (at times, we shall find it easier to denote these as $\mathbf{X}^Q_{A^{1,2,3},B}$, since they are the same for both up- and downlike quarks),
\begin{align}\label{FlavorGaugeXd}
    &\mathbf{X}^{d}_{A^1} = \frac{1}{2}\begin{pmatrix}
    0 & c_d & 0 \\
    c_d & 0 & s_d \\
    0 & s_d & 0
    \end{pmatrix}, & \mathbf{X}^{d}_{A^2} = \frac{1}{2}\begin{pmatrix}
    0 & -i c_d & 0 \\
    i c_d & 0 & i s_d \\
    0 & -i s_d & 0
    \end{pmatrix}, \\
    &\mathbf{X}^{d}_{A^3} = \frac{1}{2}\begin{pmatrix}
    c_d^2 & 0 & c_d s_d \\
    0 & -1 & 0 \\
    c_d s_d & 0 & s_d^2
    \end{pmatrix}, & \mathbf{X}^{d}_{B} = \frac{1}{2}\begin{pmatrix}
    2-3 c_d^2 & 0 & -3 s_d c_d \\
    0 & -1 & 0 \\
    -3 s_d c_d & 0 & -1+3c_d^2
    \end{pmatrix} \nonumber
\end{align}
for the down-like $SU(2)_L$ singlet quarks, and
\begin{align}\label{FlavorGaugeXu}
    &\mathbf{X}^u_{A^1} = \frac{1}{2}\begin{pmatrix}
    2 s_H c_H c_u^2 & c_u (c_H^2-s_H^2) & 2 s_H c_H s_u c_u \\
    c_u (c_H^2-s_H^2) & -2 s_H c_H & s_u (c_H^2-s_H^2) \\
    2 s_H c_H s_u c_u & s_u(c_H^2-s_H^2) & 2 s_H c_H s_u^2
    \end{pmatrix}, & \mathbf{X}^{u}_{A^2} = \frac{1}{2}\begin{pmatrix}
    0 & -i c_u & 0 \\
    i c_u & 0 & i s_u \\
    0 & -i s_u & 0
    \end{pmatrix} \\
    &\mathbf{X}^{u}_{A^3} = \frac{1}{2}\begin{pmatrix}
    c_u^2(c_H^2-s_H^2) & -2 s_H c_H c_u & s_u c_u (c_H^2-s_H^2) \\
    -2 s_H c_H c_u & -(c_H^2-s_H^2) & -2 s_H c_H s_u \\
    s_u c_u (c_H^2-s_H^2) & -2 s_H c_H s_u & s_u^2 (c_H^2-s_H^2)
    \end{pmatrix}, & \mathbf{X}^{u}_B = \frac{1}{2} \begin{pmatrix}
    2-3 c_u^2 & 0 & -3 s_u c_u \\
    0 & -1 & 0 \\
    -3 s_u c_u & 0 & -1+ 3 c_u^2
    \end{pmatrix} \nonumber
\end{align}
for the up-like $SU(2)_L$ singlet quarks. Using these expressions, we can now compute the various terms $(\mathbf{\Omega}^{A^1,A^2,A^3,B}_{L,R})^{(U,u,D,d)}_{nm}$ which encapsulate the piece of the coefficient in Eq.(\ref{Vector4Fermion}) that depends on the bulk profiles.

At this point we can begin to elucidate the behavior of these couplings by working in the ZMA. To begin, we take Eq.(\ref{Vector4Fermion}) and make the substitutions outlined in Eqs.(\ref{ZMAProfileSubstitutions}) and (\ref{ZMAMixingSubstitutions}) for the bulk quark profiles. Omitting terms arising from the odd quark profiles $\mathbf{S}^{U,u,D,d}_n(\phi)$, which are suppressed by ratios of the squares of fermion masses to the KK scale $M_{KK}$ anyway, we find specifically that Eq.(\ref{Vector4Fermion}) suggests we must evaluate integrals of the form,
\begin{align}\label{ThetaFIntegral}
    \Theta_F(\eta_A,\eta_B,\eta_C,\eta_D,\delta) \equiv \int_0^{\pi} d \phi_1 \int_{0}^{\pi} d \phi_2 \, \frac{(kr_c)^2}{4}F(\eta_A)F(\eta_B)F(\eta_C)F(\eta_D)\epsilon^{\frac{1}{2}(\eta_A+\eta_B+\eta_C+\eta_D)} \\
    \times\bigg\{e^{(\sigma_1+\sigma_2)}e^{\frac{1}{2}(\eta_A+\eta_B-2)\sigma_1}e^{\frac{1}{2}(\eta_C+\eta_D-2)\sigma_2}\bigg( \sum_{n} \frac{\chi^{A^{1,2,3},B}_n(\phi_1)\chi^{A^{1,2,3},B}_n(\phi_2)}{(m^{A^{1,2,3},B}_n)^2} \bigg)\bigg\}, \nonumber
\end{align}
where the funciton $F(\eta)$ is first defined in Eq.(\ref{QuarkApproxProfiles}), and $\delta$ is the parameter (only influencing the sum over the gauge fields' KK modes) related to the gauge field's bulk mass, first defined in Eq.(\ref{VectorGaugeFlavorSolution}). We can now substitute Eq.(\ref{FlavorGaugeTowerExchange}), which gives an analytical expression for the sum over all KK modes for a vector gauge field with a bulk mass term, into Eq.(\ref{ThetaFIntegral}) and evaluate the integral in full, obtaining
\begin{align}\label{ThetaFDefinition}
    &\Theta_F (\eta_A,\eta_B,\eta_C,\eta_D, \delta) = \frac{kr_c}{4 M_{KK}^2 \sinh[kr_c \pi \delta]} \frac{F(\eta_A)F(\eta_B)F(\eta_C)F(\eta_D)}{(\tilde{\eta}_{AB}^2-4 \delta^2)(\tilde{\eta}_{CD}^2-4 \delta^2)(\delta^2-1)} \\
    &\times \bigg\{ 8(\cosh[kr_c \pi \delta]-\epsilon^{\frac{1}{2}\tilde{\eta}_{AB}}-\epsilon^{\frac{1}{2}\tilde{\eta}_{CD}}+ \cosh[kr_c \pi \delta]\epsilon^{\frac{1}{2}(\tilde{\eta}_{AB}+\tilde{\eta}_{CD})})(2+\tilde{\eta}_{AB})(2+\tilde{\eta}_{CD})\delta \nonumber\\
    &+4\sinh[kr_c \delta](-1+\epsilon^{\frac{1}{2}(\tilde{\eta}_{AB}+\tilde{\eta}_{CD})})\bigg(2 \tilde{\eta}_{AB} \tilde{\eta}_{CD}+4(2+\tilde{\eta}_{AB}+\tilde{\eta}_{CD})\delta^2 \nonumber\\
    &\qquad \qquad \qquad \phantom{+4\sinh[kr_c \delta](-1+\epsilon^{\frac{1}{2}(\tilde{\eta}_{AB}+\tilde{\eta}_{CD})})\bigg(} -\frac{(4\tilde{\eta}_{AB}\tilde{\eta}_{CD}+16 \delta^2)(-1+\delta^2)}{(\tilde{\eta}_{AB}+\tilde{\eta}_{CD})} \bigg)\bigg\}, \nonumber
\end{align}
where
\begin{align}
    &\tilde{\eta}_{AB,CD} \equiv 2+ \eta_{A,C} + \eta_{B,D}.
\end{align}
We can now use this function in our expressions for the various Wilson coefficients in Eq.(\ref{Vector4Fermion}). In general, we obtain
\begin{align}\label{VectorFlavor4FermionCoefficients}
    &(C^{q q'}_{LL})^{A^p,B}_{nm;rs} = 2 \pi g_{A,B}^2 \sum_{a,b,c,d=1}^{3} \Theta_F(\eta^Q_a, \eta^Q_b, \eta^Q_c, \eta^Q_d, \delta)(\mathbf{X}^{Q}_{A^{p},B})_{ab}(\mathbf{X}^{Q'}_{A^{p},B})_{cd}(\mathbf{U}^{q}_L)^\dagger_{na} (\mathbf{U}^{q}_L)_{bm} (\mathbf{U}^{q'}_L)^\dagger_{rc} (\mathbf{U}^{q'}_L)_{ds}, \nonumber\\
    &(C^{q q'}_{LR})^{A^p,B}_{nm;rs} = 2 \pi g_{A,B}^2 \sum_{a,b,c,d=1}^{3} \Theta_F(\eta^Q_a, \eta^Q_b, \eta^{q'}_c, \eta^{q'}_d, \delta)(\mathbf{X}^{Q}_{A^{p},B})_{ab}(\mathbf{X}^{q'}_{A^{p},B})_{cd}(\mathbf{U}^{q}_L)^\dagger_{na} (\mathbf{U}^{q}_L)_{bm} (\mathbf{U}^{q'}_R)^\dagger_{rc} (\mathbf{U}^{q'}_R)_{ds},\\
    &(C^{q q'}_{RL})^{A^p,B}_{nm;rs} = 2 \pi g_{A,B}^2 \sum_{a,b,c,d=1}^{3} \Theta_F(\eta^q_a, \eta^q_b, \eta^Q_c, \eta^Q_d, \delta)(\mathbf{X}^{q}_{A^{p},B})_{ab}(\mathbf{X}^{Q}_{A^{p},B})_{cd}(\mathbf{U}^{q}_R)^\dagger_{na} (\mathbf{U}^{q}_R)_{bm} (\mathbf{U}^{q'}_L)^\dagger_{rc} (\mathbf{U}^{q'}_L)_{ds},\nonumber \\
    &(C^{q q'}_{RR})^{A^p,B}_{nm;rs} = 2 \pi g_{A,B}^2 \sum_{a,b,c,d=1}^{3} \Theta_F(\eta^q_a, \eta^q_b, \eta^{q'}_c, \eta^{q'}_d, \delta)(\mathbf{X}^{q}_{A^{p},B})_{ab}(\mathbf{X}^{q'}_{A^{p},B})_{cd}(\mathbf{U}^{q}_R)^\dagger_{na} (\mathbf{U}^{q}_R)_{bm} (\mathbf{U}^{q'}_R)^\dagger_{rc} (\mathbf{U}^{q'}_R)_{ds},\nonumber,
\end{align}
with
\begin{align}
    &\delta \equiv \begin{cases}
    \sqrt{1+\frac{g_A^2 v_F^2}{2 k^2}}, & A^p \\
    \sqrt{1+\frac{(-3 g_B)^2 v_F^2}{2 k^2}}, & B
    \end{cases}.
\end{align}
In Eq.(\ref{VectorFlavor4FermionCoefficients}), $q$ and $q'$ can refer either to up-like ($u$) or down-like ($d$) quarks. We also remind the reader that the matrices $\mathbf{U}^{q,q'}_{L,R}$ are the unitary matrices that diagonalize the ZMA quark mass matrices, as first defined in Eq.(\ref{UnitaryMatrices}).

\begin{figure}
    \centerline{\includegraphics[width=3.5in]{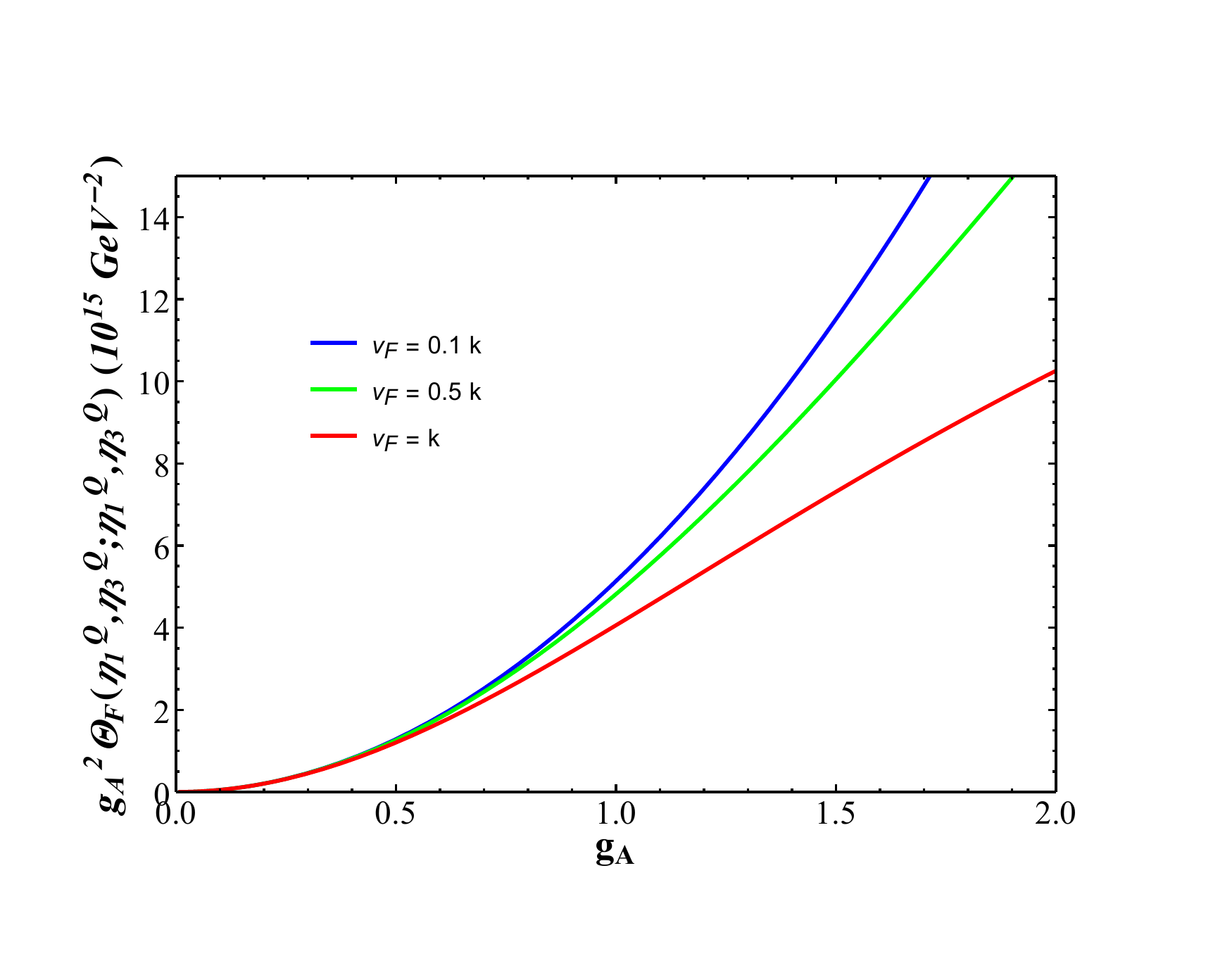}
    \hspace{-0.75cm}
    \includegraphics[width=3.5in]{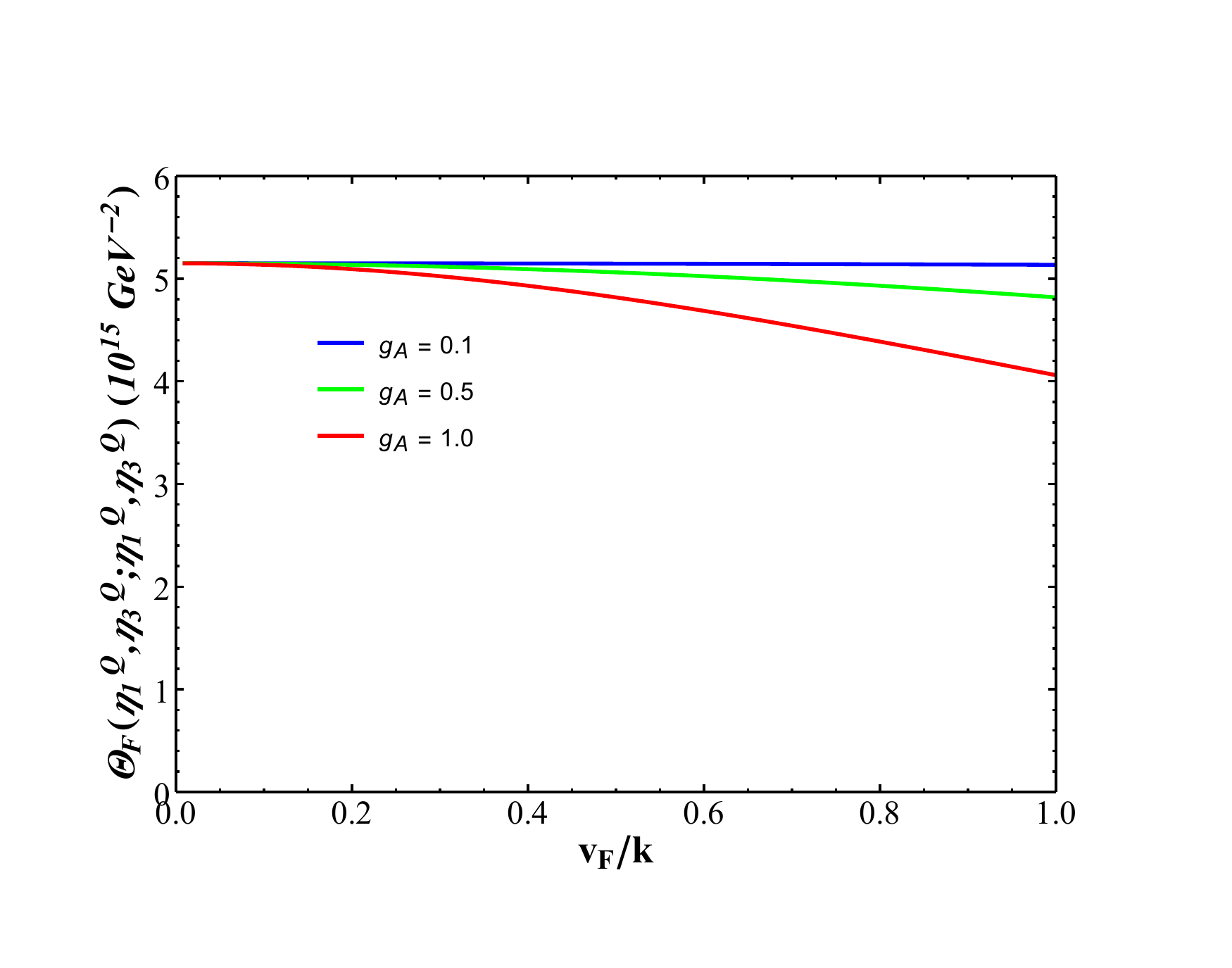}}
    \caption{Dependence of $g_A^2 \Theta_F (\eta^Q_1,\eta^Q_3,\eta^Q_1,\eta^Q_3)$ on $g_A$ (left) and dependence of $\Theta_F (\eta^Q_1,\eta^Q_3,\eta^Q_1,\eta^Q_3)$ on $v_F$ (right), for a sample point in parameter space}
    \label{fig:ThetaFVariationPlots}
\end{figure}

It should be noted that the coefficients in Eq.(\ref{VectorFlavor4FermionCoefficients}) display somewhat non-trivial behavior as functions of the coupling constant $g_A$ and the vacuum expectation value parameter $v_F$, due to the dependence of $\Theta_F$ on the parameter $\delta$. Inspection of the functional form of $\Theta_F$ allows us to make several observations on this behavior-- in particular, that $\Theta_F$ will generally decrease in magnitude with increasing $\delta$. Physically, this can be thought of as a consequence of increasing the gauge field's bulk mass, increasing the masses of the gauge field's KK modes and decreasing each field's contribution to a low-energy 4-fermion operator. So, since $\delta$ increases with an increase in the coupling strength $g_A$, we anticipate that the 4-fermion coefficients in Eq.(\ref{VectorFlavor4FermionCoefficients}) will not simply increase quadratically with increasing $g_A$, but rather have this increase at least partially offset by decreases in $\Theta_F$. In particular, for large $\delta$ (corresponding to large $g_A$ and/or large $v_F$), we see that $\Theta_F \sim \delta^{-2} \sim g_A^{-2} (v_F/k)^{-2}$, indicating that for sufficiently large $g_A$, the coefficients in Eq.(\ref{VectorFlavor4FermionCoefficients}) should actually become asymptotically flat. Numerically, the coupling strength at which these coefficients begin to exhibit this behavior varies significantly depending on the value of $v_F$-- the lower $v_F$, the larger the value $g_A$ must be before the 4-fermion coefficients effectively stop increasing in magnitude with increasing $g_A$; in practice, for $v_F < k$, we generally do not observe the asymptotic behavior even for $g_A^2/(4 \pi) \sim \alpha_s(m_Z)$, namely, even when we make the coupling constant for the flavor gauge bosons comparable to that of the strong force. In Figure \ref{fig:ThetaFVariationPlots}, we plot the dependence of $g_A^2 \Theta_F(\eta^Q_1,\eta^Q_3,\eta^Q_1,\eta^Q_3)$ (relevant in, for example, computations of the flavor gauge boson contribution to $\bar{B}^{0}-B^{0}$ mixing parameters) on the coupling $g_A$ for a particular point in our model's parameter space at different values of the bulk vacuum expectation value $v_F$, as well as the dependence of $\Theta_F(\eta^Q_1,\eta^Q_3,\eta^Q_1,\eta^Q_3)$ (note the lack of a $g_A^2$ factor) on $v_F$ for various choices of $g_A$.

Notably, for the $A^{1,2,3}_\mu$, the asymptotic behavior of the coupling with increasing $g_A$ isn't realistically achievable for $O(1)$ $g_A$ and $v_F$, however, the increase of $g_A^2 \Theta_F$ with increasing $g_A$ is significantly affected by the value of $v_F$; the larger $v_F$, the less $g_A^2 \Theta_F$ increases with increasing $g_A$. For $B$ exchanges, the beginnings of asymptotically flat behavior for $g_B^2 \Theta_F$ is even visible, due to the enhancement to the $B$ bulk mass from the $-\frac{3}{2}$ $U(1)_F$ charge of the bulk scalars. Similarly, we note that for even modest $g_A$, such as $g_A \sim 0.5$, $g_A^2 \Theta_F$ falls significantly with increasing $v_F/k$; for $g_A=1.0$, for example, $g_A^2 \Theta_F$ with $v_F/k=1$ can be as much as 30\% smaller than $g_A^2 \Theta_F$ when $v_F=0.5$.

\subsection{Four-Quark Operators: $H_{1-4}$ Exchanges}\label{FlavorScalar4FermionSection}
Having discussed the effects of the exchange of gauge bosons in our model, we continue our discussion of four-quark operators with those that arise from exchanges of the flavor scalar particles, $H_{1-4}$, with the particles defined in the mass eigenstate basis of Eq.(\ref{HBulkMasses}). Inspecting the action, we note that these fields have Yukawa interactions with the quarks in the 4-dimensional effective theory of the form,
\begin{align}
    &\int d^4 x \sqrt{2 \pi} \int_{-\pi}^{\pi} d \phi \chi^{H_i}_p(\phi) \bigg\{ \\
    &\bigg( (\Vec{a}^{(U)}_n)^\dagger \mathbf{C}^{(U)}_n(\phi) \mathbf{X}^{U}_{H_i} \mathbf{S}^{(U)}_m(\phi) \Vec{a}^{(U)}_m + (\Vec{a}^{(u)}_n)^\dagger \mathbf{S}^{(u)}_n(\phi) \mathbf{X}^{u}_{H_i} \mathbf{C}^{(u)}_m(\phi) \Vec{a}^{(u)}_m \bigg)(\bar{u}^{(n)}_L H^{(p)}_i u^{(m)}_R)+h.c. \bigg\}, \nonumber
\end{align}
with an analogous expression for down-like quarks. Note that here, unlike for the gauge coupling constant $g_E$ in Eq.(\ref{GaugeKKCouplingTerms}), the the Yukawa coupling constants $y^{Q,d}$ and $y^{u}_{1,2}$ are incorporated into the  bulk coupling matrices $\mathbf{X}^{U,u,D,d}_{H_i}$; this is simply a matter of notational convenience. By inspecting the action, we find the coupling matrices are given by
\begin{align}\label{XQHi}
    &\mathbf{X}^{U,D}_{H_1} =\frac{k c_Q s_Q (\eta^Q_3-\eta^Q_1)}{2 v_F} \begin{pmatrix}
    0 & -s_Q & 0 \\
    -s_Q & 0 & c_Q \\
    0 & c_Q & 0
    \end{pmatrix}, \nonumber\\
    &\mathbf{X}^{U,D}_{H_2} =\frac{i k c_Q s_Q (\eta^Q_3-\eta^Q_1)}{2 v_F} \begin{pmatrix}
    0 & -s_Q & 0 \\
    s_Q & 0 & -c_Q \\
    0 & c_Q & 0
    \end{pmatrix}, \\
    &\mathbf{X}^{U,D}_{H_3} =\frac{k c_Q s_Q (\eta^Q_3-\eta^Q_1)}{2 v_F} \begin{pmatrix}
    2 c_Q s_Q & 0 & -(c_Q^2-s_Q^2) \\
    0 & 0 & 0 \\
    -(c_Q^2-s_Q^2) & 0 & -2 c_Q s_Q
    \end{pmatrix}, \nonumber\\
    &\mathbf{X}^{U,D}_{H_4} =\frac{k c_Q s_Q (\eta^Q_3-\eta^Q_1)}{2 v_F} \begin{pmatrix}
    -2 c_Q s_Q & 0 & (c_Q^2-s_Q^2) \\
    0 & 0 & 0 \\
    (c_Q^2-s_Q^2) & 0 & 2 c_Q s_Q
    \end{pmatrix}\nonumber
\end{align}
for the $SU(2)_L$ doublet quarks,
\begin{align}\label{XdHi}
    &\mathbf{X}^{d}_{H_1} = \frac{k c_d s_d(\eta^d_3-\eta^d_1)}{2 v_F}\begin{pmatrix}
    0 & s_d & 0 \\
    s_d & 0 & -c_d \\
    0 & -c_d & 0
    \end{pmatrix}, \nonumber \\
    &\mathbf{X}^{d}_{H_2} = \frac{i k c_d s_d(\eta^d_3-\eta^d_1)}{2 v_F}\begin{pmatrix}
    0 & s_d & 0 \\
    -s_d & 0 & c_d \\
    0 & -c_d & 0
    \end{pmatrix}, \\
    &\mathbf{X}^{d}_{H_3} = \frac{k c_d s_d(\eta^d_3-\eta^d_1)}{2 v_F}\begin{pmatrix}
    -2 c_d s_d & 0 & (c_d^2-s_d^2) \\
    0 & 0 & 0 \\
    (c_d^2-s_d^2) & 0 & 2 c_d s_d
    \end{pmatrix}, \nonumber \\
    &\mathbf{X}^{d}_{H_4} = \frac{k c_d s_d(\eta^d_3-\eta^d_1)}{2 v_F}\begin{pmatrix}
    2 c_d s_d & 0 & -(c_d^2-s_d^2) \\
    0 & 0 & 0 \\
    -(c_d^2-s_d^2) & 0 & -2 c_d s_d
    \end{pmatrix}, \nonumber
\end{align}
for the down-like $SU(2)_L$ singlet quarks, and
\begin{align}\label{XuHi}
    &\mathbf{X}^{u}_{H_1} = \frac{k c_u s_u(\eta^u_3-\eta^u_1)}{2 v_F}\begin{pmatrix}
    4 c_H s_H c_u s_u & s_u (c_H^2-s_H^2) & -2 c_H s_H (c_u^2-s_u^2) \\
    s_u (c_H^2-s_H^2) & 0 & -c_u (c_H^2-s_H^2)\\
    -2 c_H s_H (c_u^2-s_u^2) & -c_u(c_H^2-s_H^2) & -4 c_H s_H c_u s_u 
    \end{pmatrix}, \nonumber\\
    &\mathbf{X}^{u}_{H_2} = \frac{ik c_u s_u(\eta^u_3-\eta^u_1)}{2 v_F}\begin{pmatrix}
    0 & s_u & 0 \\
    -s_u & 0 & c_u\\
    0 & -c_u & 0 
    \end{pmatrix},\\
    &\mathbf{X}^{u}_{H_3} = \frac{k c_u s_u}{2 v_F} \begin{pmatrix}
    -2 (c_H^2-s_H^2)c_u s_u & 2 c_H s_H s_u & (c_H^2-s_H^2)(c_u^2-s_u^2) \\
    2 c_H s_H s_u & 0 & -2 c_H s_H c_u \\
    (c_H^2-s_H^2)(c_u^2-s_u^2) & -2 c_H s_H c_u & 2 (c_H^2-s_H^2) c_u s_u
    \end{pmatrix}, \nonumber \\
    &\mathbf{X}^{u}_{H_4} = \frac{k c_u s_u (\eta^u_3-\eta^u_1)}{2 v_F} \begin{pmatrix}
    2 c_u s_u & 0 & -(c_u^2-s_u^2) \\
    0 & 0 & 0 \\
    -(c_u^2-s_u^2) & 0 & -2 c_u s_u
    \end{pmatrix} \nonumber
\end{align}
for up-like $SU(2)_L$ singlet quarks. Notably, in Eqs.(\ref{XQHi}-\ref{XuHi}), we have used expressions for the bulk Yukawa couplings $y^{Q,d}$ and $y^{u}_{1,2}$ in terms of our bulk localization parameters and rotation angles $\eta^{Q,u,d}_{1,2,3}$ and $\theta_{H,Q,u,d}$, given originally in Eq.(\ref{BulkYukawaExpressions}). We also note that while $H_1$, $H_3$, and $H_4$ have ordinary scalar couplings to quarks, the field $H_2$, being $CP$-odd, has a pseudoscalar coupling to fermions.

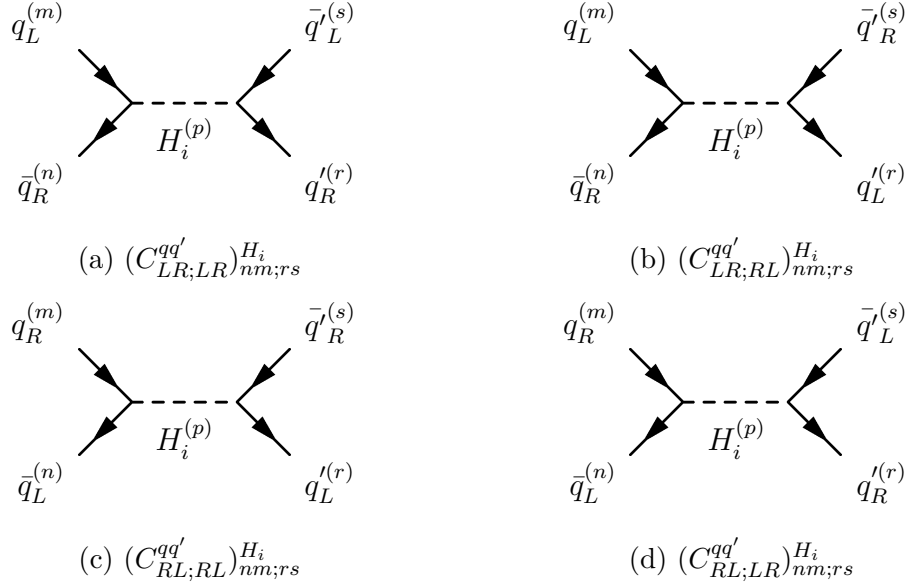
\begin{figure}
  \centering
  \begin{subfigure}[b]{0.45\textwidth}
    \centering
    \fmfframe(10,25)(10,25){
    \begin{fmfgraph*}(100,40)
        \fmfleft{i1,i2}
        \fmfright{o1,o2}
        \fmf{fermion}{i2,v1,i1}
        \fmf{dashes,label=$H_i^{(p)}$}{v1,v2}
        \fmf{fermion}{o2,v2,o1}
        \fmflabel{$q^{(m)}_L$}{i2}
        \fmflabel{$\bar{q}^{(n)}_R$}{i1}
        \fmflabel{$\bar{q'}^{(s)}_L$}{o2}
        \fmflabel{$q'^{(r)}_R$}{o1}
    \end{fmfgraph*}
    }
    \caption{$(C^{qq'}_{LR;LR})^{H_i}_{nm;rs}$}
  \end{subfigure}
  \begin{subfigure}[b]{0.45\textwidth}
    \centering
    \fmfframe(10,25)(10,25){
    \begin{fmfgraph*}(100,40)
        \fmfleft{i1,i2}
        \fmfright{o1,o2}
        \fmf{fermion}{i2,v1,i1}
        \fmf{dashes,label=$H_i^{(p)}$}{v1,v2}
        \fmf{fermion}{o2,v2,o1}
        \fmflabel{$q^{(m)}_L$}{i2}
        \fmflabel{$\bar{q}^{(n)}_R$}{i1}
        \fmflabel{$\bar{q'}^{(s)}_R$}{o2}
        \fmflabel{$q'^{(r)}_L$}{o1}
    \end{fmfgraph*}
    }
    \caption{$(C^{qq'}_{LR;RL})^{H_i}_{nm;rs}$}
  \end{subfigure}
  \begin{subfigure}[b]{0.45\textwidth}
    \centering
    \fmfframe(10,25)(10,25){
    \begin{fmfgraph*}(100,40)
        \fmfleft{i1,i2}
        \fmfright{o1,o2}
        \fmf{fermion}{i2,v1,i1}
        \fmf{dashes,label=$H_i^{(p)}$}{v1,v2}
        \fmf{fermion}{o2,v2,o1}
        \fmflabel{$q^{(m)}_R$}{i2}
        \fmflabel{$\bar{q}^{(n)}_L$}{i1}
        \fmflabel{$\bar{q'}^{(s)}_R$}{o2}
        \fmflabel{$q'^{(r)}_L$}{o1}
    \end{fmfgraph*}
    }
    \caption{$(C^{qq'}_{RL;RL})^{H_i}_{nm;rs}$}
  \end{subfigure}
  \begin{subfigure}[b]{0.45\textwidth}
    \centering
    \fmfframe(10,25)(10,25){
    \begin{fmfgraph*}(100,40)
        \fmfleft{i1,i2}
        \fmfright{o1,o2}
        \fmf{fermion}{i2,v1,i1}
        \fmf{dashes,label=$H_i^{(p)}$}{v1,v2}
        \fmf{fermion}{o2,v2,o1}
        \fmflabel{$q^{(m)}_R$}{i2}
        \fmflabel{$\bar{q}^{(n)}_L$}{i1}
        \fmflabel{$\bar{q'}^{(s)}_L$}{o2}
        \fmflabel{$q'^{(r)}_R$}{o1}
    \end{fmfgraph*}
    }
    \caption{$(C^{qq'}_{RL;LR})^{H_i}_{nm;rs}$}
  \end{subfigure}
    \caption{
    The Feynman diagrams that give rise to the effective 4-fermion interactions discussed in Eq.(\ref{HVertices}). Here, $H_i^{(p)}$ denotes the $p^{th}$ KK mode of the scalar $H_i$ bosons, however, with the substitution $H^{(p)}_i \rightarrow a^{(p)}_i$, these diagrams also depict the exchanges which give rise to the interactions discussed in Eq.(\ref{aVertices}) from exchanges of the physical pseudoscalar particles discussed in Section \ref{FlavorGoldstone4FermionSection}.}
    \label{fig:HFeynmanDiagrams}
\end{figure}

Similarly to Section \ref{FlavorGauge4FermionSection}, we now find it convenient to define quantities
\begin{align}\label{ScalarOmega}
    &(\mathbf{\Omega}^{H_i}_{LR})^{(U,D)}_{nm}(\phi) \equiv (\Vec{a}^{(U,D)}_n)^{\dagger}\mathbf{C}^{(U,D)}_n (\phi) \mathbf{X}^{U,D}_{H_i} \mathbf{S}^{(U,D)}_m(\phi) \Vec{a}^{(U,D)}_m, \nonumber\\
    &(\mathbf{\Omega}^{H_i}_{LR})^{(u,d)}_{nm}(\phi) \equiv (\Vec{a}^{(u,d)}_n)^{\dagger}\mathbf{S}^{(u,d)}_n (\phi) \mathbf{X}^{u,d}_{H_i} \mathbf{C}^{(u,d)}_m(\phi) \Vec{a}^{(u,d)}_m, \\
    &(\mathbf{\Omega}^{H_i}_{RL})^{(U,u,D,d)}_{nm}(\phi) = (\mathbf{\Omega}^{H_i}_{LR})^{(U,u,D,d)*}_{mn}(\phi). \nonumber
\end{align}
The four-fermion operators emerging in the Hamiltonian from these scalar exchanges then have the form,
\begin{align}\label{HVertices}
    &(C^{qq'}_{LR;LR})^{H_i}_{nm;rs} (\bar{q}^{(n)}_L q^{(m)}_R)((\bar{q'}^{(r)}_L q'^{(s)}_R)), \nonumber \\
    &(C^{qq'}_{LR;RL})^{H_i}_{nm;rs} (\bar{q}^{(n)}_L q^{(m)}_R)((\bar{q'}^{(r)}_R q'^{(s)}_L)), \\
    &(C^{qq'}_{RL;RL})^{H_i}_{nm;rs} (\bar{q}^{(n)}_R q^{(m)}_L)((\bar{q'}^{(r)}_R q'^{(s)}_L)), \nonumber \\
    &(C^{qq'}_{RL;LR})^{H_i}_{nm;rs} (\bar{q}^{(n)}_R q^{(m)}_L)((\bar{q'}^{(r)}_L q'^{(s)}_R)), \nonumber
\end{align}
where $q$ and $q'$ can denote up-like ($u$) or down-like ($d$) quarks. As in Section \ref{FlavorGauge4FermionSection}, we depict the Feynman diagrams that contribute to each of these vertices in Figure \ref{fig:HFeynmanDiagrams}, noting that these diagrams are also valid for the other scalar exchanges in our model, those of the new physical pseudoscalars $a_i$, which we shall discuss in Section \ref{FlavorGoldstone4FermionSection}.

\begin{fmffile}{diagram2}
\end{fmffile}

The value of the coefficients in Eq.(\ref{HVertices}) are given by
\begin{align}\label{FlavorScalar4FermionCoefficients}
    (C^{qq'}_{LR;LR})^{H_i}_{nm;rs} &= (2 \pi) \int_{-\pi}^{\pi} \int_{-\pi}^{\pi} d\phi_1 d\phi_2 \bigg\{ \bigg( \sum_{p} \frac{\chi^{H_i}_p(\phi_1)\chi^{H_i}_p(\phi_2)}{(m^{H_i}_p)^2} \bigg) \nonumber\\
    &\times[(\mathbf{\Omega}^{H_i}_{LR})^{(Q)}_{nm}(\phi_1)+(\mathbf{\Omega}^{H_i}_{LR})^{(q)}_{nm}(\phi_1)][(\mathbf{\Omega}^{H_i}_{LR})^{(Q')}_{rs}(\phi_2)+(\mathbf{\Omega}^{H_i}_{LR})^{(q')}_{rs}(\phi_2)] \bigg\}, \nonumber \\
    (C^{qq'}_{LR;RL})^{H_i}_{nm;rs} &= (2 \pi) \int_{-\pi}^{\pi} \int_{-\pi}^{\pi} d\phi_1 d\phi_2 \bigg\{ \bigg( \sum_{p} \frac{\chi^{H_i}_p(\phi_1)\chi^{H_i}_p(\phi_2)}{(m^{H_i}_p)^2} \bigg)\\
    &\times[(\mathbf{\Omega}^{H_i}_{LR})^{(Q)}_{nm}(\phi_1)+(\mathbf{\Omega}^{H_i}_{LR})^{(q)}_{nm}(\phi_1)][(\mathbf{\Omega}^{H_i}_{RL})^{(Q')}_{rs}(\phi_2)+(\mathbf{\Omega}^{H_i}_{RL})^{(q')}_{rs}(\phi_2)] \bigg\}, \nonumber
\end{align}
with
\begin{align}
    (C^{qq'}_{RL;RL})^{H_i}_{nm;rs} &= (C^{qq'}_{LR;LR})^{H_i*}_{mn;sr}\\
    (C^{qq'}_{RL;LR})^{H_i}_{nm;rs} &= (C^{qq'}_{LR;RL})^{H_i*}_{mn;sr}. \nonumber
\end{align}
We can evaluate these integrals by inserting Eq.(\ref{FlavorScalarTowerExchange}) into Eq.(\ref{FlavorScalar4FermionCoefficients}). Unlike our treatment of the operator coefficients in Section \ref{FlavorGauge4FermionSection}, we find that numerically, the integrals in Eq.(\ref{FlavorScalar4FermionCoefficients}) \textit{cannot} be well-approximated by making the substitutions of Eqs.(\ref{ZMAProfileSubstitutions}) and (\ref{ZMAMixingSubstitutions}) for the quark bulk profiles; this is likely due to the presence of additional uncontrolled terms in the low-mass expansion of Eq.(\ref{ZMAProfileSubstitutions}) which we have neglected, as discussed briefly in Section \ref{BulkFermionSection}. However, it is still possible to evaluate these integrals numerically using the exact expressions for the quark bulk profiles. It should be noted that because the bulk masses of the fields $H_{1-4}$, given in Eq.(\ref{HBulkMasses}), are sensitive to the bulk scalar potential parameters $\lambda_{1,3,4,5}$ in Eq.(\ref{BulkV}), in order to perform numerical computations of these couplings we must specify these parameters. This can be straightforwardly accomplished by simply randomly generating a set of $\lambda_{1,3,4,5}$ parameters that satisfy the minimization conditions discussed in Section \ref{BulkScalarModelSection}; for definiteness, we specify
\begin{align}\label{SampleLambdas}
    &\lambda_1 = 0.925071, &\lambda_3 = 0.196495, \;\;\;\;\;\;\;&\lambda_4 = 0.161396, & \lambda_5 = -0.093493.&
\end{align}
This selection is purely arbitrary; to arrive at it (and generate other selections for testing), we first randomly generated $\lambda_1$ and $\lambda_4$ values, both between 0 and 2; we arrive at this range because the squared bulk masses of the scalars given in Eq.(\ref{HBulkMasses}) are only positive if $\lambda_1$ and $\lambda_4$ are both positive, and we assume they do not substantially exceed $O(1)$. Then, we randomly generated $\lambda_3 \in [-2 \lambda_1, 2 \lambda_1]$ and $\lambda_5 \in [-\lambda_4,\lambda_4]$, with these parameters bounds emerging again from the requirement that the squared bulk masses in Eq.(\ref{HBulkMasses}) be positive. We shall see that for most observables we consider, the effect of scalar exchanges is negligible, and therefore making a specific choice for the bulk scalar potential parameters has no impact on our results. For the case of $\bar{K}^{0}-K^{0}$ mixing, the only process we consider for which the scalar exchanges represent a significant contribution, we find that altering our selection of $\lambda_{1,3,4,5}$ only affects the dominant Wilson coefficients here by $\sim5-10\%$, resulting in no qualitative changes in the results of our analysis from using different $\lambda_{1,3,4,5}$. We note that some Wilson coefficients undergo greater variations (some on the level of $O(100\%)$ corrections) with different $\lambda_{1,3,4,5}$ selections, however these coefficients are invariably numerically insignificant in any of our computations of observables. As such, we make the selection of Eq.(\ref{SampleLambdas}) for all subsequent computations in this work, and do not perform a detailed exploration of scenarios with different values for these parameters. 

\begin{figure}
    \centering
    \includegraphics{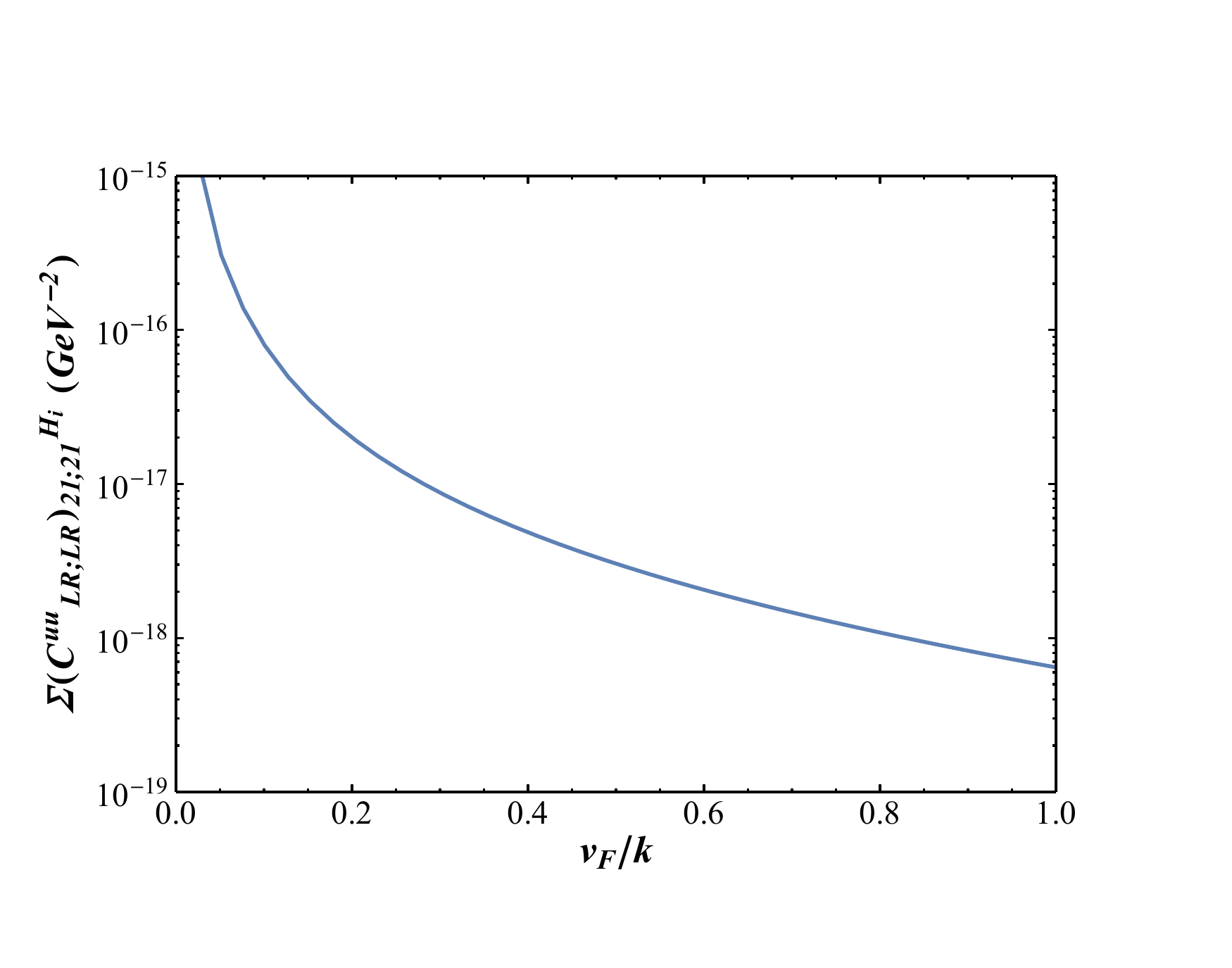}
    \caption{At a sample point in parameter space, the Wilson coefficient $\sum_{i=1}^4 (C^{uu}_{LR;LR})^{H_i}_{21;21}$, in units of $\textrm{GeV}^{-2}$.}
    \label{fig:CHuuLRuuLR}
\end{figure}

In Figure \ref{fig:CHuuLRuuLR}, we plot one flavor-changing coupling arising from scalar exchanges for a benchmark point (all other points we have sampled represent similar behavior), at different values of the bulk scalar vacuum expectation value $v_F$. Notably, the magnitude of this coupling (and others) are strongly dependent on $v_F$; as $v_F$ decreases, the 4-quark coefficients arising from exchanges of $H_{1-4}$ increase substantially; this can be intuitively understood by noting that Eq.(\ref{FlavorScalarTowerExchange}) has a rough dependence $\sim v_F^{-3}$, stemming partially from geometry and partially from the fact that a smaller $v_F$ will yield lower masses for the KK scalars, while the coupling matrices in Eqs.(\ref{XQHi}-\ref{XuHi}) add a further $\sim v_F^{-1}$ dependence to the final coupling, since the products of the bulk Yukawa couplings and the vacuum expectation value $v_F$ are fixed by the need to recreate the observed quark masses and mixings.

It should also be noted that for the light SM-like quarks, the 4-quark operators arising from the exchange of these scalars generically undergo a substantial suppression relative to those arising from the exchange of vector bosons. This can be intuitively understood by examining the form of the bulk fermion profiles in Eq.(\ref{ScalarOmega}); for the SM-like quarks, the orbifold odd profiles $\mathbf{S}^{(U,u,D,d)}_n(\phi)$ are generically approximately proportional to the ratios of the masses of these quarks to the KK scale $M_{KK}$, as can be seen in Eq.(\ref{ZMAProfileSubstitutions}) (note that although we have abandoned the ZMA here, this argument based on a ZMA profile does numerically hold in general). Because the bulk scalars $H_{1-4}$ already have their lowest-lying KK modes at masses of $O(M_{KK})$, we have that \textit{any} 4-quark operator arising from the exchange of these scalars should have a suppression of $\sim (1/M_{KK}^{2} )(m_{q_1} m_{q_2}/M_{KK}^2)$ (where $m_{q_1}$ and $m_{q_2}$ are some pair of SM quark masses), that is, \textit{the exchange of bulk scalars produces 4-quark operators that are suppressed by an additional powers of} $M_{KK}$ \textit{compared to those arising from bulk vector boson exchange.} In practice, then, we find that we can ignore the contribution of the scalars $H_{1-4}$ in most computations of tree-level flavor-changing effects; we shall discuss the sole exception to this rule of thumb in Section \ref{KKBarMixingSection}.

Before moving on, it may be of interest, given our discussion in Section \ref{BulkMatterModelSection} on the possibility that the fermion bulk masses might be the consequence of $Z_2$-odd bulk scalar vevs, which would necessarily not have a flat profile in the bulk, to remark on whether such a setup would still exhibit the substantial suppression of scalar-mediated flavor changing currents found in our model. While a detailed numerical exploration of such a modified model is not the purpose of this work, we can at least note that the principles underpinning our intuitive explanation of this suppression still hold in the case of non-constant fermion bulk masses. In particular, even for non-flat bulk masses, the $Z_2$-odd bulk fermion profiles for the SM-like quarks still vanish in the zero-mode approximation (when brane-localized mass terms arising from the SM Higgs vev are ignored) \cite{ahmed}. Hence, we can expect that once the brane masses are included in the calculation, the results will be analogous to the flat bulk mass case: The lightest $Z_2$-odd bulk fermion profiles will still bear a significant suppression factor roughly proportional to the ratio of the SM quark mass to the KK mass. Therefore, although we caution that a numerical exploration of this case would be needed to confirm any expectations, we anticipate based on intuitive arguments that a model in which our even scalars with flat vevs are replaced by odd scalars with non-constant vevs will exhibit similarly large suppression of any scalar-mediated 4-quark operators among SM-like quarks. This is in agreement with the results of \cite{ahmed}, in which the authors note similar behavior for new four-quark couplings arising from the gauge singlet odd bulk scalars in their own model.

\subsection{Four-Quark Operators: $a_{1-4}$ Exchanges}\label{FlavorGoldstone4FermionSection}
\begin{figure}
    \centering
    \includegraphics{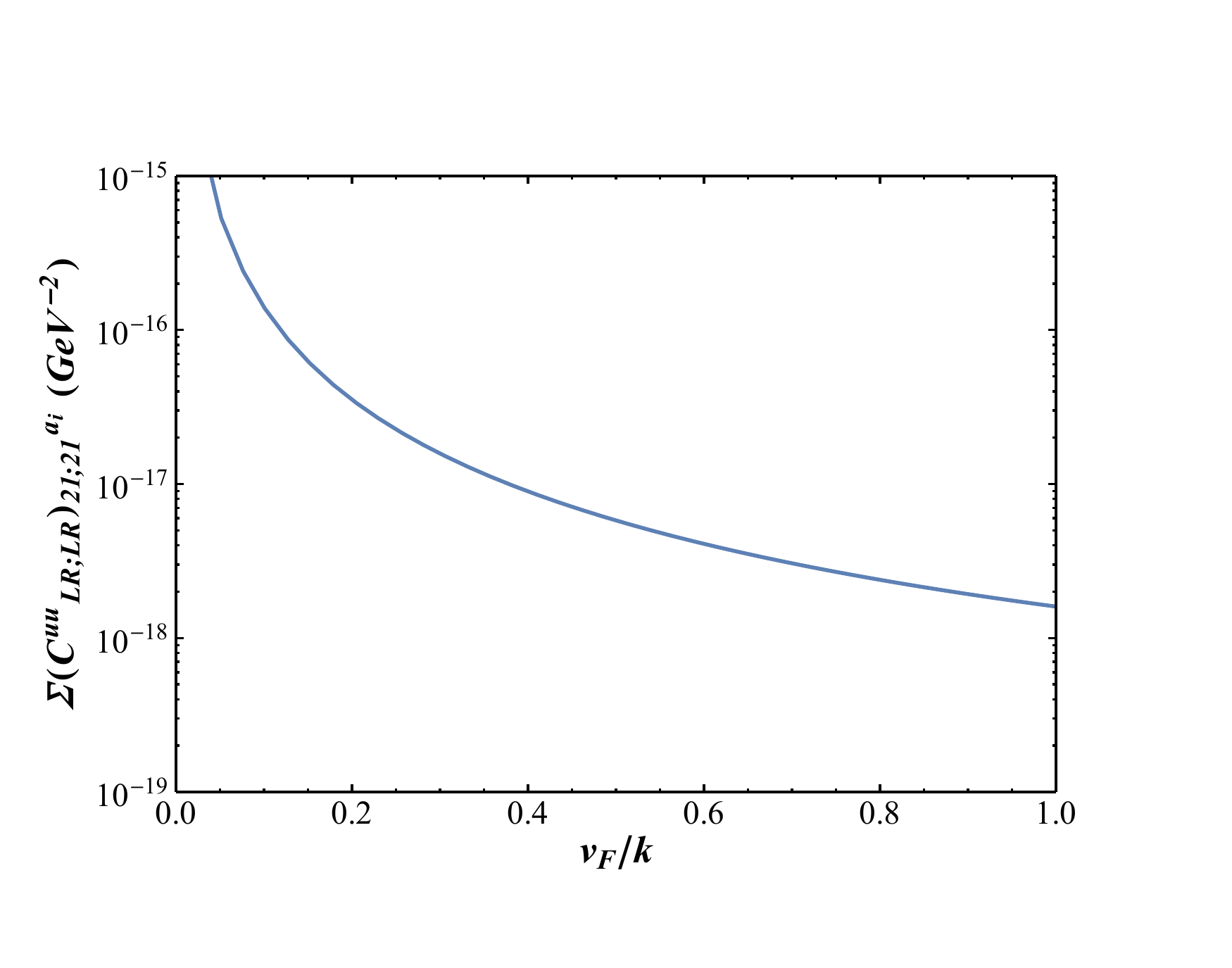}
    \caption{At a sample point in parameter space, the Wilson coefficient $\sum_{i=1}^4 (C^{uu}_{LR;LR})^{a_i}_{21;21}$ defined in Eq.(\ref{aVertices}), in units of $\textrm{GeV}^{-2}$.}
    \label{fig:CauuLRuuLR}
\end{figure}

We now discuss the four-quark operators arising from the other set of new scalars arising in our theory, namely those stemming from the orbifold-odd scalar fields $a_{1-4}$ that emerge as mixtures of  the components of the bulk scalars that lack a bulk mass and the fifth components of the flavor gauge fields. Because these fields are mixtures of bulk scalars and bulk gauge fields, the expressions for their coupling terms to quarks are dramatically more complicated than those of other sectors of the theory, consisting of both gauge couplings from the fifth components of the bulk gauge fields and bulk Yukawa couplings from the bulk scalars. Before beginning, it is useful to introduce some notation. First, throughout this section we shall denote the scalar fields $H_{5-8}$ (in the basis defined in Eq.(\ref{HGoldstones})), which remain without a bulk mass after bulk SSB and mix with the bulk gauge field components $A^{1-3}_\phi$ and $B_\phi$, by the gauge field with which they mix. So, we denote $H_5$ by $B_H$, $H_6$ by $A^3_H$, $H_7$ by $A^2_H$, and $H_8$ by $A^1_H$. Each of these pairs of fields corresponds to a single \textit{physical} scalar field $a_{1-4}$; we denote the $A^1_{\phi}-A^1_H$ mixture by $a_1$, the $A^2_{\phi}-A^2_H$ mixture by $a_2$, the $A^3_\phi-A^3_H$ mixture by $a_3$, and the $B_\phi-B_H$ mixture by $a_4$. Furthermore, in writing the coupling expressions, we find it convenient to define
\begin{align}
    &(\mathbf{\Omega}^{A^i_\phi,B_\phi}_{LR})_{nm}^{(U,D)}(\phi) \equiv (\Vec{a}^{(U,D)}_n)^\dagger \mathbf{C}^{(U,D)}_n (\phi) (i \mathbf{X}^{U,D}_{A^i,B} )\mathbf{S}^{(U,D)}_m \Vec{a}^{(U,D)}_m, \nonumber \\
    &(\mathbf{\Omega}^{A^i_H,B_H}_{LR})_{nm}^{(U,D)}(\phi) \equiv (\Vec{a}^{(U,D)}_n)^\dagger \mathbf{C}^{(U,D)}_n (\phi) (\mathbf{X}^{U,D}_{A^i_H, B_H})\mathbf{S}^{(U,D)}_m \Vec{a}^{(U,D)}_m, \\
    &(\mathbf{\Omega}^{A^i_\phi,B_\phi}_{LR})_{nm}^{(u,d)}(\phi) \equiv (\Vec{a}^{(u,d)}_n)^\dagger \mathbf{S}^{(u,d)}_n (\phi) (i \mathbf{X}^{u,d}_{A^{i}}) \mathbf{C}^{(u,d)}_m (\phi) \Vec{a}^{(u,d)}_m, \nonumber\\
    &(\mathbf{\Omega}^{A^i_H,B_H}_{LR})_{nm}^{(u,d)}(\phi) \equiv (\Vec{a}^{(u,d)}_n)^\dagger \mathbf{S}^{(u,d)}_n (\phi) (\mathbf{X}^{u,d}_{A^{i}_H,B_H}) \mathbf{C}^{(u,d)}_m (\phi) \Vec{a}^{(u,d)}_m, \nonumber\\
    &(\mathbf{\Omega}^{A^i_{\phi,H},B_{\phi,H}}_{RL})^{(U,u,D,d)}_{nm}(\phi) = (\mathbf{\Omega}^{A^i_{\phi,H},B_{\phi,H}}_{LR})^{(U,u,D,d)*}_{mn}(\phi), \nonumber
\end{align}
where $\mathbf{X}^{U,u,D,d}_{A^{i}_{H}, B_{H}}$ and $\mathbf{X}^{U,u,D,d}_{A^{i},B}$ are coupling matrices as we have used in Sections \ref{FlavorGauge4FermionSection} and \ref{FlavorScalar4FermionSection}; notably, recall that explicit forms of the matrices $\mathbf{X}^{U,u,D,d}_{A^{i},B}$ have been computed in Eqs.(\ref{FlavorGaugeXQ}-\ref{FlavorGaugeXu}). We find the remaining coupling matrices to be given by
\begin{align}
    &\mathbf{X}^{U,D}_{A^1_H} = \frac{i k s_Q c_Q(\eta^Q_{3}-\eta^Q_{1})}{2 v_F} \begin{pmatrix}
    0 & s_Q & 0 \\
    -s_Q & 0 & c_Q \\
    0 & -c_Q & 0
    \end{pmatrix}, \nonumber\\
    &\mathbf{X}^{U,D}_{A^2_H} = \frac{k s_Q c_Q(\eta^Q_{3}-\eta^Q_{1})}{2 v_F} \begin{pmatrix}
    0 & s_Q & 0 \\
    s_Q & 0 & -c_Q \\
    0 & -c_Q & 0
    \end{pmatrix}, \\
    &\mathbf{X}^{U,D}_{A^3_H} =\mathbf{X}^{U,D}_{B_H}= \frac{i k s_Q c_Q(\eta^Q_{3}-\eta^Q_{1})}{2 v_F} \begin{pmatrix}
    0 & 0 & 1 \\
    0 & 0 & 0 \\
    -1 & 0 & 0
    \end{pmatrix}, \nonumber
\end{align}
for $SU(2)_L$ doublet quarks,
\begin{align}
    &\mathbf{X}^{d}_{A^1_H} =\frac{i k c_d s_d (\eta^d_3-\eta^d_1)}{2 v_F} \begin{pmatrix}
    0 & -s_d & 0 \\
    s_d & 0 & -c_d \\
    0 & c_d & 0
    \end{pmatrix}, \nonumber \\
    &\mathbf{X}^{d}_{A^2_H} =\frac{k c_d s_d (\eta^d_3-\eta^d_1)}{2 v_F} \begin{pmatrix}
    0 & -s_d & 0 \\
    -s_d & 0 & c_d \\
    0 & c_d & 0
    \end{pmatrix}, \\
    &\mathbf{X}^{d}_{A^3_H} =\mathbf{X}^{d}_{B_H} =\frac{i k c_d s_d (\eta^d_3-\eta^d_1)}{2 v_F} \begin{pmatrix}
    0 & 0 & -1 \\
    0 & 0 & 0 \\
    1 & 0 & 0
    \end{pmatrix}, \nonumber
\end{align}
for down-like $SU(2)_L$ singlet quarks, and
\begin{align}
    &\mathbf{X}^{u}_{A^1_H} =\frac{i k c_u s_u (\eta^u_3-\eta^u_1)}{2 v_F} \begin{pmatrix}
    0 & -(c_H^2-s_H^2)s_u & -2 c_H s_H  \\
    (c_H^2-s_H^2)s_u & 0 & -(c_H^2-s_H^2)c_u \\
    2 c_H s_H & (c_H^2-s_H^2) c_u & 0
    \end{pmatrix}, \nonumber \\
    &\mathbf{X}^{u}_{A^2_H} = \frac{k c_u s_u (\eta^u_3-\eta^u_1)}{2 v_F} \begin{pmatrix}
    0 & -s_u & 0 \\
    -s_u & 0 & c_u \\
    0 & c_u & 0
    \end{pmatrix}, \\
    &\mathbf{X}^{u}_{A^3_H} =\frac{i k c_u s_u (\eta^u_3-\eta^u_1)}{2 v_F} \begin{pmatrix}
    0 & 2 c_H s_H s_u & -(c_H^2-s_H^2)  \\
    -2 c_H s_H s_u & 0 & 2c_H s_H c_u \\
    (c_H^2-s_H^2) & -2 c_H s_H c_u & 0
    \end{pmatrix}, \nonumber\\
    &\mathbf{X}^{u}_{B_H} =\frac{i k c_u s_u (\eta^u_3-\eta^u_1)}{2 v_F} \begin{pmatrix}
    0 & 0 & -1  \\
    0 & 0 & 0 \\
    1 & 0 & 0
    \end{pmatrix}, \nonumber
\end{align}
for up-like $SU(2)_L$ singlet quarks. We remind the reader that the above formulas use the angle and bulk localization parameters that we first defined in Eq.(\ref{BulkYukawaExpressions}). We may then write the coupling terms of the field $a_1$ with two quarks as
\begin{align}
    \sqrt{2 \pi} \int_{-\pi}^{\pi} d\phi &\bigg\{g_A [ (\mathbf{\Omega}^{A^1_\phi}_{LR})^{(Q)}_{nm}(\phi)+(\mathbf{\Omega}^{A^1_\phi}_{LR})^{(q)}_{nm}(\phi)] \frac{\gamma k}{m^{a_1}_p}\chi^{a_1}_p(\phi)\\
    &+[(\mathbf{\Omega}^{A^1_H}_{LR})^{(Q)}_{nm}(\phi)+(\mathbf{\Omega}^{A^1_H}_{LR})^{(q)}_{nm}(\phi)]\frac{e^{2 \sigma}}{m^{a_1}_p}\frac{1}{r_c}\partial_\phi(e^{-2 \sigma} \chi^{a_1}_p(\phi)) \nonumber \bigg\} a_1 \,(\bar{q}^{(n)}_L q^{(m)}_R)+h.c., \nonumber \\
    &\gamma \equiv \begin{cases}
    \frac{g_A v_F}{\sqrt{2} k} & a_{1,2,3} \\
    \frac{-3 g_B v_F}{\sqrt{2} k} & a_4
    \end{cases} \nonumber
\end{align}
where $q=u$ and $Q=U$ for up-like quarks, with $q=d$ and $Q=D$ for down-like quarks, with analogous expressions for couplings with other $a$ fields. These coupling terms then yield four-fermion Hamiltonian terms (again, we specialize our discussion to $a_1$, but our arguments hold for all other $a$ fields),
\begin{align}\label{aVertices}
    &(C^{qq'}_{LR;LR})^{a_i}_{nm;rs} (\bar{q}^{(n)}_L q^{(m)}_R)(\bar{q'}^{(r)}_L q^{(s)}_R), \nonumber\\
    &(C^{qq'}_{LR;RL})^{a_i}_{nm;rs} (\bar{q}^{(n)}_L q^{(m)}_R)(\bar{q'}^{(r)}_R q^{(s)}_L), \\
    &(C^{qq'}_{RL;LR})^{a_i}_{nm;rs} (\bar{q}^{(n)}_R q^{(m)}_L)(\bar{q'}^{(r)}_L q^{(s)}_R), \nonumber\\
    &(C^{qq'}_{RL;RL})^{a_i}_{nm;rs} (\bar{q}^{(n)}_R q^{(m)}_L)(\bar{q'}^{(r)}_R q^{(s)}_L), \nonumber
\end{align}
where as before, $q$ and $q'$ denote either up-like ($u$) or down-like ($d$) quarks, depending on the specific coupling being considered. For clarity, we note that the Feynman diagrams which give rise to these terms are identical to those depicted in Figure \ref{fig:HFeynmanDiagrams}, once the substitution $H^{(p)}_i\rightarrow a^{(p)}_i$ is made. In the case of $a_1$ (with the other couplings possessing entirely analogous expressions),
\begin{align}
    &(C^{qq'}_{LR;LR})^{a_1}_{nm;rs} = -(2 \pi)\int_{-\pi}^{\pi}\int_{-\pi}^{\pi} d\phi_1 d\phi_2 \bigg\{ \\
    &[(\mathbf{\Omega}^{A^1_\phi}_{LR})^{(Q)}_{nm}(\phi_1)+(\mathbf{\Omega}^{A^1_\phi}_{LR})^{(q)}_{nm}(\phi_1)][(\mathbf{\Omega}^{A^1_\phi}_{LR})^{(Q')}_{rs}(\phi_2)+(\mathbf{\Omega}^{A^1_\phi}_{LR})^{(q')}_{rs}(\phi_2)]\bigg(\sum_{p} \frac{\gamma^2 k^2 \chi^{a_1}_{p}(\phi_1) \chi^{a_1}_{p}(\phi_2)}{(m^{a_1}_p)^4} \bigg) \nonumber \\
    &+[(\mathbf{\Omega}^{A^1_\phi}_{LR})^{(Q)}_{nm}(\phi_1)+(\mathbf{\Omega}^{A^1_\phi}_{LR})^{(q)}_{nm}(\phi_1)][(\mathbf{\Omega}^{A^1_H}_{LR})^{(Q')}_{rs}(\phi_2)+(\mathbf{\Omega}^{A^1_H}_{LR})^{(q')}_{rs}(\phi_2)]\bigg(\sum_p \frac{\gamma k^2 \tilde{\chi}^{a_1}_p(\phi_2)\chi^{a_1}_p(\phi_1)}{(kr_c)(m^{a_1}_p)^4} \bigg) \nonumber \\
    &+[(\mathbf{\Omega}^{A^1_H}_{LR})^{(Q)}_{nm}(\phi_1)+(\mathbf{\Omega}^{A^1_H}_{LR})^{(q)}_{nm}(\phi_1)][(\mathbf{\Omega}^{A^1_\phi}_{LR})^{(Q')}_{rs}(\phi_2)+(\mathbf{\Omega}^{A^1_\phi}_{LR})^{(q')}_{rs}(\phi_2)]\bigg(\sum_p \frac{\gamma k^2 \tilde{\chi}^{a_1}_p(\phi_1) \chi^{a_1}_p(\phi_2)}{(kr_c)(m^{a_1}_p)^4} \bigg) \nonumber \\
    &+[(\mathbf{\Omega}^{A^1_H}_{LR})^{(Q)}_{nm}(\phi_1)+(\mathbf{\Omega}^{A^1_H}_{LR})^{(q)}_{nm}(\phi_1)][(\mathbf{\Omega}^{A^1_H}_{LR})^{(Q')}_{rs}(\phi_2)+(\mathbf{\Omega}^{A^1_H}_{LR})^{(q')}_{rs}(\phi_2)] \bigg( \sum_p \frac{k^2 \tilde{\chi}^{a_1}_p(\phi_1) \tilde{\chi}^{a_1}_p(\phi_2)}{(kr_c)^2 (m^{a_1}_p)^4}\bigg) \bigg\} \nonumber,
\end{align}
where
\begin{align}
    &\tilde{\chi}^{a_1}(\phi) \equiv e^{2 \sigma}\partial_\phi e^{-2 \sigma} \chi^{a_1}_p(\phi).
\end{align}
Inserting Eqs.(\ref{NeededSum1}-\ref{NeededSum3}), we can evaluate these 4-fermion operators numerically, as we have already done with the $H_{1-4}$ exchanges in Section \ref{FlavorScalar4FermionSection}. In Figure (\ref{fig:CauuLRuuLR}), we plot the same flavor-changing coupling as displayed in Figure (\ref{fig:CHuuLRuuLR}), only now arising from the exchange of the scalars $a_{1-4}$ rather than $H_{1-4}$, revealing similar basic behavior for these functions. Again we observe a strong inverse dependence of the coupling with respect to the bulk vacuum expectation value parameter $v_F$, and once again we can assume extreme suppression of these couplings, due to the additional factors of the ratios of the SM-like quark masses to $M_{KK}$, as we already noted in Section \ref{FlavorScalar4FermionSection}, 

\subsection{Four-Quark Operators: SM Vector Boson Exchanges}\label{SMGauge4FermionSection}
Having discussed the effects of the gauge bosons and scalars arising in our model from the $SU(2)_F\times U(1)_F$ symmetry, it is also necessary to discuss the flavor effects of SM gauge bosons in the RS framework, which it has been noted in \cite{casagrande2,chang,ahmed,blanke} are significant in a generic anarchic model. We begin with a discussion of the neutral gauge bosons, $Z$, the photon, and the gluon, as any 4-quark interactions arising from exchanges of these particles emerge from the Feynman diagrams in Figure \ref{fig:GaugeFeynmanDiagrams}, and as such can be written using the same expression for the operators as we used for the $SU(2)_F\times U(1)_F$ gauge bosons in Eq.(\ref{Vector4Fermion}), albeit using the appropriate coupling matrices, coupling constants, and summations over the bulk profiles for the bulk SM gauge bosons, rather than the gauge bosons arising from the new flavor symmetry. Notably, because each fermion generation is put into identical representations in the SM gauge groups, all of the coupling matrices $\mathbf{X}^{U,u,D,d}_{W^{\pm},Z,\gamma,g}$ in these equations are proportional to the identity matrix $\mathbb{I}_{3\times3}$ in 3-dimensional generation space, that is,
\begin{equation}\label{DiagonalXSM}
    \mathbf{X}^{U,u,D,d}_{Z,\gamma,g} = X^{U,u,D,d}_{Z,\gamma,g} \mathbb{I}_{3\times3},
\end{equation}
where the unbolded $X$ variables are simply scalar values. For the SM gauge fields, we have
\begin{align}
    \begin{matrix}
    X^{U,u}_{\gamma} = +\frac{2}{3}, & X^{D,d}_{\gamma} = -\frac{1}{3}, & X^{U,u,D,d}_{g}= 1,\\
    X^{U}_{Z} = \frac{1}{2}-\frac{2}{3} s_W^2, & X^{D}_{Z}= -\frac{1}{2} +\frac{1}{3}s_W^2 & X^{u}_{Z} = -\frac{2}{3} s_W^2, & X^{d}_Z = \frac{1}{3}s_W^2,
    \end{matrix}
\end{align}
where $s_W$ is the sine of the Weinberg angle, as usual. Here, we have deliberately suppressed color factors in the gluon couplings; they follow the same prescription as in the SM, and we shall include them explicitly when relevant to our discussions. Given that each coupling matrix is proportional to identity, then, we find it useful when writing coupling expressions here to define the variable $(\bar{\bm{\Omega}}_{L,R})^{(U,u,D,d)}_{nm}$ as
\begin{equation}\label{OmegaId}
     (\bar{\bm{\Omega}}_{L,R})^{(U,u,D,d)}_{nm} \equiv (\bm{\Omega}^E_{L,R})^{(U,u,D,d)}_{nm}(\phi) \bigg\rvert_{\mathbf{X}^{U,u,D,d}_E = \mathbb{I}_{3\times3}},
\end{equation}
where we remind the reader that $(\bm{\Omega}^E_{L,R})^{(U,u,D,d)}_{nm}(\phi)$ is defined in Eq.(\ref{OmegaDefs}).

With our definitions in place, we now discuss the photon and gluon field exchanges, in which case the sum over all massive KK modes in Eq.(\ref{Vector4Fermion}) takes the form given in Eq.(\ref{PhotonGaugeSum}).
Deriving the 4-fermion operators for these exchanges in direct analogy to our treatment of the flavor gauge bosons in Section \ref{FlavorGauge4FermionSection}, we arrive at
\begin{align}
    &(C^{qq'}_{LL})^{\gamma,g}_{nm;rs} = 2 \pi g^2_{\textrm{em},s} \sum_{a,c=1}^3 \Theta_\gamma(\eta^Q_a,\eta^Q_c)X^{Q}_{\gamma,g} X^{Q'}_{\gamma,g} (\mathbf{U}^{q}_L)^{\dagger}_{na}(\mathbf{U}^{q}_L)_{am} (\mathbf{U}^{q'}_L)^{\dagger}_{rc} (\mathbf{U}^{q'}_L)_{cs}, \nonumber\\
    &(C^{qq'}_{LR})^{\gamma,g}_{nm;rs} = 2 \pi g^2_{\textrm{em},s} \sum_{a,c=1}^3 \Theta_\gamma(\eta^Q_a,\eta^{q'}_c)X^{Q}_{\gamma,g} X^{q'}_{\gamma,g} (\mathbf{U}^{q}_L)^{\dagger}_{na}(\mathbf{U}^{q}_L)_{am} (\mathbf{U}^{q'}_R)^{\dagger}_{rc} (\mathbf{U}^{q'}_R)_{cs},\\
    &(C^{qq'}_{RL})^{\gamma,g}_{nm;rs} = 2 \pi g^2_{\textrm{em},s} \sum_{a,c=1}^3 \Theta_\gamma(\eta^q_a,\eta^Q_c)X^{q}_{\gamma,g} X^{Q'}_{\gamma,g} (\mathbf{U}^{q}_R)^{\dagger}_{na}(\mathbf{U}^{q}_R)_{am} (\mathbf{U}^{q'}_L)^{\dagger}_{rc} (\mathbf{U}^{q'}_L)_{cs}, \nonumber\\
    &(C^{qq'}_{RR})^{\gamma,g}_{nm;rs} = 2 \pi g^2_{\textrm{em},s} \sum_{a,c=1}^3 \Theta_\gamma(\eta^q_a,\eta^{q'}_c)X^{q}_{\gamma,g} X^{q'}_{\gamma,g} (\mathbf{U}^{q}_R)^{\dagger}_{na}(\mathbf{U}^{q}_R)_{am} (\mathbf{U}^{q'}_R)^{\dagger}_{rc} (\mathbf{U}^{q'}_R)_{cs},\nonumber
\end{align}
where $g_{\textrm{em}}=e=g s_W$ is the electromagnetic coupling constant, $g_s$ is the strong coupling constant, and the function $\Theta_\gamma$ is given by
\begin{align}
    \Theta_\gamma(\eta_A,\eta_B) \equiv &\frac{|F(\eta_A)|^2|F(\eta_B)|^2}{4 k^2 \epsilon^2} \bigg[ \frac{kr_c (4+\eta_A+\eta_B)}{(2+\eta_A+\eta_B)(2+\eta_A)(2+\eta_B)}\\
    &+\frac{(1-\epsilon^{\eta_A})(1-\epsilon^{\eta_B})}{2 kr_c \pi^2 \eta_A \eta_B}-\frac{1}{2 \pi}\bigg( \frac{(1-\epsilon^{\eta_B})(4+\eta_A)}{(2+\eta_A^2)\eta_B}+\frac{(1-\epsilon^{\eta_A})(4+\eta_B)}{(2+\eta_B)^2 \eta_A}\bigg)\bigg], \nonumber
\end{align}
where again we use the function $F(\eta)$ first defined in Eq.(\ref{QuarkApproxProfiles}). Next, we consider the 4-fermion operators arising from exchanges over the entire $Z$ boson tower. Notably, because of the presence of the light, SM-like $Z$ boson in these exchanges, it is no longer reasonable to omit the contribution of the orbifold odd bulk fermion fields, $(\Omega^Z_L)^{u,d}_{nm}(\phi)$ and $(\Omega^Z_R)^{U,D}_{nm}(\phi)$ in the ZMA evaluation of Eq.(\ref{Vector4Fermion}) \cite{casagrande}; although these contributions are suppressed by $O(m_{q}^2/M_{KK}^2)$, the presence of terms of $O(1/m_Z^2)$ in the summation over $Z$ boson exchanges Eq.(\ref{WZGaugeSum}), instead of solely $O(1/M_{KK}^2)$ terms, means that the orbifold odd fields can still contribute to the resultant operator to leading order in the KK scale. Using the coupling $(C^{qq'}_{LL})^{Z}_{nm;rs}$ as an example, we shall briefly outline how these additional terms are included. In the exact expression, this coefficient would be given by
\begin{align}\label{CqqLLDelta1}
    &(C^{qq'}_{LL})^{Z}_{nm;rs} = 2 \pi \frac{g^2}{c_W^2} \int_{-\pi}^{\pi} d\phi_1 \int_{-\pi}^{\pi} d\phi_2 \bigg\{ e^{(\sigma_1+\sigma_2)}\bigg[ \frac{1}{2 \pi m_Z^2} + O\bigg(\frac{1}{M_{KK}^2}\bigg)\bigg]\\
    &\times[(T^3_q-s_W^2 Q_q)(\bar{\bm{\Omega}}_{L})^{(Q)}_{nm}(\phi_1)-s_W^2 Q_q (\bar{\bm{\Omega}}_{L})^{(q)}_{nm}(\phi_1)][(T^3_{q'}-s_W^2 Q_{q'})(\bar{\bm{\Omega}}_{L})^{(Q')}_{rs}(\phi_2)-s_W^2 Q_{q'} (\bar{\bm{\Omega}}_{L})^{(q')}_{rs}(\phi_2)]\bigg\}, \nonumber
\end{align}
where the $O(1/(M_{KK}^2)$ terms are given explicitly by Eq.(\ref{WZGaugeSum}), $T^3_{q,q'}$ are the weak isospins of the $q$ and $q'$ quarks, and $Q_{q,q'}$ are their electric charges. Focusing specifically on the leading order terms arising from the $1/m_Z^2$ term in Eq.(\ref{CqqLLDelta1}), and recalling the orthonormality condition for fermion fields, given in Eq.(\ref{FermionNormalizationConditions}), we have this piece of the four-fermion operator given by
\begin{align}
    &\frac{g^2}{c_W^2 m_Z^2} (T^3_q-s_W^2 Q_q)(T^3_{q'}-s_W^2 Q_{q'})\delta_{nm} \delta_{rs} \\
    &-\frac{g^2}{2 \pi c_W^2 m_Z^2}\int_{-\pi}^{\pi} d\phi_1 \int_{-\pi}^{\pi} d\phi_2 \, e^{(\sigma_1+\sigma_2)} [T^3_q(T^3_{q'}-s_W^2 Q_{q'}) (\bar{\bm{\Omega}}_L)^{q}_{nm}(\phi_1)(\bar{\bm{\Omega}}_L)^{(Q')}_{rs}(\phi_2)] \nonumber\\
    &-\frac{g^2}{2 \pi c_W^2 m_Z^2}\int_{-\pi}^{\pi} d\phi_1 \int_{-\pi}^{\pi} d\phi_2 \, e^{(\sigma_1+\sigma_2)} [T^3_{q'}(T^3_{q}-s_W^2 Q_{q}) (\bar{\bm{\Omega}}_L)^{(Q)}_{nm}(\phi_1)(\bar{\bm{\Omega}}_L)^{(q')}_{rs}(\phi_2)], \nonumber
\end{align}
where the integrals in the second and third lines of this equation can be approximately determined from the ZMA expressions for the orbifold odd fermion fields, given by Eqs.(\ref{ZMAProfileSubstitutions}) and (\ref{ZMAMixingSubstitutions}). Including these terms (and analogous terms for couplings such as $(C^{qq'}_{LR})^Z$ and $(C^{qq'}_{RR})^Z$), the complete ZMA expressions for the operators arising from $Z$ boson exchange will be
\begin{align}
    (C^{qq'}_{LL})^{Z}_{nm;rs} &= \bigg( \frac{g}{c_W m_Z}\bigg)^2(T^3_q-s_W^2 Q_q)(T^3_{q'}-s_W^2 Q_{q'}^2) \delta_{nm} \delta_{rs} \\ 
    &+2 \pi \bigg(\frac{g}{c_W}\bigg)^2\sum_{a,c=1}^3 \bigg\{ (T^3_q-s_W^2 Q_q)(T^3_{q'}-s_w^2 Q_{q'})\Theta_Z(\eta^Q_a,\eta^Q_c)(\mathbf{U}^q_L)^{\dagger}_{na}(\mathbf{U}^q_L)_{am}(\mathbf{U}^{q'}_L)^{\dagger}_{rc}(\mathbf{U}^{q'}_L)_{cs} \nonumber\\
    &-T^3_{q}(T^3_{q'}-s_W^2 Q_{q'})\delta_Z(\eta^Q_c,\eta^{q}_a) x^q_n x^q_m (\mathbf{U}^q_R)^\dagger_{na}(\mathbf{U}^q_R)_{am} (\mathbf{U}^{q'}_L)^{\dagger}_{rc} (\mathbf{U}^{q'}_L)_{cs} \nonumber \\
    &-T^3_{q'}(T^3_{q}-s_W^2 Q_{q})\delta_Z(\eta^Q_a,\eta^{q'}_c) x^{q'}_r x^{q'}_s (\mathbf{U}^q_L)^\dagger_{na}(\mathbf{U}^q_L)_{am} (\mathbf{U}^{q'}_R)^{\dagger}_{rc} (\mathbf{U}^{q'}_R)_{cs} \bigg\} \nonumber \\
    (C^{qq'}_{LR})^{Z}_{nm;rs} &= \bigg( \frac{g}{c_W m_Z} \bigg)^2 (T^3_{q}-s_W^2 Q_q)(-s_W^2 Q_{q'}) \delta_{nm} \delta_{rs} \\
    &+2 \pi \bigg( \frac{g}{c_W} \bigg)^2 \sum_{a,c=1}^3 \bigg\{ (T^3_q-s_W^2 Q_q)(-s_W^2 Q_{q'}) \Theta_Z(\eta^Q_a,\eta^{q'}_c)(\mathbf{U}^{q}_L)^{\dagger}_{na}(\mathbf{U}^{q}_L)_{am} (\mathbf{U}^{q'}_R)^{\dagger}_{rc} (\mathbf{U}^{q'}_R)_{cs} \nonumber \\
    &-T^3_q (-s_W^2 Q_{q'}) \delta_Z(\eta^{q'}_c,\eta^q_a) x^q_n x^q_m (\mathbf{U}^q_R)^{\dagger}_{na}(\mathbf{U}^q_R)_{am} (\mathbf{U}^{q'}_R)^{\dagger}_{rc}(\mathbf{U}^{q'}_R)_{cs} \nonumber \\
    &+T^3_{q'} (T^3_{q}-s_W^2 Q_{q}) \delta_Z(\eta^Q_a,\eta^Q_{c})x^{q'}_r x^{q'}_s (\mathbf{U}^{q}_L)^{\dagger}_{na} (\mathbf{U}^{q}_L)_{am} (\mathbf{U}^{q'}_L)^{\dagger}_{rc} (\mathbf{U}^{q'}_L)_{cs} \bigg\}, \nonumber \\
    (C^{qq'}_{RL})^{Z}_{nm;rs} &= \bigg( \frac{g}{c_W m_Z} \bigg)^2 (-s_W^2 Q_q)(T^3_{q'}-s_W^2 Q_{q'}) \delta_{nm} \delta_{rs} \\
    &+2 \pi \bigg( \frac{g}{c_W} \bigg)^2 \sum_{a,c=1}^3 \bigg\{ (-s_W^2 Q_q)(T^3_{q'}-s_W^2 Q_{q'}) \Theta_Z(\eta^q_a,\eta^{Q}_c)(\mathbf{U}^{q}_R)^{\dagger}_{na}(\mathbf{U}^{q}_R)_{am} (\mathbf{U}^{q'}_L)^{\dagger}_{rc} (\mathbf{U}^{q'}_L)_{cs} \nonumber \\
    &+T^3_q (T^3_{q'}-s_W^2 Q_{q'}) \delta_Z(\eta^{Q}_c,\eta^Q_a) x^q_n x^q_m (\mathbf{U}^q_L)^{\dagger}_{na}(\mathbf{U}^q_L)_{am} (\mathbf{U}^{q'}_L)^{\dagger}_{rc}(\mathbf{U}^{q'}_L)_{cs} \nonumber \\
    &-T^3_{q'} (-s_W^2 Q_{q}) \delta_Z(\eta^{q}_a,\eta^{q'}_{c})x^{q'}_r x^{q'}_s (\mathbf{U}^{q}_R)^{\dagger}_{na} (\mathbf{U}^{q}_R)_{am} (\mathbf{U}^{q'}_R)^{\dagger}_{rc} (\mathbf{U}^{q'}_R)_{cs} \bigg\}, \nonumber\\
    (C^{qq'}_{RR})^{Z}_{nm;rs} &= \bigg( \frac{g}{c_W m_Z} \bigg)^2 (-s_W^2 Q_q)(-s_W^2 Q_{q'}) \delta_{nm} \delta_{rs} \\
    &+2 \pi \bigg( \frac{g}{c_W} \bigg)^2 \sum_{a,c=1}^3 \bigg\{ (-s_W^2 Q_q)(-s_W^2 Q_{q'}) \Theta_Z(\eta^{q}_a,\eta^{q'}_c)(\mathbf{U}^{q}_R)^{\dagger}_{na}(\mathbf{U}^{q}_R)_{am} (\mathbf{U}^{q'}_R)^{\dagger}_{rc} (\mathbf{U}^{q'}_R)_{cs} \nonumber \\
    &+T^3_q (-s_W^2 Q_{q'}) \delta_Z(\eta^{q'}_c,\eta^{Q}_a) x^q_n x^q_m (\mathbf{U}^q_L)^{\dagger}_{na}(\mathbf{U}^q_L)_{am} (\mathbf{U}^{q'}_R)^{\dagger}_{rc}(\mathbf{U}^{q'}_R)_{cs} \nonumber \\
    &+T^3_{q'} (-s_W^2 Q_{q}) \delta_Z(\eta^{q}_a,\eta^Q_{c})x^{q'}_r x^{q'}_s (\mathbf{U}^{q}_R)^{\dagger}_{na} (\mathbf{U}^{q}_R)_{am} (\mathbf{U}^{q'}_L)^{\dagger}_{rc} (\mathbf{U}^{q'}_L)_{cs} \bigg\}, \nonumber
\end{align}
where we have the functions
\begin{align}
    \Theta_Z(\eta_A,\eta_C) &\equiv \frac{|F(\eta_A)|^2|F(\eta_C)|^2}{4 M_{KK}^2} \bigg[ \frac{-1+2 kr_c \pi}{2 kr_c \pi^2 |F(\eta_A)|^2|F(\eta_C)|^2} \\
    &-\bigg( \frac{kr_c}{|F(2+\eta_C)|^2|F(\eta_A)|^2}+\frac{kr_c}{|F(2+\eta_A)|^2|F(\eta_C)|^2}\bigg) \nonumber\\
    &+\frac{kr_c}{(2+\eta_A+\eta_C)}\bigg( \frac{1}{|F(2+\eta_A)|^2}+\frac{1}{|F(2+\eta_C)|^2}-\frac{\epsilon^{2+\eta_A}}{|F(\eta_C)|^2}-\frac{\epsilon^{2+\eta_C}}{|F(\eta_A)|^2}\bigg) \bigg] \nonumber
\end{align}
and
\begin{align}
    \delta_Z(\eta_A, \eta_C) = \frac{\epsilon^{2+\eta_C}|F(\eta_C)|^2}{2 m_Z^2 \pi \eta_C^2}\bigg(1-\epsilon^{-2}+\frac{1}{|F(\eta_C-2)|^2}+\frac{1}{|F(-\eta_C-2)|^2}\bigg).
\end{align}

\begin{figure}
  \centering
  \begin{subfigure}[b]{0.45\textwidth}
    \centering
    \fmfframe(10,25)(10,25){
    \begin{fmfgraph*}(100,40)
        \fmfleft{i1,i2}
        \fmfright{o1,o2}
        \fmf{fermion}{i2,v1,i1}
        \fmf{boson,label=$W^{(p)}$}{v1,v2}
        \fmf{fermion}{o2,v2,o1}
        \fmflabel{$d^{(m)}_X$}{i2}
        \fmflabel{$\bar{u}^{(n)}_X$}{i1}
        \fmflabel{$\bar{u}^{(s)}_Y$}{o2}
        \fmflabel{$d^{(r)}_Y$}{o1}
    \end{fmfgraph*}
    }
    \caption{$(C^{UD;DU}_{XY})^W_{nm;rs}$}
  \end{subfigure}
  \begin{subfigure}[b]{0.45\textwidth}
    \centering
    \fmfframe(10,25)(10,25){
    \begin{fmfgraph*}(100,40)
        \fmfleft{i1,i2}
        \fmfright{o1,o2}
        \fmf{fermion}{i2,v1,i1}
        \fmf{boson,label=$W^{(p)}$}{v1,v2}
        \fmf{fermion}{o2,v2,o1}
        \fmflabel{$u^{(m)}_X$}{i2}
        \fmflabel{$\bar{d}^{(n)}_X$}{i1}
        \fmflabel{$\bar{d}^{(s)}_Y$}{o2}
        \fmflabel{$u^{(r)}_Y$}{o1}
    \end{fmfgraph*}
    }
    \caption{$(C^{DU;UD}_{LR})^W_{nm;rs}$}
  \end{subfigure}
    \caption{The Feynman diagrams that give rise to the effective 4-fermion interactions discussed in Eq.(\ref{WVertices}). Here, $W^{(p)}$ denotes the $p^{th}$ KK mode of a $W^{\pm}$ vector gauge boson, and $X$ and $Y$ are chiralities (\ie, $X=L,R$ and $Y=L,R$.}
    \label{fig:WFeynmanDiagrams}
\end{figure}
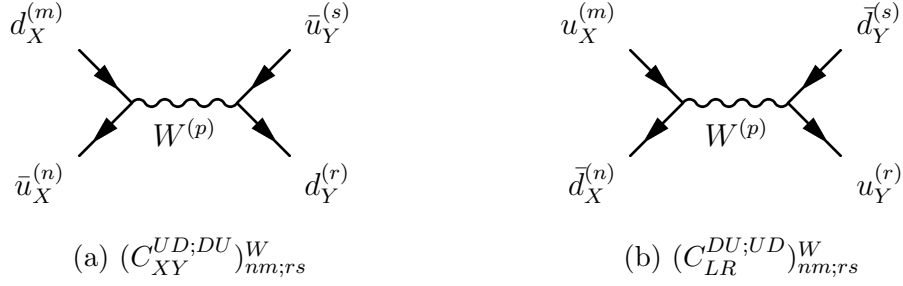

Finally, we consider the four-fermion operators arising from $W$ boson exchanges, which, while not used our analysis in Section \ref{FlavorObservablesSection}, we include for completness. Due to the fact that the $W^{\pm}$ field couples only to the $SU(2)_L$ doublet quarks, and does so via a charged current, we can modify Eq.(\ref{Vector4Fermion}) to specifically treat these operators. To do so, it is first convenient to define the quantities,
\begin{align}
    &(\bar{\bm{\Omega}}_L)^{(UD,DU)}_{nm}(\phi) \equiv (\Vec{a}^{(U,D)}_n)^\dagger \mathbf{C}^{(U,D)}_n(\phi) \mathbf{C}^{(D,U)}_m(\phi) \Vec{a}^{(D,U)}_n, \\
    &(\bar{\bm{\Omega}}_R)^{(UD,DU)}_{nm}(\phi) \equiv (\Vec{a}^{(U,D)}_n)^\dagger \mathbf{S}^{(U,D)}_n(\phi) \mathbf{S}^{(D,U)}_m(\phi) \Vec{a}^{(D,U)}_n, \nonumber
\end{align}
to more compactly write charged currents. Then, the four-fermion operators emerging from $W$ boson KK tower exchange are then given by
\begin{align}\label{WVertices}
    &(C^{(UD;DU)}_{XY})^{W}_{nm;rs} (\bar{u}^{(n)}_X \gamma^\mu d^{(m)}_X)(\bar{d}^{(r)}_Y \gamma_\mu u^{(s)}_Y),
    &(C^{(DU;UD)}_{XY})^{W}_{nm;rs} (\bar{d}^{(n)}_X \gamma^\mu u^{(m)}_X)(\bar{u}^{(r)}_Y \gamma_\mu d^{(s)}_Y),
\end{align}
where
\begin{align}
    (C^{(UD;DU)}_{XY})^{W}_{nm;rs} \equiv &2 \pi \frac{g^2}{2} \int_{-\pi}^{\pi} d \phi_1 \int_{-\pi}^{\pi} d\phi_2 \, \bigg\{ \\
    &e^{(\sigma_1+\sigma_2)}(\bar{\bm{\Omega}}_X)^{(UD)}_{nm}(\phi_1)(\bar{\bm{\Omega}}_Y)^{(DU)}_{rs}(\phi_2) \bigg( \sum_{p} \frac{\chi^W_p(\phi_1) \chi^W_p(\phi_2)}{(m^W_p)^2} \bigg)\bigg\}, \nonumber \\
    (C^{(DU;UD)}_{XY})^{W}_{nm;rs} \equiv &2 \pi \frac{g^2}{2} \int_{-\pi}^{\pi} d \phi_1 \int_{-\pi}^{\pi} d\phi_2 \, \bigg\{ \\
    &e^{(\sigma_1+\sigma_2)}(\bar{\bm{\Omega}}_X)^{(DU)}_{nm}(\phi_1)(\bar{\bm{\Omega}}_Y)^{(UD)}_{rs}(\phi_2) \bigg( \sum_{p} \frac{\chi^W_p(\phi_1) \chi^W_p(\phi_2)}{(m^W_p)^2} \bigg) \bigg\}. \nonumber
\end{align}
For clarity, we have also included the Feynman diagrams that contribute to the Wilson coefficients in Eq.(\ref{WVertices}) in Figure \ref{fig:WFeynmanDiagrams}. Applying the ZMA substitutions of Eqs.(\ref{ZMAProfileSubstitutions}) and (\ref{ZMAMixingSubstitutions}), as well as the summation identity Eq.(\ref{WZGaugeSum}) for the sum over the $W$ boson bulk profiles, we obtain
\begin{align}\label{VectorW4Fermion}
    (C^{(DU;UD)}_{LL})_{nm;rs} &= \frac{4 G_F}{\sqrt{2}} [(\mathbf{U}^{d}_L)^{\dagger}(\mathbf{U}^{u}_L)]_{nm} [(\mathbf{U}^{u}_L)^{\dagger}(\mathbf{U}^{d}_L)]_{rs} \\
    &+\frac{g^2}{2} \sum_{a,c=1}^3 \Theta_{WLL}(\eta^Q_a,\eta^Q_c)(\mathbf{U}^{d}_L)^{\dagger}_{n,a}(\mathbf{U}^{u}_L)_{am} (\mathbf{U}^{u}_L)^{\dagger}_{r,c} (\mathbf{U}^{d}_L)_{cs}, \nonumber\\
    (C^{(DU;UD)}_{LR})_{nm;rs} &= \frac{g^2}{2} \sum_{c=1}^3 \Theta_{WLR}(\eta^Q_c)(\mathbf{U}^{d \dagger}_L \mathbf{U}^{u}_L)_{nm} (\mathbf{U}^{u}_L)^{\dagger}_{rc} (\mathbf{U}^{d}_L)_{cs}, \nonumber \\
    (C^{(DU;UD)}_{RL})_{nm;rs} &= \frac{g^2}{2} \sum_{a=1}^3 \Theta_{WLR}(\eta^Q_a)(\mathbf{U}^{u \dagger}_L \mathbf{U}^{d}_L)_{rs} (\mathbf{U}^{d}_L)^{\dagger}_{ra} (\mathbf{U}^{u}_L)_{as}, \nonumber
\end{align}
where
\begin{align}
    \Theta_{WLL}(\eta_A, \eta_C)  = \frac{kr_c |F(\eta_A)|^2 |F(\eta_C)|^2}{4 M_{KK}^2(2+\eta_A+\eta_C)} \bigg( \frac{-\epsilon^{2+\eta_C}}{|F(\eta_A)|^2}+ \frac{1}{|F(2+\eta_C)|^2}-\frac{\epsilon^{2+\eta_A}}{|F(\eta_C)|^2}+\frac{1}{|F(2+\eta_C)|^2}\bigg)
\end{align}
and
\begin{align}
    \Theta_{LR}(\eta)= \frac{1}{2 \pi m_W^2} \bigg( \frac{1}{\eta-2}-\frac{1}{(\eta-2)|F(\eta_C)|^2}-\frac{|F(\eta)|^2}{(\eta-2)(\eta+2)}\bigg).
\end{align}
Other couplings can be derived straightforwardly from symmetry; for example, we can use the relation $(C^{(UD;DU)}_{LL})_{nm;rs}=(C^{(DU;UD)}_{LL})_{rs;nm}$. Terms that couple two right-handed currents, \eg $(C^{(UD;DU)}_{RR})_{nm;rs}$ can be safely ignored, due to the additional $O(v^2/M_{KK}^2)$ suppression emerging from the presence of two right-handed charged currents. In Eq.(\ref{VectorW4Fermion}), we have intentionally absorbed part of the $O(v^2/M_{KK}^2)$ corrections of the $LL$ couplings into a slight redefinition of $G_F$, as commonly done (\textit{e.g.}, in \cite{goertz}). Specifically, we absorb the flat part of the sum over the $W$ KK modes into a rescaling of $G_F$, since this alteration of the $W$ four-fermion operators is universal for all fermion species. Under this redefinition, our expression for $G_F$ becomes
\begin{align}
    \frac{4 G_F}{\sqrt{2}}=\frac{g^2}{2}\bigg[1+\frac{1}{2}\frac{m_W^2}{M_{KK}^2}\bigg(1-\frac{1}{2 kr_c \pi}\bigg)\bigg].
\end{align}
With our expressions for $W$, $Z$, photon, and gluon exchange, we now have all of the SM gauge boson contributions to flavor-changing 4-quark operators.

\begin{fmffile}{diagram3}
\end{fmffile}

\section{Flavor Observables}\label{FlavorObservablesSection}
Equipped with the computed effective four-quark operators emerging in our model, we are now prepared to discuss specific flavor observables which emerge in our model. Due to their suppression at tree level in the SM, we shall largely focus on $\Delta F = 2$ processes emerging from flavor-changing neutral currents, as these processes have no SM tree-level contribution. Before continuing our analysis, we make two points that should be remembered during the following presentation. First, it should be noted that many of the formulas in this section are sensitive to the particular phase convention used for the CKM matrix, and only hold as long as the quarks observe the usual CKM phase convention. As such, we note that any coupling expressions derived in earlier sections are then modified by introducing phase rotations to the quark fields in order to obey the usual CKM phase convention and maintain real fermion masses. Second, we note that in the flavor gauge boson sector, the two gauge coupling constants $g_A$ and $g_B$ are in principle entirely independent. However, since we find that no special cancellation or other notable behavior occurs when differing values of $g_A$ and $g_B$ are taken, while scanning our parameter space we take $g_A=g_B$ for the remainder of this paper.

\subsection{$K-\bar{K}$ Mixing}\label{KKBarMixingSection}
First, we shall address the $\Delta F = 2$ physics in the Kaon sector, that is, $\bar{K}^0-K^0$ mixing, noting that in contrast to anarchic models of RS flavor, the contribution of new physics within our model to the relevant observables is negligible. In general, the effective 4-fermion Hamiltonian for this process has the form \cite{carrasco},

\begin{equation}
    \mathcal{H}^{\Delta S = 2}_{\textit{eff}} = \sum_{i=1}^5 \mathcal{C}^K_i (\mu) \mathcal{O}^{K}_i (\mu) + \sum_{i=1}^3 \tilde{\mathcal{C}}^K_i \tilde{\mathcal{O}}_i (\mu),
\end{equation}
where $\mu$ is a renormalization scale, and we have
\begin{align}\label{KMesonOperators}
    &\mathcal{O}^K_1 = [\bar{s}_L^\alpha \gamma_\mu d^\alpha_L][\bar{s}_L^\beta \gamma_\mu d^\beta_L], &\tilde{\mathcal{O}}^K_1 = [\bar{s}_R^\alpha \gamma_\mu d_R^{\alpha}][\bar{s}_R^\alpha \gamma_\mu d_R^{\alpha}], \\
    &\mathcal{O}^K_2 = [\bar{s}_R^\alpha d^\alpha_L][\bar{s}_R^\beta d^\beta_L], &\tilde{\mathcal{O}}^K_2 = [\bar{s}_L^\alpha d_R^{\alpha}][\bar{s}_L^\beta d_R^{\beta}], \nonumber \\
    &\mathcal{O}^K_3 = [\bar{s}_R^\alpha d^\beta_L][\bar{s}_R^\beta d^\alpha_L], &\tilde{\mathcal{O}}^K_3 = [\bar{s}_L^\alpha d_R^{\beta}][\bar{s}_L^\beta d_R^{\alpha}], \nonumber \\
    &\mathcal{O}^K_4 = [\bar{s}_R^\alpha d^\alpha_L][\bar{s}_L^\beta d^\beta_R], \nonumber\\
    &\mathcal{O}^K_5 = [\bar{s}_R^\alpha d^\beta_L][\bar{s}_L^\beta d^\alpha_R], \nonumber
\end{align}
where $\alpha$ and $\beta$ denote color indices. The quantity of interest in our discussions shall be the dispersive piece of the $\bar{K}^{0}-K^{0}$ mixing matrix, which is then given by
\begin{equation}
    M^K_{12} =\frac{1}{2 m_K}\langle \bar{K}^{0} | \mathcal{H}^{\Delta S=2}_{eff} | K^{0} \rangle = \frac{1}{2 m_K} \bigg( \sum_{i=1}^5 \mathcal{C}^K_i (\mu) \langle \bar{K}^{0} | \mathcal{O}^K_i (\mu) | K^{0} \rangle +\sum_{i=1}^3 \tilde{\mathcal{C}}^K_i (\mu) \langle \bar{K}^{0} | \tilde{\mathcal{O}}^K_i (\mu) | K^{0} \rangle \bigg),
\end{equation}
where $m_K= 0.497$ GeV is the mass of the $K^0$ meson, and we note that due to the parity invariance of QCD, $\langle \bar{K}^{0} | \tilde{\mathcal{O}}_i | K^{0}\rangle = \langle \bar{K}^{0} | \mathcal{O}_i | K^{0} \rangle$. The expectation values of the various operators may be obtained from lattice QCD calculations; we use the values obtained in \cite{carrasco}, evaluated at the renormalization scale of 3 GeV at which they are evaluated in that source. To compute our mixing matrix element, then, we must derive the coefficients $\mathcal{C}^K_i$ and $\tilde{\mathcal{C}}^K_i$ from our expressions for the various four-quark coefficients that we have derived in Section \ref{4QuarkOperatorSection}; in the language of that section, we then need to compute the coefficients $(C^{dd}_{LL,LR,RL,RR})^{E}_{21;21}$ for any gauge boson $E$ and $(C^{dd}_{LR;LR,LR;RL,RL;LR,RL;RL})^{S}_{21;21}$ for any scalar $S$ in our model. The coefficients we compute using the expressions in Section \ref{4QuarkOperatorSection} are then Wilson coefficients evaluated at the scale $M_{KK}$; they can then be run down to lower energy scales using anomalous dimension matrices (ADMs) which can be extracted from \cite{bagger}. In our model, we note that several mechanisms conspire to eliminate or strongly suppress all of these coefficients. First, we consider the effects of exchanges of  vector bosons (namely, those of the gauge group $SU(2)_F \times U(1)_F$ and those of the SM). As only neutral currents contribute at tree-level to the operators of Eq.(\ref{KMesonOperators}), we can generically describe our various operators using Eq.(\ref{Vector4Fermion}). Meanwhile, as we have noted in Section \ref{BulkMatterModelSection}, the mass eigenvectors $\Vec{a}^{(D,d)}_2$ are both equal to $(0,1/\sqrt{2},0)$ (up to a complex phase), and $\Vec{a}^{(D,d)}_{1}$ and $\Vec{a}^{(D,d)}_{3}$ have no component in the direction of $\Vec{a}^{(D,d)}_2$, since there exists no mixing between the second generation of the $SU(2)_L$ doublet and down-like singlet quarks and the other generations. Then, we see that for any neutral current gauge boson $E$, the quantity $(\mathbf{\Omega}^{E}_{L})^{(D)}_{21}(\phi)$, as defined in Eq.(\ref{OmegaDefs}, is given by
\begin{equation} \label{KaonFCNCOmega}
    (\mathbf{\Omega}^{E}_{L})^{(D)}_{21}(\phi) \sim \frac{1}{\sqrt{2}} \sum_{i=1,3} (\mathbf{C}^{(D)}_2(\phi))_{22} (\mathbf{X}^{D}_E)_{2i} (\mathbf{C}^{(D)}_1(\phi))_{ii} (\Vec{a}^{(D)}_1)_i,
\end{equation}
with completely analogous expressions for right-handed quarks and the $SU(2)_L$ down-like singlet quarks. Notably, we see that the only gauge bosons which have a coupling matrix with $\mathbf{X}^{D,d}_{21}$ and/or $\mathbf{X}^{D,d}_{23}$ unequal to zero can have a non-zero contribution to any neutral-current tree-level four-quark operator coupling the $d$ quark to the $s$ quark. So, we see from Eqs.(\ref{FlavorGaugeXQ}), (\ref{FlavorGaugeXd}), and (\ref{DiagonalXSM}) that all of the 4-quark operators arising from the exchanges of the $A^3$, $B$, and SM gauge bosons have vanishing contributions to this process. This leaves only the couplings from $A^1$ and $A^2$ exchanges. Inserting Eqs.(\ref{FlavorGaugeXQ}) and (\ref{FlavorGaugeXd}) into Eq.(\ref{KaonFCNCOmega}) then, we immediately see that 
\begin{equation}
    (\mathbf{\Omega}^{A^2}_{L,R})^{(D,d)}_{21} = i (\mathbf{\Omega}^{A^1}_{L,R})^{(D,d)}_{21}.
\end{equation}
Using this identity with Eq.(\ref{Vector4Fermion}), and noting that $A^1$ and $A^2$ have identical bulk masses (and hence identical bulk profiles, $\chi^{A^{1}}_n (\phi)=\chi^{A^{2}}_n (\phi)$), we see that the Wilson coefficients of the 4-quark operators arising from $A^1$ and $A^2$ obey the relation,
\begin{equation}
    (C^{dd}_{XY})^{A^1}_{21;21}=-(C^{dd}_{XY})^{A^2}_{21;21}
\end{equation}
for any set of chiralities $X$ and $Y$. So, the only non-trivial tree-level $\Delta S = 2$ four-quark operators for $\bar{K}^0-K^0$ mixing arising from gauge boson exchange cancel one another completely. In fact, this cancellation can be seen as a consequence of our choice of bulk scalar vev structure; in the vacuum arrangement selected, it can be shown that the gauge boson interactions still respect a conserved $U(1)$ flavor-like charge, as though there remains an unbroken global $U(1)$ symmetry \cite{diazcruz}. Because the quark mass eigenstates themselves are not eigenstates of this flavor charge in our model (due to their mixing from their bulk scalar couplings), keeping track of these flavor quantum numbers is largely a pointless exercise, however due to the fact that the second generation of the down-like quarks remains completely unmixed (and hence a flavor charge eigenstate for both the $SU(2)_L$ doublet and down-like singlet fields), and its flavor charge eigenvalue is unequal to either of the two flavor charge eigenstates which mix to form the first- and third-generation down-like quarks, none of the operators in Eq.(\ref{KMesonOperators}) conserve flavor charge. As a result, we observe that $\textit{none}$ of these operators can have any contribution from any vector gauge bosons in our theory.

\begin{figure}
    \centering
    \includegraphics{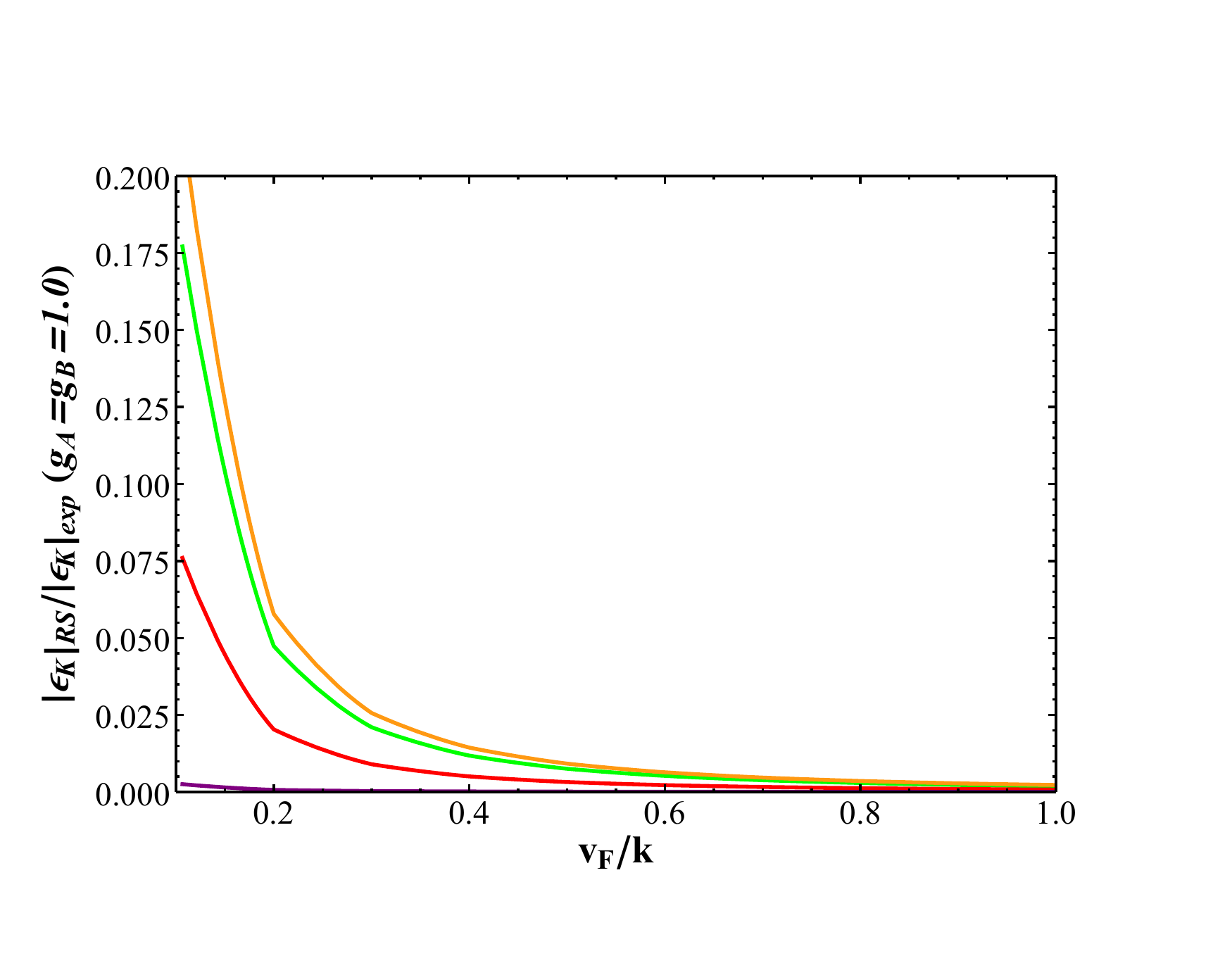}
    \caption{the contribution to $|\epsilon_K|$ emerging from new physics in our model, compared to its experimental value $(2.228 \pm 0.011)\times 10^{-3}$. Different colors represent different sample points in parameter space.}
    \label{fig:EpsilonKvF}
\end{figure}

The only processes which can contribute at tree level to $\bar{K}^0-K^0$ mixing therefore arise from exchanges of the scalars $H_{1-4}$ and $a_{1-4}$. However, as we have already seen in Sections \ref{FlavorScalar4FermionSection} and \ref{FlavorGoldstone4FermionSection}, these couplings are extremely suppressed because they link orbifold odd quark bulk profiles to orbifold even ones. Despite this, we also note that in anarchic RS flavor models, the contributions to the indirect $CP$ violation parameter $\epsilon_K$ can be as much as 3 orders of magnitude larger than its experimentally measured value. In the interest of caution, then, we proceed to complete a computation of $\epsilon_K$ within our model. We begin by recalling that the new physics contribution to $\epsilon_K$ is given by \cite{casagrande2,burasEpsilonK}
\begin{equation}
    |\epsilon_K| = \frac{1}{2 m_K}\frac{\kappa_\epsilon }{\sqrt{2}(\Delta m_K)_{\textrm{exp}}} |\textrm{Im} \langle \bar{K}^0 | \mathcal{H}^{\Delta S=2}_{eff}| K^0 \rangle|,
\end{equation}
with $\kappa_\epsilon \approx 0.92$ \cite{casagrande2,bauer,burasEpsilonK}. Evaluating the expectation value $\langle \bar{K}^{0} | \mathcal{H}^{\Delta S=2}_{eff} | K^{0} \rangle$ using the methods we have outlined earlier in this section, then, we can estimate the contribution that new physics makes to the indirect $CP$-violating parameter $\epsilon_K$. In Figure \ref{fig:EpsilonKvF}, we plot our computed value of $|\epsilon_K|$ stemming solely from our model's new physics (namely, the flavor-changing neutral currents arising from RS effects on the SM gauge sector and our new flavor gauge bosons and scalars), compared with the experimentally measured value of this parameter, for different choices of $v_F/k$ and for several different sample points in parameter space. Notably, the magnitude of $\epsilon_K$ dramatically declines with increasing $v_F$ (as might be expected given the behavior of the Wilson coefficients that contribute to the $\Delta S = 2$ Hamiltonian), but even when $v_F =0.1 k$, the correction does not, for any of our sample points, represent an unreasonable fraction of total observed indirect $CP$ violation. Even if there were a point in parameter space such that the $v_F=0.1 k$ contribution to $\epsilon_K$ were unacceptably large, this contribution decreases so rapidly as $v_F/k$ increases that we can easily evade such constraints by increasing $v_F$ only slightly, to $0.2k$ or higher. As such, indirect CP violation in the $\bar{K}^{0}-K^{0}$ imposes negligible, if any, constraints on our model; even conservatively all we can really assert is that $v_F$ must not be unreasonably small, or in other words, $v_F \gsim 0.1 k$.

We conclude our remarks on $\bar{K}^{0}-K^{0}$ mixing with one final note: We have found numerically that the contribution to the $\Delta S = 2$ Hamiltonian of $a_{1-4}$ exchanges is identically zero, making the exchange of the bulk scalars $H_{1-4}$ the sole contributor to $CP$ violation in this sector. Intuitively, this can be seen as a manifestation of the flavor charge conservation that protects this mixing from contributions from the vector gauge bosons, however, as a detailed exploration of this phenomenon would involve discussing suppression mechanisms for an already suppressed (by powers of the quark masses) operator, we omit any further investigation of this phenomenon here.

\subsection{$B^0-\bar{B}^0$ Mixing}\label{BBBarMixingSection}
\begin{figure}
    \centerline{\includegraphics[width=3.5in]{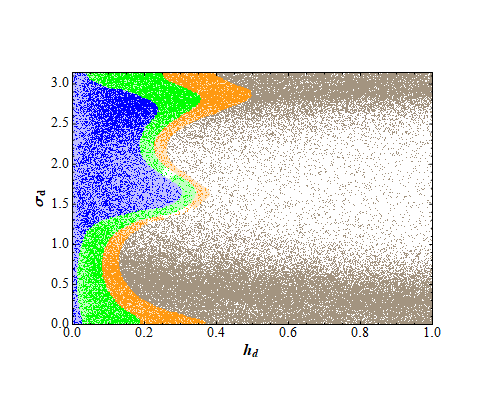}
    \hspace{-0.75cm}
    \includegraphics[width=3.5in]{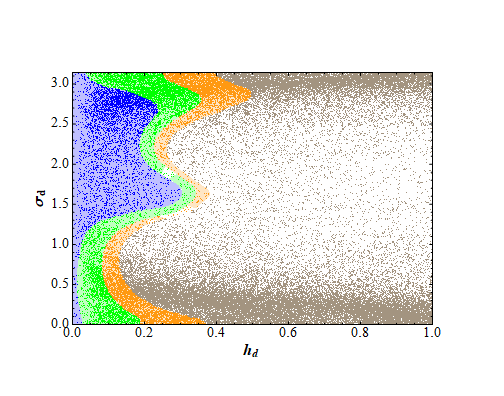}}
    \vspace*{-0.25cm}
    \centerline{\includegraphics[width=3.5in]{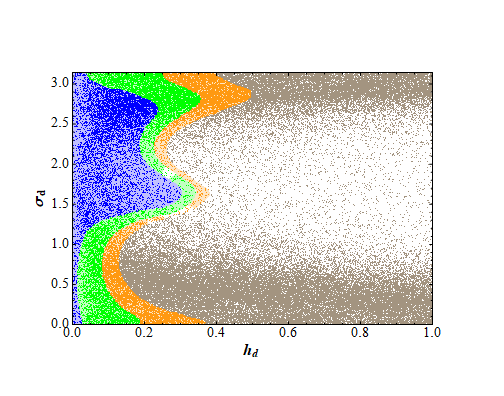}
    \hspace{-0.75cm}
    \includegraphics[width=3.5in]{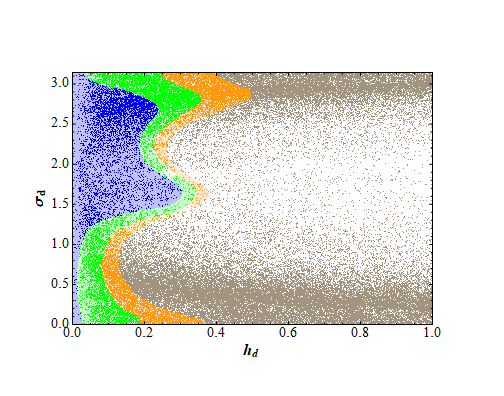}}
    \caption{$h_d-\sigma_d$ scatter plots for the full sample of parameter space points, for $g_A=g_B=0.1$ and $v_F/k=0.1$ (top left), $g_A=g_B=1.0$ and $v_F/k=0.1$ (top right), $g_A=g_B=0.1$ and $v_F/k=1.0$ (bottom left), and $g_A=g_B=1.0$ and $v_F/k=1.0$ (bottom right). The blue represents the 68\% CL experimental allowed region, the green represents the 95\% CL region, and the orange represents the 99.7\% CL region \cite{ckmfitterBBBar}}
    \label{fig:BBBarMixing}
\end{figure}

We now discuss the constraints on new physics stemming from $\bar{B}^{0}-B^{0}$ mixing within our model. Notably, we find that any 4-quark operators in $\bar{B}^{0}_s-B^{0}_s$ mixing stemming from gauge boson exchanges vanish in our model, by the same reasoning as we applied in Section \ref{KKBarMixingSection} when discussing $\bar{K}^{0}-K^{0}$ mixing. Unlike the case of $\bar{K}^{0}-K^{0}$ mixing, however, there is no mixing parameter sensitive enough to the highly suppressed operators emerging from scalar exchanges in order to merit even a somewhat quantitative investigation of these effects; rather, we simply ignore them.  Turning our attention instead to $\bar{B}^{0}-B^{0}$ mixing, we see that because the first- and third-generation down-like quarks mix freely in our model, the contributions to the mixing parameters in this sector from both SM and $SU(2)_F \times U(1)_F$ gauge bosons are unsuppressed. We can then compute $\langle \bar{B}^{0} | \mathcal{H}^{\Delta B = 2}_{eff} | B^{0} \rangle$ in much the same manner as we computed $\langle \bar{K}^{0} | \mathcal{H}^{\Delta S=2}_{eff} | K^{0} \rangle$ in Section \ref{KKBarMixingSection}, the only differences being our substitution of $B$ meson hadronic matrix elements (extracted from \cite{bazavov}) rather than the matrix elements used in \ref{KKBarMixingSection}, and the operator definitions in Eq.(\ref{KMesonOperators}) undergoing the substitution, $s \rightarrow b$. In terms of the four-fermion operators computed in Section \ref{4QuarkOperatorSection}, we determine the non-zero Wilson coefficients to be
\begin{align}\label{BBBarWilsonCoeffs}
    &\mathcal{C}^{B}_1(M_{KK}) = \frac{1}{3} (C^{dd}_{LL})^g_{31;31} + (C^{dd}_{LL})^\gamma_{31;31}+(C^{dd}_{LL})^Z_{31;31} +(C^{dd}_{LL})^{B}_{31;31}+ \sum_{i=1}^3 (C^{dd}_{LL})^{A^i}_{31;31},\\
    &\tilde{\mathcal{C}}^{B}_1(M_{KK}) = \frac{1}{3} (C^{dd}_{RR})^g_{31;31} + (C^{dd}_{RR})^\gamma_{31;31}+(C^{dd}_{RR})^Z_{31;31} +(C^{dd}_{RR})^{B}_{31;31}+ \sum_{i=1}^3 (C^{dd}_{RR})^{A^i}_{31;31}, \nonumber \\
    &\mathcal{C}^{B}_4(M_{KK}) = -2 (C^{dd}_{LR})^g_{31;31}, \nonumber \\
    &\mathcal{C}^{B}_5(M_{KK}) = \frac{2}{3} (C^{dd}_{LR})^g_{31;31} -4 (C^{dd}_{LR})^\gamma_{31;31} - 4 (C^{dd}_{LR})^Z_{31;31} - 4 (C^{dd}_{LR})^{B}_{31;31} - 4 \sum_{i=1}^3 (C^{dd}_{LR})^{A^i}_{31;31}.\nonumber
\end{align}
Running these coefficients down to the scale $m_b=4.18$ GeV of the hadronic matrix elements given in \cite{bazavov}, we can now evaluate $\langle \bar{B}^{0} | \mathcal{H}^{\Delta B = 2}_{eff} | B^{0} \rangle$. In order to determine how the RS contributions to this process constrain the theory, we now divide the dispersive mixing amplitude we have derived above, (namely, just the expectation value $M^B_{12} = \langle \bar{B}^{0} | \mathcal{H}^{\Delta B = 2}_{eff} | B^{0} \rangle/(2 m_B)$, where $m_B=5.28$ GeV is the mass of the $B^0$ meson) by one-loop electroweak SM contribution \cite{burasBBBar},
\begin{equation}
    (M^B_{12})_{SM} = \frac{G_F^2 m_W^2 \eta_B m_B B_B f^2_B}{12 \pi^2} S_0(m_t^2/m_W^2) (V_{td}^* V_{tb})^2,
\end{equation}
where $\eta_B \approx 0.84$ is a QCD correction, $S_0(x)$ is the Inami-Lim function \cite{burasBBBar,inami}, $B_B$ is a QCD bag parameter extracted from the same hadronic matrix elements as were used in evaluating the RS contribution $M^B_{12}$, and $f_B$ is the $B$ meson decay constant. We now follow \cite{ckmfitterBBBar} and parameterize the ratio of $(M^B_{12})_{RS}$ to $(M^B_{12})_{SM}$ as
\begin{equation}
    \frac{(M^B_{12})_{RS}}{(M^B_{12})_{SM}} = h_d e^{2i \sigma_d},
\end{equation}
where $(M^B_{12})_{RS}$ is the contribution to $M^B_{12}$ stemming entirely from tree-level RS contributions, and $h_d$ and $\sigma_d$ are real parameters. We can then compute $h_d$ and $\sigma_d$ for the points in parameter space that we have numerically sampled, and compare the results to existing constraints based on a global fit of meson mixing observables to new physics \cite{ckmfitterBBBar}. A scatter plot of our numerically sampled parameter points in the $h_d-\sigma_d$ plane is  is presented in Figure \ref{fig:BBBarMixing}, for different values of the bulk $SU(2)_F \times U(1)_F$ coupling constants $g_{A,B}$ and the bulk vev $v_F$.

Notably, in the general case $h_d$ can be unacceptably large. In fact, we find that $h_d$ can even in some cases obtain values of as high as $\sim 10$. Since our sampling of points is by no means guaranteed to be unbiased, however, this does not necessarily indicate a fine-tuning problem. However, even if we do assume that our sampling is at least close to representative of our parameter space as a whole, we note that a qualitative examination still suggests that the fine-tuning situation is likely significantly better than the tuning necessary for the flavor-anarchic model (assuming that the parameters we have assumed to vanish in Section \ref{BulkMatterModelSection} do not present an additional naturalness problem, which we posited in that section they do not with minimal modification of the model). In our sample, the \textit{largest} values we obtain for $h_d$ are roughly an order of magnitude greater than the $h_d$ constraints. Specifically, the $95\%$ quantile of the $h_d$ values for the most restricted scenario we consider, in which $g_A=g_B=1.0$ and $v_F/k=0.1$, is roughly 7.5, or approximately 25 times the maximum allowed $h_d$ (at 95\% CL), which we can (very roughly) estimate as approximately 0.3 from the allowed region depicted in Figure \ref{fig:BBBarMixing}. The median $h_d$ value for this data set is approximately 1.4, or just several times the maximum allowed value. In \cite{casagrande2} at the scale $M_{KK}= 5$ TeV, the authors find that the median $\epsilon_K$ value their sampling of anarchic models yields exceeds experimental constraints by roughly an order of magnitude, and at worst, roughly $5\%$ of their sample points exceed the SM result by $O(100)$. While hardly a precise comparison between the models, this simple reckoning suggests that the $\epsilon_K$ prediction in the anarchic model tends to be further divorced from experimental constraints than the $h_d$ prediction in our model (since the discrepancy between its median predictions and the experimentally allowed region are greater), and that points in the anarchic parameter space which satisfy the $\epsilon_K$ constraints are likely less numerically stable, since the anarchic $\epsilon_K$ predictions span a much broader range of proportional values than our model's $h_d$ predictions. We leave a detailed probe of the fine-tuning present in our model, and a quantitative comparison with the anarchic case (or to other models with some form of flavor symmetry), to future work.

\begin{figure}
    \centerline{\includegraphics[width=3.5in]{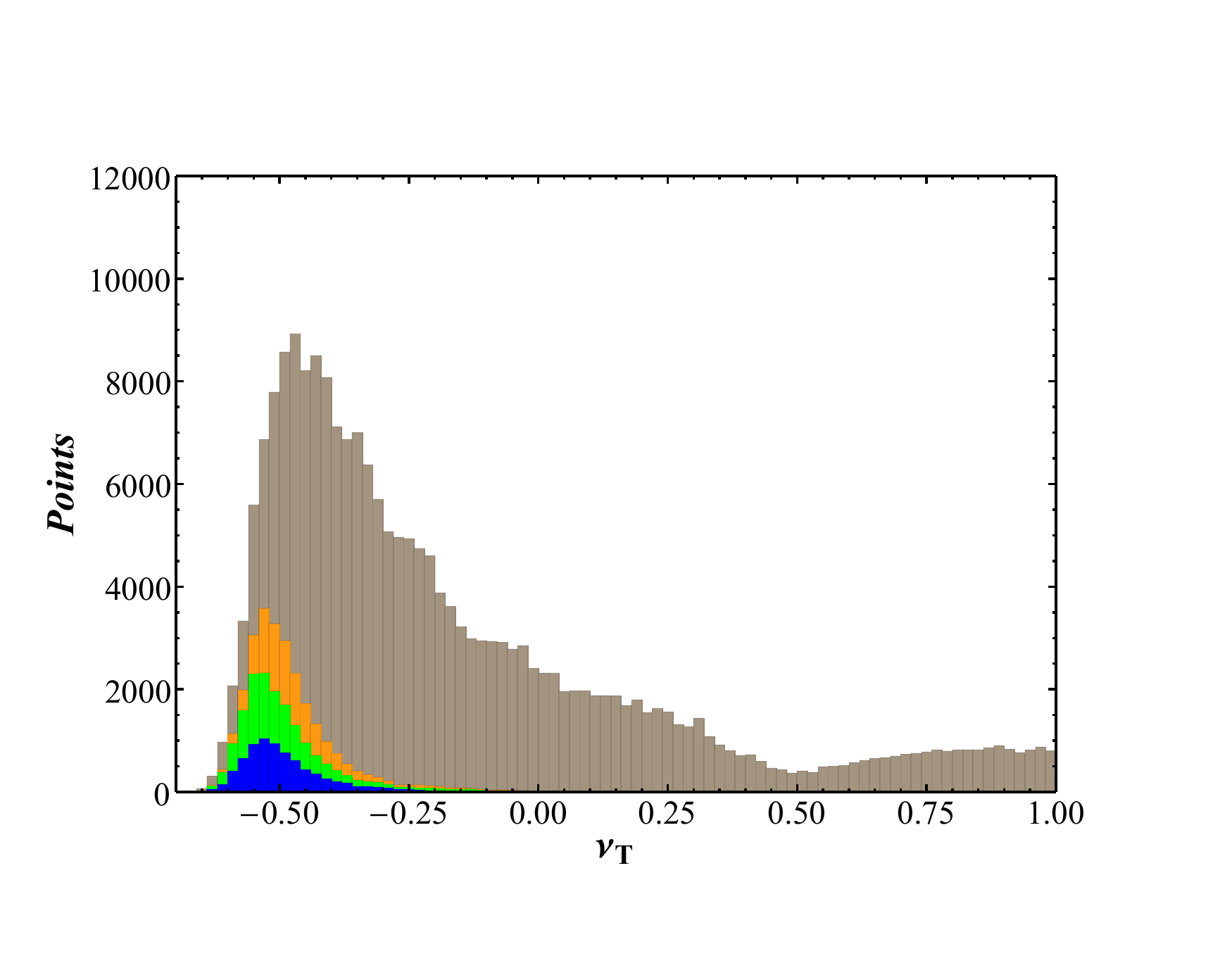}
    \hspace{-0.75cm}
    \includegraphics[width=3.5in]{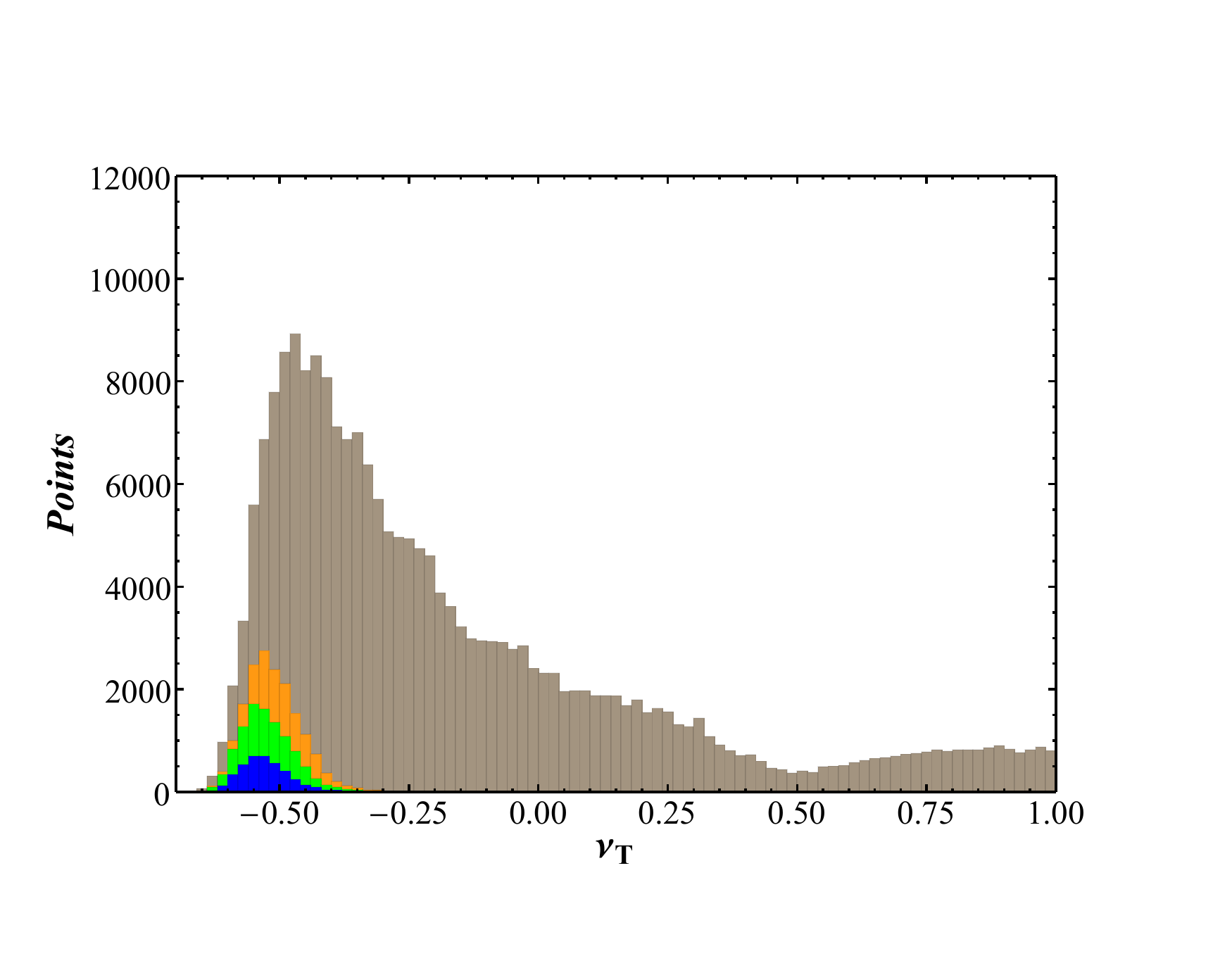}}
    \vspace*{-0.25cm}
    \centerline{\includegraphics[width=3.5in]{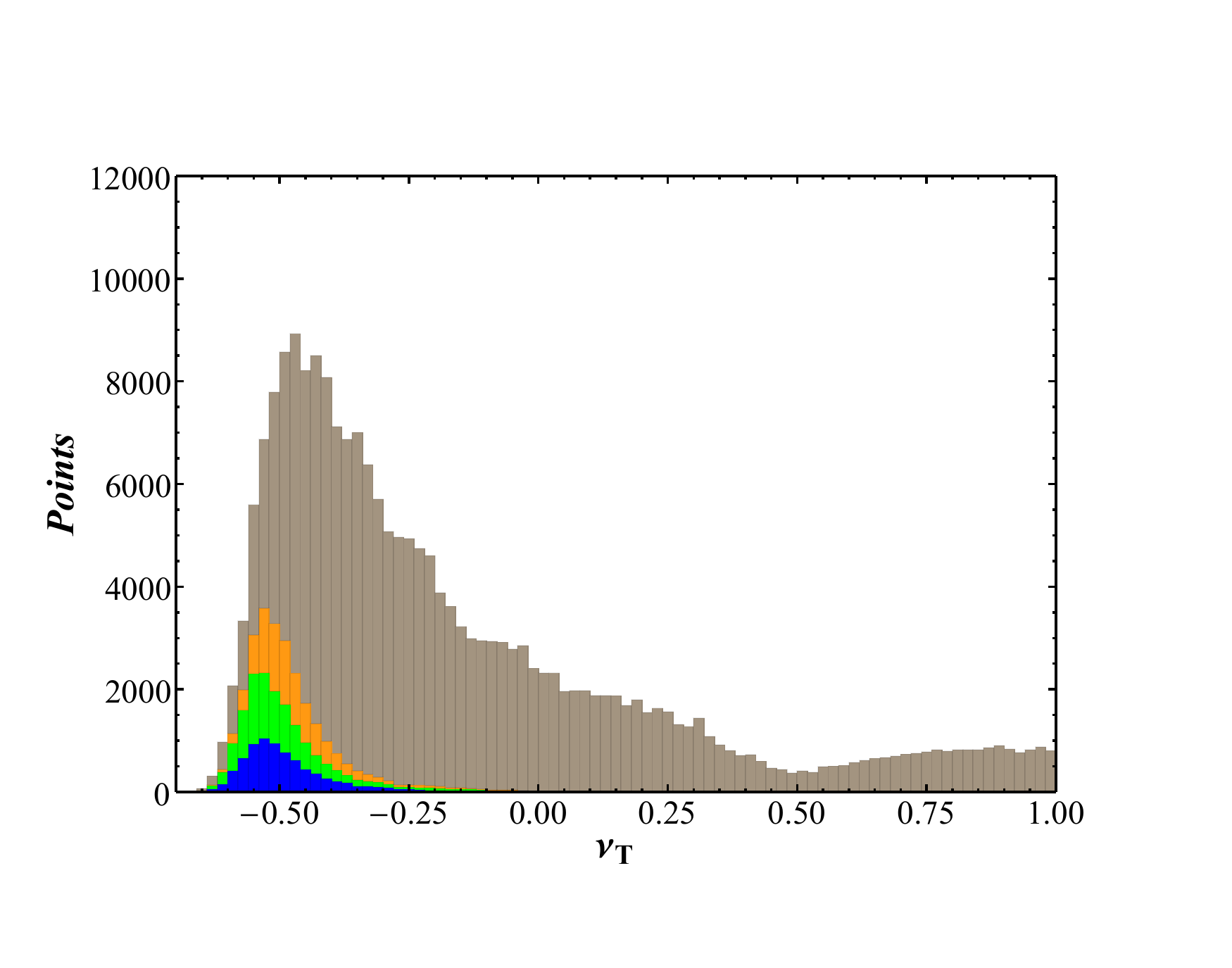}
    \hspace{-0.75cm}
    \includegraphics[width=3.5in]{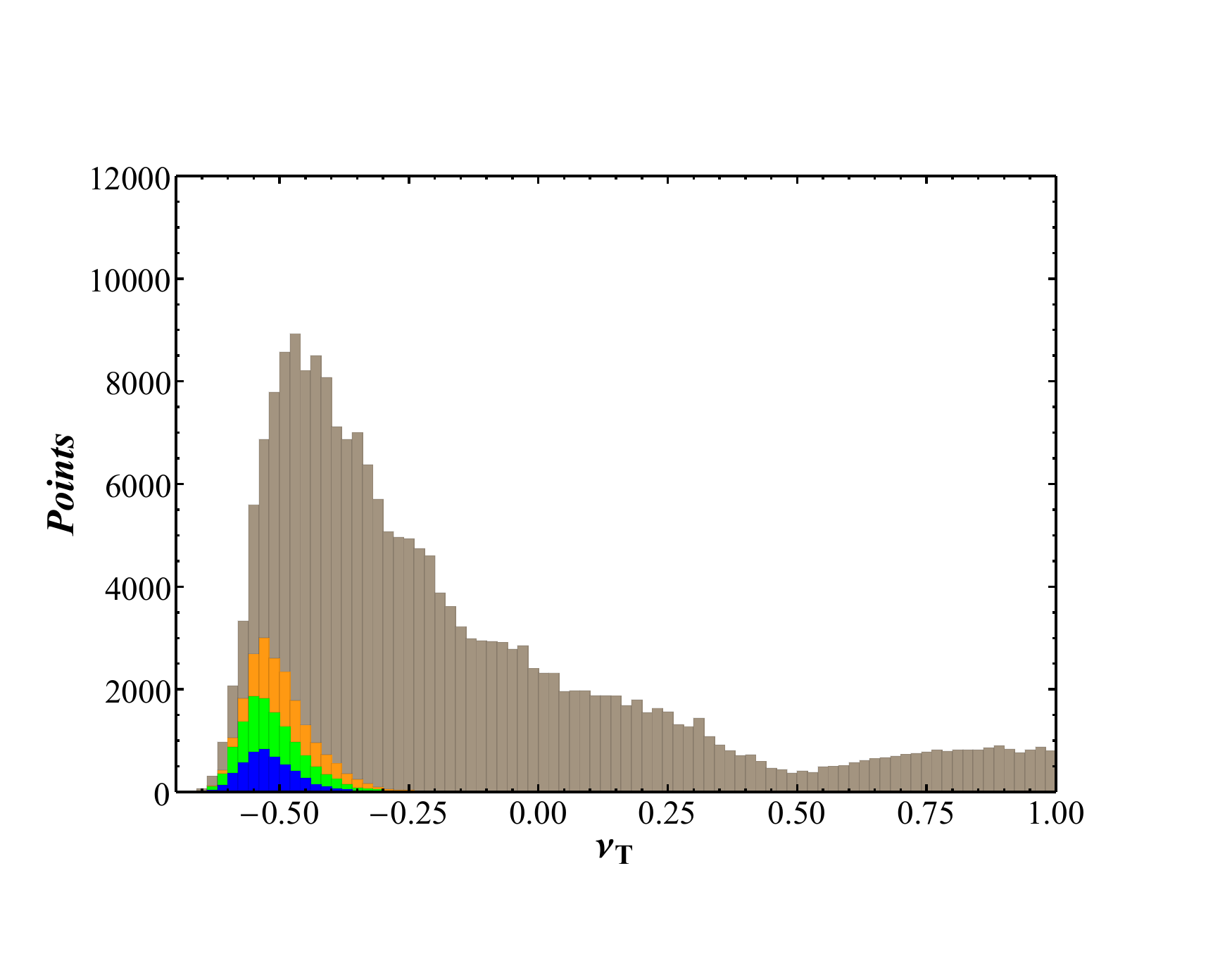}}
    \caption{Histograms of the parameter $\nu_T$, for $g_A=g_B=0.1$ and $v_F/k=0.1$ (top left), $g_A=g_B=1.0$ and $v_F/k=0.1$ (top right), $g_A=g_B=0.1$ and $v_F/k=1.0$ (bottom left), and $g_A=g_B=1.0$ and $v_F/k=1.0$ (bottom right). The blue represents points which satisfy the 68\% CL experimental limit in the $h_d-\sigma_d$ plane, the green represents the 95\% CL region, the orange represents the 99.7\% CL region, and the beige includes all points in our sample pool, with no $h_d-\sigma_d$ constraints.}
    \label{fig:BBBarnus}
\end{figure}

Fine-tuned or not, the harsh $\bar{B}^{0}-B^{0}$ constraints give us significant information about the region of model parameter space that are acceptable. In particular, we notice that the $\bar{B}^{0}-B^{0}$ constraints are particularly sensitive to the value of the bulk localization parameter $\nu_T$ (as described in Section \ref{BulkMatterModelSection}). In Figure \ref{fig:BBBarnus}, we depict histograms of this parameter for several values of $g_{A,B}$ and the bulk vev $v_F$. Clearly from the Figure, the points in parameter space that recreate the quark masses and mixings with a $\nu_T$ above $\sim-0.25$ are generally excluded by the $\bar{B}^{0}-B^{0}$ mixing constraints. Given its strong influence on the bulk localization of the third-generation $SU(2)_L$ doublet quarks, this is unsurprising: There exist a number of points in our parameter space that recreate the CKM matrix and the observed quark masses with an $O(1)$ and positive $\nu_T$, indicating highly TeV-brane localized $b_L$ and $t_L$ quarks, however, it is well-known that a more TeV-brane localized fermion field is more strongly coupled to the KK modes of bulk gauge fields. As a result, these points in parameter space are more subject to large new flavor-changing neutral currents featuring $b_L$ than those which have a more negative $\nu_T$. 

\begin{figure}
    \centerline{\includegraphics[width=3.5in]{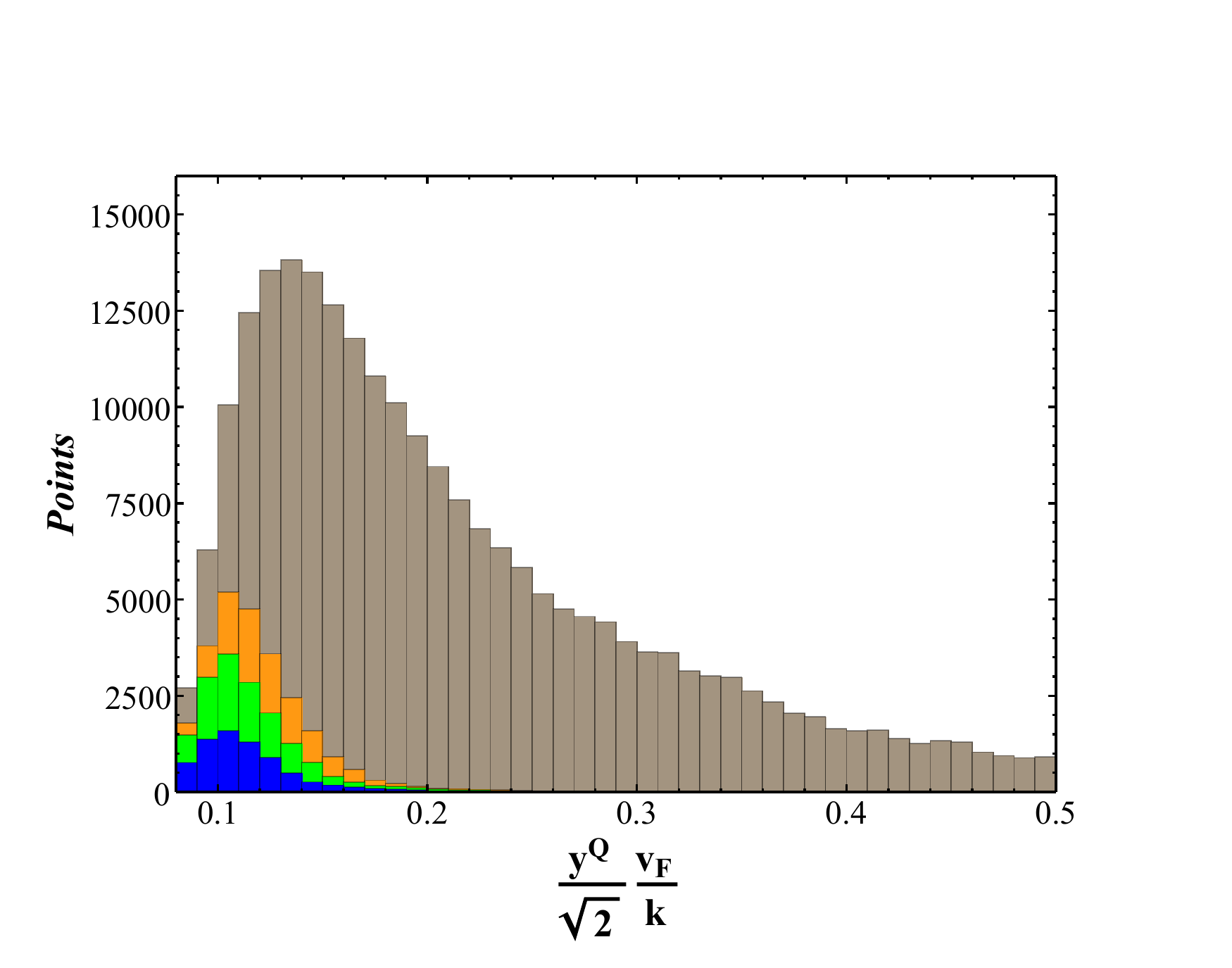}
    \hspace{-0.75cm}
    \includegraphics[width=3.5in]{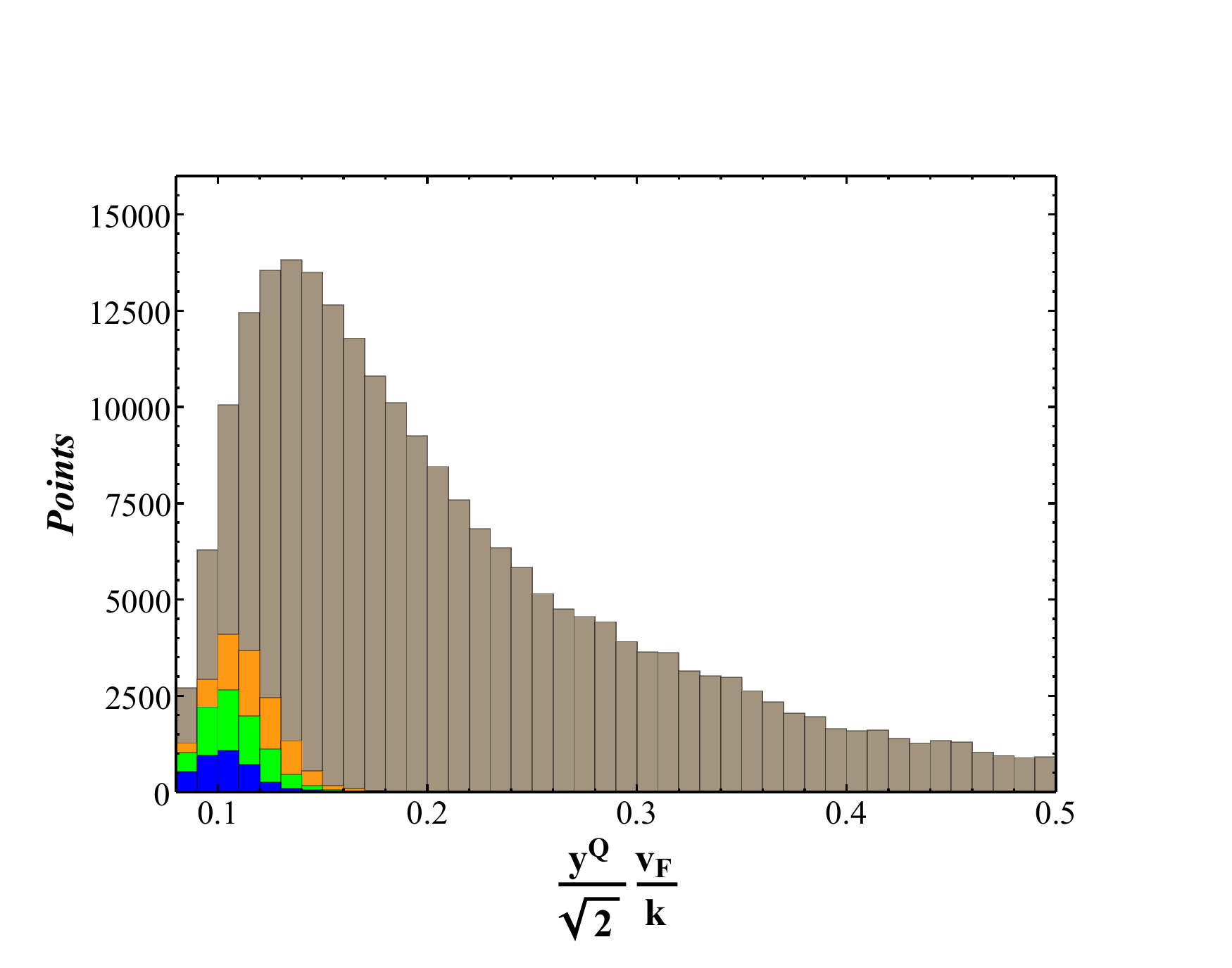}}
    \vspace*{0.0cm}
    \centerline{\includegraphics[width=3.5in]{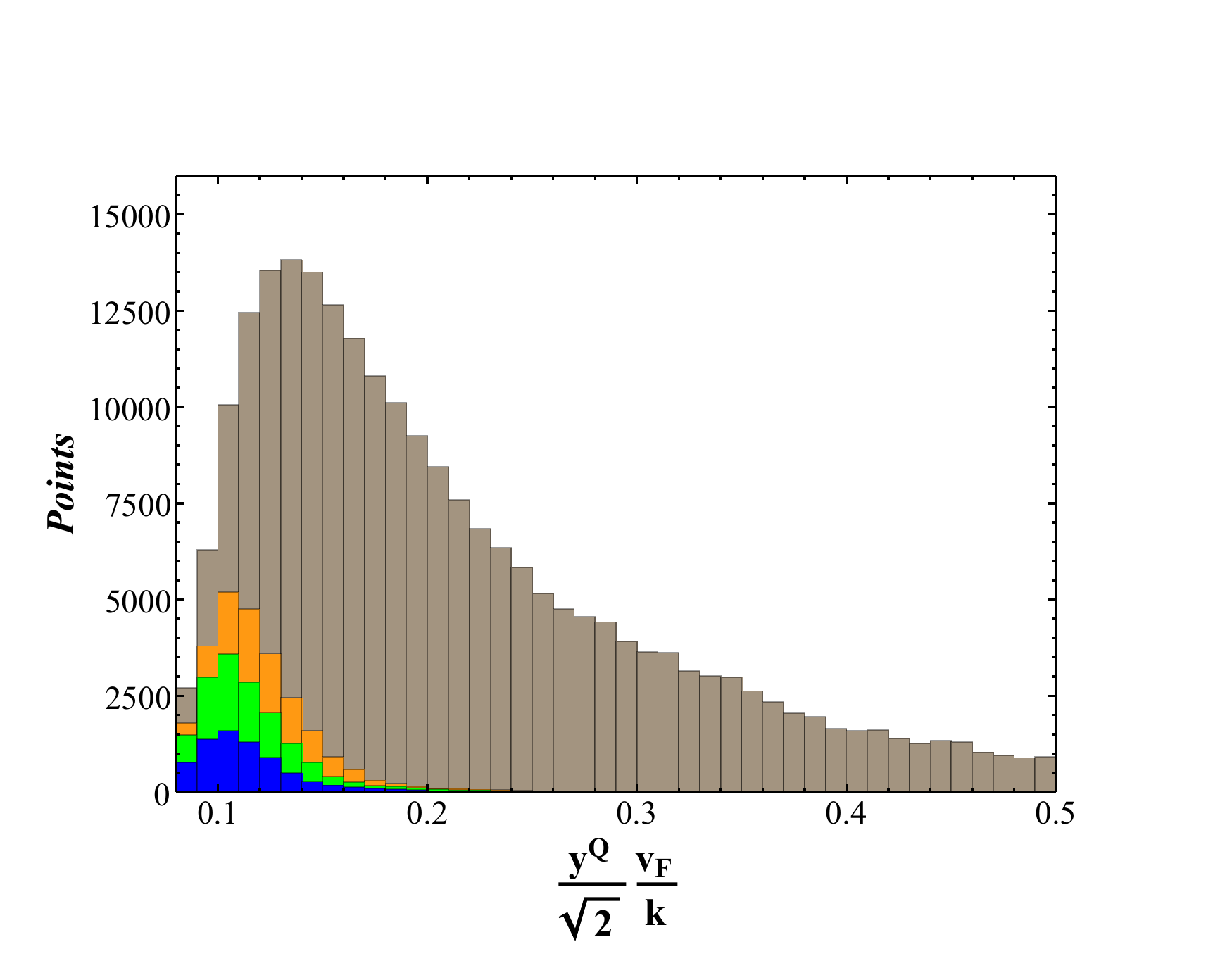}
    \hspace{-0.75cm}
    \includegraphics[width=3.5in]{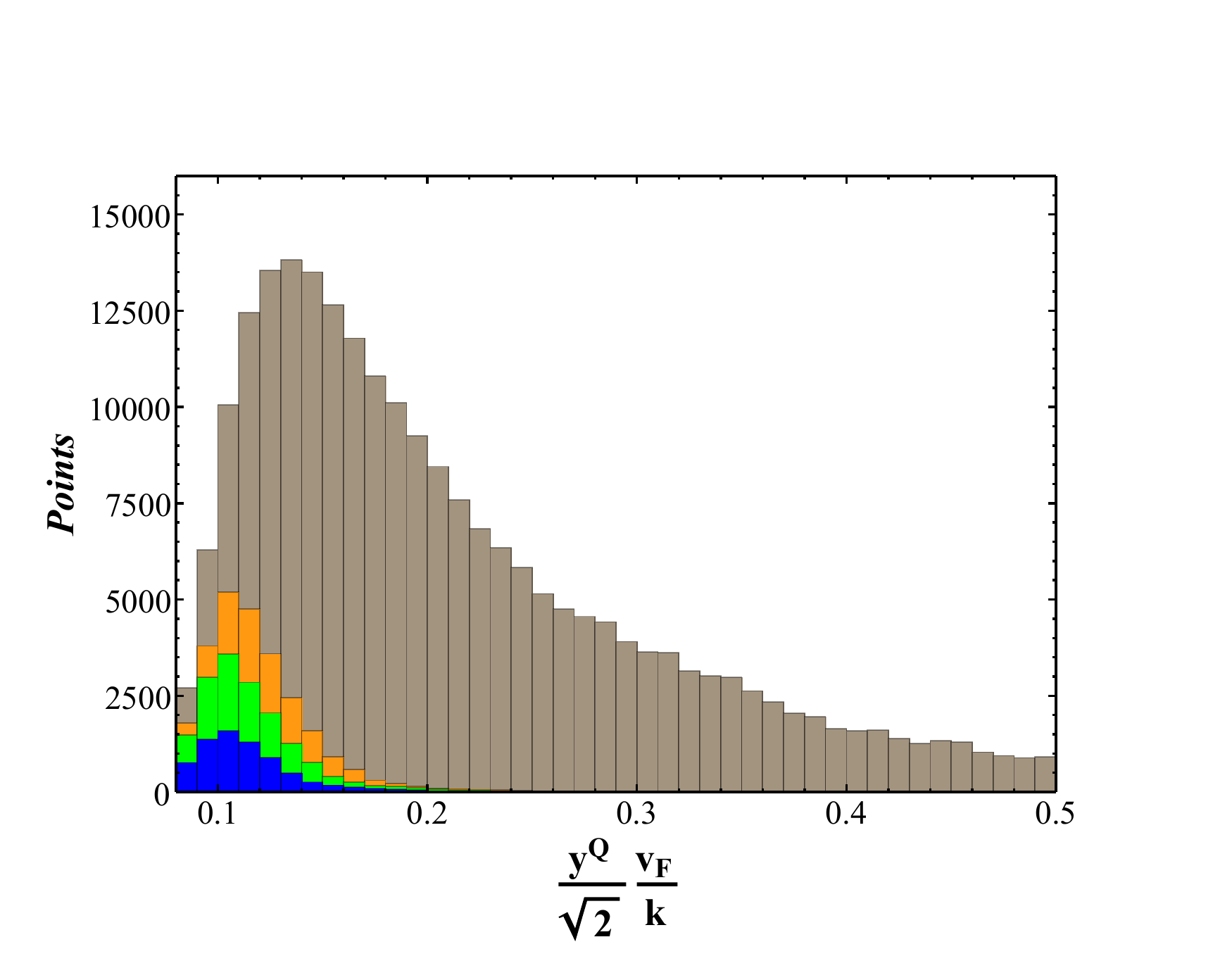}}
    \caption{Histograms of the $y^Q$, for $g_A=g_B=0.1$ and $v_F/k=0.1$ (top left), $g_A=g_B=1.0$ and $v_F/k=0.1$ (top right), $g_A=g_B=0.1$ and $v_F/k=1.0$ (bottom left), and $g_A=g_B=1.0$ and $v_F/k=1.0$ (bottom right). The blue represents points which satisfy the 68\% CL experimental limit in the $h-\sigma_d$ plane, the green represents the 95\% CL region, the orange represents the 99.7\% CL region, and the beige includes all points in our sample pool, with no $h_d-\sigma_d$ constraints.}
    \label{fig:BBBarMixingys}
\end{figure}

We also note that the $\bar{B}^{0}-B^{0}$ constraints have a significant effect on the various bulk Yukawa coupling terms permitted in our model, which we depict in histograms in Figure \ref{fig:BBBarMixingys}. In particular, we see that the bulk Yukawa coupling $y^Q$, which parameterizes the bulk mixing between the third and first generations of the $SU(2)_L$ doublet fields, is visibly favored to be smaller in parameter space points which satisfy the constraints on $B^0$ meson mixing. This is, again, unsurprising; as $y^Q$ is smaller, the mixing between the first and third generation states in the final theory diminishes, which in turn results in smaller flavor-changing neutral currents between these generations.

\begin{figure}
    \centerline{\includegraphics[width=3.5in]{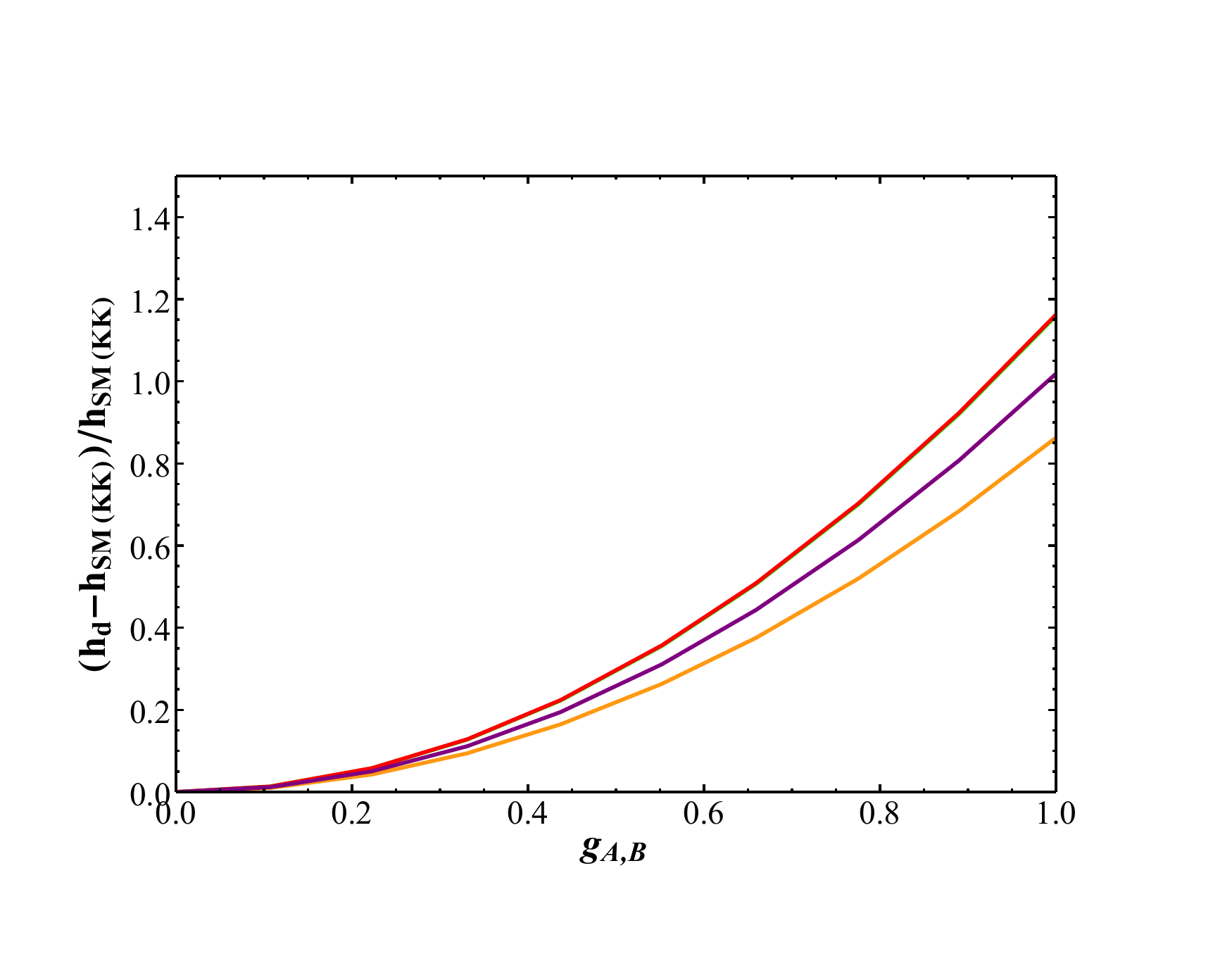}
    \hspace{-0.75cm}
    \includegraphics[width=3.5in]{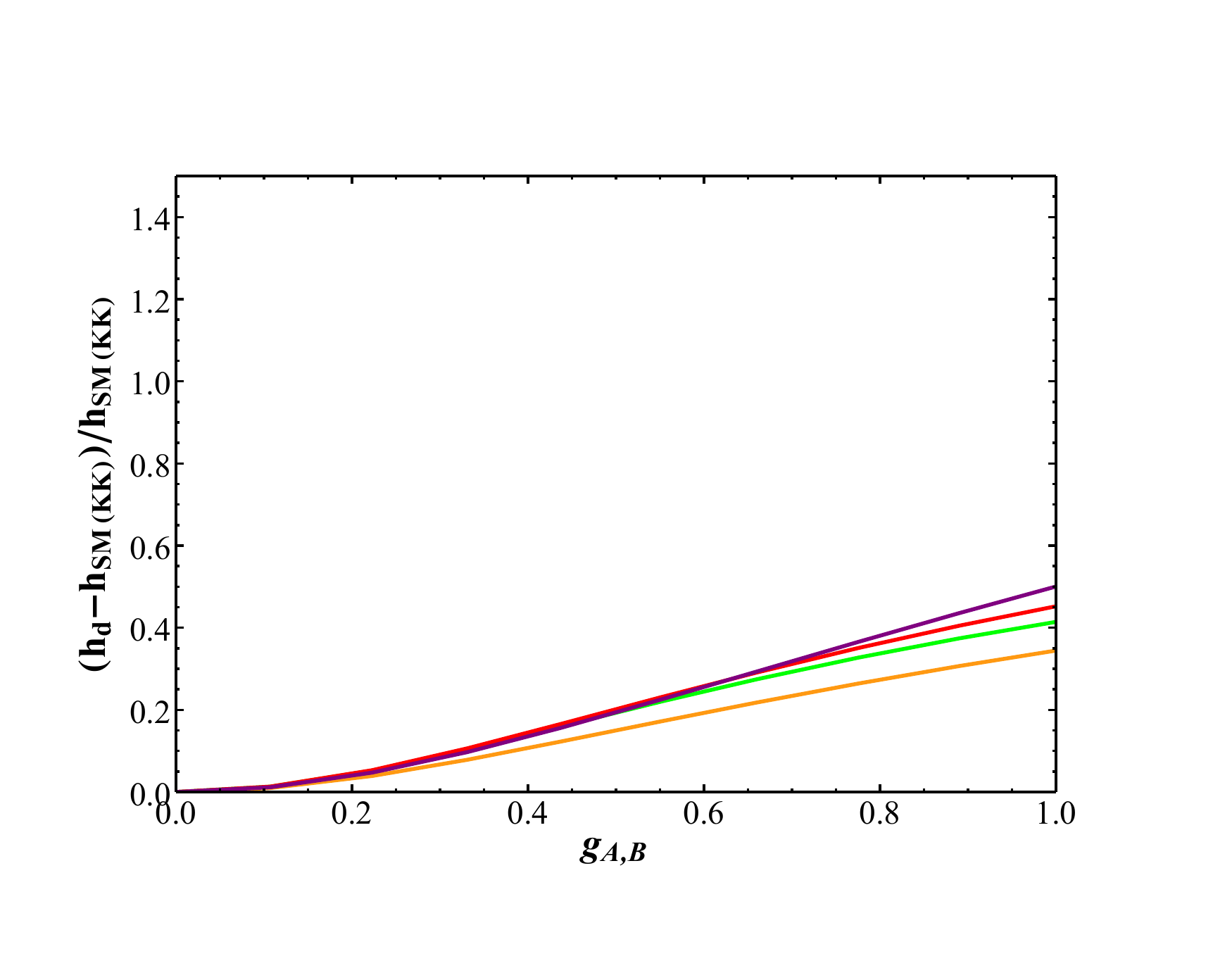}}
    \vspace*{-0.25cm}
    \centering
    \includegraphics[width=3.5in]{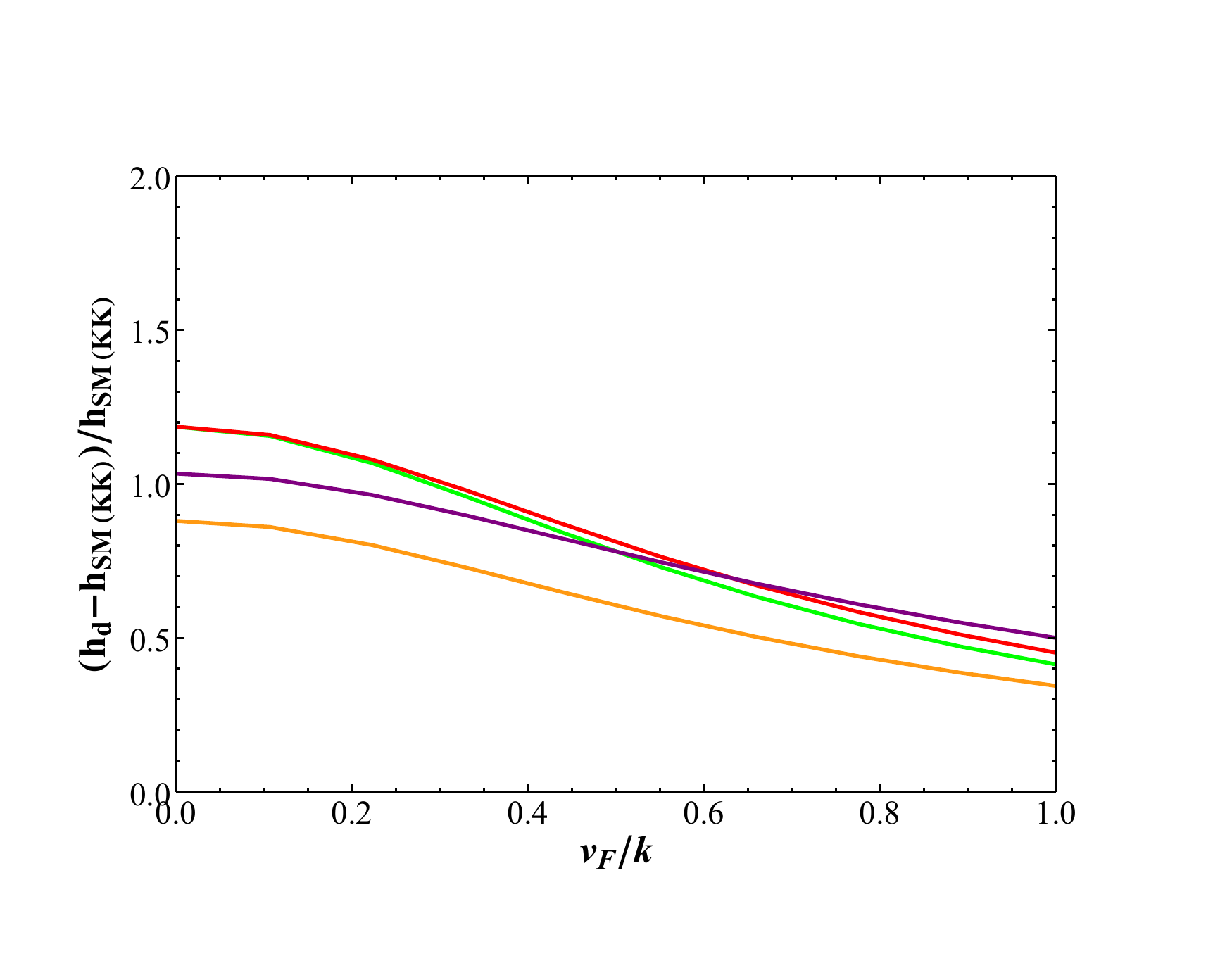}
    \caption{The dependence of the parameter $h_d$ on $g_{A,B}$ for $v_F=0.1k$ (top left), on $g_{A,B}$ for $v_F=k$ (top right), and on $v_F$ for $g_{A,B}=1.0$ (bottom), for different points in our sample parameter space, as represented by curves of different colors.}
    \label{fig:BBBarMixingGaugeDependence}
\end{figure}

Next, we comment briefly on the sensitivity of our constraints to the $SU(2)_F \times U(1)_F$ gauge coupling constants and the bulk vev $v_F$. In Figure \ref{fig:BBBarMixingGaugeDependence}, we attempt to explore the relative contribution of the new flavor gauge boson to the $\bar{B}^{0}-B^{0}$ mixing parameters compared to that of the KK modes of the SM gauge bosons by plotting the quantity,
$(h_{\textrm{full}}-h_{\textrm{SM(KK)}})/h_{\textrm{SM(KK)}}$, where $h_{\textrm{full}}$ is the computed value of $h_d$ where all contributions are included, while $h_{\textrm{SM(KK)}}$ is $h_d$ where only contributions from KK towers emerging from SM gauge boson fields ($Z$, the photon, and the gluon) are included in the computation. Notably, we see that for stronger couplings ($g_{A,B} \sim 1$), the effects of the exchange of flavor gauge bosons can rival that of the KK modes of SM bosons. Intuitively, this is to be expected, since at $g_{A,B} \sim 1$, the gauge coupling strength of the $SU(2)_F \times U(1)_F$ bosons is roughly equivalent to that of the strong force, which should \cite{casagrande} dominate the SM gauge boson contribution to these parameters. If $v_F$ is low, we find that the contribution from the flavor gauge bosons can even significantly exceed the contribution from the SM gauge fields, representing more than two thirds of the computed value of $h_d$ for some parameter space points with $v_F=0.1$, $g_A=g_B=1$. We also note that in the majority of cases, the effects of the flavor gauge bosons interfere constructively with the effects of the SM gauge bosons; with the flavor gauge bosons included, $h_d$ is almost invariably larger than the value computed with just the $Z$, photon, and gluon KK tower contributions. This can be easily seen from the number of points which are within the 95\% CL allowed region of $h_d-\sigma_d$ space with various choices of the flavor gauge couplings: For $g_A=g_B=0.1$ and $v_F=0.1 k$, 17173 of our sampled points have $h_d$ and $\sigma_d$ within this range, however, for the choice $g_A=g_B=1.0$, only 10286 points of the same sample pass the cut.

Before concluding our discussion of the $\bar{B}^0-B^{0}$ mixing parameters we note that our choice of normalization of various dimensionful bulk parameters in Section \ref{BulkSSBSection} in fact has a significant impact on our qualitative conclusions here. For example, if in lieu of writing the bulk vev as $v_F/\sqrt{2 \pi r_c}$, we were instead to write the equally valid $v'_F k^{1/2}$, this would have the effect of increasing our effective vev in all of our computations by a factor of $\sqrt{2 \pi k r_c} \approx 8.43$, and as a result, the parameter space that would seem "natural" with $v'_F \sim O(1)$ (although it should be noted that naturalness is not a good guide for the values of parameters here) differs dramatically from that which has $v_F \sim O(1)$. In particular, for $v'_F \sim O(1)$, we find that the contribution from the flavor gauge bosons to $h$ is so suppressed that it is more than two orders of magnitude smaller than the contribution from the SM gauge bosons. However, as the results of these computations become nearly equivalent to the regime where $g_{A,B}$ is small (for example, $g_A=g_B=0.1$), we do not further explore the effects of altering our normalization choices for $v_F$ or any other parameters.

\subsection{$D-\bar{D}$ Mixing}\label{DDBarMixingSection}
We now discuss constraints on our model arising from $D-\bar{D}$ mixing. Our treatment here closely resembles that of our treatment of $\bar{K}^{0}-K^{0}$ and $\bar{B}^{0}-B$ mixing in Sections \ref{KKBarMixingSection} and \ref{BBBarMixingSection}, in that we evaluate the $\Delta C=2$ Wilson coefficients in the Hamiltonian at the scale $M_{KK}$ and run them down to the scale of 3 GeV, at which scale hadronic matrix elements for the general set of $\Delta C=2$ operators are given in \cite{bazavovDDBar}. Ignoring the effects of the scalars again, the non-zero Wilson coefficients $\mathcal{C}^D_i$ are easily found in analogous expressions to those of Eq.(\ref{BBBarWilsonCoeffs}); we arrive at
\begin{align}\label{DDBarWilsonCoeffs}
    &\mathcal{C}^{D}_1(M_{KK}) = \frac{1}{3} (C^{uu}_{LL})^g_{21;21} + (C^{uu}_{LL})^\gamma_{21;21}+(C^{uu}_{LL})^Z_{21;21} +(C^{uu}_{LL})^{B}_{21;21}+ \sum_{i=1}^3 (C^{uu}_{LL})^{A^i}_{21;21},\\
    &\tilde{\mathcal{C}}^{D}_1(M_{KK}) = \frac{1}{3} (C^{uu}_{RR})^g_{21;21} + (C^{uu}_{RR})^\gamma_{21;21}+(C^{uu}_{RR})^Z_{21;21} +(C^{uu}_{RR})^{B}_{21;21}+ \sum_{i=1}^3 (C^{uu}_{RR})^{A^i}_{21;21}, \nonumber \\
    &\mathcal{C}^{D}_4(M_{KK}) = -2 (C^{uu}_{LR})^g_{21;21}, \nonumber \\
    &\mathcal{C}^{D}_5(M_{KK}) = \frac{2}{3} (C^{uu}_{LR})^g_{21;21} -4 (C^{uu}_{LR})^\gamma_{21;21} - 4 (C^{uu}_{LR})^Z_{21;21} - 4 (C^{uu}_{LR})^{B}_{21;21} - 4 \sum_{i=1}^3 (C^{uu}_{LR})^{A^i}_{21;21}.\nonumber
\end{align}
The renormalization-group running can then be performed with the same ADMs as used in Section \ref{BBBarMixingSection}, after which we can use the evolved coefficients to evaluate $\langle \bar{D} | \mathcal{H}^{\Delta C = 2}_{eff} | D \rangle$. 

To extract constraints from this calculation, we refer to \cite{hflavDDBar}, which includes the results of a global fit for the parameter,
\begin{equation}
    x_{12} \equiv \frac{|\langle \bar{D}^{0} | \mathcal{H}^{\Delta C = 2}_{eff}| D^{0} \rangle|}{m_D \Gamma_D},
\end{equation}
where $\Gamma_D=2.438 \; \textrm{ps}^{-1}$ is the $D$ meson's measured decay width. In order to determine whether a point is phenomenologically acceptable, we then simply follow \cite{ligeti} and require that our computed $x_{12}$ stemming from new physics does not exceed the maximum value allowed by the results of the \cite{hflavDDBar} fit; specifically, we require our computed $x_{12}$ to be below the upper edge of the 95\% CL interval for the best fit of this parameter, 0.63\%. 

\begin{figure}
    \centerline{\includegraphics[width=3.5in]{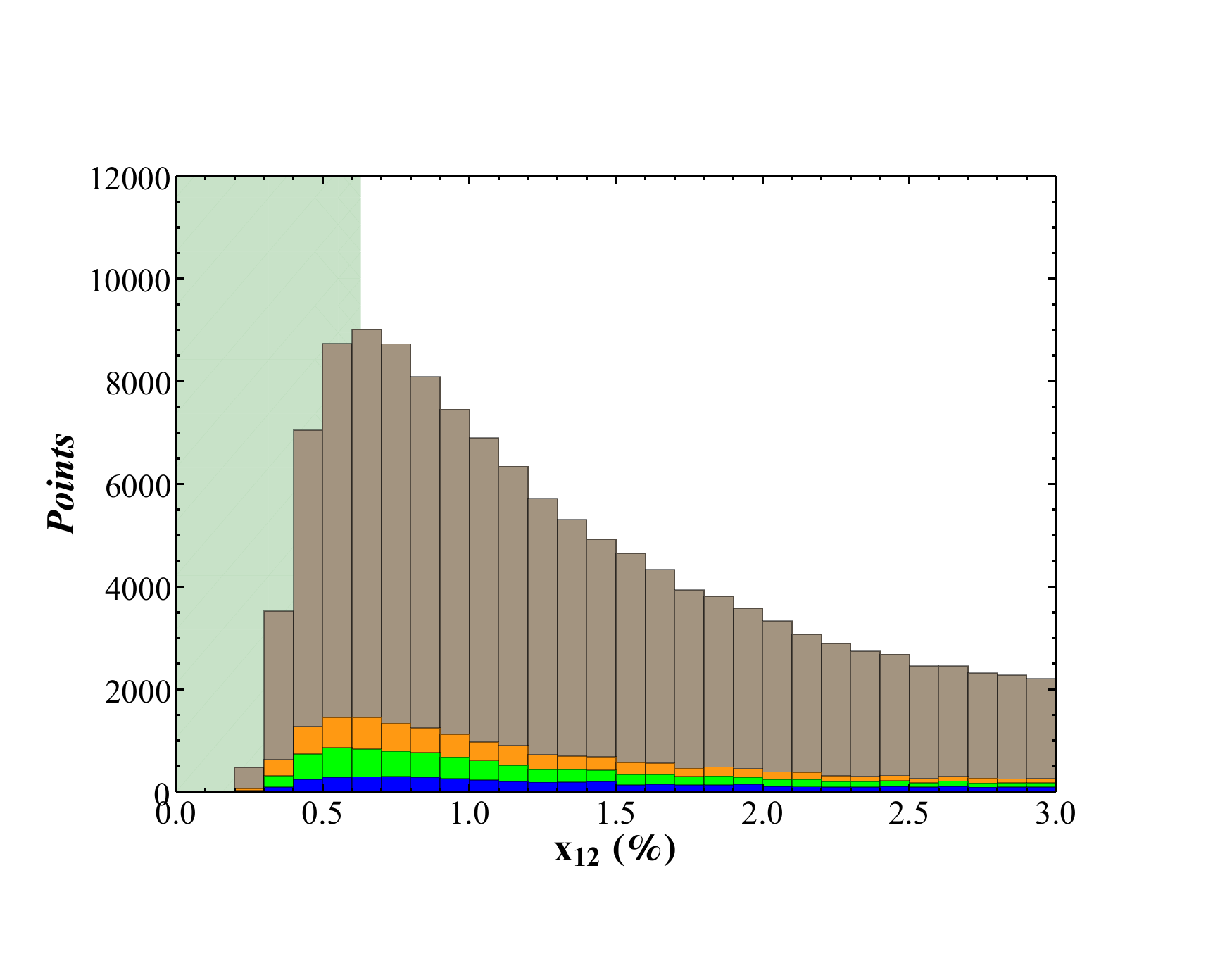}
    \hspace{-0.75cm}
    \includegraphics[width=3.5in]{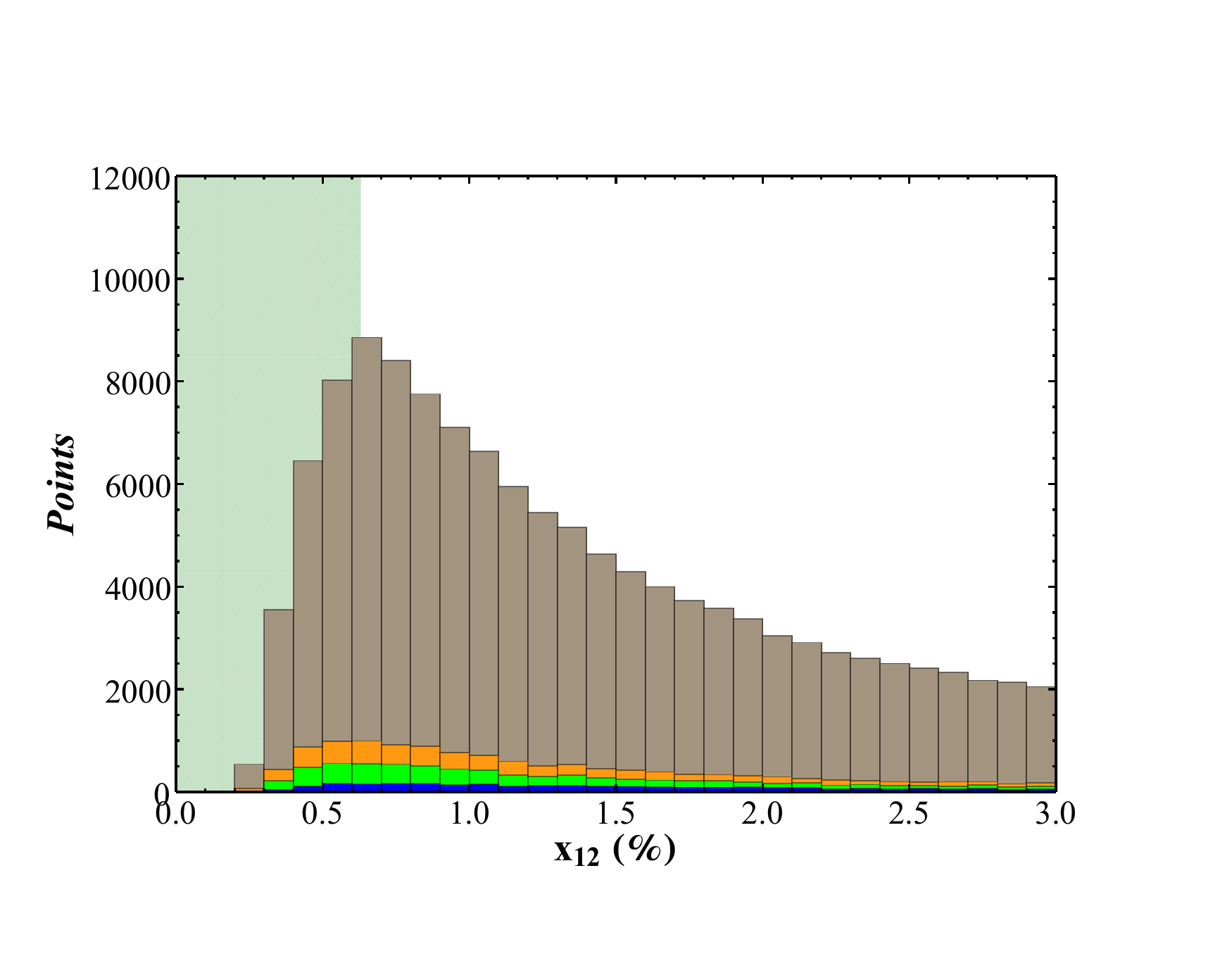}}
    \vspace*{-0.25cm}
    \centerline{\includegraphics[width=3.5in]{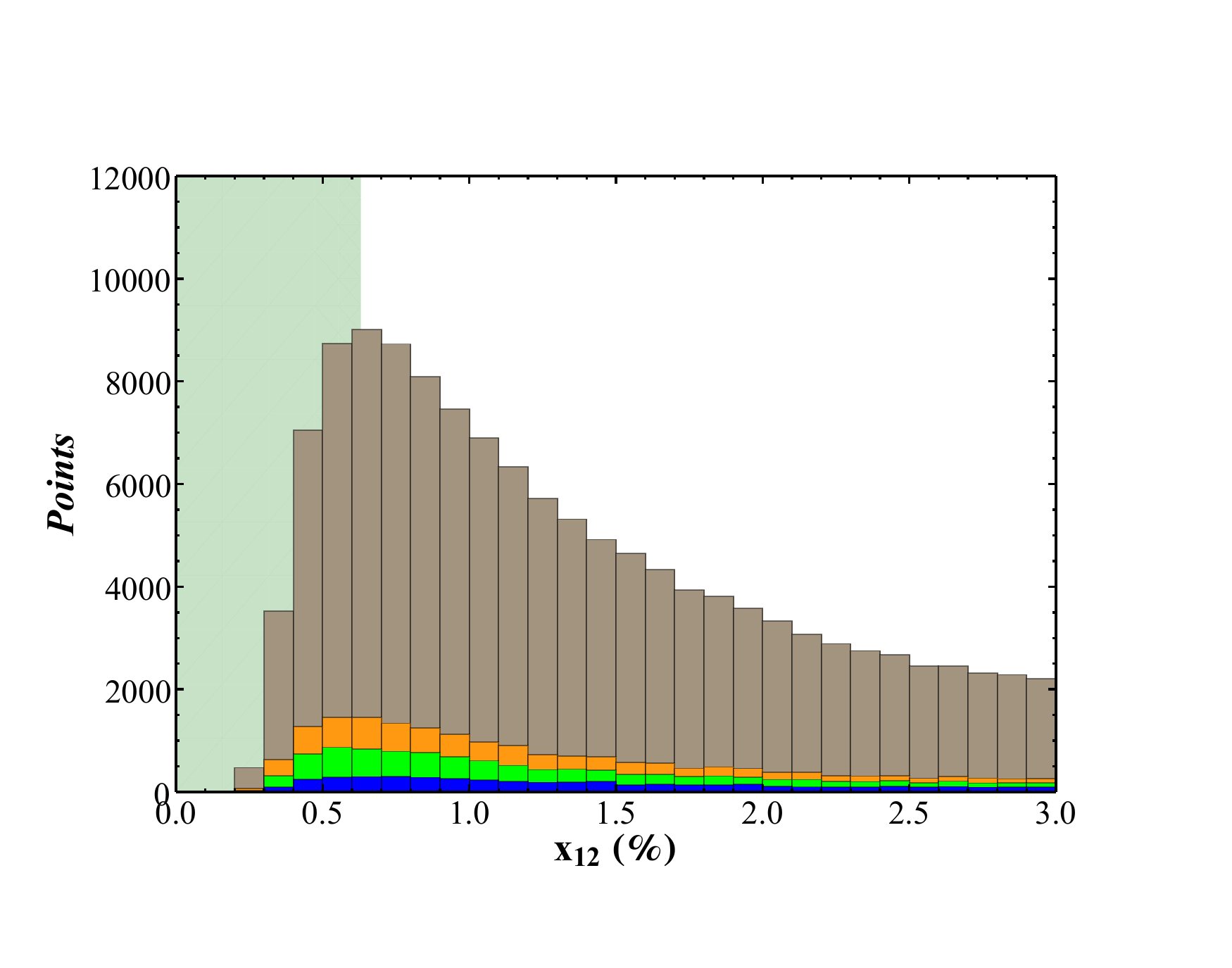}
    \hspace{-0.75cm}
    \includegraphics[width=3.5in]{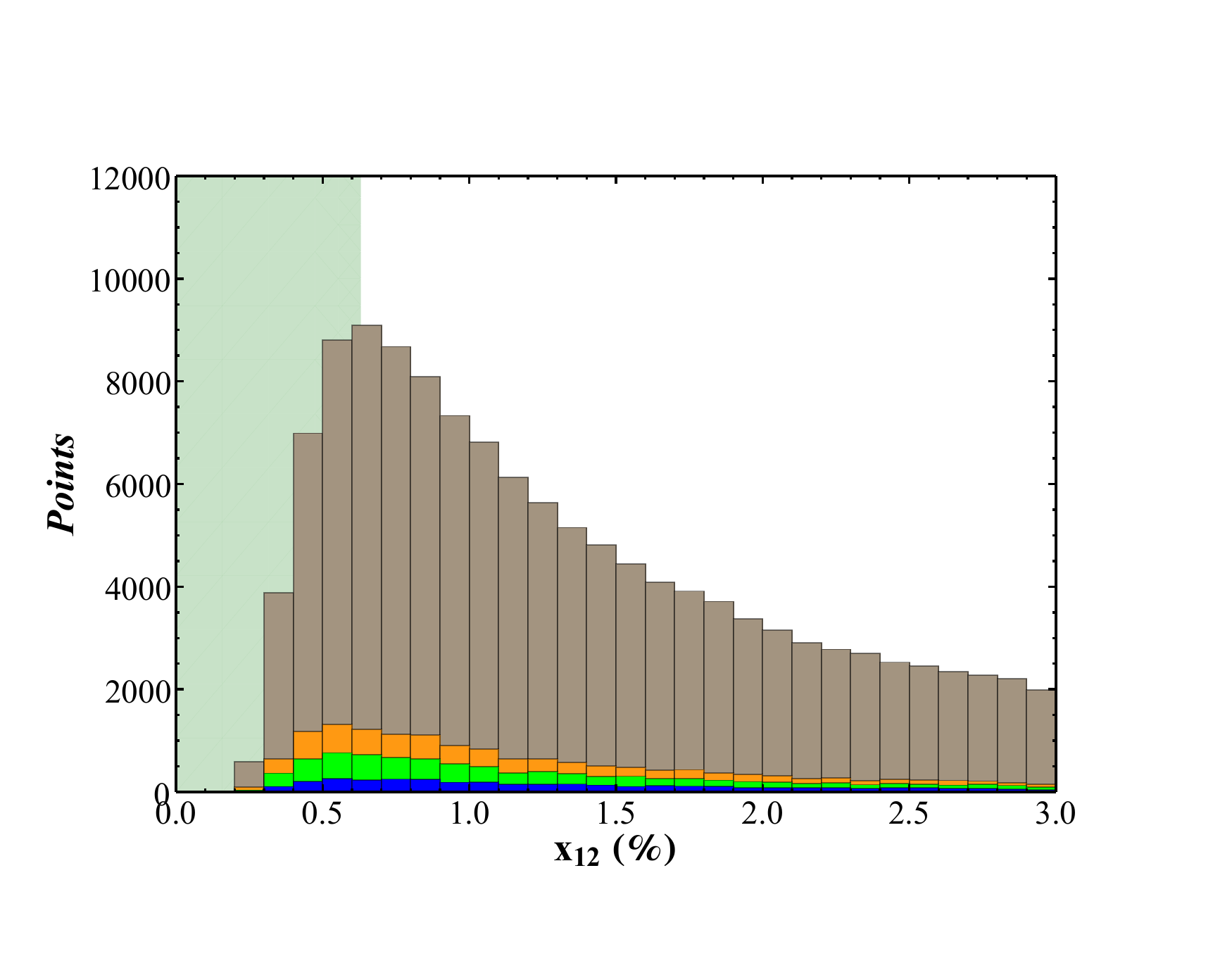}}
    \caption{Histograms of the $x_{12}$, for $g_A=g_B=0.1$ and $v_F/k=0.1$ (top left), $g_A=g_B=1.0$ and $v_F/k=0.1$ (top right), $g_A=g_B=0.1$ and $v_F/k=1.0$ (bottom left), and $g_A=g_B=1.0$ and $v_F/k=1.0$ (bottom right). The blue represents points which satisfy the 68\% CL experimental limit in the $h_d-\sigma_d$ plane, the green represents the 95\% CL region, the orange represents the 99.7\% CL region, and the beige includes all points in our sample pool, with no $h_d-\sigma_d$ constraints.}
    \label{fig:DDBarMixing}
\end{figure}

In Figure \ref{fig:DDBarMixing}, we include histograms of the computed $x_{12}$ values among our sampled points in parameter space at various values of $v_F$ and $g_{A,B}$, with sets of points within the 68\%, 95\%, and 99.7\% CL for the $\bar{B}^{0}-B^{0}$ observables plotted separately and color coded. There are two immediately salient characteristics of this plot: First, that even after applying cuts based on the $\bar{B}^{0}-B^{0}$ observables, the cut on $x_{12}$ is quite stringent, eliminating most remaining points, and second, that there is little in the way of apparent correlation between points which satisfy the $\bar{D}^{0}-D^{0}$ constraint and those which satisfy the $\bar{B}^{0}-B^{0}$ one; the histogram exhibits very little shape change, other than an obvious change in normalization, when progressively harsher $\bar{B}^{0}-B^{0}$ requirements are imposed.

The model's computed contributions to $x_{12}$ also exhibit one significantly different behavior from the contributions to $\bar{B}^{0}-B^{0}$ mixing parameters-- namely, that the \textit{minimum} contribution to $x_{12}$ among points in our parameter space is still non-negligible. Due to this fact, and the fact that any RS contributions to this observable will exhibit a roughly inverse squared dependence on the KK mass $M_{KK}$ (up to $RG$ effects on fermion masses and the Wilson coefficients of new physics operators), we can use the $x_{12}$ constraint to roughly estimate a lower bound on $M_{KK}$ in this model, even if bounds from electroweak precision measurements are ignored (perhaps ameliorated by a custodial symmetry). We find that at $M_{KK} \approx 3.3$ TeV, \textit{none} of our parameter space points can satisfy the $x_{12}$ constraint any longer, and hence, we can roughly estimate that $M_{KK} \gsim 3.3$ TeV can be thought of as an extreme lower bound of $M_{KK}$ within our model, arising from $\bar{D}^{0}-D^{0}$ meson mixing constraints.

\begin{figure}
    \centerline{\includegraphics[width=3.5in]{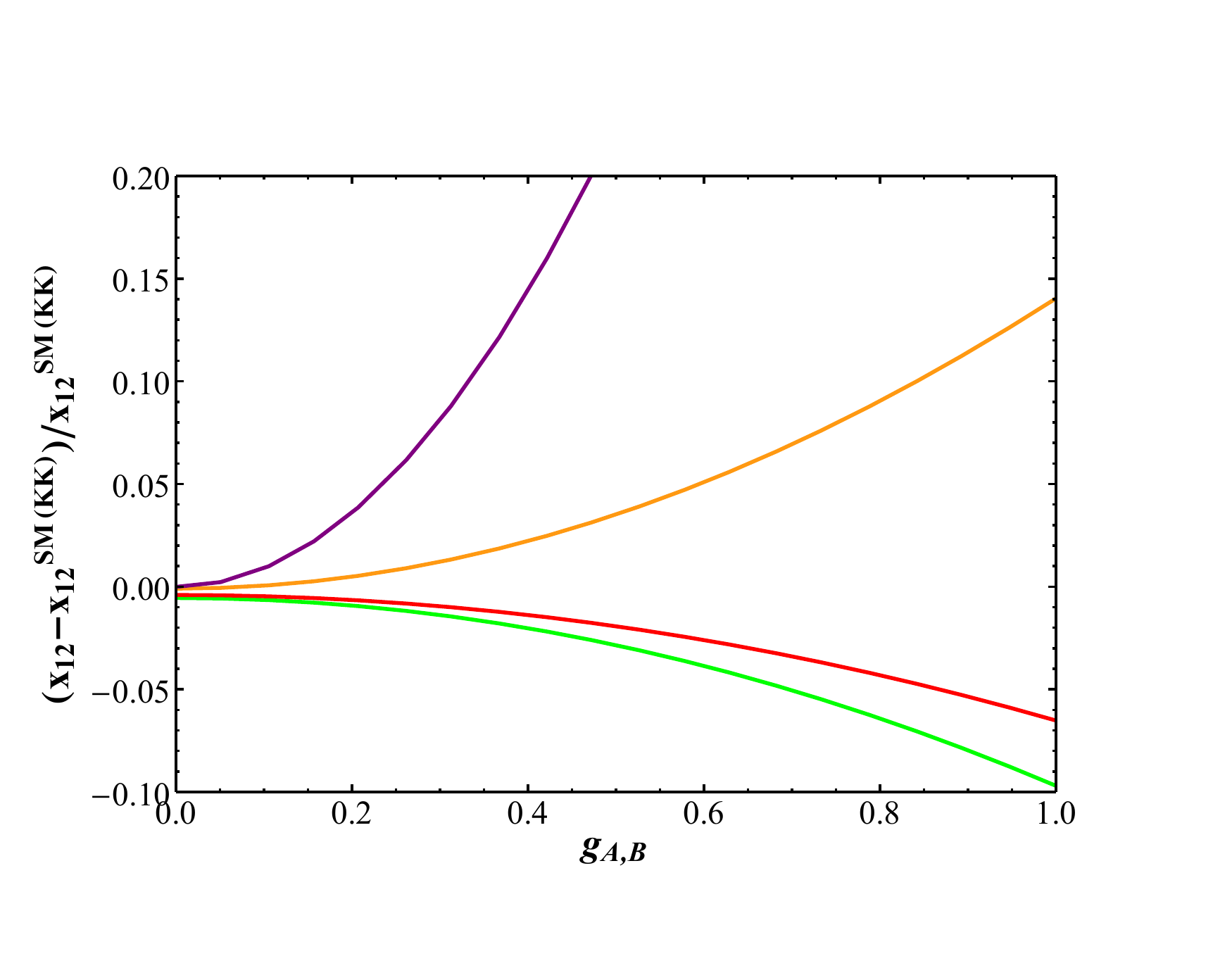}
    \hspace{-0.75cm}
    \includegraphics[width=3.5in]{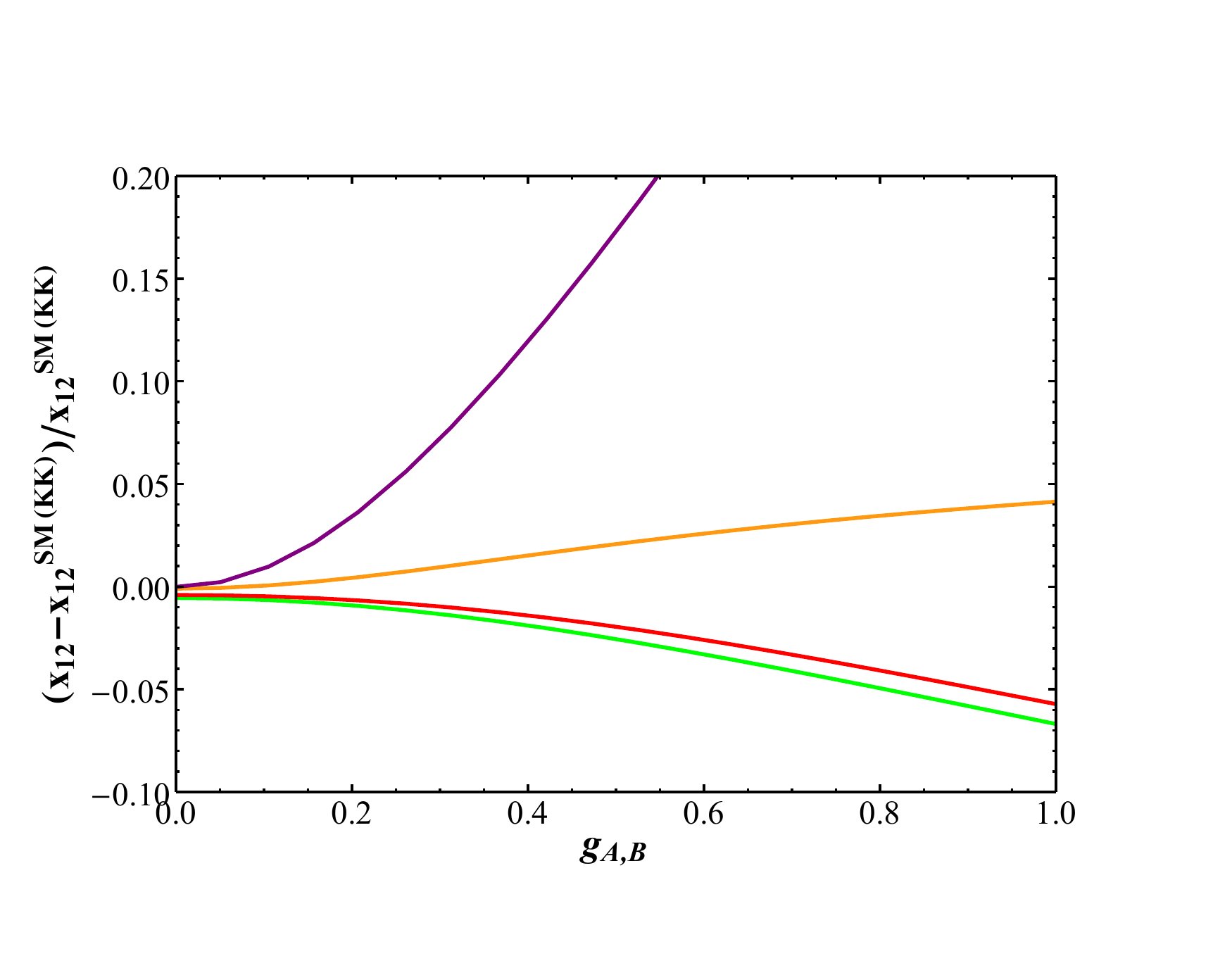}}
    \vspace*{-0.25cm}
    \centering
    \includegraphics[width=3.5in]{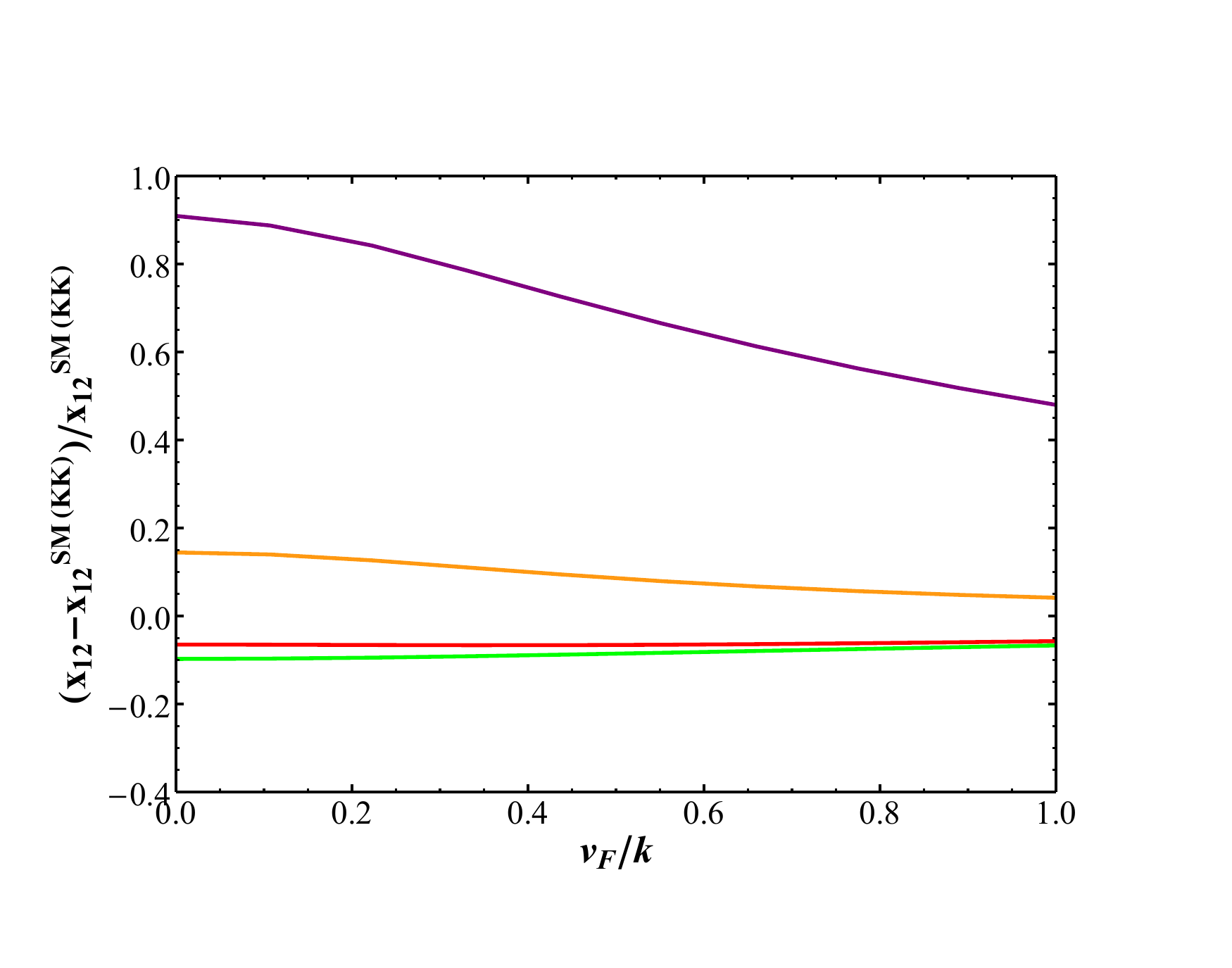}
    \caption{The dependence of the parameter $x_{12}$ on $g_{A,B}$ for $v_F=0.1k$ (top left), on $g_{A,B}$ for $v_F=k$ (top right), and on $v_F$ for $g_{A,B}=1.0$ (bottom), for several different points in our sample parameter space.}
    \label{fig:DDBarMixingGaugeDependence}
\end{figure}

Meanwhile, it is readily apparent from Figure \ref{fig:DDBarMixing} that differing choices of the parameters $v_F$ and $g_{A,B}$ have a limited effect on the fraction of points in parameter space that are of phenomenologically viable $x_{12}$. To explore the effect of the flavor gauge bosons further, we plot the quantity $(x_{12}-x_{12}^{SM(KK)})/x_{12}^{SM(KK)}$ (in analogy to our discussion of $h$ in Section \ref{BBBarMixingSection}) in Figure \ref{fig:DDBarMixingGaugeDependence}, where $x_{12}$ is the complete computed contribution of new physics to this parameter and $x_{12}^{SM(KK)}$ includes only the contribution from KK towers of SM gauge bosons, omitting those of the flavor gauge bosons, for a small sample of points at various choices of $v_F$ and $g_{A,B}$. We find (from more robust sampling than depicted in Figure \ref{fig:DDBarMixingGaugeDependence}) that in contrast to the case of $\bar{B}^{0}-B^{0}$ mixing, the flavor gauge bosons' contributions to $x_{12}$ will negatively interfere with the SM gauge bosons' contributions in approximately as many points in parameter space as they positively interfere with them. We also note that the overall level of the contribution is smaller; within the space of $g_{A,B}$ we consider, the flavor gauge boson contribution does not frequently exceed $O(10\%)$ of the contribution from the SM gauge bosons. In support of the hypothesis that the flavor gauge bosons' contributions to $x_{12}$ are as likely to interfere destructively with the SM gauge bosons' contributions as positively, we find that unlike the case of the $\bar{B}^{0}-B^{0}$ constraints, the total number of points of the parameter space which pass the $\bar{D}^{0}-D^{0}$ constraints is largely agnostic to the magnitude of the coupling constants $g_{A,B}$. For example, in the case where $v_F=0.1 k$, where due to our low vev we would expect sensitivity of the flavor gauge boson operators to changes in $g_{A,B}$ to be maximized, 14.5\% of our sample points that lie within the 99.7\% CL range of the $\bar{B}^{0}-B^{0}$ parameters also satisfy our $\bar{D}^{0}-D^{0}$ constraints if $g_A=g_B=0.1$, however, if instead $g_A=g_B=1.0$, we find that the same figure is 15.6\%; compared to the fact that the number of points which pass the 99.7\% CL constraint on the $\bar{B}^{0}-B^{0}$ parameters at $g_A=g_B=0.1$ is nearly double that of the case where $g_A=g_B=1.0$, the overall sensitivity of the $\bar{D}^{0}-D^{0}$ sector to changes in $g_{A,B}$ is practically insignificant.

\begin{figure}
    \centerline{\includegraphics[width=3.5in]{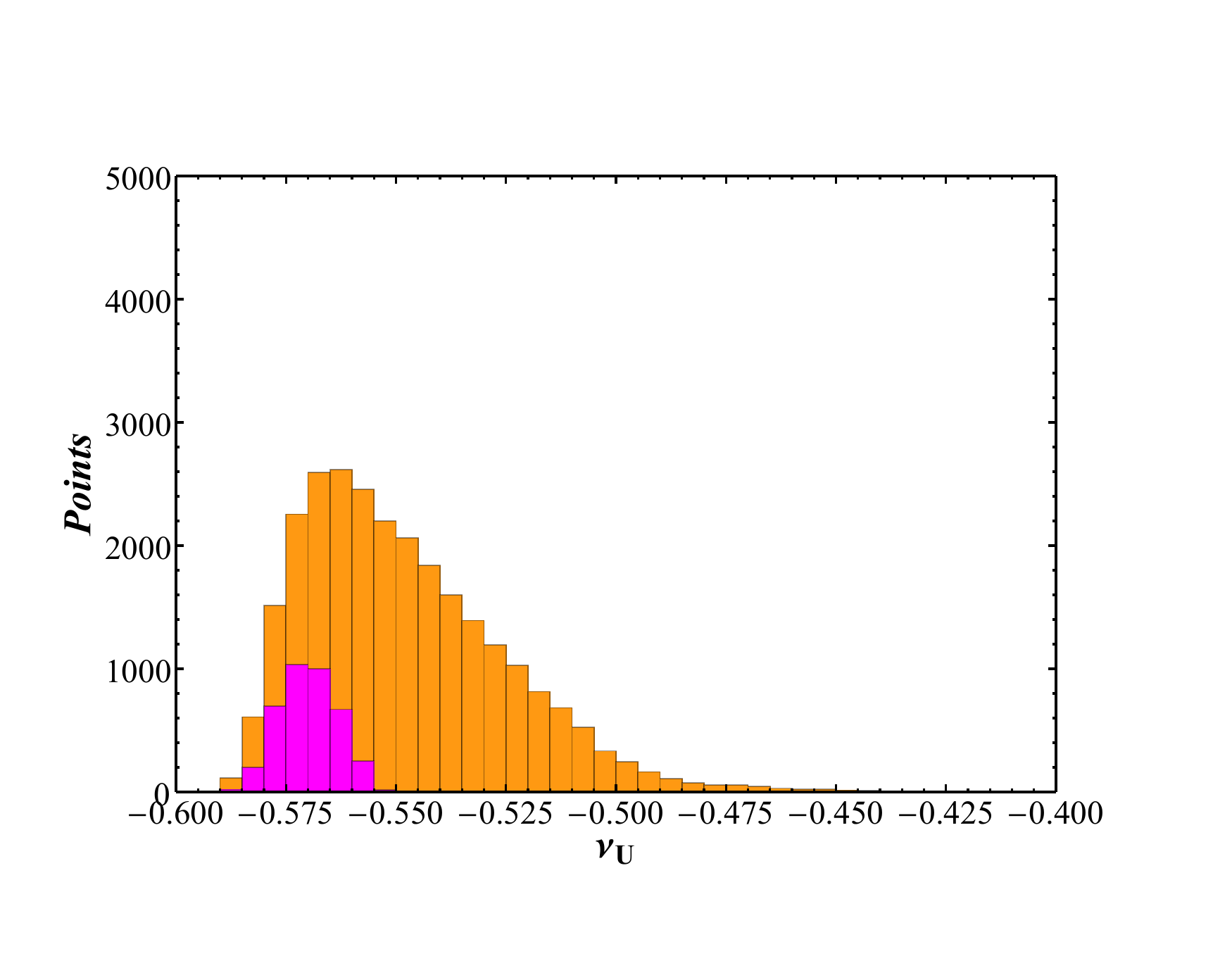}
    \hspace{-0.75cm}
    \includegraphics[width=3.5in]{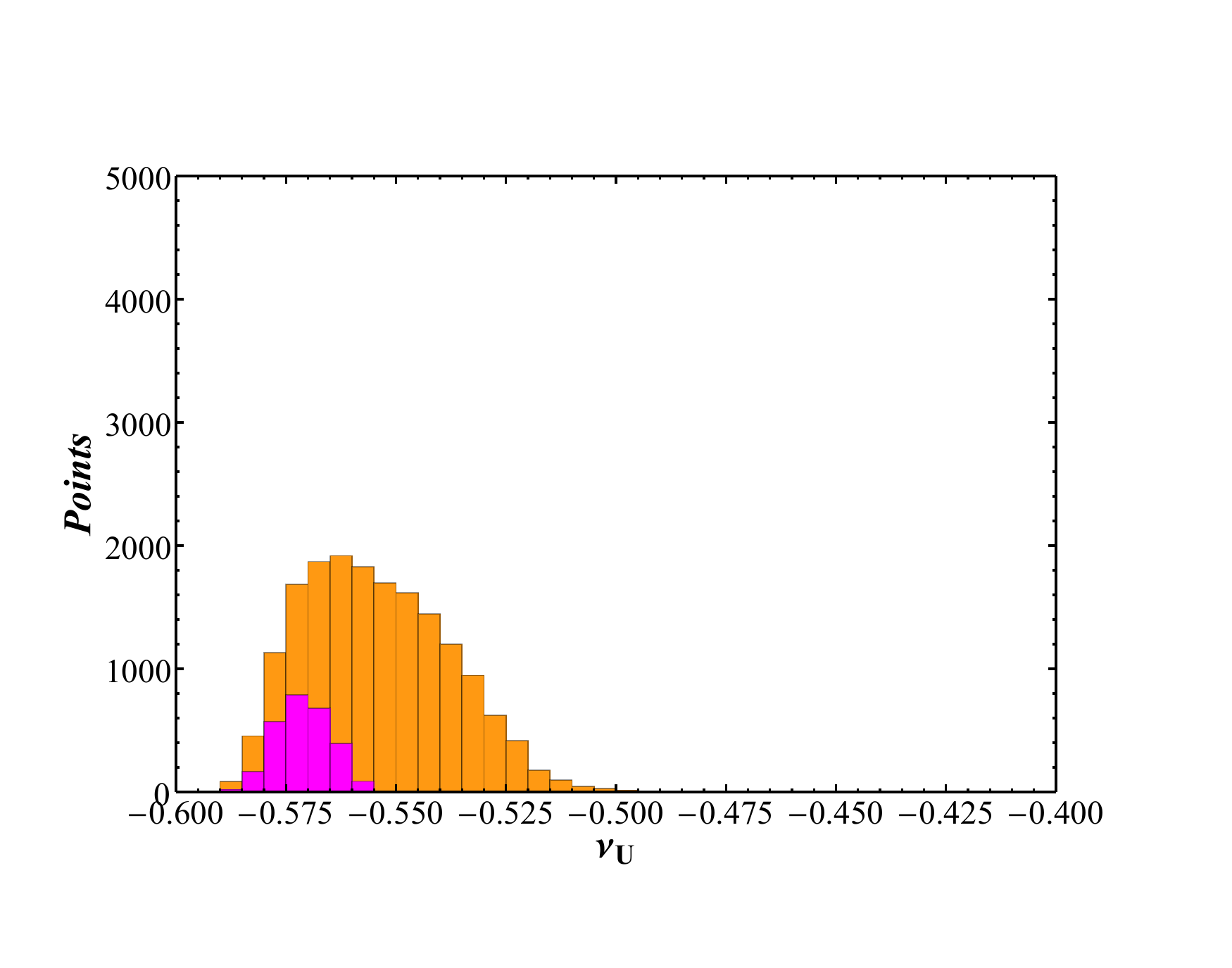}}
    \vspace*{-0.25cm}
    \centerline{\includegraphics[width=3.5in]{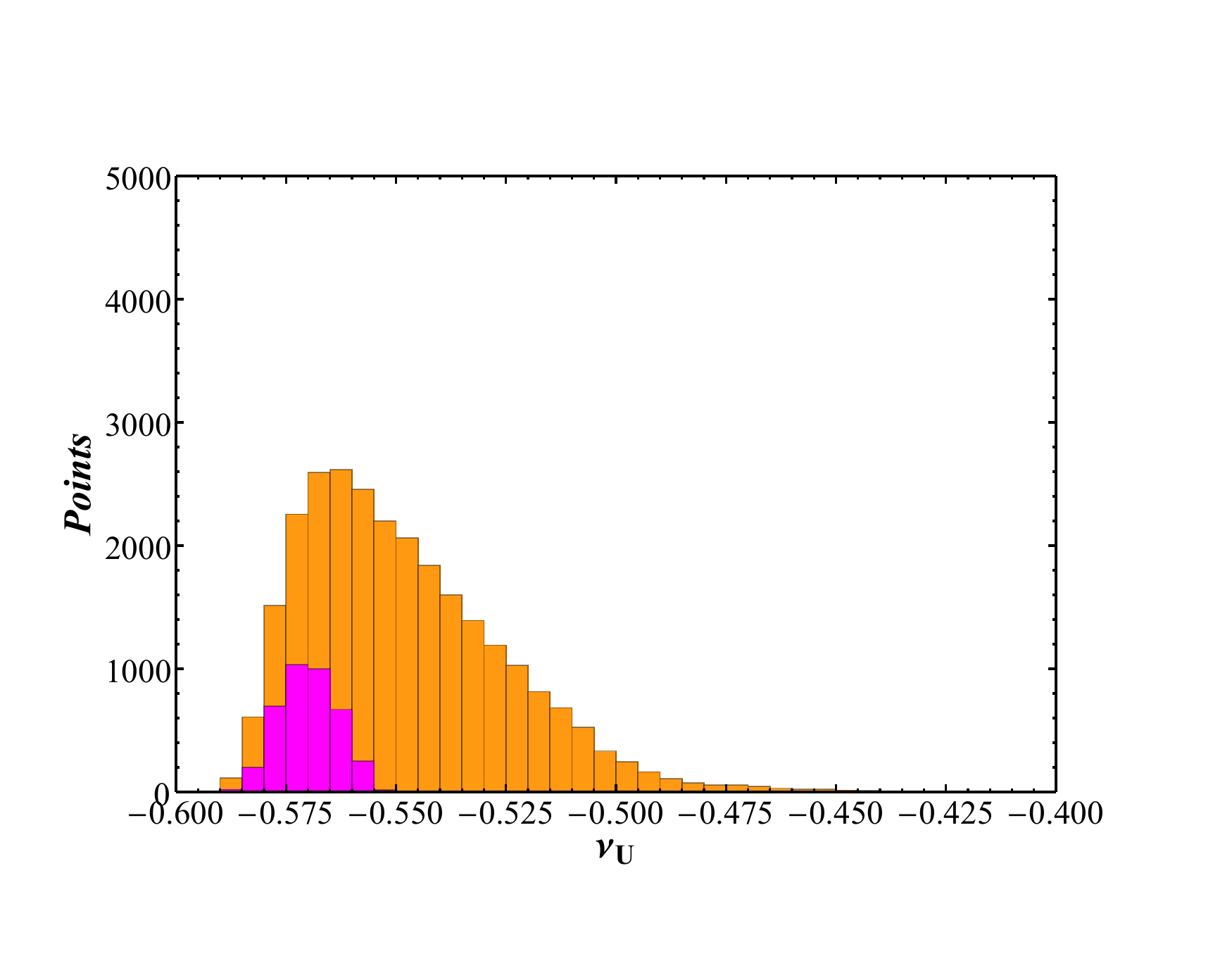}
    \hspace{-0.75cm}
    \includegraphics[width=3.5in]{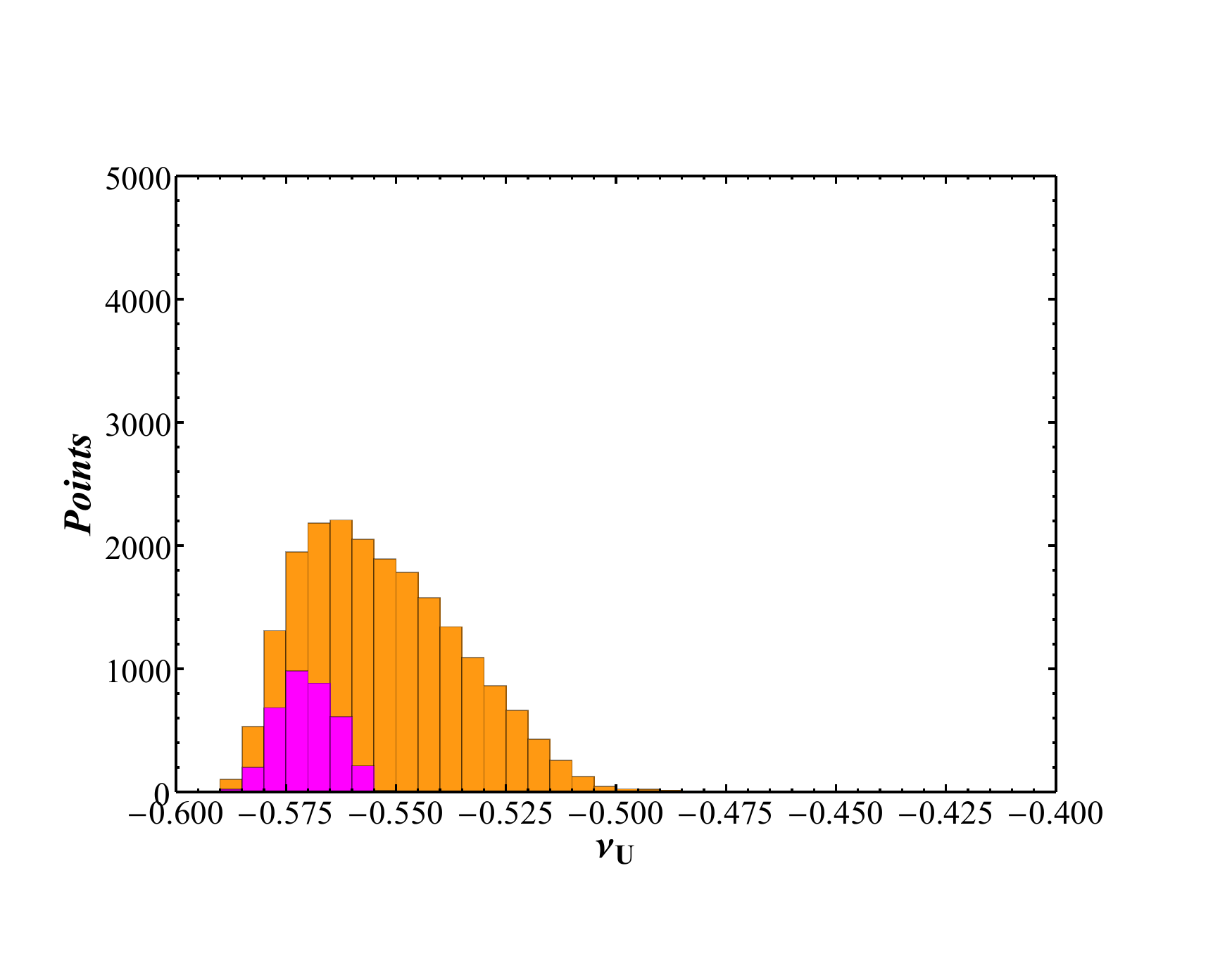}}
    \caption{Histograms of the $\nu_U$ for $g_A=g_B=0.1$ and $v_F/k=0.1$ (top left), $g_A=g_B=1.0$ and $v_F/k=0.1$ (top right), $g_A=g_B=0.1$ and $v_F/k=1.0$ (bottom left), and $g_A=g_B=1.0$ and $v_F/k=1.0$ (bottom right). The orange represents points which satisfy the 99.7\% CL experimental limit in the $h_d-\sigma_d$ plane, and the magenta represents the subset of these points which also satisfy the constraint $x_{12}<0.63\%$.}
    \label{fig:DDBarnuQ}
\end{figure}

\begin{figure}
    \centerline{\includegraphics[width=3.5in]{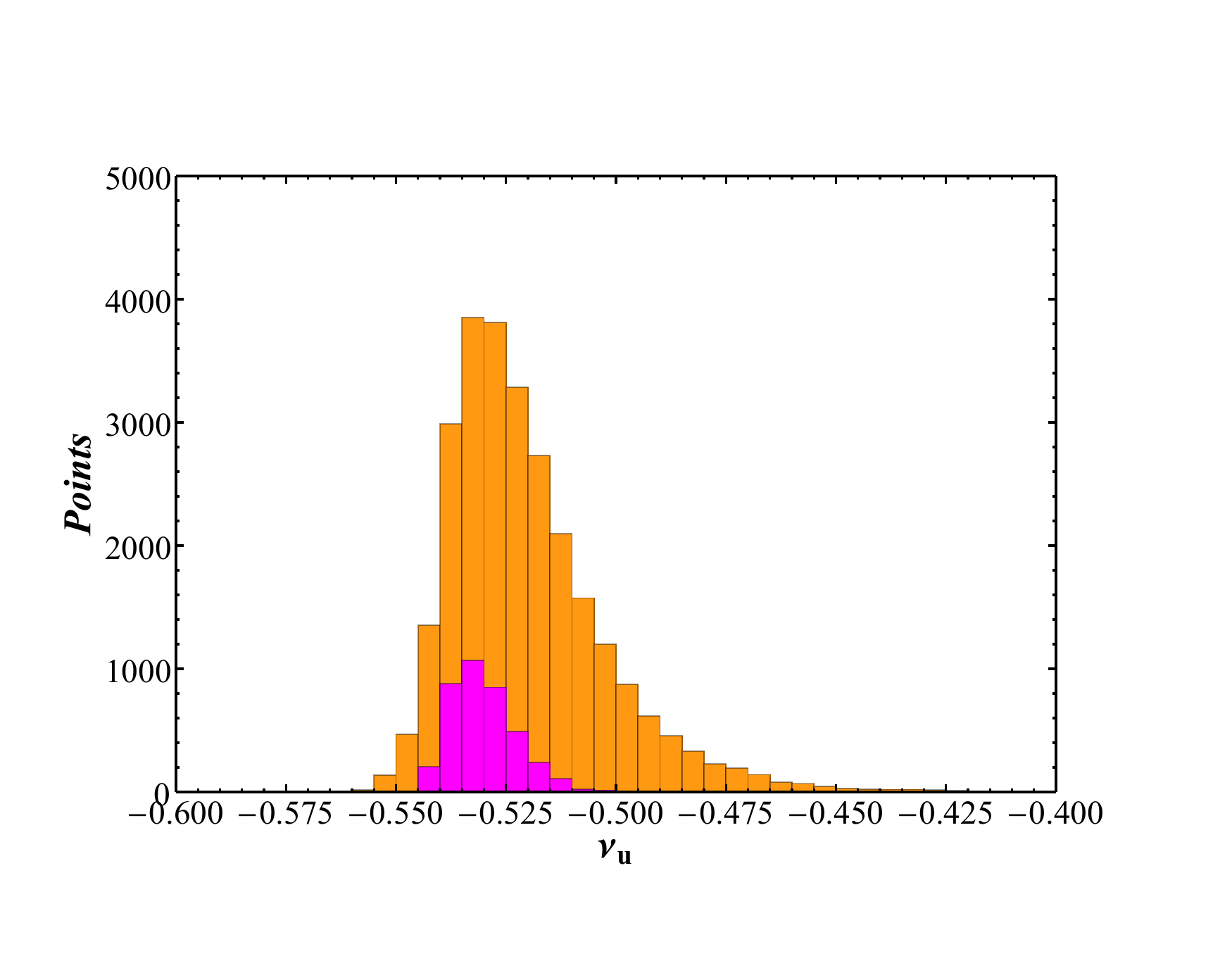}
    \hspace{-0.75cm}
    \includegraphics[width=3.5in]{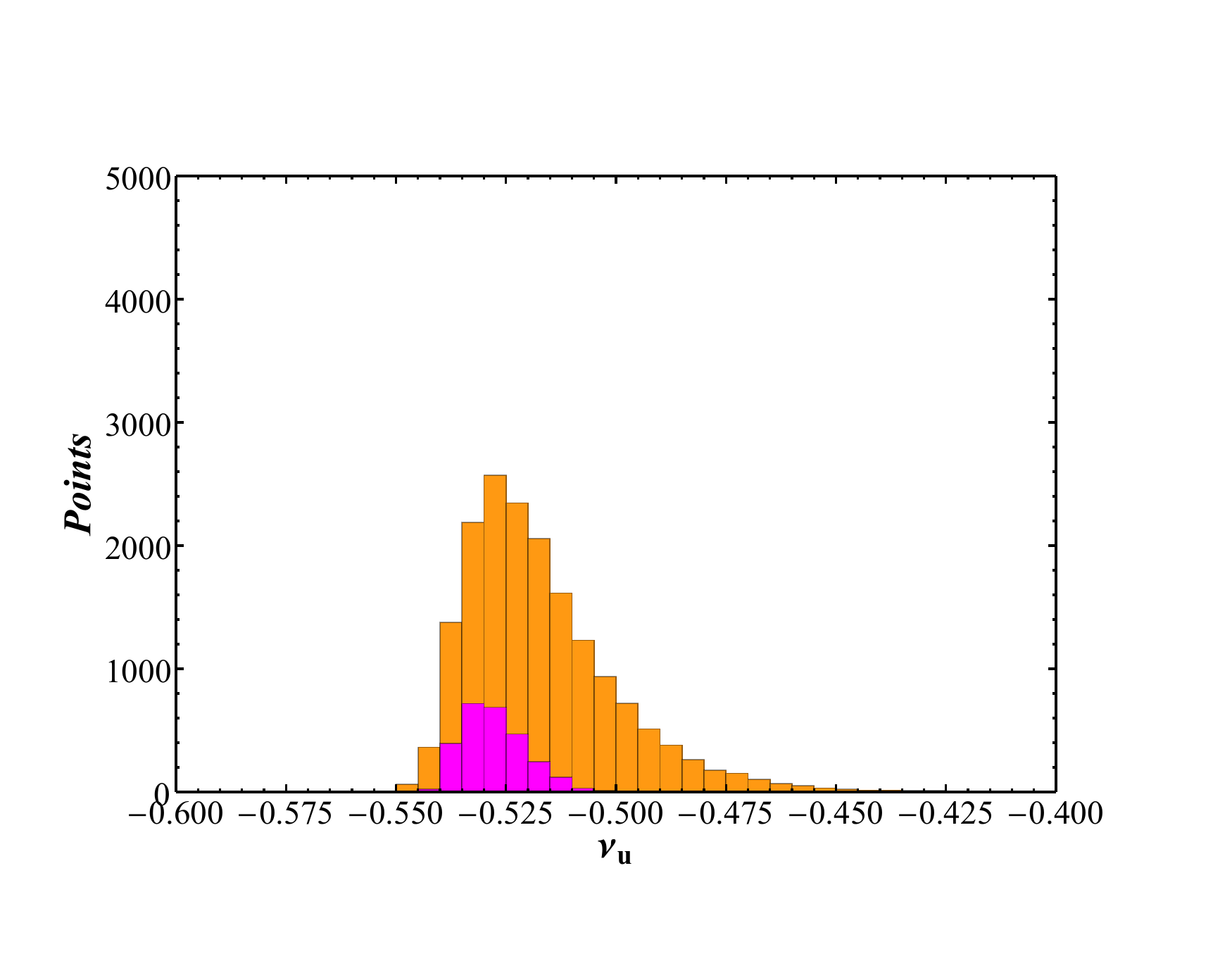}}
    \vspace*{-0.25cm}
    \centerline{\includegraphics[width=3.5in]{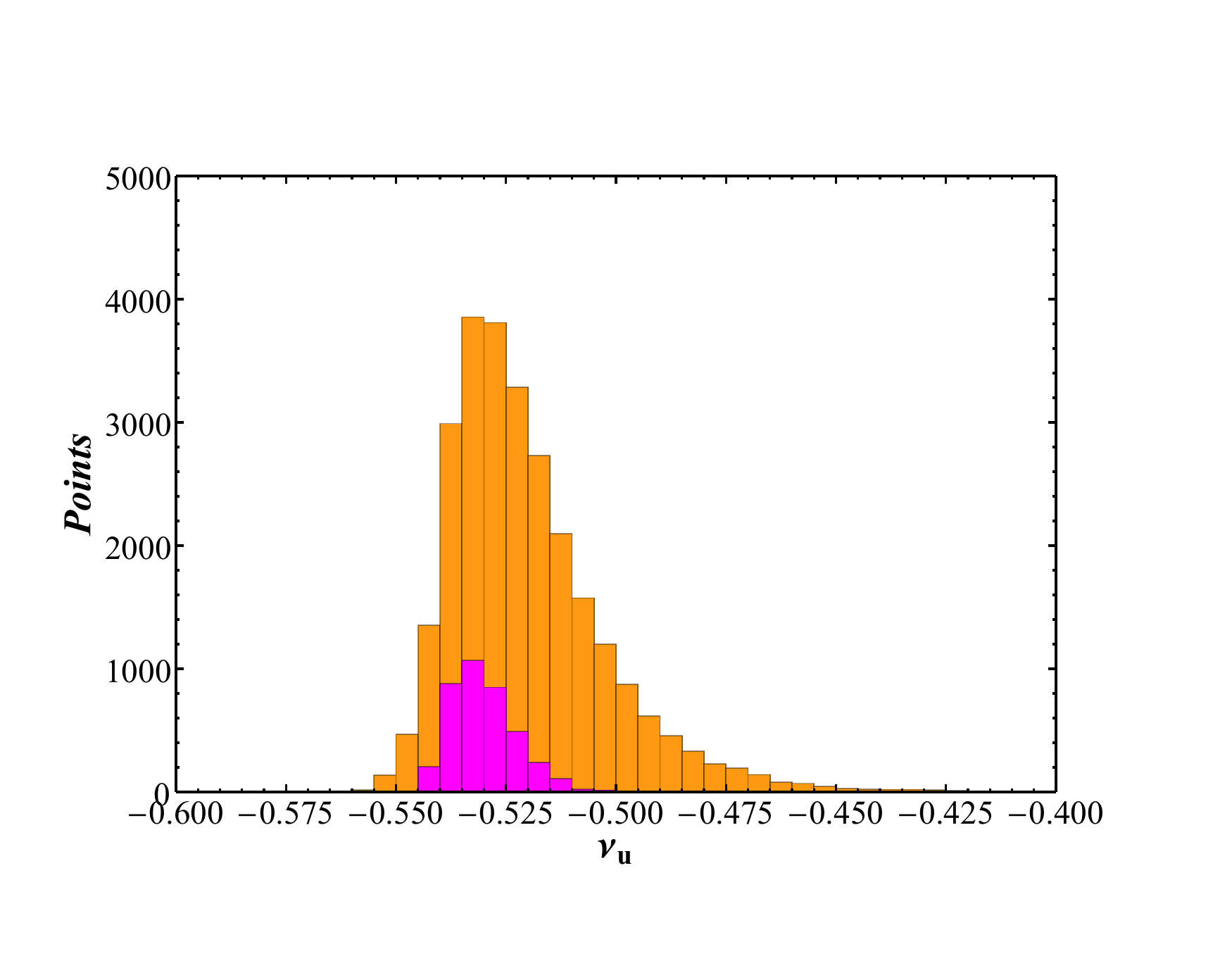}
    \hspace{-0.75cm}
    \includegraphics[width=3.5in]{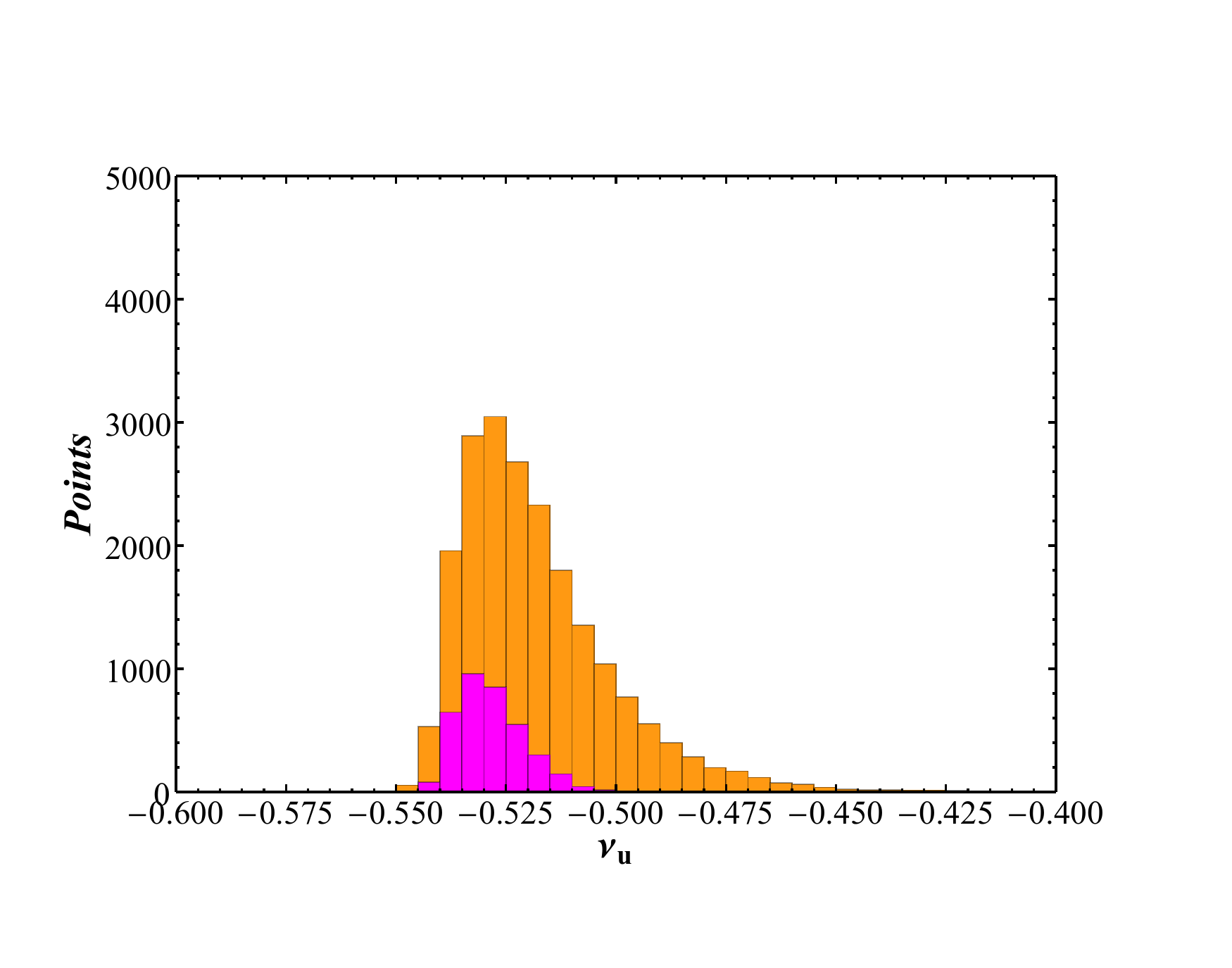}}
    \caption{Histograms of the $\nu_u$ for $g_A=g_B=0.1$ and $v_F/k=0.1$ (top left), $g_A=g_B=1.0$ and $v_F/k=0.1$ (top right), $g_A=g_B=0.1$ and $v_F/k=1.0$ (bottom left), and $g_A=g_B=1.0$ and $v_F/k=1.0$ (bottom right). The orange represents points which satisfy the 99.7\% CL experimental limit in the $h_d-\sigma_d$ plane, and the magenta represents the subset of these points which also satisfy the constraint $x_{12}<0.63\%$.}
    \label{fig:DDBarnuu}
\end{figure}

In terms of favored regions of the parameter space, the most salient pattern among the parameter space points that satisfy the new $\bar{D}^{0}-D^{0}$ requirement is that the parameters $\nu_U$ and $\nu_u$ are pushed further negative, indicating, perhaps unsurprisingly, that the $\bar{D}^{0}-D^{0}$ constraints strongly favor regions in which the first two up-like quark generations are localized closer to the Planck brane, and as such are less strongly coupled to KK gauge bosons. This pattern is easily depicted via histograms of the relevant parameters; we do so in Figures \ref{fig:DDBarnuQ} and \ref{fig:DDBarnuu}.

\subsection{Other Constraints: $Z b_L \bar{b}_L$, $t\rightarrow H c$, and $t\rightarrow Z c$}\label{OtherFlavorObsSection}

We conclude our discussion of various constraints on our model parameters by discussing several additional physics parameters that can in principle affect our model, but in practice give no further constraints and are entirely independent of the effects of the new flavor gauge bosons and bulk scalars. The first of these is the modification to the coupling of the light SM-like $Z$ boson to $b$ quarks. In general RS models of flavor, the vertex $Z \bar{b}_L b_L$ can be particularly problematic, because by necessity the third-generation $SU(2)_L$ doublet quark field must be localized closer to the TeV-brane than the first- and second- generation, and frequently must be  close enough to experience significant corrections to its SM couplings due to interaction with the KK gauge boson sectors. The correction to the $Z \bar{b}_L b_L$ coupling is approximately given by \cite{casagrande}
\begin{align}\label{ZbbFormula}
    \delta g^{bb}_{L} = \frac{m_Z^2}{M_{KK}^2}(-\frac{1}{2}+\frac{1}{3}s_W^2) \Delta_{Z\bar{b}_L b_L}+\frac{(m_b)^2}{2 M_{KK}^2} \delta_{Z \bar{b}_L b_L}
\end{align}
where
\begin{align}
    \Delta_{Z \bar{b}_L b_L} \equiv \frac{1}{4}\bigg(1-\frac{1}{kr_c \pi}\bigg)-\sum_{i=1}^{3}\frac{|F(\eta^Q_i)|^2}{2}\bigg( \frac{kr_c \pi(U^d_L)^{\dagger}_{3i}(U^d_L)_{i3}}{2+\eta^Q_i}+\frac{(U^d_L)^\dagger_{3i}(4+\eta^Q_i)(U^d_L)_{i3}}{2(2+\eta^Q_i)^2} \bigg)
\end{align}
and
\begin{align}
    \delta_{Z \bar{b}_L b_L} \equiv \sum_{i=1}^3 \frac{(U^d_R)^\dagger_{3i}(U^d_R)_{i3}}{2-\eta^{d}_i} \bigg(\frac{1}{|F(\eta^d_i)|^2}-1+\frac{|F(\eta^d_i)|^2}{2+\eta^d_i}\bigg)
\end{align}

\begin{figure}
    \centering
    \includegraphics{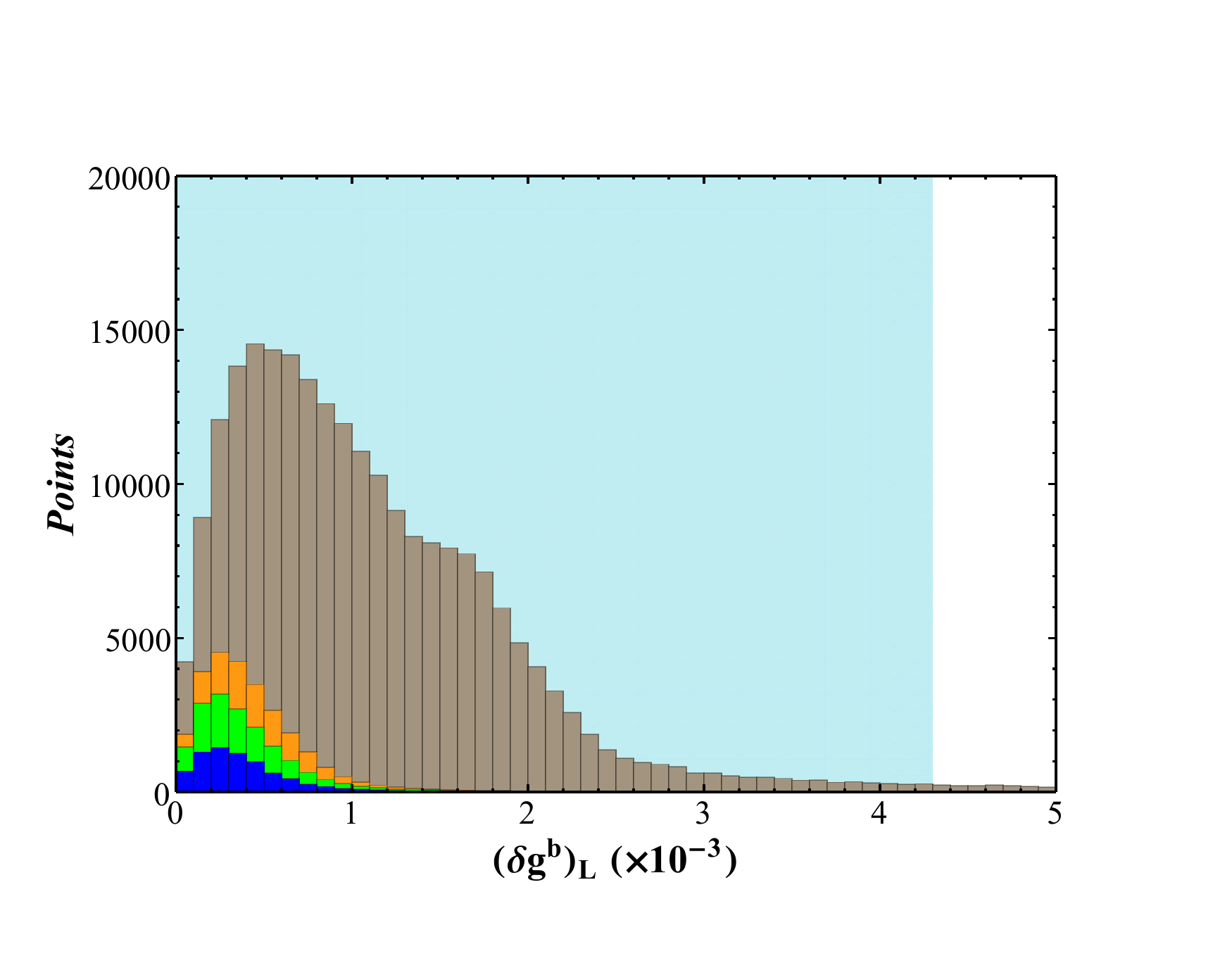}
    \caption{A histogram of $\delta g^{bb}_L$ for all parameter space points (beige), those within the 99.7\% CL region of the $h-\sigma_d$ plane (orange), within the 95\% CL region (green), and within the 68\% CL region (blue). The light blue region denotes the region $\delta g^{bb}_L<0.0043$, or the region in which the contribution to this parameter from our model is no greater than the maximum bound of the 68\% confidence interval of the best fit of $\delta g^{bb}_L$ in \cite{ciuchini}.}
    \label{fig:Zbb}
\end{figure}

Notably, the $m_b$ employed here is $m_b$ evaluated at the scale $M_{KK}$, and is therefore subject to a small correction from RGE. In Figure \ref{fig:Zbb}, we depict a histogram of the computed $Z \bar{b}_L b_L$ couplings, both before and after constraints arising from $\bar{B}^{0}-B^{0}$ mixing are applied. To estimate a constraint on the $Z \bar{b}_L b_L$ correction, we note that \cite{ciuchini} derives (from a fit of electroweak observables) a best-fit deviation of $g^{bb}_L$ from its SM value of $0.0029 \pm 0.0014$.  Notably, once even a comparatively weak (CL 99.7\%) constraint from the $\bar{B}^{0}-B^{0}$ mixing sector is applied, the result is the complete elimination of any $Z \bar{b}_L b_L$ coupling correction even beginning to approach our constraints.

The final flavor observables that we consider here are the FCNC decays, $t\rightarrow c Z$ and $t \rightarrow c H$. The $Z \bar{t}_L c_L$ and $Z \bar{t}_R c_R$ couplings are given (in the ZMA) by \cite{casagrande}
\begin{align}
    g^{tc}_L &= \label{ZtcL}\\
    &-\frac{g}{c_W}\frac{m_Z^2}{2M_{KK}^2}\bigg( \frac{1}{2} - \frac{2}{3} s_W^2 \bigg)\sum_{i=1}^3 \bigg[ \frac{kr_c \pi (U^u_L)^\dagger_{3i} |F(\eta^Q_i)|^2 (U^u_L)_{i1}}{2+\eta^Q_i}-\frac{(U^u_L)^\dagger_{3i}|F(\eta^Q_i)|^2(4+\eta^Q_i)(U^u_L)_{i1}}{2(2+\eta^Q_i)^2}\bigg] \nonumber \\
    &-\frac{g}{c_W}\frac{m_t m_c}{2 M_{KK}^2} \sum_{i=1}^3 \frac{(U^u_R)^\dagger_{3i} (U^u_R)_{i1}}{2-\eta^u_i}\bigg[\frac{1}{|F(\eta^u_i)|^2}-1+\frac{|F(\eta^u_i)|^2}{2+\eta^u_i} \bigg] \nonumber\\
    g^{tc}_R &= \label{ZtcR}\\
    &\frac{g}{c_W} \frac{m_Z^2}{2M_{KK}^2}\bigg( \frac{2}{3} s_W^2 \bigg)\sum_{i=1}^3 \bigg[ \frac{kr_c \pi (U^u_R)^\dagger_{3i} |F(\eta^u_i)|^2 (U^u_R)_{i1}}{2+\eta^u_i}-\frac{(U^u_R)^\dagger_{3i}|F(\eta^u_i)|^2(4+\eta^u_i)(U^u_R)_{i1}}{2(2+\eta^u_i)^2}\bigg] \nonumber \\
    &+\frac{g}{c_W}\frac{m_t m_c}{2 M_{KK}^2} \sum_{i=1}^3 \frac{(U^u_L)^\dagger_{3i} (U^u_L)_{i1}}{2-\eta^Q_i}\bigg[\frac{1}{|F(\eta^Q_i)|^2}-1+\frac{|F(\eta^Q_i)|^2}{2+\eta^Q_i} \bigg], \nonumber
\end{align}
where $g$ is the $SU(2)_L$ gauge coupling constant in the SM. The branching ratio $\mathcal{B}(t\rightarrow c Z)$ is then given by \cite{casagrande}
\begin{align}
    &\mathcal{B}(t\rightarrow c Z) = \frac{2(1-r_Z^2)^2(1+2 r_Z^2)c_W^2}{g^2 (1-r_W^2)^2(1+2 r_W^2)} \bigg\{|g^{tc}_L|^2+|g^{tc}_R|^2-\frac{12 r_c r_Z^2}{(1-r_Z^2)(1+2 r_Z^2)}\textrm{Re}[(g^{tc}_L)^* g^{tc}_R] \bigg\},\\
    &r_i \equiv \frac{m^{\textrm{pole}}_i}{m^{\textrm{pole}}_t}. \nonumber
\end{align}
We find that the effects of $t\rightarrow cZ$ are negligible. Among the sample of points that pass the constraints from $\bar{D}^{0}-D^{0}$ mixing and the 99.7\% CL constraints from $\bar{B}^{0}-B^{0}$ mixing when $g_A=g_B=0.1$ and $v_F=k$, the least restrictive scenario we consider, we find that the $\textit{maximum}$ branching ratio here is $4.7 \times 10^{-5}$, and the average among this population of points in parameter space is three orders of magnitude below this. Given that the current 95\% CL upper limit on $t\rightarrow Zq$ for $q=u,c$ is $\sim 5\times 10^{-4}$ \cite{pdgTopQuark}, this is entirely insignificant.

Meanwhile, the flavor-changing Higgs couplings of $H\bar{t}_L c_R$ and $H\bar{c}_L t_R$ are given in the ZMA by \cite{casagrande}
\begin{align}
    g^{t_L c_R}_H &= \frac{(m_t)^2 m_c}{v M_{KK}^2}\sum_{i=1}^3 \frac{(U^u_L)^{\dagger}_{3i} (U^u_L)_{i2}}{2-\eta^Q_i}\bigg( \frac{1}{|F(\eta^Q_i)|^2}-1+\frac{|F(\eta^Q_i)|^2}{2+\eta^Q_i}\bigg)\\
    &+\frac{(m_c)^2 m_t}{v M_{KK}^2}\sum_{i=1}^3 \frac{(U^u_R)^{\dagger}_{3i} (U^u_R)_{i2}}{2-\eta^u_i}\bigg( \frac{1}{|F(\eta^u_i)|^2}-1+\frac{|F(\eta^u_i)|^2}{2+\eta^u_i}\bigg), \nonumber \\
    g^{c_L t_R}_H &= \frac{(m_c)^2 m_t}{v M_{KK}^2}\sum_{i=1}^3 \frac{(U^u_L)^{\dagger}_{2i} (U^u_L)_{i3}}{2-\eta^Q_i}\bigg( \frac{1}{|F(\eta^Q_i)|^2}-1+\frac{|F(\eta^Q_i)|^2}{2+\eta^Q_i}\bigg)\\
    &+\frac{(m_t)^2 m_c}{v M_{KK}^2}\sum_{i=1}^3 \frac{(U^u_R)^{\dagger}_{2i} (U^u_R)_{i3}}{2-\eta^u_i}\bigg( \frac{1}{|F(\eta^u_i)|^2}-1+\frac{|F(\eta^u_i)|^2}{2+\eta^u_i}\bigg) \nonumber
\end{align}
and the branching ratio $\mathcal{B}(t\rightarrow cH)$ is given by \cite{casagrande}
\begin{align}
    \mathcal{B}(t\rightarrow cH) = \frac{\sqrt{2} (1-r_H^2)^2 r_W^2}{(1-r_W^2)^2(1+2 r_W^2)4 m_W^2 G_F} \bigg( |g^{c_L t_R}_H|^2+|g^{t_L c_R}|^2+ \frac{4 r_c}{1-r_H^2}\textrm{Re}[g^{t_L c_R}_H g^{c_L t_R}_H]\bigg)
\end{align}
We find, just as in the case of the $t\rightarrow c Z$ branching fraction, there is no significant contribution to the $t\rightarrow cH$ process from the parameter points in our model. Among those points in parameter space which pass our least stringent constraints from $\bar{B}^{0}-B^{0}$ and $\bar{D}^{0}-D^{0}$ mixing, we find that the maximum branching fraction we observe is $7.4\times 10^{-6}$, which is about 30 times larger than the mean value among this set of points. The current 95\% CL upper limit on $t\rightarrow H c$, meanwhile, is $1.6\times10^{-3}$ \cite{pdgTopQuark}. Hence, we can ignore this effect as well.

\section{Discussion and Conclusions}\label{ConclusionSection}
Within this paper, we have developed an $SU(2)_F \times U(1)_F$ gauged quark flavor symmetry in the framework of a warped extra dimension. To avoid flavor physics constraints, we have broken this symmetry entirely in the extra dimensional bulk, producing new KK towers for the flavor gauge bosons as well as for the bulk scalar fields which break the symmetry. Our use of spontaneous symmetry breaking in the bulk necessitates the inclusion of KK towers not just of the components of the bulk scalars that acquire a non-zero bulk mass, but also those of physical fields arising from a mixture of the fifth components of the gauge fields and the bulk Goldstone bosons created after SSB. We derive exact expressions for the bulk profiles and masses of the members of these latter towers in the unitary gauge, and find agreement with \cite{Falkowski:2008fz}. In addition to fully breaking the gauge symmetry in the bulk, our choice of bulk scalar vacuum expectation values maintains a flavor charge conserved by interactions with the flavor gauge bosons, and as a result, the most troublesome flavor observable in RS models, $\epsilon_K$, is insulated from unacceptably large corrections: The only tree-level new physics contributions to this process arise from the exchange of bulk scalars, and we have seen that these exchanges are highly suppressed for the SM-like quarks.

To probe the characteristics of our model further, we have performed a numerical probe of possible points in parameter space by randomly generating 8000 sets of input brane-localized Yukawa parameters, followed by performing a broad scan of remaining model parameters in order to produce $\approx 2.3 \times 10^5$ points in parameter space for us to examine. By probing the tree-level contributions of the model to $\Delta F = 2$ neutral meson mixing processes, which the absence of a tree-level SM contribution renders particularly sensitive to new physics, we find that there still exist a significant number of parameter points which remain phenomenologically viable even if the KK scale $M_{KK}=k \epsilon$ is set as low as 5 TeV. The dominant constraints on the model stem from $\bar{B}^{0}-B^{0}$ and $\bar{D}^{0}-D^{0}$ mixing. In spite of being designed as a mechanism to protect certain sensitive flavor observables from unacceptably large RS contributions (specifically $\epsilon_K$), our model does possess some specific predictive power: In particular, our model leads to significant corrections of $O(10\%)$ to $\bar{B}^{0}-B^{0}$ mixing observables even for points in parameter space not currently ruled out by experiment, but does $\textit{not}$ anticipate similar corrections to any other down-like neutral meson mixing observables, in dramatic contrast to models of RS without flavor symmetry \cite{casagrande2}, and so is likely falsifiable in the near future with LHCb and Belle-II data \cite{ckmfitterBBBar}.

While a complete quantitative comparison of this model to a generic RS model with flavor anarchy would require a more detailed probing of the model parameter space than this work presents, we do note that our present model construction likely suffers from some degree of fine-tuning. For example, in Section \ref{BulkMatterModelSection} we have set two bulk coupling terms that are allowed by symmetries to zero. However, as discussed in that section, there very likely exist slightly modified constructions that preclude these couplings from emerging but preserve the model's other results, and as such, we do not consider this to present an insurmountable obstacle to the model's naturalness. The harsh $\bar{B}^{0}-B^{0}$ constraints discussed in Section \ref{BBBarMixingSection}, meanwhile, also suggest a degree of fine-tuning (with the caveat that our sampling of parameter space is not necessarily unbiased), however in that section, we present qualitative arguments which indicate that the degree of fine-tuning to meet these constraints is likely less extreme than the tuning required to produce a phenomenologically viable anarchic model.

In relation to other attempts to ameliorate the RS flavor problem with bulk gauge symmetries, meanwhile, this work hopes to elucidate model-building options in this direction that have not, until now, been seriously explored. In particular, we find that it is possible to construct a model in which the additional gauge fields associated with the new symmetry mediate significant flavor-changing interactions, and that therefore requiring either the corresponding gauge couplings to nearly vanish, as in \cite{csakiU1}, or requiring particular alignments to preclude any flavor gauge boson contributions to flavor-changing currents, as in \cite{csakiShining}. Furthermore, this work hopes to elucidate the model-building utility of gauge symmetries spontaneously broken by scalars in a warped extra dimensional bulk, and leave the possibility open to a more robust exploration of its applications to flavor physics (or other extended bulk gauge groups) in the future. In particular, the gauge bosons and scalars arising from such symmetries are naturally hidden at the KK scale, even if they possess identical boundary conditions to those of the SM gauge bosons in the bulk, and the form of scalar couplings in the bulk will, in most realizations of any fermion-coupled scalar sector, naturally provide substantial suppression to couplings of the scalar fields to the SM-like fermions. In short, the use of spontaneously broken bulk gauge symmetries provides an additional straightforward and effective method of restricting new physics in RS models with extended gauge sectors to higher energy scales.

$\textbf{Acknowledgements: }$

We would like to thank JoAnne Hewett for advice and discussions related to this work. This work was supported by the Department of Energy, Contract DE-AC02-76SF00515.

\appendix
\section{Subleading Corrections to Wolfenstein Parameters}\label{DeltaTermsAppendix}
In this section, we reproduce the subleading corrections (in hierarchies of the quark bulk profiles) to the Wolfenstein parameters mentioned in Section \ref{CKMSection}, specifically Eq.(\ref{WolfensteinExpansion}).
\begin{align}\label{DeltaTermsExplicit}
    &\Delta_{\lambda^2} \equiv -\frac{|\bar{Y}_t-Y_u|^2 s_H^2}{|\bar{Y}_t|^2 c_Q^2 c_H^2} \bigg\lvert \frac{F(\eta^Q_1)}{F(\eta^Q_2)} \bigg\rvert^2- \frac{2 Re[\bar{Y}_t^*-Y_u^* s_H^2]}{c_u^2 c_H^2 |\bar{Y}_t|^2} \bigg\lvert \frac{F(\eta^u_1)}{F(\eta^u_2)} \bigg\rvert^2, \nonumber \\
    &\Delta_{A^2 \lambda^4} \equiv \bigg( 1 - 2 Re \bigg[ \frac{(Y_b Y_u c_d c_Q + Y_d Y_u s_d s_Q - Y_d \bar{Y}_t s_d s_Q)(\bar{Y}_t-Y_u s_H^2)}{s_H^2(Y_b c_d c_Q+Y_d s_d s_Q)(\bar{Y}_t-Y_u)Y_u}\bigg]\bigg)\frac{|\bar{Y}_t-Y_u|^2 s_H^2}{|\bar{Y_t}|^2 c_Q^2 c_H^2} \bigg\lvert \frac{F(\eta^Q_1)}{F(\eta^Q_2)} \bigg\rvert^2 \nonumber\\
    &\;\;\;\;\;\;\;\;\;-\frac{|Y_u|^2 s_H^2 c_H^2}{|\bar{Y}_t-Y_u s_H^2|^2 s_Q^2} \bigg\lvert \frac{F(\eta^Q_2)}{F(\eta^Q_3)} \bigg\rvert^2-\frac{2 Re[Y_u^* \bar{Y}_t]c_H^2}{s_u^2 |\bar{Y}_t-Y_u s_H^2 |^2} \bigg\lvert \frac{F(\eta^u_2)}{F(\eta^u_3)} \bigg\rvert^2,\\
    &\Delta_{\bar{\rho}-i\bar{\eta}} \equiv \frac{(\bar{Y}_t^*-Y_u^*)}{\bar{Y}_t^*} \frac{Y_u Y_b c_Q c_d c_H^2-(\bar{Y}_t-Y_u)Y_d s_Q s_d}{Y_u(Y_b c_Q c_d+Y_d s_Q s_d)c_H^2 c_Q^2} \bigg\lvert \frac{F(\eta^Q_1)}{F(\eta^Q_2)} \bigg\rvert^2 \nonumber \\
    &\;\;\;\;\;\;\;\;\;+\frac{(\bar{Y}_t^*-Y_u^*)}{Y_t^*}\frac{Y_b Y_u c_d c_Q c_H^2-(\bar{Y_t}-Y_u)Y_d s_d s_Q}{c_H^2 c_u^2 (Y_u(Y_b c_d c_Q+Y_d s_d s_Q)-Y_d \bar{Y}_t s_Q s_d)} \bigg\lvert \frac{F(\eta^u_1)}{F(\eta^u_2)} \bigg\rvert^2 \nonumber \\
    &\;\;\;\;\;\;\;\;\;-\frac{\bar{Y}_t Y_u^* c_H^2}{(\bar{Y}_t-Y_u)}\frac{Y_b Y_u c_d c_Q c_H^2-(\bar{Y}_t-Y_u)Y_d s_Q s_d}{Y_u (Y_b c_Q c_d+Y_d s_Q s_d)|\bar{Y}_t-Y_u s_H^2|^2 s_Q^2} \bigg\lvert \frac{F(\eta^Q_2)}{F(\eta^Q_3)} \bigg\rvert^2+\frac{\bar{Y}_t Y_u^* c_H^2}{|\bar{Y}_t-Y_u s_H^2|^2 s_u^2} \bigg\lvert \frac{F(\eta^u_2)}{F(\eta^u_3)} \bigg\rvert^2 \nonumber
\end{align}
The most salient feature of these expansions is the absence of any terms of the form $|F(\eta^d_i)|^2/|F(\eta^d_{i+1})|^2$; this is easily understandable by considering the form of the down-like quarks' brane-localized Yukawa matrix, given in Eq.(\ref{YukawaMatrices}). First, we note that any subleading corrections to the Wolfenstein parameters in the ZMA ultimately stem from corrections to the quark diagonalization matrices $U^{u,d}_{L,R}$. In particular, because the only off-diagonal terms in the down-like quarks' Yukawa matrix occur between the first and third generations, we see that any subleading corrections to the matrix $U^d_{L,R}$ must include only the ratios $|F(\eta^{Q,d}_1)|^2/|F(\eta^{Q,d}_3)|^2$, which, based on our assumption of hierarchical bulk profiles (that is, $|F(\eta^{Q,u,d}_1)|<<|F(\eta^{Q,u,d}_2|<<|F(\eta^{Q,u,d}_3)|$) must be doubly suppressed, as it is equal to the product of $|F(\eta^{Q,d}_1)|^2/|F(\eta^{Q,d}_2)|^2$ and $|F(\eta^{Q,d}_2)|^2/|F(\eta^{Q,d}_3)|^2$. As a result, we find that we can omit \textit{all} correction terms to the Wolfenstein parameters stemming from $U^d_{L,R}$; they are numerically insignificant.


    



\end{document}